\documentstyle[12pt,epsfig]{article}
\input psfig.tex
\def\slashchar#1{\setbox0=\hbox{$#1$}  % set a box for #1
   \dimen0=\wd0     % and get its size
   \setbox1=\hbox{/} \dimen1=\wd1  % get size of /
   \ifdim\dimen0>\dimen1   % #1 is bigger
      \rlap{\hbox to \dimen0{\hfil/\hfil}} % so center / in box
      #1     % and print #1
   \else     % / is bigger
      \rlap{\hbox to \dimen1{\hfil$#1$\hfil}} % so center #1
      /      % and print /
   \fi}      %

\def\disp{\displaystyle }
\def\qeqs#1#2{ (\ref{#1}) and (\ref{#2}) }
\def\sumi{ \sum_{i=1}^3 }
\def\suma{ \sum_{a=4}^7 }
\def\zeichen{\stackrel{\disp\sim}{ T\to\infty} }
\def\ba{\begin{array} }
\def\ea{\end{array} }
\def\linie{\ \vrule height 14pt depth 7pt \ }
\def\bea{\begin{eqnarray}}
\def\eea{\end{eqnarray}}
\def\nn{\nonumber \\ }
\def\queq#1{(\ref{#1})}

\def\vx{ {\vec x}  }
\def\qeq#1{ (\ref{#1})  }
\def\({\Bigl(}
\def\){\Bigr)}
\def\[{\Bigl[}
\def\]{\Bigr]}
\def\half{\ifinner {\scriptstyle {1 \over 2}}
   \else {1 \over 2} \fi}
\def\dsl{\rlap{/}{\partial}}
\def\frac#1#2{{#1 \over #2}}

\def\piv{\vec\pi}

\def\Sp{{\rm Sp}}
\def\Tr{ {\rm Tr}}
\def\ddmu{\partial_\mu}
\def\dumu{\partial^\mu}

\def\sld{\slashchar{\partial}}

\def\Sesp{S (\sigma, \vec\pi)}

\def\beq{ $$ }
\def\eeq{  $$ }
\def\Im{ {\rm Im} }
\def\Re{ {\rm Re} }
\def\bra#1{\langle #1 \,\vert}
\def\ket#1{\vert\, #1 \rangle}
\def\d{\mbox{d}}
\def\e{\mbox{e}}
\def\D{{\cal D}}
\def\Tr{\mbox{Tr}}
\def\Sp{\mbox{Sp}}
\def\fm{\mbox{fm}}
\def\MeV{\mbox{MeV}}

\def\sign{\mbox{sign}}
\def\Sesp{S(\sigma,\vec \pi)}
\def\mbar{m_0}
\def\barM{M_u}

\def\nl{\hfil\break}   % \nl is newline

\def\xslide#1#2#3#4#5#6{\centerline{\psfig
{figure=#1,height=#2,bbllx=#3bp,bblly=#4bp,bburx=#5bp,bbury=#6bp,clip=}}}
\parskip=\normalbaselineskip
\begin{document}
\vglue2.5cm

\centerline{\large\bf BARYONS AS NON-TOPOLOGICAL CHIRAL SOLITONS}   
\vskip8mm
\centerline{\large Chr.V. Christov$^{a,f}$, A.  Blotz$^{a,b}$, H.-C. Kim$^a$, 
P. Pobylitsa$^{a,c}$, T. Watabe$^a$} 
\centerline{\large Th. Meissner$^d$, E. Ruiz Arriola$^e$ and K. Goeke$^a$}
\vskip4mm
\centerline{$^{a}$ Ruhr-Universit\"at Bochum, Institut f\"ur Theoretische  
Physik  II, D-44780 Bochum, Germany}
\centerline{$^{b}$ Department of Physics, State University of New York, Stony
Brook, NY 11790, USA}
\centerline{$^{c}$ St.Petersburg Nuclear Physics Institute, Gatchina, 
St.Petersburg 188350, Russia}
\centerline{$^{d}$ Department of Physics and Astronomy, University of South 
Carolina, Columbia, SC 29208, USA}
\centerline{$^{e}$ Departamento de F\'{\i}sica Moderna, Universidad de 
Granada, E-18071 Granada, Spain}      
\centerline{$^{f}$ Institute for Nuclear Research and Nuclear Energy, 1178 
Sofia, Bulgaria}

\vspace {-9cm}
\begin{flushright}{RUB-TPII-32/95\\ December, 95\\ {\it Prog.Part.Nucl.Phys.}
{\bf 37}, 1996}
\end{flushright}
\vspace {6cm}

%\vskip4cm
%\centerline{\epsfysize=6cm\epsffile{rub-logo.eps}}%
%\vskip4pt

%\epsfxsize = 6 cm
%\centerline{\epsfbox{rub-logo.eps}}
%\vspace{0mm}
\vskip6cm
\begin{center}
{\Large Ruhr-Universit\"at Bochum} \\
{\large Institut f\"ur Theoretische Physik II}\\
{\large Teilchen- und Kernphysik}
\end{center}

%\newpage

%\vglue3cm
%\hskip4cm{\Large \it ``Though this be madness, yet there is method in it.''}
%\vskip-0.4cm
%\hskip11cm{\large Shakespeare, Hamlet, II-2}
%\vskip0.5cm

\newpage
\setcounter{page}{1}
\vglue1cm

\centerline{\large\bf BARYONS AS NON-TOPOLOGICAL CHIRAL SOLITONS}   
\vskip8mm
\centerline{\large Chr.V. Christov$^{a,f}$, A.  Blotz$^{a,b}$, H.-C. Kim$^a$, 
P. Pobylitsa$^{a,c}$, T. Watabe$^a$} 
\centerline{\large Th. Meissner$^d$, E. Ruiz Arriola$^e$ and K. Goeke$^a$}
\vskip4mm
\centerline{$^{a}$ Ruhr-Universit\"at Bochum, Institut f\"ur Theoretische  
Physik  II, D-44780 Bochum, Germany}
\centerline{$^{b}$ Department of Physics, State University of New York, Stony
Brook, NY 11790, USA}
\centerline{$^{c}$ St.Petersburg Nuclear Physics Institute, Gatchina, 
St.Petersburg 188350, Russia}
\centerline{$^{d}$ Department of Physics and Astronomy, University of South 
Carolina, Columbia, SC 29208, USA}
\centerline{$^{e}$ Departamento de F\'{\i}sica Moderna, Universidad de 
Granada, E-18071 Granada, Spain}      
\centerline{$^{f}$ Institute for Nuclear Research and Nuclear Energy, 1178 
Sofia, Bulgaria}

%\vspace {-9cm}
%\begin{flushright}{RUB-TPII-32/95\\ December, 95}
%\end{flushright}
%\vspace {6cm}
\vspace {1cm}

\begin{abstract}
The present review gives a survey of recent developments and applications
of the Nambu--Jona-Lasinio model with $N_f=2$ and $N_f=3$ quark flavors for 
the structure of baryons. The model is an effective 
chiral quark theory which incorporates
the SU(N$_f$)$_L\otimes$SU(N$_f$)$_R\otimes$U(1)$_V$ approximate symmetry  of 
Quantum chromodynamics. The approach describes the spontaneous chiral 
symmetry breaking and dynamical quark mass generation. 
Mesons appear as quark-antiquark excitations and baryons arise as 
non-topological solitons with three valence quarks and a polarized Dirac sea. 
For the evaluation of the baryon properties the present review concentrates 
on the non-linear Nambu--Jona-Lasinio model
with quark and Goldstone degrees of freedom which 
is identical to the Chiral quark soliton model 
obtained from the instanton liquid model of the QCD vacuum.  In this 
non-linear model, a wide variety of observables of baryons of the octet 
and decuplet is considered.
These include, in particular, electromagnetic, axial, pseudoscalar and  
pion nucleon form factors and the related static properties like magnetic 
moments, radii and coupling constants of the nucleon as well as the mass 
splittings and electromagnetic form factors of hyperons. 
Predictions are given for the strange form factors, the scalar form factor 
and the tensor charge of the nucleon.

\end{abstract}

\hskip4cm{\Large \it ``Though this be madness, yet there is method in it.''}
\vskip-0.4cm
\hskip11cm{\large Shakespeare, Hamlet, II-2}
\vskip0.5cm
-------------------------------\nl
E-mail:\nl 
christov@hadron.tp2.ruhr-uni-bochum.de \nl
blotz@nuclear.physics.sunysb.edu\nl
kim@hadron.tp2.ruhr-uni-bochum.de\nl
pavelp@hadron.tp2.ruhr-uni-bochum.de\nl
watabe@hadron.tp2.ruhr-uni-bochum.de\nl
meissner@nuc003.psc.scarolina.edu\nl
earriola@ugr.es \nl
goeke@hadron.tp2.ruhr-uni-bochum.de\nl

\newpage
\begin{center}
\begin{tabular}{lll}
1.&{\bf Introduction}\dotfill& 4\\  
2.&{\bf SU(2) NJL model: vacuum sector and meson properties}\dotfill& 8\\  
 2.1& Bosonization of the NJL model\dotfill& 8\\  
 2.2& Divergences and regularization\dotfill& 12\\ 
 2.3& Spontaneously broken chiral symmetry and constituent quark mass
\dotfill& 14\\
 2.4& Mesonic properties\dotfill& 15\\
 2.5& Fixing of the parameters in the vacuum 
sector \dotfill& 18\\
 2.6 &Restriction to Goldstone degrees of freedom: 
the chiral circle \dotfill& 19\\
 2.7 & Generalization to higher flavor groups and relation to the
topological approach \dotfill& 21\\
3. & {\bf Nucleon as a non-topological soliton in the NJL model}
\dotfill& 23\\
 3.1 & Nucleon correlation function\dotfill& 24\\ 
 3.2 & ``Classical'' soliton \dotfill& 25\\ 
 3.3 & Baryon number of the soliton \dotfill& 27\\
 3.4 & Stationary hedgehog meson-field configuration\dotfill& 
29\\ 
 3.5 & Selfconsistent soliton solution\dotfill& 32\\
 3.6 & Semiclassical quantization of the soliton \dotfill& 35\\
 3.7 & Nucleon matrix elements of quark current \dotfill& 41\\ 
4.& {\bf Nucleon properties in the SU(2) NJL
model}\dotfill& 47\\  
 4.1 & Electromagnetic properties\dotfill& 47\\ 
 4.2 & $E2/M1$ ratio for the $\gamma N \rightarrow \Delta$
transition \dotfill& 53\\ 
 4.3 & Axial properties \dotfill& 55\\ 
 4.4 & Pion nucleon form factor \dotfill& 61\\  
 4.5 & PCAC and the Goldberger-Treiman relation \dotfill& 63\\ 
 4.6 & Tensor charges \dotfill& 65\\  
 4.7 & Electric polarizability of the nucleon  \dotfill& 66\\ 
5. & {\bf SU(3)--Flavor NJL model} \dotfill& 68\\ 
 5.1 & Extension of the NJL model to SU(3) flavors\dotfill& 68\\ 
 5.2 & Vacuum solutions and fixing of parameters \dotfill& 69\\  
 5.3 & Restriction to Goldstone modes \dotfill& 71\\ 
 5.4 & Trivial embedding of SU(2) hedgehog soliton into SU(3)\dotfill& 72\\
 5.5 & Expansion in angular velocity \dotfill& 72\\ 
 5.6 & Strange mass terms of the collective lagrangian\dotfill& 75\\ 
 5.7 & Hyperon  splittings in linear order - Sum rules\dotfill& 77\\ 
%\end{tabular}
%\end{center}
%\newpage
%\begin{center}
%\begin{tabular}{llr}
 5.8 & Hyperon splitting in second order - wave functions\dotfill& 78\\ 
 5.9 & Yabu-Ando diagonalization method \dotfill& 80\\ 
 5.10 & Results: mass splittings in the SU(3) NJL model\dotfill& 81\\ 
6.& {\bf Baryon properties in the SU(3) NJL model} \dotfill& 84\\ 
 6.1 & Electromagnetic properties \dotfill& 84\\ 
 6.2 & Strange vector form factors of the nucleon and related observables
\dotfill& 87\\ 
 6.3 & Scalar form factor and the sigma term $\pi N$ \dotfill& 92\\ 
 6.4 & Axial charges of the nucleon \dotfill& 92\\ 
 6.6 & Gottfried sum rule \dotfill& 95\\ 
7.& {\bf Summary}\dotfill& 97\\ 
& {\bf References}\dotfill& 98\\ 
\end{tabular}
\end{center}
\newpage
\section{Introduction}

The structure and the dynamics of hadrons are generally believed to be
described by  Quantum chromodynamics (QCD). 
A prominent feature of QCD is the asymptotic freedom which allows 
the processes at high energies to be 
described perturbatively in terms of these degrees of freedom. 
However, at energies comparable to the low-lying hadron masses 
QCD shows non-perturbative phenomena such as confinement of quarks and 
gluons and the spontaneous breakdown of chiral symmetry. It makes the
description in terms of quarks and gluons enormously complicated and the only 
direct source of information from QCD is the lattice Monte Carlo simulations. 
In this situation it is natural to
invoke effective models of the strong interaction which incorporate the
relevant degrees of freedom in the low energy regime. Those are all
related to chiral symmetry and its dynamical breaking. In fact, the
spontaneous breakdown of the  chiral symmetry has been
phenomenologically recognized to be an essential feature of the strong
interaction long before the advent of QCD. One has by now
accumulated a large number of empirical facts and, following from this,
somewhat generally accepted features of low energy hadronic dynamics,
which have to be respected by effective models:
\begin{itemize}
\item
The existence of light pseudoscalar mesons (pions) and the fact that
there is no degenerate partner of the nucleon with negative parity
suggests that the pions are Goldstone bosons associated with the
dynamical breaking of SU(2)${}_L \otimes$SU(2)${}_R$ chiral flavor
symmetry down to SU(2)${}_V$. This picture can be extended to SU(3),
although obviously the larger masses of kaons and $\eta$ indicate that
the explicit chiral symmetry breaking is larger in this case.
\item
The relatively large mass of $\eta^\prime$ suggests that the flavor 
$U_A$(1)-current is not conserved even in chiral limit.
\item
The hadron mass differences in the SU(3)-flavor multiplets appear as
being governed by a linear breaking of the SU(3)${}_V$-symmetry in a rather
simple fashion and 
follow phenomenologically the Gell-Mann-Okubo mass formula.
\item
The masses of the low lying mesons and baryons, apart from the
Goldstone bosons, are quite well described in terms of the constituent
quark model whose basic ingredients are massive constituent
up and down ($M_{u,d}~\approx 350$~MeV) and strange
($M_s~\approx 500$~MeV) quarks. 
\end{itemize}

Actually the above empirical facts can be understood from the
viewpoint of QCD. Its lagrangian is given by
\begin{equation}
{\cal L}_{\rm QCD} = \bar \Psi ( i \gamma^\mu \partial_\mu + g 
\gamma^\mu A_\mu  - \hat m) \Psi - \frac12 {\rm Sp} F_{\mu\nu} F^{\mu\nu}\,.
\label{eq:LQCD}
\end{equation}

Here $\Psi$ comprises a quark field with $N_c=3$ colors and $N_f=6$ flavors,
and  $A_\mu$ and $F_{\mu\nu}$ are the corresponding SU(3)-color gauge
field and field strength tensor, respectively. 

The current quark masses are represented in (\ref{eq:LQCD})
by the matrix $\hat m= {\rm diag}(m_u, m_d, m_s, m_c, m_b, m_t)$. Since up-,
down- and strange
current quark masses are much smaller than those of other quarks,
there is a natural separation between ``light quark physics'', relevant
for energies $\sim~1$~GeV, and ``heavy quark physics'' for higher
energies. Thus for systems composed of up-, down- and strange quarks one
can ignore the heavy quarks and work with $N_f =2$ or 3. Accordingly it
is physically reasonable to consider the limit of vanishing (light)
current quark masses and if needed the current quark masses can be
treated as perturbations. This limit corresponds
to exact chiral symmetry. The latter is broken spontaneously which results in 
a non-zero chiral condensate $\langle 0 | \bar \Psi\Psi | 0 \rangle$.
The breakdown of chiral symmetry leads to the appearance of Goldstone
bosons and to a dynamical generation of a constituent quark mass of
a few hundred MeV.
Although for many observables
one can probably neglect $m_u$ and $m_d$, the finite value of
$m_s~\approx~175$~MeV is too large to be ignored. In particular, the 
relatively large current strange mass $m_s$ is responsible for 
the baryon mass splitting summarized in the Gell-Mann-Okubo mass formula.

The number of colors $N_c$ plays a special role in QCD. The inverse 
number of colors $1/N_c$ can be considered ('t Hooft, 1974) as a
small parameter of QCD. Although in the real world $N_c=3$, the 
$1/N_c$ expansion is a very useful tool to analyze non-perturbative phenomena 
in the QCD. We will use this technique also in the effective theory 
considered below. In fact, concepts like that of constituent quarks and the 
idea of describing baryons as chiral solitons with a polarized Dirac sea emerge
naturally from the large $N_c$ limit of QCD and have proven to be 
phenomenologically quit useful.   

In this report we shall concentrate on theory and applications of an
effective model, which on one hand is formulated in terms of quarks
interacting via a simple and tractable quark-antiquark interaction, and
on the other hand fulfills basically the above mentioned
phenomenological demands, in particular the dynamical chiral symmetry
breaking and the explicit symmetry breaking due to current quark
masses. The model has first been proposed by Nambu and Jona-Lasinio (1961a,b)
in a different context. It has become quite
popular and successful in a form where the fermion fields are re-interpreted
as quark fields. In the meson sector it has already been the subject of several
review articles (Vogl and Weise, 1991, Klevansky, 1992, Hatsuda and
Kunihiro, 1994). In the baryon
sector the most interesting aspect is the applications to physical
observables of nucleon and hyperons and this is the
subject of the  present report. Previous reviews cover only partially
these topics (Meissner et al., 1993, Alkofer et al., 1995).
The Nambu -- Jona-Lasinio (NJL) model provides an excellent account of the 
today's experimental data if both
the collective quantization and the perturbative expansion in $m_s$ are
performed properly. The model describes the physics
in a way quite different from the other well known baryonic effective
model, i.e. the Skyrme model (Skyrme 1961, 1962). In the Skyrme-approach
(Adkins et al., 1993, Holzwarth and Schwesinger, 1986, Zahed and
Brown, 1986) one does not have quarks but meson fields and the
baryons are described as topological solitons. In fact, the NJL model
contains this physical picture as a limiting case, though being a
genuine non-topological model. The relationship between these two
approaches will also be a subject of discussion of our review.

The NJL model in its simplest SU($N_f=2$)-form is given by
\begin{equation}
{\cal L} = \bar\Psi  ( i \gamma^\mu \partial_\mu - m_0) \Psi
+ \frac G2 [(\bar \Psi\Psi)^2 + (\bar\Psi  i \gamma_5 \vec\tau \Psi)^2 ] \,.
\label{eq:LNJL}
\end{equation}
Here $\Psi$ are quarks fields, which are flavor doublets with $N_c=3$ colors, 
and $m_0$ is the average up and down quark current mass. The 
corresponding semibosonized lagrangian is
\begin{equation}
{\cal L}^\prime= {\bar{\Psi}}(i\sld -\sigma-i\vec\pi\cdot \vec\tau\,
\gamma_5) \Psi - {1\over 2G}(\sigma^2+\vec\pi^2) + {m_0\over G} \sigma\,,
\label{LNJLboson}
\end{equation}
where $\sigma$ and $\vec\pi$ are auxiliary meson fields.  The model
is not renormalizable and lacks confinement. Although these features
were for many years a reason to ignore the NJL model, nowadays the view
is different. For an effective model in the low energy regime, the
problem of non-renormalizability can be avoided if
one has a reasonable means to fix a suitable ultraviolet cut-off.
The lack of confinement provides in NJL model unphysical 
$\bar\Psi\Psi$-thresholds.  
However, there are strong arguments that the basic properties 
of light hadrons can be reproduced without invoking confinement and one of them
is the recent lattice QCD results for the correlation functions of hadronic 
currents in the cooled vacuum (Chu {\it et al.,} 1994). This point of view 
will be also supported by the results reviewed in the present paper.

In the NJL model, baryons appear as large $N_c$ solutions consisting of
valence quarks bound to the Dirac sea via classical (mean-field) meson fields 
in accordance with the arguments of Witten (1979).  
This physical picture was proposed by Kahana and Ripka (1984), and by
Diakonov {\it et al.} (1986). However, the mean-field solutions exist only
if the classical meson fields are subjected to the condition of the
chiral circle
\begin{equation}
\sigma^2+\piv^2=M^2\,,
\label{chcircle}
\end{equation}
where $M$ is the dynamically generated constituent quark mass. Under this
constraint the NJL lagrangian (\ref{LNJLboson}) reduces to an effective
lagrangian
\begin{equation}
{\cal L}^\prime = \bar \Psi ( i \gamma^\mu \partial_\mu - m_0 
- MU^{\gamma_5} )\Psi\,,
\label{eq:LNJLpion}
\end{equation}
in which the quarks interact 
only with Goldstone fields and the SU(2) matrix 
$U$ corresponds to a non-linear representation of the pion
field $MU^{\gamma_5}=\sigma+i\gamma_5\vec\pi\vec\tau$. 
For SU(3) the chiral field
configuration $U$ also encompasses kaon and eta fields. In the actual
calculations $U$ will be determined selfconsistently by minimizing the
effective action which corresponds to ${\cal L}^\prime$ (\ref{eq:LNJLpion}). 
The model with constraint (\ref{chcircle}) is usually referred to as the 
non-linear NJL model or, when applied to baryons, the Chiral quark soliton 
model.

The assumption of the chiral circle (non-linear pion field
configuration) is not an {\it ad hoc} device to allow for solitonic
solutions of NJL. Physically, it is a means of separating low-energy mesonic 
degrees of freedom (Goldstone bosons) from heavier ones (sigma meson) and to 
``freeze'' the latter. The physics, which apparently goes beyond 
the scope of the NJL model, can be understood in the framework of the 
underlying theory, the low-energy QCD. 

There are various attempts (McKay and 
Munczek, 1985, Cahill and Roberts, 1985, Diakonov and Petrov, 1986, 
Ball, 1989, Schaden {\it et al.,} 1990) to relate 
the NJL model to some low-energy limit of QCD. Further, arguments to 
illuminate the chiral structure of the NJL model are given by Ebert and Volkov
 (1983), Volkov (1984), Dhar and Wadia (1984) and Ebert and Reinhardt (1986). 
The relationship to the chiral perturbation theory (Gasser and Leutwyler, 1982)
is studied by Hansson {\it et al.} (1990), Bernard and Meissner (1991), 
Ruiz Arriola (1991), Sch\"uren {\it et al.} (1992) and Bijnens {\it et al.} 
(1993).  
In fact, the physics behind the present model can be well understood in the 
framework of the instanton liquid model of QCD (Shuryak 1982, 1983, 
Diakonov and Petrov, 1984, 1986) which is also supported by recent 
lattice QCD calculations (Chu {\it et al.,} 1994). The assumption that the 
non-perturbative ground 
state of QCD is governed by a dilute liquid of interacting instantons and 
anti-instantons allows to obtain in large $N_c$ limit an effective low-energy 
theory (Diakonov and Petrov, 1986)
which can be used to justify the NJL lagrangian on the chiral circle, i.e.
eq.~(\ref{eq:LNJLpion}). In fact, by assuming the instanton model of the QCD 
vacuum, Diakonov and Petrov (1984, 1986) derived an effective 
low-energy QCD lagrangian in terms of a 't~Hooft-like $2N_f$-fermion 
interaction. The crucial quantity in the instanton vacuum is the
ratio $\bar\rho/\bar R$ of the average size $\bar\rho$ of instantons to the 
average distance $\bar R$ between them. Both the phenomenological analysis 
(Shuryak 1982, 1983) and the variational estimates 
(Diakonov and Petrov, 1984) suggested a relatively small value
$\bar\rho / \overline R \approx 1/3$ which has been confirmed by the 
lattice calculations (Chu {\it et al.,} 1994). Thus, besides the inverse 
number of colors $1/N_c$ this small ratio plays also an important role of a 
small parameter in the 
theory. If one places light quarks in the instanton vacuum, one observes
that chiral symmetry is dynamically broken due to the delocalization of
the would-be fermion zero modes related to the individual instantons.
Spontaneous chiral symmetry breaking manifests itself in a non-zero chiral
condensate $\langle 0 |\bar\Psi\Psi| 0\rangle$, massless Goldstone bosons, 
and in a momentum dependent
dynamical quark mass $M(k)$ with $M(0)\approx 350$~MeV and $M(k)$ vanishing
for momenta $k\gg1/\bar \rho$. Due to the instanton-induced 't Hooft-like 
interaction the 
$\eta^\prime$ is not a Goldstone boson, so that the $U_A(1)$ problem is 
properly solved. The momentum dependence of $M(k)$ provides a
natural cut-off $1/\bar \rho$ which makes the resulting theory finite. 
It is important that the dynamical quark mass is parametrically small
$M\bar\rho=O(\bar\rho^2 / \bar R^2)\ll 1$ compared to the cutoff
$1/\bar \rho$. As a final result, due to the diluteness parameter 
$\bar\rho / \overline R$ in this instanton model of 
QCD vacuum, at low momenta $k<1/\bar \rho$ and large $N_c$ the QCD is 
reduced to an effective quark meson theory with a lagrangian 
which includes only constituent quarks of mass $M(k)$ interacting 
with Goldstone 
pion fields. If both quark and pion momenta are smaller than $1/\bar \rho$
the dynamical mass $M(k)$ can be approximated by
$M=M(0)$. In this approximation one reproduces the semibosonized NJL 
lagrangian (\ref{eq:LNJLpion}). Since the inverse size of
nucleons corresponds to $\sim 300$~MeV this can be considered as ``much'' less
than the cutoff $1/\bar\rho$ and the effective lagrangian  (\ref{eq:LNJLpion})
can be also applied to describe baryons. Hence a systematic treatment of the 
nucleons and hyperons involves only quarks and Goldstone bosons, and the 
implementation of the chiral circle in (\ref{LNJLboson}) is justified. 
It should be noted that the only relevant degrees of 
freedom are pions and quarks and excitations connected with other degrees of 
freedom like vector mesons have $k \sim 1/\bar\rho$ and do not contribute 
essentially  to the dynamics at momenta $k \ll 1/\bar\rho$. Following this 
logic one realizes that for a proper description of 
the baryon ground-state properties the only relevant degrees of freedom are 
those of quarks and pions, and  heavier meson degrees of freedom 
like scalar and vector mesons are not important. 

The present paper concentrates on the results for the baryon properties  
in the non-linear NJL model. 
The paper reviews applications in flavor SU(2) and SU(3) versions to  
electromagnetic, axial, scalar and strange vector 
form factors and the related static properties like magnetic moments, radii 
and coupling constants of the nucleon, and to the mass splittings and 
electromagnetic properties of the hyperons of the octet and
decuplet. Calculations of the spin properties are discussed in
connection with the ``spin crisis''. 
Predictions are reviewed also for nucleon strange form factors
and tensor charges. For all observables, the calculations reported are 
performed with one and the same set of parameter values. These values are 
fixed 
in the meson sector by fitting the physical values of the pion mass 
and the pion decay constant, and additionally to the kaon mass in the case of 
SU(3). The only parameter, which is free, is the constituent mass $M$. 
However, 
as we will show for the constituent mass of about $M=420$ MeV the model is 
rather successful in describing the baryon properties. 
Actually, the comparison of the model results with the 
experimental data of the nucleon and also of the hyperons in the octet and 
decuplet shows that the NJL model on the chiral circle 
indeed is able to reproduce most of the data 
within 15~\%. This includes critical observables like the axial vector 
coupling constant, $g_A$, and the magnetic moments of  proton and neutron. 
Only a few observables are less successfully described (within 30 \%) as 
e.g. the neutron electric 
form factor and the corresponding charge radius.  

Since the present theory yields a rather good agreement with the experimental 
data one can use the model picture to obtain some insight into the low-energy
structure 
of the nucleon and the other octet and decuplet baryons. The basic mechanism 
is obviously the interaction of quarks with pseudoscalar 
Goldstone bosons, the latter being themselves quark-antiquark excitations of 
the spontaneously broken chiral vacuum. Heavier mesons like scalar 
mesons or vector mesons are apparently not needed in this scenario. In 
addition, the confining forces seem also not to be much relevant in
describing the ground-state properties of low mass mesons and baryons. 
Hence, the only important 
mechanism seems to be chiral symmetry breaking. The calculations support 
the physical picture of baryons as non-topological solitons of $N_c=3$ valence 
quarks bound to a 
polarized Dirac sea via classical (mean-field) meson fields and being well 
separated from the negative Dirac continuum. The polarized 
Dirac sea phenomenologically corresponds to a meson cloud. Most of the 
observables receive about 30 \% contributions from the Dirac sea quarks. 
This valence picture of 
the baryon in the present model differs from the one of the topological 
solitons favored in the Skyrme model.  In principle, the NJL model can be 
reduced to a topological approach  similar to the Skyrme model 
by an expansion of the effective action in terms of gradients of the 
meson fields (gradient expansion). This expansion does only converge if the 
valence level is close to the negative Dirac continuum. This, however, 
requires in the NJL model unphysically large values of the constituent mass 
for which the baryon properties are not reproduced.
Hence, {\it within} the NJL model there is no way 
to recover Skyrme type models. 

%%%%%%%%%%%%%%%%%%%%%%%%%%%%%%%%%%%%%%%%%%%%%%%%%%%%%%%%%%%%%%%%%%%%%%%%%%%%%
%
%        review: chapter 2
%
%%%%%%%%%%%%%%%%%%%%%%%%%%%%%%%%%%%%%%%%%%%%%%%%%%%%%%%%%%%%%%%%%%%%%%%%%%%%%%
%
%
%     review2.tex
%
%
%%%%%%%%%%%%%%%%%%%%%%%%%%%%%%%%%%%%%%%%%%%%%%%%%%%%%%%%%%%%%%%%%%%%%%%%%%%%%%%
%\input review2

\section{SU(2) NJL model: vacuum sector and meson properties}

In this chapter we deal with the Nambu -- Jona Lasinio model
with two lightest quark flavors.
To begin with, we study the NJL lagrangian and its symmetries.
Using the functional integral approach we bosonize the theory
and solve the effective bosonized theory in the mean-field approximation
which corresponds to the leading order of the expansion
in inverse number of quark colors.
In the functional integral approach the mean-field solution
naturally appears as a saddle point of the bosonized action.
We analyze the structure of the ultraviolet divergences of the
model and introduce the regularization scheme.
We show that in the mean-field treatment of the model chiral symmetry is 
spontaneously broken,
which leads to a generation of the constituent quark mass.
An important consequence of the spontaneous breakdown of chiral
symmetry is the appearance of light Goldstone mesons (pions). We discuss 
shortly the description of mesons in the model
and sketch the derivation of the meson propagators, masses and
of the meson decay constant needed to fix the model parameters.

\vskip1cm
2.1 \underline{Bosonization of the NJL model}
\vskip4mm

\underline{NJL lagrangian and its symmetries}.
The NJL lagrangian contains a local
four-fermion interaction. In the simplest SU(2)
version it includes only scalar and pseudoscalar couplings:
\begin{equation}
{\cal L}=\bar{\Psi}(x) \[i\slashchar\partial - {\hat m}\] \Psi(x)
+ {G\over 2} \[ \(\bar{\Psi}(x) \Psi(x)\)^2 +
\(\bar{\Psi} (x) i\gamma_5 {\vec \tau}\Psi(x) \)^2 \]
\label{njlsu2}
\end{equation}
where one recognizes the free and the interaction parts
\begin{eqnarray}
&& {\cal L}_{free} = \bar{\Psi}(x) \[i\slashchar\partial
 - {\hat m}\] \Psi(x) \,,
\nonumber\\
&& {\cal L}_{int} = {G\over 2} \[ \(\bar{\Psi}(x) \Psi(x)\)^2 +
\(\bar{\Psi} (x) i\gamma_5 {\vec \tau}\Psi(x) \)^2 \]
\label{e22}
\end{eqnarray}
respectively. Here $\Psi(x)$ denotes a quark field with $u$ and $d$ flavors
and $N_c$ colors. The matrices $\tau^a$ $(a=1,2,3)$ are usual Pauli matrices.
The $\hat m$ represents the current quark mass matrix given by
\begin{equation}
{\hat m} = \left( \matrix{ m_u  &  0  \cr
                            0   & m_d } \right)=
                            m_0 {\bf 1} + \tau_3 m_3\,.
\label{e23}
\end{equation}
It is useful to divide it in
the average mass $m_0$ and in the isospin mass difference $m_3$
defined as
\begin{equation}
m_0 = {1\over 2} ( m_u + m_d ), \qquad  m_3 = {m_u - m_d\over 2}\,.
\label{e24}
\end{equation}
In (\ref{njlsu2}) $G$ is the coupling constant of the four fermion
interaction.

Lagrangian (\ref{njlsu2}) possesses a number of symmetries
which we also have in QCD. In particular, it
is symmetric under the global $U(1)$ transformation
\begin{equation}
\Psi(x) \rightarrow \Psi^\prime(x)= \exp (i\delta) \Psi(x)\,.
\label{e26}
\end{equation}

If one neglects the quark isospin mass difference $m_3$ and considers
$m_u = m_d=m_0$, which corresponds to exact isospin symmetry,
lagrangian (\ref{njlsu2}) is also invariant  under the
transformation
\begin{equation}
 \Psi(x) \to \Psi'(x)= \exp (i \vec\tau\cdot\vec\rho) \Psi(x)\,.
\label{e27a}
\end{equation}
If we additionally neglect the current mass $m_0$,
it becomes invariant under global chiral transformations:
\begin{equation}
\Psi(x) \to \Psi'(x)= \exp (i \gamma_5 \vec\tau\cdot\vec\kappa)\Psi(x)\,.
\label{e27b}
\end{equation}
Thus, in chiral limit $m_u=m_d=0$,
the NJL lagrangian has an
$SU_L(2)\otimes SU_R(2) \otimes U_V(1)$ symmetry.
Therefore, many well-known results obtained
within the current algebra can be reproduced in the NJL model.
Later we will show that chiral symmetry is spontaneously
broken which leads to such fundamental consequences as the appearance
of Goldstone bosons (pions) and to the generation of the dynamical quark
mass.

The above transformations correspond to the following Noether currents
\begin{equation}
B_\mu (x) = \bar \Psi(x)\gamma_\mu \Psi(x)
\qquad \hskip0.85cm{\rm ( baryon \, current)}\,,
\label{e28a}
\end{equation}
\begin{equation}
V_\mu^a (x) = \bar \Psi(x) \gamma_\mu {\tau^a\over 2} \Psi(x)
\qquad \hskip0.4cm{\rm ( vector \, current)} \,,
\label{e28b}
\end{equation}
\begin{equation}
A_\mu^a (x) = \bar \Psi(x) \gamma_\mu \gamma_5 {\tau^a\over 2}
\Psi(x) \qquad  \hskip0.3cm{\rm ( axial \, current)}\,,
\label{e28c}
\end{equation}
whose divergences are given by
\begin{equation}
\partial^\mu B_\mu = 0\,,
\label{e29a}
\end{equation}
\begin{equation}
\partial^\mu  V_\mu^a =
 i \bar \Psi [ {\hat m}, {\tau^a\over 2} ] \Psi \,,
\label{e29b}
\end{equation}
\begin{equation}
\qquad \partial^\mu  A_\mu^a
= i \bar \Psi\gamma_5
\{{\hat m}, {\tau^a\over 2} \} \Psi\,.
\label{e29c}
\end{equation}
The baryon current and the third component of the vector isotopic current
$V_\mu^3$ are conserved for any values of the quark masses.
For the conservation of $V_\mu^1$, $V_\mu^2$ we need equal
quark masses $m_u=m_d$,
whereas the axial current $A_\mu^a$ is conserved
only in chiral limit $m_u=m_d=0$, so that the last equation
(\ref{e29c})  expresses the partial conservation of the axial
current (PCAC).

Below for simplicity we work with the NJL model in the approximation
$m_u=m_d=m_0$. However, most of our formalism can be easily generalized to 
the case $m_u\ne m_d$ at least in the linear approximation in
small quark masses.

The NJL lagrangian is assumed to mimic the low energy effective theory
of QCD with gluon degrees of freedom integrated out.
However the quark fields carry a color index
that runs from $1$ to $N_c$ ($N_c=3$ in the real world).
Instead of the $SU(N_c)$ local gauge invariance of QCD,
in the NJL lagrangian, we have only the global $SU(N_c)$ invariance.
Although the color degrees of freedom enter the NJL lagrangian
in a rather trivial way, their role will be very important
for the justification of the mean-field approximation to the NJL model.
Later we shall show that the mean-field approach becomes asymptotically
exact in the limit of large number of quark colors $N_c$.

\vskip1cm
\underline{Auxiliary boson fields}.
The NJL lagrangian (\ref{njlsu2}) contains only quark degrees of freedom.
On the other hand, the light mesons which we are going to describe
by using this effective lagrangian are pions. Therefore, it
is desirable to find an equivalent formulation of the theory that
involves meson fields and not only quarks. Another reason why we need
the bosonization is that we are going to solve the model by using the mean
field approximation. In terms of the functional integral approach the mean
field solution corresponds to the saddle-point approximation.  The
saddle-point solution can be obtained only in terms of boson
fields so that the bosonization of the NJL lagrangian is a natural step
towards the mean-field solution of this model.

Historically, the bosonization has first been formulated in solid state
physics where it is known as Hubbard-Stratononovich transformation.  In the
present theory it has been first applied by Eguchi (1976), Kikkawa (1976) and
Kleinert (1976).

For simplicity we describe the bosonization of NJL
on the level of the partition function
\begin{equation}
Z = \int \D \Psi \D \bar \Psi
\e^{i \int {d}^4 x {\cal L}(x)}
\label{e211}
\end{equation}
keeping in mind that the generalization to other quantities is
straightforward.

Under proper normalization of the integration measure one can write the 
following identity
\begin{equation}
1=\int \D \sigma \D \vec\pi \exp \biggl \{ -i\int d^4 x{1\over 2G}
\left[
\left(\sigma
+ G \left(\bar{\Psi}\Psi - {m_0 \over G}\right)\right )^2
+ \left( \vec \pi  + G \bar{\Psi} i \vec\tau\,\gamma_5  \Psi \right)^2
\right ] \biggr \}
\label{e210}
\end{equation}
with a gaussian path integral over auxiliary boson fields
$\sigma$ and $\vec\pi$ on the rhs. Inserting this identity into
the partition function $Z$ we obtain:
\begin{equation}
 Z =\int \D \bar \Psi  \D \Psi \D \sigma \D \vec\pi
 \e^{ i \int \d^4 x {\cal L}^\prime (x)  }\,,
\label{e212}
\end{equation}
where the path integral runs over both the quark and
auxiliary meson fields.
The corresponding semibosonized lagrangian is
\begin{equation}
{\cal L}^\prime= {\bar{\Psi}}(i\sld -\sigma-i\vec\pi\cdot \vec\tau\,
\gamma_5) \Psi - {1\over 2G}(\sigma^2+\vec\pi^2) + {m_0\over G} \sigma\,.
\label{e213}
\end{equation}
One should keep in mind that on this stage the introduced auxiliary
fields $\sigma$ and $\piv$ are non-dynamical fields and no kinetic term
$\half \left [ \ddmu\sigma \dumu\sigma + \ddmu\piv \dumu\piv \right ]$
appears in lagrangian (\ref{e213}). However, after integrating out the quarks
in (\ref{e212}), i.e. after including the quark loop
effects, the fields will be ``dressed'' and can describe the
physical mesons. It should be stressed that since the bosonization procedure
is exact, the semibosonized lagrangian
${\cal L^\prime}$ has the same symmetries as the initial lagrangian
(\ref{njlsu2}). Note that the meson fields $\sigma$, $\vec\pi$
transform under the chiral rotation (\ref{e27b}) as follows 
\begin{equation}
(\sigma + i \vec\pi \cdot\vec\tau) \to
(\sigma'  + i \vec\pi{}' \cdot\vec\tau) =
\exp (- i \vec\tau\cdot\vec\kappa)
(\sigma + i \vec\pi \cdot\vec\tau)
\exp (- i \vec\tau\cdot\vec\kappa)\,.
\label{sigma-pi-chiral}
\end{equation}

Diagrammatically, the described bosonization procedure corresponds to
rearrangement and resummation of the graphs of the 4-fermion point interaction
into {\it quark-meson} vertices of the Yukawa type by introducing collective
scalar-isoscalar ($\sigma$) and pseudoscalar-isovector ($\piv$)
auxiliary boson fields, both carrying the
quantum numbers of the composite operators $({\bar{\Psi}} \Psi)$ and
$({\bar{\Psi}}i\vec\tau\gamma_5\Psi)$ but not color. The
idea has a long history and traces back to the sixties (see for instance
(Coleman, 1985) and references therein).

\vskip1cm
\underline{Effective action}.
Now instead of the four-fermion interaction we
have a theory of quarks interacting with boson fields.
Next we notice that the integral over quark fields in
eq.~(\ref{e212}) is gaussian and can be done exactly, which leads
to a determinant of the Dirac operator
\begin{eqnarray}
 \int \D \bar \Psi  \D \Psi
 \exp \left[ i \int \d^4 x
 {\bar{\Psi}}(i\sld -\sigma-i\vec\pi\cdot \vec\tau\,\gamma_5) \Psi  \right]
&=& \mbox{Det}^{N_c}
(i\sld -\sigma-i\vec\pi\cdot \vec\tau\,\gamma_5)
\nonumber
\\
&=& \exp\left[ N_c \Tr\, \log
(i\sld -\sigma-i\vec\pi\cdot \vec\tau\,\gamma_5) \right]
\,.
\label{quarks-out}
\end{eqnarray}
Here the power of $N_c$ appeared due to $N_c$ colors of the quark fields.

The Dirac operator can be rewritten in the form
\begin{equation}
i\sld -\sigma-i\vec\pi\cdot \vec\tau\,\gamma_5
= \gamma^0[i\partial_t - h(\sigma,\piv)]\,.
\end{equation}
where $h(\sigma,\piv)$ is the one-particle Dirac hamiltonian
\begin{equation}
h(\sigma,\piv) = - i \gamma^0\gamma^k \nabla_k
+ \gamma^0(\sigma+i\vec\pi\cdot\vec\tau \gamma_5)\,.
\label{Dirac-covariant}
\end{equation}

Below we prefer to work with the euclidean time $\tau$
and perform the Wick rotation
\begin{equation}
\tau = it\,.
\end{equation}
In the euclidean case the following change should be done in
(\ref{quarks-out})
\begin{equation}
\mbox{Det} (i\sld -\sigma-i\vec\pi\cdot \vec\tau\,\gamma_5)
= \mbox{Det} [i\partial_t - h(\sigma,\piv)] \to
\mbox{Det} [\partial_\tau + h(\sigma,\piv)]\,.
\end{equation}
Integrating out the quarks as it is done in (\ref{quarks-out})
we arrive at the following completely bosonized euclidean effective action
\begin{equation}
\Sesp = -N_c \Tr\, \log
 \, D(\sigma,\piv)+{1 \over 2G} \int \d^4 x
(\sigma^2 + \vec\pi^2)- {m_0 \over G} \int d^4 x \sigma\,.
\label{bosonized-action}
\end{equation}
Here we use the notation
\begin{equation}
D(\sigma,\piv)=\partial_\tau + h(\sigma,\piv)\,.
\label{e217d}
\end{equation}

\vskip1cm
\underline{Large $N_c$ and mean-field approximation}.
In terms of the effective action (\ref{bosonized-action})
the partition function of the model is given by equation
\begin{equation}
Z = \int \D \sigma \D \vec\pi \e^{-\Sesp}\,.
\label{e216}
\end{equation}

All our formulas, which have been derived so far, are exact. However,
in order to evaluate the functional integral
(\ref{e216}) we need some approximation. Note
that the fermion determinant contribution to the effective action
(\ref{bosonized-action}) has a prefactor of $N_c$.
For a consistent large $N_c$ treatment of the effective action
(\ref{bosonized-action}) we take the coupling constant of the NJL model
to be of order
\begin{equation}
G = O(N_c^{-1})\,.
\end{equation}
Then all terms in (\ref{bosonized-action})
will be of order $O(N_c)$ and in large $N_c$ limit
the functional integral (\ref{e216}) can be evaluated in the saddle-point
approximation. The stationary meson configuration $\sigma_c,\piv_c$
minimizing the effective action $\Sesp$
can be found by solving the saddle
point equations
\begin{equation}
{\delta \Sesp \over \delta \sigma}
\Bigr|_{\sigma=\sigma_c \atop \pi=\pi_c}  = 0 \,,
\qquad
{\delta \Sesp \over \delta \vec\pi}
\Bigr|_{\sigma=\sigma_c \atop \pi=\pi_c} = 0\,.
\label{e220a}
\end{equation}
Thus, in the leading order of the $1/N_c$ expansion the functional integral
(\ref{e216}) is simply given by the value of the integrand
at the saddle point. One can show that this approximation
is nothing else but the Hartree mean-field approximation in the original
$4$-fermion version, which has been used repeatedly
(Bernard {\it et al.,} 1984, Bernard, 1986,  Ferstl {\it et al.,} 1986, Bernard
{\it et al.}, 1987, Providencia {\it et al.,} 1987, Bernard {\it et al.}, 
1988, Klimt {\it et al.,} 1990) and has been reviewed by Vogl and Weise (1991) 
and Klevansky (1992).

\newpage
\vskip1cm
2.2 \underline{Divergences and regularization}
\vskip4mm

Like the Fermi theory of weak interaction the Nambu--Jona-Lasinio model
is non-renormalizable, because the coupling constant of the 4-fermion
interaction $G$ has the dimension $ [G]=  mass^{-2}$.
This means that in each increasing order in $G$ new graphs with
a higher degree of ultraviolet divergence appear.
In order to get a well defined theory it is therefore necessary to
specify  how the infinities of the model have to be treated.

One faces the ultraviolet divergences in the effective action
(\ref{bosonized-action}) already in the leading order
in $N_c$ which are due to the Dirac determinant
$\mbox{Det} \, D(\sigma,\piv)$. It is convenient to consider separately
the divergences of the real part
\begin{equation}
\Re \, \Tr \, \log \, D
=  \frac{1}{2} \,  \Tr \, \log \, (D^{\dag} D)\,,
\label{e217a}
\end{equation}
as well as of the imaginary part
\begin{equation}
\Im \, \Tr \, \log \, D
=  \frac{1}{2i} \, \Tr \, \log \, ( D/ D^{\dag})\,.
\label{e217b}
\end{equation}

\vskip1cm
\underline{UV divergences}.
Let us start from the real part of the effective action (\ref{e217a}).
The functional determinant $\mbox{Det} (D^{\dag} D)$ is a complicated
non-local functional of fields $\sigma,\piv$ and the simplest way to
study its divergences is to consider the case of slowly changing fields
$\sigma,\piv$ and to expand this functional determinant in a series
of integrals of local polynomials in derivatives of fields
$\sigma,\piv$ (Aitchison and Fraser, 1984, 1985a, 1985b). It should be noted 
that in the case of the proper-time regularized effective action 
(see next subsection eq.~(\ref{e229})) for a
systematic expansion it is more convenient to use the
heat kernel method (Kleinert, 1976, Nepomechie, 1985, Ebert and Reinhardt,
1986). However, the structure of divergent terms can be determined by a
simple dimensional analysis without explicit
calculations. Since the effective action is
dimensionless and the divergent terms should contain a positive power
or the logarithm of the cutoff, the local integrals of the fields and their
derivatives in the divergent terms should have a dimension of $mass^{m}$ 
where $m\geq 0$. Combining this dimensional argumentation with the 
chiral symmetry it is easy to see that 
the divergent terms are of the form
\begin{eqnarray}
\Re\, \Tr \, \log \,  \, D
&=& c_1 \int d^4 x (\sigma^2+\vec\pi^2)
+ c_2 \int d^4 x (\sigma^2 + \vec\pi ^2)^2
\nonumber\\
&+& c_3 \int d^4 x [(\partial^\mu \sigma)
(\partial_\mu \sigma) +(\partial^\mu \vec\pi) (\partial_\mu \vec\pi) ]
+ {\tilde S} (\sigma,\vec\pi)\,.
\label{e226}
\end{eqnarray}
Here $c_1$, $c_2$, $c_3$ are some divergent coefficients and
${\tilde S} (\sigma,\vec\pi)$ stands for the ultraviolet finite part. 
Apparently, in
order to examine the particular divergent structure of coefficients
$c_i$ one should perform explicitly a gradient expansion of the real part of
the effective action (\ref{e217a}).
Note that among the divergent terms there appeared the kinetic
term for meson fields  $\sigma,\vec\pi$. Later we will introduce a
regularization so that this kinetic term will be finite
but non-vanishing. The presence of this kinetic term means that
the mesonic fields, which  initially have been introduced as auxiliary
unphysical fields, become dynamical fields after taking into account
the quark loop effects.

\vskip1cm
\underline{Imaginary part of the effective action.}
In the case of the $SU(2)$ flavor group the imaginary part of the effective
action vanishes. It is connected to the fact that operators
$D(\sigma,\piv)$ and $D^\ast(\sigma,\piv)$
(the asterisk stands for usual complex, not hermitian, conjugation)
are unitary equivalent in the $SU(2)$ case
\begin{equation}
D^\ast(\sigma,\piv) = V\tau^2 D(\sigma,\piv) (V\tau^2)^\dagger\,.
\label{Dast}
\end{equation}
Here $V$ is some unitary spin matrix defined by 
\begin{equation}
\gamma_\mu^\ast = V \gamma_\mu V^{-1}  \qquad(\mu=1,2,3,4,5)\,,
\label{V}
\end{equation}
where the asterisk stands for complex conjugation and  $\gamma_\mu$ are 
euclidean hermitian Dirac matrices.

Using  equality (\ref{Dast}) one immediately concludes that
\begin{equation}
 \mbox{Det} \, ( D/ D^{\dag}) = 1\,,
\end{equation}
which means that the imaginary part of the effective action vanishes exactly
in the $SU(2)$ case.

In the case of higher flavor groups the imaginary part of the effective
action is not necessarily vanishing. However, it still
remains finite. The simplest way to check it is again to consider
the gradient expansion.

\vskip1cm
\underline{Regularization scheme.}
The NJL model is a non-renormalizable
effective theory which should be considered as a low-energy approximation to 
QCD and hence is physically relevant only up to some momentum
scale. Actually, in the instanton liquid model (Diakonov and Petrov, 1986) the 
inverse size of the instantons $1/\bar\rho$ plays the role of a natural 
cutoff in the theory free from the UV divergences. 
Therefore, we have to regularize the present theory introducing a cutoff at 
this scale.
There are many ways to do it and
there is no reason to prefer some scheme in favor of the other
provided certain constraints such as  Lorentz invariance, gauge
invariance, current conservation, integer baryon number,
are obeyed (Ball, 1989). We have already shown that only the real part of the 
effective action contains ultraviolet divergences and hence needs  
regularization. In general, one is 
free to regularize the finite imaginary part as well. However, it would lead 
to additional complications related for instance to the definitions of the 
baryon number and the charge. Some insights into the problem can be gained 
from the 
underlying non-local theory (Diakonov and Petrov, 1986) which because of the 
momentum dependent dynamical mass $M(k)$ is finite. In fact, Diakonov and 
Petrov (1986), and
Ball and Ripka (1994) have shown that the results for the amplitude of the 
anomalous processes $\pi^0\to 2\gamma$ in the non-local 
theory are very similar to those of the NJL model with non-regularized 
imaginary part of the effective action. 

In this review we regularize only the divergent real part 
and keep the finite imaginary part unaffected. We will discuss the results 
obtained in the proper-time scheme based on the identity
\begin{equation}
\Tr[\log \,K - \log \,K_0]
=-\int_0^\infty {\d u\over u}\Tr \Bigl(\e^{-uK}-\e^{-uK_0}
\Bigr)\,.
\label{PT}
\end{equation}
In the case of our interest $K=D^\dagger D$ and
the ultraviolet divergence appears as a singularity in the integration
over small $u$. We make this integral finite by simply introducing a
cutoff $\Lambda^{-2}$ in the lower limit
\begin{equation}
\half\Tr\,\log\,(D^\dagger D)_{reg}-\half \Tr\,\log\,(D^\dagger_0 D_0)_{reg} =
-\half\Tr\int_{1/\Lambda^2}^\infty {\mbox{d}u \over u}
\Bigl[\e^{-u D^\dagger D} - \e^{-u D_0^\dagger D_0} \Bigr]\,.
\label{e229}
\end{equation}
Here we subtract the fermion determinant of the "free" Dirac operator
$D_0=D(\sigma_0,\piv_0)$ corresponding to some fixed
fields $\sigma_0,\piv_0$. One can generalize the proper-time
regularization in the form
\begin{equation}
\half\Tr\,\log\,(D^\dagger D)_{reg}-\half \Tr\,\log\,
(D^\dagger_0 D_0)_{reg} =
-\half\Tr\int_0^\infty {\mbox{d}u \over u} \phi(u,\Lambda)
\Bigl[\e^{-u D^\dagger D} - \e^{-u D_0^\dagger D_0} \Bigr]\,,
\label{e229g}
\end{equation}
where $\phi(u,\Lambda)$ is a function properly chosen to make
the integral finite. In particular, for the specific choice
$\phi (u)= \theta ( {1\over \Lambda^2} - u )$ one recovers
(\ref{e229}).
Different cutoff schemes have been considered in detail by Meissner 
Th. {\it et al.} (1990) for the vacuum, and by Blotz {\it et al.} (1990) and 
D\"{o}ring {\it et al.} (1992a) in the soliton sectors. In both regimes the 
results of using other regularization schemes, Pauli-Villars and both sharp 
and soft O(3)- and O(4)-cuttofs, were rather similar to those of the 
proper-time scheme if the real part was only regularized. 

\vskip1cm
2.3 \underline{Spontaneously broken chiral symmetry and constituent quark
mass}
\vskip4mm

The  main observation of Nambu and Jona-Lasinio (1961a,b) in their original
work is that if the coupling constant $G$ is larger than a certain
critical value, the four fermion interaction leads to a non-trivial vacuum
solution in which chiral symmetry is spontaneously broken and in this way
fermions acquire a dynamical mass. This was done in the canonical
formalism using a Bogoliubov-Valatin transformation from bare massless 
fermions to constituent massive fermions as quasiparticles. This phenomenon is
called dynamical mass generation. In the present section we will
see explicitly how the spontaneous chiral symmetry breaking is realized within
the path-integral approach to the NJL model in large $N_c$ limit.

We look for the vacuum saddle-point solution of
eqs.(\ref{e220a}) in the form
\begin{equation}
\sigma_c \neq 0, \hskip2cm \piv_c=0\,.
\label{VEVmeson}
\end{equation}
where the expectation value $\sigma_c$ is a constant field.
Using non-regularized effective action~(\ref{bosonized-action})  the
saddle-point equation~(\ref{e220a}) reads
\begin{equation}
{\delta \Sesp \over \delta \sigma}
{\Biggr\vert}_{\sigma=\sigma_c}  =
{1\over G}\sigma_c - N_c\Tr{1\over D(\sigma_c)}-{m_0\over G}
={1\over G}\sigma_c-8N_c  \sigma_c I_1(\sigma_c)
-{m_0\over G}=0\,,
\label{gap}
\end{equation}
In the case of the proper-time regularization of the fermion determinant
(\ref{e229}) the divergent integral $I_1(M)$ in the saddle-point
eq.~(\ref{gap}) should be replaced by:
\begin{equation}
I_1(M)= \int{\d^4 k\over (2\pi)^4}  {1\over \left( {k^2 + M^2}\right)} \to
I_1^\Lambda(M)={1\over (4\pi)^2}\int_{\Lambda^{-2}}^\infty {\d u\over
u^2}\e^{-u M^2}\,.
\label{gapint}
\end{equation}
Eq.~(\ref{gap}) is a non-linear equation
for $\sigma_c$ which is known as {\it gap equation}.
In fact, it corresponds to the Schwinger-Dyson
equation in the mean-field Hartree approximation.

The non-zero value $\sigma_c$ enters the Dirac operator (\ref{Dirac-covariant})
in the same way as the quark current mass and as such plays a role of a quark
mass which is usually called constituent quark mass
\begin{equation}
M=\sigma_c
\label{constituent-mass}
\end{equation}

\begin{figure}[t]
%\vspace{1cm}
\centerline{\epsfysize=1.in\epsffile{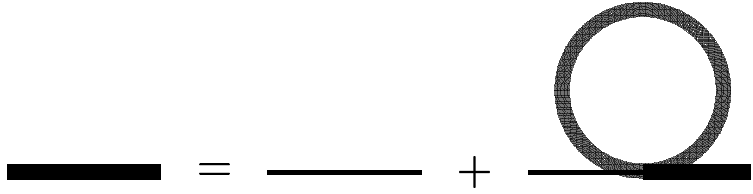}}%\vskip4pt
\caption{Diagram representation of the gap equation. The thin line corresponds
to ``bare'' quarks whereas the thick line represents the constituent 
``dressed'' quarks.}
\label{f21}
\end{figure}

The gap equation is illustrated in fig.~\ref{f21}. Because of the
quark-loop contribution the quark line gets ``dressed'' and the quarks
become massive even in chiral limit.
As a result in the Dirac quark spectrum there appears a mass gap of $2M$
between the positive and negative energy continua.

Note that the non-vanishing saddle-point expectation value $\sigma_c\ne 0$
is not invariant under the chiral transformation
(\ref{sigma-pi-chiral}) even in chiral limit. This means that in the
saddle-point solution chiral symmetry is spontaneously broken and the 
non-zero value for the constituent quark mass
(\ref{constituent-mass}) is the consequence of this spontaneous breakdown of
chiral symmetry.

Analogous to the magnetization in
ferromagnets an order parameter for the non-trivial vacuum can be introduced.
In fact, it is the quark condensate $\langle\bar \Psi\Psi\rangle=\langle\bar u u\rangle+\langle\bar d d\rangle$, which
characterizes the strength of the spontaneous breakdown of chiral
symmetry, and as such plays the role of the chiral order parameter.
In the leading order of the large $N_c$ approximation we obtain
\begin{equation}
<\bar \Psi\Psi> \>  =  - \frac{1}{V_4}
\frac{\partial}{\partial m_0} \, \log Z
= -{\partial \over \partial m_0}
\left[ \frac{1}{V_4} S(\sigma_c,\vec\pi_c) - \frac{m_0^2}{2G} \right]
\end{equation}
Here $V_4$ is the four-dimensional euclidean space-time volume.
Note that the additional term $m_0^2/(2G)$
appearing here should have been written already in the semibosonized
lagrangian (\ref{e213}) but we have dropped it there since it is a field
independent constant. Differentiating the bosonized action
$S(\sigma_c,\vec\pi_c)$ (\ref{bosonized-action}) with respect to $m_0$ at the 
saddle point $\sigma_c\equiv M,\vec\pi_c=0$ we get:
\begin{equation}
<\bar \Psi\Psi> \>
= - \, {1\over G}(\sigma_c-m_0) = - \, {1\over G}(M-m_0)\,.
\label{qq}
\end{equation}
Using explicitly the gap equation (\ref{gap}) we can rewrite this
result in the form
\begin{equation}
<\bar \Psi\Psi> \>=-8N_c M I_1(M) \to -8N_cM {1\over (4\pi)^2}
\int_{\Lambda^{-2}}^\infty {\d u\over u^2}\e^{-u M^2}\,.
\label{qqI1}
\end{equation}
One realizes that the condensate $<\bar \Psi\Psi>$ is a quadratically
divergent quantity  and depends strongly on details of the regularization
scheme.

\vskip1cm
2.4 \underline{Mesonic properties}
\vskip4mm

The evaluation of the mesonic two-point functions within the NJL model
has been undertaken by several authors, both in a pure fermionic theory
using a Bethe-Salpeter formalism in the ladder approximation
(Blin {\it et al.,} 1988, Bernard {\it et al.,} 1988, Bernard and 
Meissner U.-G., 1988,
Klimt {\it et al.,} 1990, Takizawa {\it et al.,} 1990)  and in a bosonized 
version (Jaminon {\it et al.,} 1989, 1992) in the path integral formalism. 
In this section
we sketch the calculation of mesonic spectra in the
bosonized version of the model as it has been done by Jaminon {\it et al.}
The advantage of their calculations is that using path integrals one is
able to start directly from the regularized theory.

The mesonic two-point function $K_{ab}$  can be obtained from the
generating functional
\begin{equation}
Z(j) = \int \D \phi_a e^{-S+j\cdot\phi}\,
\label{e216a}
\end{equation}
with explicit meson source $j_a$ included
\begin{equation}
K_{ab}(x-y)={\delta^2 \ln Z\over \delta j_a(x)
\delta j_b(y)}\Bigg\vert_{j=0}\,.
\label{defmesprop}
\end{equation}
Here we use an abbreviation $j\cdot\phi=\int\d^4x\, j_a(x)\phi_a(x)$
where $\phi_a\equiv (\sigma,\piv)$ stands for both meson fields.

In the model the mesons appear as low-lying collective $\bar \Psi\Psi$
excitations. In order to describe these modes one should consider meson
fluctuations around the stationary meson field configuration:
\begin{equation}
\phi_a=\phi^0_a + \tilde\phi_a\,.
\label{mesfluct}
\end{equation}
In the leading order in $N_c$, the saddle-point value $\phi^0_a$, 
which minimizes the effective action (\ref{bosonized-action}), coincides with
the ``classical'' value $<\phi_a>$
\begin{equation}
<\phi_a(x)>= {\delta \ln Z\over \delta
j_a(x)}\Bigg\vert_{j=0}\equiv\phi_a^0(x)\,.
\label{classmesvalue}
\end{equation}
Since the meson fluctuations are small (lowest
excitations) we can expand the effective action (\ref{bosonized-action}) up
to the second order in the fluctuating meson fields 
(small amplitude approximation):
\begin{equation}
Z(j)  = \e^{- S(\phi_a^0)+j\cdot \phi_0}
\int \D \tilde\phi \e^{-{1\over 2}\tilde\phi{\delta^2 S\over\delta
\phi \delta \phi}\tilde\phi+j\cdot\tilde\phi}\,,
\label{genfunctmes}
\end{equation}
where we use the short notation
\begin{equation}
\tilde\phi{\delta^2 S\over\delta\phi \delta\phi}\tilde\phi
=\int\d^4x\d^4y \tilde\phi_a(x){\delta^2 S\over\delta\phi_a(x)
\delta\phi_b(y)}\Bigg\vert_{\phi_a^0}\tilde\phi_b(y)\,,
\label{genfunctmesnot}
\end{equation}
In the exponent we have a bilinear form and the integral can be easily
evaluated 
\begin{equation}
\ln Z(j)= -S(\phi^0)+j\cdot\phi^0
-{1\over 2}\Tr\log\Bigl[{\delta^2 S\over\delta\phi\delta \phi}
\Bigr] +{1\over 2}j\Bigl[{\delta^2 S\over\delta\phi\delta \phi}
\Bigr]^{-1}j\,.
\label{lnZ}
\end{equation}
For the last term in (\ref{lnZ}) we use the same short notation as 
(\ref{genfunctmesnot}) and  $\Tr$ is the functional trace over the meson 
degrees of freedom. The first term in rhs is
the effective action in leading order in $N_c$ (\ref{bosonized-action}).
The third one is the one-meson loop contribution. Compared to the leading term
$S(\phi_a^0)\sim N_c$ it is suppressed by $1/N_c$. It means that it is
parametrically small (large $N_c$) and we do not include it in our
considerations (zero-meson-loop approximation). From eq.(\ref{lnZ}) one
also realizes that the inverse meson propagators in leading order in
$N_c$ are given by the second variation of the effective action $S$
with respect to the meson fields at the stationary point $\phi_0$:
\begin{equation}
K_{\phi}^{-1}(x-y)={\delta^2 S\over\delta\phi(x)\delta\phi(y)}
\Bigg\vert_{\phi_0}\,.
\label{mesprop}
\end{equation}
The particular calculations of the r.h.s. of the above equation are
straightforward but quite lengthy and we skip it here. We refer the
reader to (Jaminon et al., 1989, 1992) for details. After Fourier
transform to the momentum space the
final result for the meson propagator in leading order in $N_c$ is
\begin{equation}
K_{\phi}(q^2)={1\over Z_p(q^2)}\ {1\over q^2+
4M^2\delta_{\phi\sigma}+{m_0\over GM  Z_p(q^2)}}\,.
\label{Kphi}
\end{equation}
The function $Z_p(q^2)$ corresponds to a quark-loop with two
pseudoscalar-isovector insertions ($i\gamma_5\tau_a$). Here we present
only the proper-time regularized expression for it:
\begin{equation}
Z_p(q^2)={4N_c \over (4\pi)^2}\int\limits_0^{1}\mbox{d} \beta
\int\limits_{\Lambda^{-2}}^\infty {\mbox{d} u\over
u}\mbox{e}^{-u\biggl[M^2+{q^2\over 4}(1-\beta^2)\biggr]}\,.
\label{Zpq2}
\end{equation}
The Pauli-Villars regularized expression can be found in
(Sch\"uren {\it et al.,} 1993).
The on-shell meson mass corresponds to the pole of the meson propagator
\begin{equation}
K_{\phi}^{-1}\(q^2=-m^2_\phi\)=0\,.
\label{mesonmass}
\end{equation}
From (\ref{Kphi}) one can see that in chiral limit $m_0 \to 0$, the pions 
become massless Goldstone bosons whereas $m_\sigma=2M$.
The meson coupling constants are given by the residue of
the propagator at the pole
\begin{equation}
g_{\phi}^2=\lim\limits_{q^2\rightarrow -m^2_\phi}(q^2+m^2_\phi)
K_{\phi}(q^2)\,,
\label{couplconst}
\end{equation}
and as usual the physical meson fields are defined as
\begin{equation}
\pi^a_{ph}={\pi^a\over g_{\pi}} \qquad \mbox{and} \qquad
\sigma_{ph}={\sigma\over g_{\sigma}}\,.
\label{physmeson}
\end{equation}
Another important quantity, which we need to fix our
parameters, is the pion decay constant. It is defined as the matrix
element of the axial current between the physical vacuum and a physical
pion state of a four momenta $q$:
\begin{equation}
<0\mid A_\mu^a(z)\mid\pi^b(q)>=i q_\mu f_\pi\e^{-iqz}\delta^{ab}\,.
\label{fpidef}
\end{equation}
Using the Lehmann-Symanzik-Zimmermann reduction formula one can write
the pion decay constant as
\begin{equation}
f_\pi=f( q^2=-m^2_\pi)\,.
\label{fpidef1}
\end{equation}
which is given by
\begin{equation}
i\delta^{ab}\int
{\mbox{d}^4 q\over(2\pi)^4} q_\mu {f(q^2)\over q^2+m_\pi^2}\e^{-iqz}=
<0\mid{\cal T}[ A_\mu^a(z)\pi^b_{ph}(0)]\mid 0>\,,
\label{fpidef2}
\end{equation}
where  ${\cal T}$ stands for time ordering.
In order to calculate the rhs of eq.(\ref{fpidef2})
we couple an external axial source $J_\mu^a$
\begin{equation}
D(\sigma,\vec\pi)\longrightarrow D(\sigma,\vec\pi)+J
_\mu^a(z)\gamma^\mu\gamma_5{\tau_a\over 2}\,.
\label{axialmatr}
\end{equation}
in the effective action (\ref{bosonized-action}) and in eq.(\ref{e216a}).
Using the generating functional $Z$ which now explicitly depends on $J_\mu^a$
we can write
\begin{equation}
<0\mid{\cal T}[ A_\mu^a(z)\pi^b_{ph}(0)]\mid 0>={1\over g_\pi}
{\delta^2 \ln Z\over\delta J^a_\mu(z) \delta j^b(0)}\,.
\label{axialmatr1}
\end{equation}
From (\ref{lnZ}) one can calculate directly
\begin{equation}
{\delta^2 \ln Z\over\delta J^a_\mu(z) \delta j_b(0)}=\int \d^4 x
{\delta^2 S\over\delta J^a_\mu(z) \delta \pi_b(x)}K_\pi(x)\,,
\label{dlnZdj}
\end{equation}
and combining eqs.(\ref{fpidef2}) and (\ref{dlnZdj}) one gets
\begin{equation}
q_\mu f(q^2)\delta^{ab}={1\over g_\pi}(q^2+m_\pi^2)K_\pi(q^2)\int\mbox{d}^4z
{\delta^2 S\over\delta J^a_\mu(z) \delta \pi^b(0)}\e^{-iqz}\,.
\label{fpiq2}
\end{equation}
The calculation of ${\delta^2 S\over\delta J^a_\mu \delta \pi_b}$
is almost identical to one of the inverse pion propagator (\ref{mesprop}).
The final result for the physical pion decay constant reads
\begin{equation}
f_\pi=M g_\pi Z_p(q^2=-m_\pi^2)\,,
\label{fpi-final}
\end{equation}
where the regularized function $Z_p(q^2)$ is defined in (\ref{Zpq2}).
In chiral limit the pion coupling constant (\ref{couplconst})
is given by
\begin{equation}
g^2_\pi={1\over Z_p(0)}\,.
\label{gpi-chiral-limit}
\end{equation}
Inserting this result in (\ref{fpi-final})  one recovers the
Goldberger-Treiman relation on the quark level:
\begin{equation}
f_\pi g_\pi=M\,.
\label{GTR}
\end{equation}
In chiral limit using the above relation
and the explicit expression for $Z_p(0)$ (\ref{Zpq2}) we get for the pion
decay constant a simple expression
\begin{equation}
f_\pi^2=M^2 {N_c \over 4\pi^2}
\int\limits_{\Lambda^{-2}}^\infty {\mbox{d} u\over u}\mbox{e}^{-uM^2}\,.
\label{fpi-chiral-limit}
\end{equation}

\newpage
2.5 \underline{Fixing of the parameters in the vacuum sector }
\vskip4mm

The SU(2) NJL model, treated in the leading  order in $N_c$ (no meson loops),
contains three free parameters. Two of them, the coupling constant
$G$ and the quark current mass $m_0$, are presented in the NJL
lagrangian (\ref{njlsu2}). The third one, the cutoff  $\Lambda$ is
needed to make the theory finite. The parameters $m_0$ and $\Lambda$
are fixed to reproduce the physical pion mass $m_\pi=139$ MeV and the
physical pion decay constant $f_\pi=93$ MeV. As usual, using the gap
equation (\ref{gap}), the last model parameter, the coupling constant
$G$, is eliminated in favor of the constituent quark mass $M$.

Thus, starting with a particular value for $M$ one fixes the cutoff
$\Lambda$ reproducing the physical value of the pion decay constant
(\ref{fpi-final}). For the pion mass we use the equation obtained from
(\ref{Kphi}) and (\ref{mesonmass}) and also the gap equation
(\ref{gap}):
\begin{equation}
m_\pi^2 = { m_0 \over G M Z_p(q^2=-m_\pi^2)}={ m_0 <\bar \Psi\Psi>
\over f_\pi^2}+O(m_0^2) \,,
\label{2004c}
\end{equation}
where the quark condensate is given by (\ref{qqI1}).
In fact, we recover the
Gell-Mann -- Oakes -- Renner (GMOR) relation:
\begin{equation}
m_\pi^2 f_\pi^2=-m_0< \bar \Psi \Psi >\,.
\label{GMOR}
\end{equation}
Now, we are left with only one free parameter, the constituent quark
mass $M$. In principle, the latter can be related to the phenomenological
value of the quark condensate $\langle \bar \Psi \Psi \rangle$. We remind 
that we work in an effective non-renormalizable theory with a finite  cutoff, 
so that the quantities which in QCD depend on the normalization point
should be compared to the particular values of these quantities in our model 
at the scale of order of the model cutoff. Analysis of 
(Gasser and Leutwyler, 1982) shows 
$\langle \bar \Psi \Psi \rangle=(-280\pm 30\mbox{MeV})^3$ at
$\overline{\mbox{MS}}$ renormalization scale $\mu=1$ GeV. 
(Gasser and Leutwyler, 1982).

\begin{figure}
\centerline{\epsfysize=2.55in\epsffile{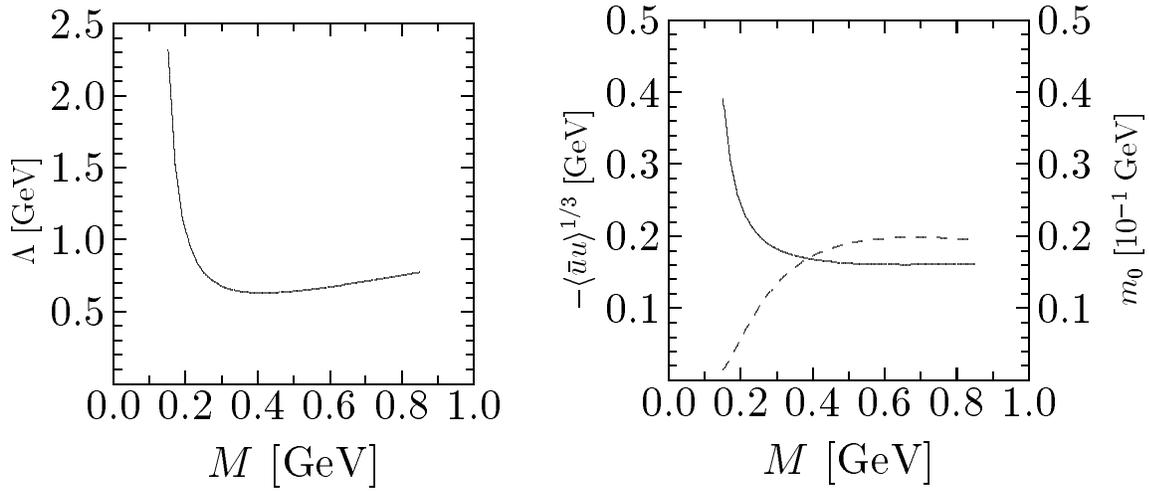}}%\vskip4pt
\caption{Cutoff $\Lambda$ on the left and the quark
condensate $\langle\bar uu\rangle$ (solid line) 
and current quark mass $m_0$ (dashed line) on the right in the
case of the proper-time regularization as a function of the
constituent quark mass $M$. }
\label{f22}
\end{figure}

The dependence of $\Lambda$ on $M$ can be seen on fig.\ref{f22}.
The numerical value of the cutoff $\Lambda$ depends noticeably on
the regularization scheme chosen.
The results for the quark condensate $\langle\bar uu\rangle$
and the current quark mass $m_0$ can be also seen on the right side of 
fig.\ref{f22} as a function of $M$. For both observables there is a plateau 
for $M$ larger than 350 MeV.  

\newpage
%\vskip1cm
2.6 \underline{Restriction to Goldstone degrees of freedom: The chiral circle}
\vskip4mm

Up to now we have considered the meson sector of the model bosonized in 
terms of independent fields $\sigma$ and $\vec\pi$ (linear NJL model). 
However, our main aim is to
apply the model to baryons. Although it is successful in describing
the meson sector (for review see Klevansky 1992, Vogl and Weise, 1991,
Hatsuda and Kunihiro, 1994), the version with independent $\sigma$ and 
$\vec\pi$ does not provide a soliton solution with non-zero baryon number and 
therefore cannot be applied to baryons. In fact, it has been shown
(Sieber {\it et al.}, 1992, Watabe and Toki, 1992) that this solution 
collapses to an infinitely bound point-like solution in
the linear NJL-model . Apparently,
the success of the NJL model in the meson sector shows that the model
incorporates some essential physics
and should be not discarded but rather modified. In order to understand
which ingredient of the NJL model should be revised let us recall that
we are looking for a model describing the low-energy baryon properties.
The particle spectrum of the NJL model with independent fields $\sigma$,
$\vec\pi$ includes quarks, pions and sigma mesons,
and the sigma meson appears to be much heavier than the others.
Furthermore, it is precisely the sigma field which drives the collapse of
the soliton in the linear NJL model. Thus, we argue that the NJL model
in its linear formulation does not correctly describe the degrees of
freedom related to the sigma meson. Since we are
interested in the low-energy properties of pions and baryons
this degree of freedom seems to be not relevant. Therefore, a natural 
solution  to this problem is simply to freeze the sigma-degree
of freedom. This will be done by invoking the chiral circle condition. The
corresponding model is called non-linear NJL model which applied to
the baryon sector is equivalent to the Chiral quark soliton model of
Diakonov {\it et al.,}  (1988).

In order to separate the degrees of freedom corresponding to pions
from the sigma meson degrees of freedom, let us note that
in chiral limit we have a whole family of vacuum solutions
connected by the chiral transformations (\ref{sigma-pi-chiral})
\begin{equation}
\sigma + i \vec\pi\cdot\vec\tau = M U\,,
\label{u}
\end{equation}
where $U$ is a constant $SU(2)$ flavor matrix,
and the meson fields fulfill the chiral circle constraint 
(\ref{chcircle})
\begin{equation}
\sigma^2+\piv^2=M^2\,.
\end{equation}
The pions are massless Goldstone particles and in order to
construct the effective theory of soft pions it is sufficient to consider only
slowly changing $SU(2)$ fields $U(x)$ of the form (\ref{u}).
With meson fields of the form (\ref{u}) the corresponding semibosonized 
lagrangian (\ref{e213}) can be written in the form
\begin{equation}
L=\Psi^\dagger[\partial_\tau + h(U)] \Psi\,,
\label{U-lagrangian}
\end{equation}
where the current quark mass $m_0$ is included in the
one-particle hamiltonian
\begin{equation}
h(U) = - i\gamma^0\gamma^k \nabla_k+M\gamma^0 U^{\gamma_5}+m_0\gamma^0\,.
\label{hU}
\end{equation}
In fact, this treatment of $m_0$ corresponds to a redefinition of
auxiliary scalar field $\sigma$ used in the identity (\ref{e210}).
Here we use notation
\begin{equation}
U^{\gamma_5} = \frac{1+\gamma_5}2 U
+ \frac{1-\gamma_5}2 U^\dagger\,.
\label{u1}
\end{equation}

Integrating out the quarks we arrive at the bosonized non-local
effective action
\begin{equation}
S(U)=-N_c\Tr\log D(U)\,,
\label{SeffU}
\end{equation}
where
\begin{equation}
D(U)= \partial_\tau + h(U)\,.
\label{D-U-definition}
\end{equation}

From now on we will work with this effective action,
i.e. with the non-linear version of NJL model.

Although, as we will see, the solutions of the non-linear 
and linear version of the NJL model are qualitatively different in the 
soliton sector, they 
are equivalent in the meson sector as far as soft pions are concerned. This
has the important consequence that we do not have to change the way in which
the parameters of the model are fixed.
The pion decay constant is given by the same expression
(\ref{fpi-chiral-limit}) and the GMOR relation (\ref{GMOR}) also
remains. In fact, the GMOR formula is valid with accuracy $O(m_0^2)$ in
any model where chiral symmetry is spontaneously  broken.

As explained above, the lagrangian (\ref{U-lagrangian})
is suitable for the description of soft pions. In the case of baryons, however,
the pions are no longer ``soft'', although their momenta are low (of the order 
of the inverse baryon size) compared to the baryon mass. This leads us to ask:
can one use the lagrangian (\ref{U-lagrangian}) for the description of 
baryons? Apparently the answer to this question goes beyond the
model itself. It depends on the dynamics of the underlying theory,
which in our case is QCD, related to the low-energy baryon properties.
The physical aspect of the problem is to what extent the chiral
lagrangian (\ref{U-lagrangian}) can mimic the "exact" effective
low-energy lagrangian of QCD. Indeed, we understand
that the NJL model is only a model and that the "true"
effective lagrangian of QCD has a much more complicated structure
including non-local multi-fermion interactions.
Various attempts (McKay and Munczek, 1985, Cahill and Roberts, 
1985, Diakonov and Petrov, 1986, Ball, 1989, 
Schaden {\it et al.,} 1990), have been made to "derive" the lagrangian 
(\ref{U-lagrangian}) from QCD.
In the instanton liquid model (Diakonov and Petrov, 1984, 1986)
it has been shown that indeed the lagrangian (\ref{U-lagrangian}) can be used 
to describe the pion dynamics relevant for the low-energy baryon structure.

The instanton liquid model assumes that the QCD vacuum is described by a
gluon field configuration corresponding to interacting instantons and
anti-instantons. The model successfully describes the
spontaneous breakdown of chiral symmetry
which can be understood already on the level
of quarks interacting with instantons before the reformulation
of this interaction in terms of the effective
quark-pion interaction. Due to the spontaneous breakdown of chiral
symmetry quarks acquire a dynamical mass $M(k)$. In contrast to the NJL model,
however, this dynamical quark mass depends on the quark
momentum $k$ and the effective lagrangian describing the
quark interactions induced by instantons is non-local.
Therefore, again in contrast to the NJL model, which due to the local four
fermion interaction is not renormalizable and has to be
regularized, the instanton liquid model is ultraviolet finite.
The typical momentum scale above which the non-locality
of the effective quark action becomes important is
the inverse average size of the instanton $1/\overline\rho$.
In particular, the dynamical quark mass $M(k)$ can be treated as a constant
at momenta $k\ll {1/\overline\rho}$ and vanishes at momenta
$k \gg 1/\overline\rho$.
Thus the scale ${1/\overline\rho}$ effectively plays a role of a cutoff.

Besides the average instanton size $\overline\rho$ there is
another important parameter
of the instanton model, namely the average distance $\overline R$ between
instantons. The instanton medium is rather dilute:
$\overline\rho/\overline R \approx 1/3$, so that one can develop a
systematic expansion in this small parameter. Since the dynamical quark
mass $M(k)$ appears as a collective effect of the whole instanton
medium, it is parametrically small for a
dilute medium. Indeed, a detailed analysis (Diakonov and 
Petrov, 1984) shows that
for $\overline\rho\ll \overline R$
\begin{equation}
M(0)\overline\rho =O\Biggl({\overline\rho^2\over \bar R^2}\Biggr)\ll 1
\label{M-small}
\end{equation}
which means that the dynamical quark mass $M(k)$ is much less than the cutoff
${1/\overline\rho}$.
On the other hand, as mentioned above,
$M(k)$ remains constant as long as $k$ is less than
this cutoff.
Thus the presence of the small parameter $\overline\rho/\overline R$
in the instanton model allows to approximate this non-local effective theory by
a theory with a local (non-renormalizable) lagrangian and with a cutoff 
$\Lambda$ of the order of ${1/\overline\rho}$.

A detailed analysis of the non-local four quark interaction of the
instanton model leads to the conclusion 
that in the case of the small
parameter $\overline\rho/\overline R$, the pion and quark
degrees of freedom can be consistently treated within the local
lagrangian (\ref{U-lagrangian}) up to momenta $k\sim M\ll
{1/\overline\rho}$.
The degrees of freedom, which lead to the violation of
the chiral circle condition, correspond to excitations with
momentum $k$ of order of ${\overline\rho\,}^{-1}$.
Indeed, the specific characteristics of the pion and quark degrees of
freedom are that the pions are light Goldstone
bosons and the quarks have a parametrically small dynamical mass $M$.

It is important to note that in the lagrangian (\ref{U-lagrangian})
and the effective action (\ref{SeffU}), the only interactions are those of
quarks with Goldstone degrees of freedom. These are pions in SU(2) and in
addition, kaons and $\eta$-mesons in SU(3). Furthermore, there is no need to
include heavier mesons like e.g. vector mesons $\rho$, $\omega$ and $A_1$ in
the lagrangian (\ref{U-lagrangian}) on the chiral circle. 

Altogether, the instanton model not only naturally leads to the lagrangian
(\ref{U-lagrangian}) with meson fields restricted to the chiral circle, it
also determines the region of the applicability of this lagrangian.
In particular, it can be used to 
describe the baryons since the characteristic pion momenta are low compared 
to the cutoff.

\vskip1cm
2.7 \underline{Generalization to higher flavor groups and relation to the
topological approach}
\vskip4mm

The NJL lagrangian (\ref{njlsu2}) has a structure which suggests a trivial 
generalization to three and more flavors. In such a case, however, the 
resulting lagrangian is invariant under $\mbox{U}(3)_L\otimes\mbox{U}(3)_R$ 
and some terms have to be added in order to break the $U_A(1)$ symmetry 
(Vogl and Weise, 1991, Klevansky, 1992, Hatsuda and Kunihiro, 1994). 
On the other hand, if one works 
in the instanton liquid model (Diakonov and Petrov, 1986) one arrives at a 
SU(3) effective quark lagrangian of 't Hooft-like form
which breaks the 
$\mbox{U}_A(1)$ symmetry explicitly. Further, after bosonization of this 't 
Hooft-like terms one ends up after a couple of approximations with a 
lagrangian describing the interaction of quarks with SU(3) Goldstone bosons. 
This lagrangian has then exactly the structure (\ref{U-lagrangian}) after 
simply replacing the SU(2) matrix $U$ by a SU(3) one in it. In this way, the 
lagrangian (\ref{U-lagrangian}) can be immediately generalized to any number 
of flavors. In fact, in section 5 we will eventually apply it to hyperons.

Although the present model with the meson fields restricted to the
$SU(N_f)$ has the same meson degrees of freedom as the
Skyrme model, in contrast to it our bosonized action (\ref{SeffU}) is a
complicated non-local functional. For slowly changing fields $U(x)$ we can
perform a gradient expansion for the action (\ref{SeffU}). From the
regularized real part one obtains
(in chiral limit $m_0=0$)
\begin{equation}
\Re S(U) = \frac{f_\pi^2}4 \int d^4 x
\Sp (\partial_\mu U)(\partial_\mu U^\dagger) - {N_c\over192 \pi^2}
\int\d^4x\bigl[2\Sp(\partial_\mu L_\mu)^2 +\Sp(L_\mu L_\nu L_\mu
L_\nu)\bigr]+\ldots \,,
\end{equation}
where $L_\mu=iU^\dagger\partial_\mu U$ and the ellipsis stands for the
terms with higher derivatives. One recognizes the kinetic term and the four 
derivative terms similar to those of the Skyrme lagrangian although the latter 
differ in signs and coefficients.

The analogy with the Skyrme model gets closer if one considers the
imaginary part of the effective action. As we already stated before,
the imaginary part of the effective action exactly vanishes in the $SU(2)$
case. However, for higher flavor groups $SU(N_f)$ the effective action
$S(U)$ is generally speaking complex. In contrast to the real part,
the imaginary part of the effective action is free of the ultraviolet
divergences.
A straightforward way to show this is to perform a gradient expansion
for this imaginary part. In the leading order of the gradient expansion
one finds (Dhar and Wadia, 1984, Dhar {\it et al.,} 1985)
the famous Wess-Zumino term (Wess and Zumino, 1971, Wess, 1972)
\begin{equation}
\Im S(U) \Big\vert_{grad} =-{{i N_c}\over {240 \pi^2}} \int
d^5 x \epsilon_{\mu_1\mu_2\mu_3\mu_4\mu_5}
\Sp \left [\left ( {U^{\dag}} \partial_{\mu_1} U \right )
\left ( {U^{\dag}} \partial_{\mu_2} U \right )
\left ( {U^{\dag}} \partial_{\mu_3} U \right )
\left ( {U^{\dag}} \partial_{\mu_4} U \right )
\left ( {U^{\dag}} \partial_{\mu_5} U \right )
\right ]\,.
\label{e2wzw}
\end{equation}

Note that for time-independent  fields $U(x)$ the imaginary
part $\Im \log \mbox{Det} D(U)$
vanishes, because the one-particle hamiltonian $h(U)$
is hermitian. The same is also valid for the Wess-Zumino term.

In the Skyrme model the Wess-Zumino term generates the baryon current
\begin{equation}
 B^\mu (x)  =- {{1} \over 24\pi^2} \epsilon^{\mu\nu\alpha\beta}
\Sp \( U^\dagger \partial_\nu U U^\dagger \partial_\alpha U
U^\dagger \partial_\beta U \)\,.      \label{bartopo}
\end{equation}
The corresponding baryon charge
\begin{equation}
B = \int d^3x B^0(x)=Q_T
\label{bartopo1}
\end{equation}
is exactly the winding number $Q_T$ of the chiral field $U(x)$:
\begin{equation}
Q_T =- {{1} \over 24\pi^2} \epsilon^{ijk} \int d^3x
\Sp \( U^\dagger \partial_i U U^\dagger \partial_j U
U^\dagger \partial_k U \)\,.
\label{windnumber}
\end{equation}

Thus, we see that using the gradient expansion one can trace
a rather interesting relation between the model with the non-local
action (\ref{SeffU}) and a local Skyrme-type lagrangian with the 
Wess-Zumino term.
However, although this relation is rather attractive from the mathematical
point of view, we have the problems of the higher order terms as well
as with the validity of the gradient expansion itself.
We remind that the above gradient expansions of both the real and
the imaginary parts of the effective action are valid only for slowly
changing fields $U(x)$. However, as we shall see in the next part, in
the case of the nucleon the corresponding saddle-point field $U(x)$
varies rather fast, so that the gradient expansion generally
cannot be used. Therefore, in the present model, the
relation between the baryon and topological charges used in the Skyrme
model is not valid.

\newpage
%\input review3
%%%%%%%%%%%%%%%%%%%%%%%%%%%%%%%%%%%%%%%%%%%%%%%%%%%%%%%%%%%%%%%%%%%%%%%%%%%%%
%
%        review: chapter 3
%
%%%%%%%%%%%%%%%%%%%%%%%%%%%%%%%%%%%%%%%%%%%%%%%%%%%%%%%%%%%%%%%%%%%%%%%%%%%%%%
%
%
%       review3.tex
%
%
%%%%%%%%%%%%%%%%%%%%%%%%%%%%%%%%%%%%%%%%%%%%%%%%%%%%%%%%%%%%%%%%%%%%%%%%%%%%%%%
%
%
\section{Nucleon as a non-topological soliton in SU(2) NJL model}

This chapter is devoted to the description of the nucleon in the NJL
model. Using the functional integral approach we show that in the large
$N_c$ approximation the nucleon naturally appears as a many-body bound
state of quarks in the selfconsistent mean chiral field $U$.
The spectrum of the Dirac one-particle hamiltonian with this
field $U$ contains a discrete valence level.
The nucleon state corresponds to the Hartree picture where
all negative Dirac sea levels and the valence level are occupied
by quarks. In fact, a similar physical picture of the
nucleon has been first considered by
(Kahana {\it et al.}, 1984) in the context of the sigma model.

Although the mean chiral field $U$ resembles to some extent the
soliton of the Skyrme model, the important difference
is that, whereas in the Skyrme model the baryon number of the nucleon
is identified with the winding number of the chiral mean field $U$,
in the NJL model the nucleon baryon number $B=1$ is explicitly given by
the presence of the valence level occupied by $N_c$ quarks.
In order to distinguish the NJL mean-field solution
from the chiral soliton of the Skyrme model, it is commonly called a
non-topological soliton.

Depending on constituent mass $M$ in the NJL model we have two different and 
to some extent controversial physical pictures of the nucleon:
\begin{itemize}
\item At $M \approx 400$ MeV a clear valence picture prevails. The valence 
quarks occupy a bound state in the gap between the positive and negative 
continuum and its single-particle energy is around 200 MeV.
The number of quarks in this valence level determines the baryon number and
the polarization of the Dirac sea plays the role of the pion cloud.
The mean field is noticeably varying over the size of the soliton and the 
gradient expansion of the action does not converge.
\item At $M > 1000$ MeV the valence level has come asymptotically close to the
negative continuum and its  single-particle energy is slightly above $-M$. 
Still the number of quarks in this valence level determines the baryon number 
and the polarization of the Dirac sea plays the role of the pion cloud.
However, the gradient expansion in this situation works well, and
the non-local effective action of the NJL model can be well approximated by a 
local one. Hence, a Skyrme-type picture emerges in which the winding number of the meson field equals the baryon number of the system. 
\item Obviously, since $M$ is a free model parameter, only the agreement
with the experiment can give a preference to one of these physically different
pictures. In fact, all
calculations within the present model, which reproduce the experimental data, 
support the picture of valence quarks with polarized Dirac sea and not the 
Skyrme picture.
\end{itemize}

Actually, there has been a debate in the literature whether the 
valence quark picture is still valid if vector mesons are explicitly taken 
into account in the NJL lagrangian. In fact, several theoretical and numerical 
investigations have been performed to clarify the role of $\rho$, $\omega$ and 
$A_1$ mesons in the solitonic sector (Alkofer and Reinhardt, 1990, Alkofer 
{\em et al.,} 1992, D\"oring {\em et al.,} 1992b, 1993, 1995, Sch\"uren 
{\em et al.,} 1992, 1993, Z\"uckert {\em et al.,} 1994, Weigel {\em et al.,} 
1995). Unfortunately, no 
clear conclusion on the formalism and on the effects of the vector mesons on 
the nucleon observables has been obtained and none of the above approaches is beyond criticism. In the present paper, we do not 
consider vector quark couplings or vector meson fields at all. 

The picture of bound valence quarks interacting with a polarized Dirac sea 
bears some similarity to the idea of chiral quark models where the Dirac sea 
is replaced by the dynamical meson clouds and the quarks are confined by a bag
(Theberge {\it et al.}, 1980, Thomas, 1993). In the chiral sigma model one does
not have confining walls but the system is bound by sigma and pion fields  
respecting the spontaneous breaking of chiral symmetry
(Gell-Mann and L\'evy, 1960, Birse and Banerjee, 1984, 1985, 
Birse 1990). This model allows to use the Peierls-Yoccoz projection 
method for the quantization of collective degrees of freedom (Birse, 1986, 
Fiolhais {\em et al.,} 1988, Neuber {\em et al.,} 1993). There are also
hybrid models which combine valence quarks with meson clouds. This 
is e.g. the chiral bag model of Brown and Rho (1979). In the model, the pion 
field outside the bag carries a fractional baryon number, and it is connected 
to the interior by the axial current. The linear chiral sigma model of Kahana 
and Ripka (1984) involves both sea quarks and dynamical pion and sigma fields, 
and is renormalizable. In fact, these authors were the first to treat
the polarization of the Dirac sea in the context of the valence picture for the
baryon (Kahana and Ripka, 1984, Soni, 1987, Li {\em et al.,} 1988). The model
shows, however, a vacuum instability and hence has never been applied in a
selfconsistent way to baryons.

One should also mention some papers in which the nucleon is not 
described as a soliton but as a three-body bound state as a solution of
Faddeev-type equations (Johii {\em et al.,} 1993, Huang and Tjon, 
1994, Buck {\em et al.,} 1993, Weiss {\em et al.,} 1993b, Hellstern and 
Weiss, 1995).

In order to describe the nucleon in the NJL model, we consider the nucleon
correlation
function at large euclidean time separation. First, we concentrate on the
solution in strict large $N_c$ limit - the ``classical'' soliton solution. We
discuss the classical part of the soliton mass and the baryon number. The
selfconsistent hedgehog soliton is presented and the results are discussed.
As a second step, the quantization of the hedgehog soliton is done.
The rotational zero modes are considered and the related moment of inertia and
the energy of the quantized soliton are given. The translational zero modes 
are also shortly discussed. In the end, we present the evaluation of the 
nucleon matrix element of a quark current in this quantization scheme.

\vskip1cm
3.1 \underline{Nucleon correlation function}
\vskip4mm

The correlation functions are a general tool to study the structure of any kind
of matter and in particular, they provide a powerful method to study the
structure of the vacuum and the hadronic spectrum in the QCD.
The main idea behind this is that the correlation function
of a given current $\hat O$
\begin{equation}
\Pi(t)=<0|\hat O(t) \hat O^\dagger(0)|0>
\label{Pidef}
\end{equation}
can be written in terms of intermediate physical states
\begin{equation}
\Pi(t)=\sum_n|<0|\hat O|n>|^2 \e^{-iE_nt}\,,
\label{Pi}
\end{equation}
which in euclidean space-time ($\tau=it$) becomes a sum of decreasing
exponents. Hence, from the behavior of the correlator at very large euclidean
separation $\tau$,
$$\Pi(\tau)\sim\exp(-m\tau)\,,$$
one is able to extract the energy (mass) $m$ of the lowest state with
quantum numbers of the current $\hat O$.

In the case of the nucleon, the correlation function is given by
\begin{equation}
\Pi_N(T)=<0|J_N(0,T/2)J_N^\dagger(0,-T/2)|0>\,,
\label{PiN}
\end{equation}
where the $N_c$-quark current $J_N$ carries the nucleon quantum numbers
$J J_3, T T_3$ and no color (Ioffe, 1981):
\begin{equation}
J_N(\vec x,t) = \frac{1}{N_c !} \
\varepsilon^{\beta_1 \cdots \beta_{N_c}} \
\Gamma^{ \{ f\} }_{J J_3, T T_3} \
\Psi_{\beta_1 f_1}(\vec x,t)\cdots\Psi_{\beta_{N_c} f_{N_c}}(\vec x,t)\,.
\label{jn}
\end{equation}
Here $\beta_i$ are color indices and $\Gamma^{f_1\cdots f_{N_c}}_{
JJ_3,TT_3}$ is a matrix with $f_i$ standing for both flavor and
spin indices. $J$ and $T$ denote the nucleon spin and isospin,
respectively.

The asymptotics of the correlation
functions at large euclidean time separations determines the nucleon mass $M_N$
\begin{equation}
\lim_{T \to+\infty }\Pi_N(T)\sim\e^{-M_N T}\,.
\label{PiNasym}
\end{equation}

The nucleon correlation function can be written as a path integral
with the action (\ref{SeffU})
\begin{equation}
\Pi_N(T)=<0|J_N(0,T/2)J_N^\dagger(0,-T/2)|0>={1\over Z}\int \D \Psi^\dagger
\D \Psi \D U  J_N(0,T/2)J_N^\dagger(0,-T/2)
\e^{-\int {\rm d}^4 x L   }\,,
\label{PiNPI}
\end{equation}

The effective lagrangian (\ref{U-lagrangian}) is
quadratic in the quark fields and a gaussian integration can be
performed. In this we apply the Wick theorem
contracting $\Psi$ and $\Psi^\dagger$ from the currents
which leads to a product of $N_c$ quark propagator (diagonal in color
indices) in the background of meson field $U$:
\begin{equation}
\Pi_N(T)= \Gamma^{\{f\}}_{N}\Gamma^{\{g\}*}_N \frac 1Z
\int\D U\prod_{i=1}^{N_c}\langle 0,T/2 |
\frac{1}{D(U)} |0,-T/2 \rangle_{f_i g_i}\e^{-S(U)}\,.
\label{PiNPIU}
\end{equation}

\vskip1cm
3.2 \underline{``Classical'' soliton}
\vskip4mm

In order to integrate over the chiral field $U$,
as in chapter 2, we use the large $N_c$
argumentation and perform the integration in the
saddle-point approximation which is exact in large $N_c$ limit.
In this we make use of the fact that in the limit of large time $T$ the
saddle-point field configuration is time independent which allows us to
consider only static fields $U$ below. Here we concentrate on the properties of
the ``classical'' soliton which is
given by this large $N_c$ saddle-point solution.

\vskip1cm
\underline{Saddle-point solution and the nucleon mass in the leading order of 
$N_c$.}
A specific feature of the nucleon case is that the functional integral
contains now not only the action exponential  $\e^{-S(U)}$
with $S(U)=O(N_c)$ but also a product of $N_c$ quark propagators
which also depend on the chiral field $U$. Thus, in the leading order
of $N_c$ these quark propagators also contribute to the saddle-point
solution.

We start with the pre-exponential factor in eq.(\ref{PiNPIU}). For the quark
propagator in the case of the static field $U$ we use its spectral 
representation
\begin{eqnarray}
\langle {\vec x}^\prime ,x_4^\prime | \frac{1}{D(U)}
|{\vec x},x_4 \rangle&=&
\theta(x_4^\prime-x_4) \sum\limits_{\epsilon_n>0}
\e^{-\epsilon_n(x_4^\prime-x_4)}\,\Phi_n({\vec x}^\prime)  \,
\Phi_n^\dagger(\vec x) \nonumber\\
&&- \theta(x_4-x_4^\prime) \sum\limits_{\epsilon_n<0}
\e^{-\epsilon_n(x_4^\prime-x_4)}\,\Phi_n({\vec x}^\prime)  \,
\Phi_n^\dagger(\vec x)
\label{spectral-representation}
\end{eqnarray}
where $\Phi_n$ and $\epsilon_n$ are the eigenfunctions and eigenvalues of
the one-particle hamiltonian $h(U)$ (\ref{hU})
in the static background field $U$
\begin{equation}
h(U)\Phi_n=\epsilon_n\Phi_n\,.
\label{Diracequation}
\end{equation}
In the case $x_4^\prime-x_4=T>0$, we have
\begin{equation}
\langle 0,T/2 | \frac{1}{D(U)} |0,-T/2 \rangle=\sum\limits_{\epsilon_n>0}
\e^{-\epsilon_n T}\,\Phi_n({\vec 0})  \,
\Phi_n^\dagger(\vec 0)
\label{spectral-representation1}
\end{equation}
where the sum goes over all levels of positive energy.
At $T\to \infty$ only the level with the lowest positive energy will
survive in the sum and hence, for the product of $N_c$ quark propagators
we have
\begin{equation}
\prod_{i=1}^{N_c}\langle 0,T/2 | \frac{1}{D(U)} |0,-T/2
\rangle\mathop{\sim}_{T\to\infty} \e^{-TN_c\epsilon_{val}(U)}\,.
\label{eval}
\end{equation}
Here, we refer to this lowest energy level as
valence level $\epsilon_{val}$.

Now, we turn to the fermion determinant. Although this determinant contains
ultraviolet divergences, we start with the non-regularized version,
because it will clarify the physics behind the saddle-point
approximation. For a static field $U$ following (\ref{SeffU}) we can write 
\begin{eqnarray}
&&S(U) - S(U_0)
=-N_c\Tr\log [D(U)/D(U_0)]\nonumber\\
&&=-N_c\Tr[\log(\partial_\tau+h)-\log(\partial_\tau+h_0)]
=-N_c T\int\limits_{-\infty}^{+\infty}{\mbox{d}\omega\over 2\pi}\int \d^3 x\,
\Sp\bra x\log(i\omega+h)-\log(i\omega+h_0)\ket x\nonumber\\
&&=N_c T
\int_{-\infty}^{+\infty}{\mbox{d}\omega\over 2\pi}\int \d^3 x\,\Sp \bra x
{i\omega\over i\omega+h}-{i\omega\over i\omega+h_0}\ket x=T N_c
\sum_{{\epsilon_\alpha<0}\atop {\epsilon_\alpha^0<0}}
(\epsilon_\alpha-\epsilon_\alpha^0)\equiv T E_{sea}(U)\,.
\label{Diracseaenergy}
\end{eqnarray}
Here $D(U_0)$ is the Dirac operator corresponding to the vacuum
saddle point solution $U_0\equiv1$.
The vacuum part $N_c\Tr\log D(U_0)$ appears due to the normalization
$1/Z$ factor in (\ref{PiNPIU}). $h_0$ is the Dirac hamiltonian corresponding
to the vacuum solution. In (\ref{Diracseaenergy})
we used that $\Sp(h-h_0)=0$ and integrated by parts over $\omega$. The energies
$\epsilon_\alpha (\epsilon_\alpha^0)$ are the eigenvalues of $h$ ($h_0$).

We see that $S(U)$ is proportional to the sum of all
negative energies, i.e. to the energy of the Dirac sea
from which the analogous vacuum contribution is subtracted.

Combining (\ref{eval}) and (\ref{Diracseaenergy}) we see that
\begin{equation}
\frac 1Z
\prod_{i=1}^{N_c}\langle 0,T/2 | \frac{1}{D(U)} |0,-T/2
\rangle \e^{-S(U)}
\mathop{\sim}_{T\to\infty} \e^{-[N_c\epsilon_{val}(U)+E_{sea}(U)]T}\,.
\end{equation}
Now the functional integral over $U$ in (\ref{PiNPIU}) can be performed
in the saddle point approximation justified in the large $N_c$ limit.
The saddle-point meson configuration $U_c$ can be found from the 
stationary condition
\begin{equation}
\delta_{U} [N_c\epsilon_{val}(U)+E_{sea}(U)]\Big\vert_{U_c}=0\,
\label{statcond}
\end{equation}
and the large $T$ asymptotics of $\Pi_N(T)$ in the saddle-point
approximation is
\begin{equation}
\Pi_N(T)\mathop{\sim}_{T\to\infty}\e^{-[N_c\epsilon_{val}(U_c)+
E_{sea}(U_c)]T}\,.
\label{PINsolenergy}
\end{equation}
In contrast to the vacuum case, the stationary meson field configuration $U_c$
is a static localized solution which is commonly called a non-topological
soliton. Comparing (\ref{PINsolenergy})
with (\ref{PiNasym})
we see that in leading order in $N_c$ the nucleon mass is
\begin{equation}
M_{cl} = \min_U[N_c\epsilon_{val}(U)+E_{sea}(U)]
=N_c\epsilon_{val}(U_c)+E_{sea}(U_c)
\label{solenergy}
\end{equation}

This result has a simple physical meaning. In our model the nucleon
appears as a bound state of quarks in a mean field $U_c$
and the energy (mass) of the nucleon state is a sum of
one-particle energies of all occupied states. These states are eigenstates of
the Dirac hamiltonian $h(U_c)$. In order to get a colorless state, each level
is occupied by $N_c$ quarks. Therefore, the contribution of each level to
the total energy of the nucleon (\ref{solenergy}) contains a common factor
$N_c$.

Thus, we see that in the path integral approach, the saddle-point approximation
for the correlation function of nucleon currents
agrees with the Hartree picture of the nucleon as a bound state
of quarks in the selfconsistent mean field $U_c$.

In the above euclidean treatment, the lowest positive level in the spectral
representation (\ref{spectral-representation1}) plays a privileged
role and we called it a valence level. Indeed, as we shall see later,
in the case of physically relevant parameter values in the
NJL model the valence energy is positive,
so that the euclidean calculation
with $E_{sea}$ including all negative levels and with
$\epsilon_{val}$ understood as the lowest positive level
gives correct result. However, let us assume that under
some change of the parameters of the model
the valence level crosses the zero energy and becomes
negative. Formally, in the euclidean treatment we face a problem, since in this
case the large time asymptotics of the euclidean quark propagator
(\ref{eval}) would be dominated by another lowest positive level. On the
other hand, from the physical point of view,
there is nothing special about crossing zero and the nucleon
still remains the bound state of the Dirac sea and valence quarks
although the valence level is negative. Apparently not the
sign of the valence level but the way in which the levels are occupied
is important. In order to overcome this problem, we can generalize the
euclidean treatment identifying the valence level with the bound
occupied level which does not belong to the polarized Dirac sea,
i.e. which originates from the upper positive continuum.
In principle, the valence level can have any sign of
the energy. Similarly,  the Dirac sea energy $E_{sea}$
should be understood as the total energy of all
levels which belong to the polarized Dirac sea.
Actually, the case in which the valence level crosses zero and becomes
negative is important for understanding the relation between the
current approach and the Skyrme-like models.

\vskip1cm
3.3 \underline{Baryon number of the soliton}
\vskip4mm

Now we turn to the baryon number of the soliton.
We remind that starting from the correlation function
of two $N_c$ quark currents $J_N$ (\ref{jn}), simply by
construction, the state with baryon number $B=1$ is created.
The saddle-point treatment described above does not destroy this feature.
It is easy to see that the physical picture associated with the
saddle-point (large $N_c$)
approximation really agrees with $B=1$. The nucleon in the model appears as a
bound state of the Dirac sea and valence quarks. Since the sea quarks
originate from the lower Dirac continuum, they carry the same
baryon number $B=0$ as the vacuum Dirac sea. Hence,
the baryon number $B=1$ is given by the $N_c$ valence quarks,
each of them carrying the baryon charge $1/N_c$, i.e. due to the
presence of the valence quarks the nucleon has a
baryon number $B=1$. Explicit proof will be given later when we will
evaluate the nucleon isoscalar electric form factor.

We conclude that the quark configuration determines the baryon number of the
NJL soliton and it is carried by the valence quarks. In particular, in order
to consider solutions with a larger, for instance $B=2$,
baryon number one should consider a correlation function of two
currents made of $2N_c$ quarks. It corresponds to a
quark configuration in which besides the filled Dirac sea the first two
levels (coming from upper continuum) are additionally occupied.

Apparently, this physical picture differs from the one of the Skyrme model
where the baryon charge is identified with the winding number of the meson
field and
the field configurations with different
winding numbers are isolated from each other from the very beginning.
The effective action of the NJL model is non-local
and the transitions between different topological sectors
are allowed.
Therefore, in the NJL model there is no {\em a priori} connection between
the winding number and the baryon charge. On the other hand, in section 2
we stated that in the leading order of the gradient expansion the
imaginary part of the effective action of the NJL
model coincides with the Wess-Zumino term which generates the
topological current in the Skyrme model. As we will see later, indeed
the baryon charge, which we discussed above in terms of occupied levels,
originates from the imaginary part of the effective action.
Hence, one expect that if the gradient expansion
is valid the baryon charge $B$ could be related to the winding number $Q_T$.
In order to examine this connection let us consider the following
quantity
\begin{equation}
\tilde B(U)
={1\over N_c}\int \d^3 x \Bigl\{ <\Psi^\dagger \Psi>_{U}
-<\Psi^\dagger \Psi>_{U_0}\Bigr\}\,,
\label{barnumsol}
\end{equation}
where the brackets mean the euclidean functional integral over the
fermion field in a classical background chiral field $U$:
\begin{equation}
<\Psi^\dagger \Psi>_{U}=\frac{\int D\Psi D\Psi^\dagger \Psi^\dagger \Psi
\exp\left\{-\int d^4 z\Psi^\dagger[\partial_t + h(U)] \Psi \right\}}
{\int D\Psi D\Psi^\dagger\exp\left\{-\int d^4 z\Psi^\dagger[\partial_t
+ h(U)] \Psi \right\}}\,.
\end{equation}
Formally $\tilde B(U)$ looks like the baryon charge.
However, the baryon
charge of the soliton depends not on the mean field
$U_c$ itself but on the way in which the quark levels of the hamiltonian
$h(U_c)$ are occupied, whereas the quantity $\tilde B(U)$
depends only on the background field $U$.
%and does not know what quark levels are occupied in the physical
%soliton state
Indeed, for the time-independent field $U$
\begin{eqnarray}
\tilde B(U)
&=&-\int\d^3 x\Sp<x|{1\over\partial_\tau+h(U)}
-{1\over \partial_\tau+h(U_0)}|x>
\nonumber\\
&=& -
\int_{-\infty}^{+\infty}{\mbox{d}\omega\over 2\pi}\int \d^3 x \Sp
\bra x{1\over i\omega+h(U)}-{1\over i\omega+h(U_0)}\ket x=
\Tr\left[\theta[-h(U)]-\theta[-h(U_0)]\right]\,
\label{barnumbersol}
\end{eqnarray}
we see that $\tilde B(U)$ counts the number of {\em negative} states
of the hamiltonian $h(U)$ from which the number of 
negative vacuum quark states is subtracted.
We rewrite (\ref{barnumsol}) in the form
\begin{equation}
\tilde B(U)
=\int_{-\infty}^{+\infty}{\mbox{d}\omega\over 2\pi }\int \d^3 x \Sp
\bra x {h(U)\over\omega^2+h^2(U)}-
{h(U_0)\over\omega^2+h^2(U_0)}\ket x\,,
\label{barnumbersol2}
\end{equation}
where
\begin{equation}
h^2(U)
= -\partial_k^2+M^2+iM\gamma_k\partial_k U^{\gamma_5}\quad \mbox{and}\quad
h^2(U_0)= -\partial_k^2+M^2\,,
\label{hgradexp}
\end{equation}
and perform a local expansion in derivatives $\partial_k U/M$.
The leading term of the gradient expansion
\begin{equation}
\tilde B(U) = - {1 \over 24\pi^2} \epsilon_{ikl}
\int \d^3 x \Sp ( U^\dagger \partial_i U) (U^\dagger \partial_k U)
(U^\dagger \partial_l U )+\cdots\,
\label{B-Q-equal}
\end{equation}
gives exactly the winding number $Q_T(U)$ (\ref{windnumber}) of the field $U$.
The higher order corrections in the gradient
expansion vanish. Indeed, from (\ref{barnumbersol}) follows that
$\tilde B(U)$ is an integer number and any higher order correction would
destroy this feature.
However, the equality (\ref{B-Q-equal}) exists only if the gradient
expansion is valid, i.e. for chiral fields $U(x)$ with
characteristic size $RM \gg 1$. We remind that for such fields
the non-local action can be also approximated by the first terms of the
gradient expansion which are similar to those of the Skyrme lagrangian.
Thus, we see that the NJL model exhibits a certain similarity
with local Skyrme-like lagrangians in the case of solitons
with $MR\gg 1$. In this limit:
\begin{itemize}
\item the baryon charge coincides with the topological charge (if all and only
the negative-energy levels are occupied);
\item the non-local action can be approximated by the first terms of the
gradient expansion which are similar to those of the Skyrme lagrangian.
\end{itemize}

\vskip1cm
\underline{Regularization.}
As we already know from the
vacuum sector the fermion determinant is divergent and must be regularized.
We start from the proper time regularization formula (\ref{e229}).
In the case of static fields $U(\vec x)$ it gives
\begin{equation}
 \half \Tr\,\log\,(D^\dagger D)_{reg}-\half \Tr\,\log\, (D^\dagger_0 D_0)_{reg} =
-\half T
\int_{-\infty}^{\infty} \frac{d\omega}{2\pi}
\int_{1/\Lambda^2}^\infty {\mbox{d}u \over u}
\Tr
\Bigl[\e^{-u(h^2+ \omega^2) } - \e^{-u(h_0^2+ \omega^2) } \Bigr]\,,
\label{D-reg-static}
\end{equation}
where $D_0\equiv D(U_0)$ corresponds to the vacuum.
The regularized analog of the Dirac sea energy $E_{sea}$
(\ref{Diracseaenergy}) is given by
\begin{equation}
E_{sea}^\Lambda(U) =
\frac{N_c}{2} \Tr\,\log\,(D^\dagger D)_{reg}
-\frac{N_c}{2} \Tr\,\log\,(D^\dagger_0 D_0)_{reg}\,.
\end{equation}
We perform the integral over $\omega$
in (\ref{D-reg-static}) and express the trace
through the sum over the eigenstates of the Dirac operators $h(h_0)$.
Finally, we get
\begin{equation}
E_{sea}^\Lambda=N_c\sum_n R^\Lambda_1(\epsilon_n)-N_c\sum_n
R^\Lambda_1(\epsilon_n^v)\,,
\label{regseaenergy}
\end{equation}
where the regularization function $R^\Lambda_1(\epsilon_n)$ is given by
\begin{equation}
R^\Lambda_1(\epsilon_n)={1\over 4\sqrt{\pi}}\int_{\Lambda^{-2}}^\infty
{\d u\over u^{3/2}} \e^{-u\epsilon_n^2}\,.
\label{energyreg}
\end{equation}
In the limit of the large cutoff $\Lambda\to\infty$ the regularized
sea energy $E_{sea}^\Lambda$ reduces to the non-regularized sum over
negative energies (\ref{Diracseaenergy}). If the
valence level is positive its energy should be added to this sum
but if the valence level energy is negative its contribution
is already contained in $E_{sea}^\Lambda$, so that the total regularized
energy of the soliton is
\begin{equation}
M_{cl}= N_c\epsilon_{val} \theta(\epsilon_{val})
+E_{sea}^\Lambda\,.
\label{classolenergy}
\end{equation}

\vskip1cm
3.4 \underline{Stationary hedgehog meson field configuration.}
\vskip4mm

Even in the simplest SU(2) meson field configuration $U$ (\ref{u}),
in which the meson fields are constrained on the chiral circle
(\ref{chcircle}), one needs additional assumptions in order to solve
eq.~(\ref{statcond}) numerically. Usually for the stationary meson field
configuration a hedgehog structure is used
\begin{equation}
U_c(\vec r)= \e^{i \Theta(r) (n^a\tau^a)}\,.
\label{Uhedgehog}
\end{equation}
The hedgehog is the most symmetric meson field configuration in which
\begin{equation}
\pi^a(\vec r)\equiv n^a \pi(r)\,
\label{pihedgehog}
\end{equation}
and the classical
meson fields are parametrized in terms of the profile function of the soliton
$\Theta(r)$ as follows
\begin{equation}
\sigma(r)=M\cos\Theta(r)\quad \mbox{and} \quad
\pi(r)=M\sin\Theta(r)\,.
\label{sigpihedgehog}
\end{equation}
The chiral profile function satisfies the boundary condition
\begin{equation}
\Theta(r)\mathop{\sim}_{r\to \infty} 0\,,
\label{hgboundcond}
\end{equation}
which ensures that the physical vacuum $U=1$ is recovered at
$r\to \infty$. In the Skyrme model, additionally to condition 
(\ref{hgboundcond}), it is imposed that
\begin{equation}
\Theta(r=0)\,=\,-\pi Q_T\,,
\label{hgboundcond0}
\end{equation}
where the number $Q_T$ (winding number) is identified with the baryon
number and as such it should be integer. In the present NJL model the $Q_T$ is
not necessarily an integer number, since due to the non-locality of the
effective action we could have in principle singular saddle-point solutions. In
addition, even an integer $Q_T$ in the NJL model is not necessary identical to
the baryon number.

At first glance the assumption of the hedgehog structure (\ref{Uhedgehog})
appears arbitrarily in the NJL model and, indeed, there is no general proof
that it corresponds to a stationary solution of (\ref{statcond}). However, such
a proof exists in the Gell-Mann
and Levy sigma model with valence quarks (Ruiz Arriola  {\it et al.}, 1989).
Additionally, there is a general argument in favor of the hedgehog choice.
In particular, one expects that the classical solution should have the most
symmetric structure which is the hedgehog one. A deviation from the hedgehog
configuration would lead to a rather complicated spectrum of rotational
excitations of the nucleon
which is difficult to be classified and in fact it is not observed in nature.

For the hedgehog meson fields the stationary condition
(\ref{statcond}) can be written in terms of the profile function
$\Theta(r)$
\begin{equation}
\delta_{\Theta} [N_c\epsilon_{val}(\Theta)+E_{sea}(\Theta)]
\Biggr\vert_{\Theta=\Theta_c}=0\,.
\label{statcondtheta}
\end{equation}
In particular, it leads to the following
classical equations of motion written in terms of $\Theta(r)$
\begin{equation}
\sin \Theta (r) S(\vec r) - \cos \Theta (r) P(\vec r) =0\,,
\label{eqsmotion}
\end{equation}
where the scalar $S(\vec r)$ and pseudoscalar $P(\vec r)$ densities read
\begin{equation}
S(\vec r)=N_cM\left [ \sum_n R_2^\Lambda (\epsilon_n)\bar \Phi_n(\vec r)
\Phi_n(\vec r) + \theta(\epsilon_{val})
\bar\Phi_{val}(\vec r) \Phi_{val}(\vec r) \right ]\,,
\label{scalar-density}
\end{equation}
\begin{equation}
P(\vec r)=N_cM\left [ \sum_n R_2^\Lambda (\epsilon_n)\bar \Phi_n(\vec r)
i\gamma_5 (\tau^a\hat n^a) \Phi_n(\vec r) + \theta(\epsilon_{val})
\bar\Phi_{val}(\vec r)i\gamma_5(\tau^a\hat n^a) \Phi_{val}(\vec r) \right ]\,,
\label{pseudoscalar-density}
\end{equation}
respectively.
The regularization function $R_2^\Lambda (\epsilon_n)$ is given by
(Reinhardt and W\"unsch, 1988,  Meissner Th. {\it et al.}, 1988)
\begin{equation}
R^\Lambda_2(\epsilon_n)={1\over 4\sqrt{\pi}}\int_{\Lambda^{-2}}^\infty
{\d u\over u^{1/2}}\epsilon_n \e^{-u\epsilon_n^2}\,.
\label{eqmotionreg}
\end{equation}

\vskip1cm
\underline{One-particle hamiltonian $h$ and Dirac spectrum.}
Due to the hedgehog ansatz (\ref{Uhedgehog}), the one-particle
hamiltonian $h$ (\ref {hU}) is invariant only with
respect to simultaneous space-isospin rotations. It means that it commutes
with the grand spin $\vec G = \vec j + \vec t$, but not separately with the
total momentum $\vec j=\vec l +\vec s$ and the isospin $\vec t$ of the
one-particle eigenstates of $h$. Since it also
commutes with the parity operator $\Pi$,
the hamiltonian $h$, the grand spin $G$, its $z$-component $G_z$ and
the parity operator $\Pi$ form a complete set of commuting operators.
Therefore, the eigenstates $\Phi_n$ of
$h$:
\begin{equation}
h\Phi_n(\vec r)=\epsilon_n\Phi_n(\vec r)
\label{eigenstate}
\end{equation}
can be characterized by the energies $\epsilon_n$ and three quantum numbers,
which correspond to $G$, $G_z$ and $\Pi$.

Following  Kahana and Ripka (1984) the eigenvalue problem
(\ref{eigenstate}) can be solved numerically in a finite quasi--discrete
basis. The basis is made discrete by putting
system in a large box of radius $D$ and imposing boundary conditions at $r=D$.
Also, it is made finite by restricting momenta of the basis states
to be smaller than the numerical cutoff $K_{max}$. Both quantities have no
physical meaning and the results should not depend on them.
In particular, $K_{max}$ has nothing to do with the model cutoff $\Lambda$
(\ref{D-reg-static}).
The typical values used are $D \sim 20/M$ and $K_{max} \sim 7 M$.

\begin{figure}
\centerline{\epsfysize=3.7in\epsffile{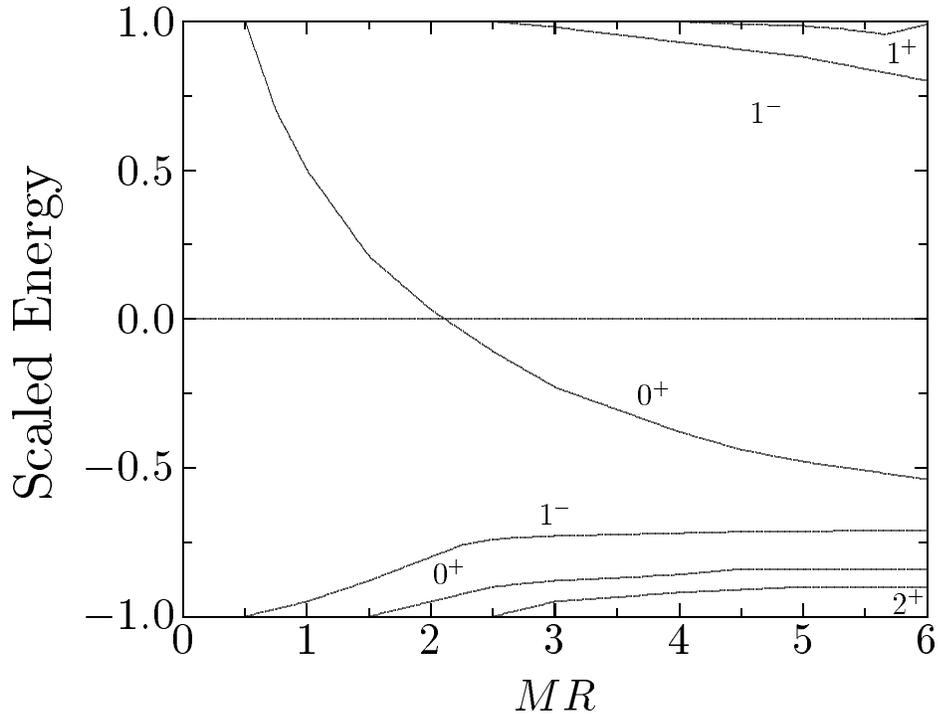}}%\vskip4pt
\caption{Quark spectrum of the one-particle hamiltonian $h$ for an
exponential profile form $\Theta (r) = - \pi \exp (- r/ R )$
as a function of the profile size $R$. Both the energies $\epsilon_n$
and $R$ are given in scaled units: ($\epsilon_n \over M$ and $MR$).}
\label{f31}
\end{figure}

As an illustration a typical spectrum of $h$ for a simple meson profile
function $\Theta (r) = - \pi \exp (- r/ R )$ are shown in Fig.~\ref{f31}.
This $\Theta$ is not a selfconsistent solution and it is purely chosen for
didactic purposes. The parameter $R$ is the size of the meson profile. 
Together with the constituent quark mass $M$, which describes the strength 
coupling between quarks and mesons (or equally the magnitude of the
chiral field), the product $RM$ characterizes the deviation of the actual field
configuration from the vacuum ($MR=0$) one. One can vary $MR$ either changing 
$M$ at a fixed profile size $R$ or vice verse. In most of our discussions, we 
prefer to use a language in which $M$ is kept fixed.
  
\begin{figure}
\centerline{\epsfysize=3.7in\epsffile{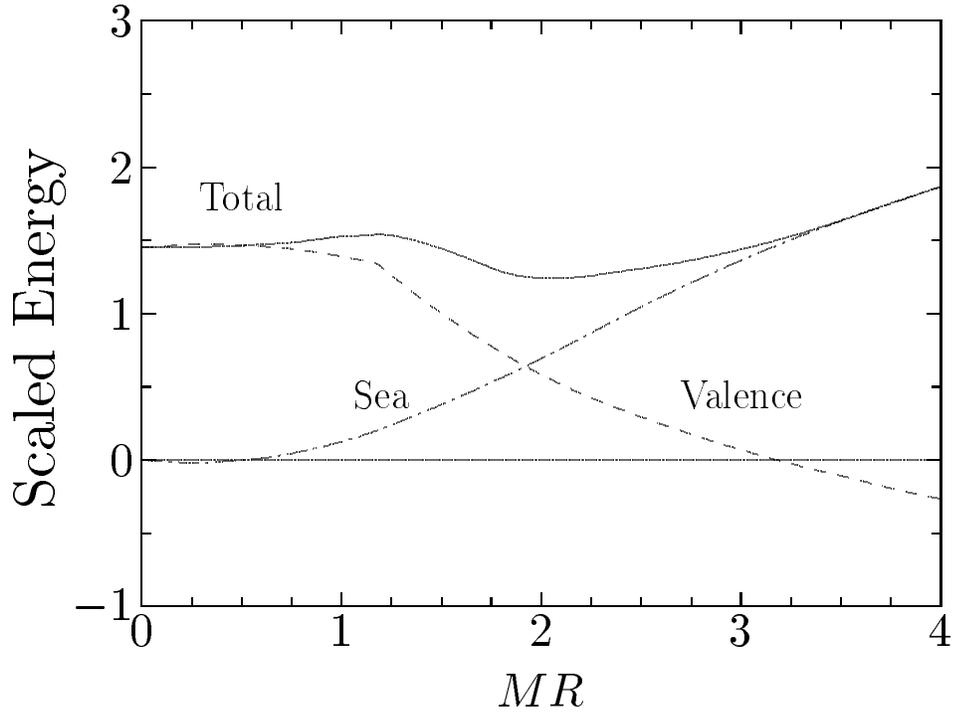}}%\vskip4pt
\caption{Classical soliton energy 
$M_{cl}$ split in Dirac sea and valence parts in the case of an exponential 
profile form. The energies are given in unit of the constituent quark mass
$M$ and the size of the soliton $R$ is given in terms of the scaled unit 
$MR$.}
\label{f32}
\end{figure}

At some $R$ the lowest quark level $G^\Pi= 0^+$
starts to deviate from the positive continuum and at large enough
($R \geq 0.5/M$) one finds a bound level which  we call the valence level
${\ket n}\equiv \ket {val}$. Its single-particle energy $\epsilon_{val}$
decreases with increasing $R$, changes sign and at very large $R$ ($RM\gg 1$)
this level approaches the negative continuum (Dirac sea). Thus, for large
$R$ the valence level approaches the negative continuum. In this cases the
chiral field is slow varying ($RM\gg 1$) and the equality (\ref{B-Q-equal})
is valid,
i.e. the baryon number can be related to the topological winding number.
Similar considerations hold for higher winding numbers (Kahana {\it et al.}, 
1984). At sufficiently large $RM\gg 1$ one therefore gets close to 
the picture of
the {\it topological soliton models} like e.g. the Skyrme model which contains
no quark degree of freedom and the winding number of the chiral field is
related to the baryon number of the soliton. At moderate values of $R$
($RM\approx 1$), however, the NJL model supports a
different physical (i.e. valence) picture: there is a positive valence level
which originates from the positive continuum, and it gets bound due to the
interaction of the valence quarks with the polarized negative continuum
(Dirac sea). This mechanism creates the non-topological soliton as a bound
many-quark state. In order to illustrate it, we plot the soliton energy as well
as the separated valence and Dirac sea parts for the same meson profile
as a function of the profile size $R$ in fig.\ref{f32}.
One clearly recognizes that $M_{cl}$ has a local minimum at
$R\,M\approx 1$ which corresponds to a solitonic solution of $B=1$. This
minimum is due to the competing behavior of the valence energy $E_{val}$ and of
the sea contribution $E_{sea}$. At very small $RM\ll 1$ we have actually the 
vacuum
quark spectrum with both unpolarized continua which means that $E_{sea}=0$
(no polarization). With increasing $R$ due to the
interaction between the valence quarks and the Dirac sea, the latter gets
more and more polarized and the polarization energy
$E_{sea}$ increases. The valence level becomes more bound and the
valence energy decreases. Thus, at small $R$ the valence part dominates over
the sea one and it changes to the opposite at large $R$. The $B=1$ soliton
solution which minimizes the soliton energy occurs at intermediate 
$RM\approx 1$ and
the soliton consists of a positive valence level (well separated from the Dirac
sea) and a polarized Dirac sea, and both parts contribute almost equally to the
soliton energy.

\vskip1cm
3.5 \underline{Selfconsistent soliton solution}
\vskip4mm 

After some pioneering calculations
with fixed meson profile (Diakonov {\it et al.}, 1988, 
Meissner Th. {\it et al.}, 1988)
the selfconsistent soliton solution with baryon number one has been first
obtained by Reinhardt and W\"unsch (1988), and by Meissner Th. {\it et al.} 
(1989). Using a selfconsistent procedure,
similar to ones known from Hartree and Hartree-Fock mean-field calculations in
atomic and nuclear physics, (\ref{eqsmotion}) and (\ref{eigenstate}) have
been solved numerically. One starts with a reasonably chosen profile 
$\Theta (r)$, diagonalizes the $h$ in the quasi-discrete finite basis
(Kahana and Ripka, 1984) and using the eigenfunctions
$\Phi_n (\vec r)$ and the eigenvalues $\epsilon_n$  obtains a new profile
function $\Theta^\prime (r)$ from the equation of motion (\ref{eqsmotion}).
The procedure is repeated until a desired degree of selfconsistency is reached.

\begin{figure}
\centerline{\epsfysize=3.7in\epsffile{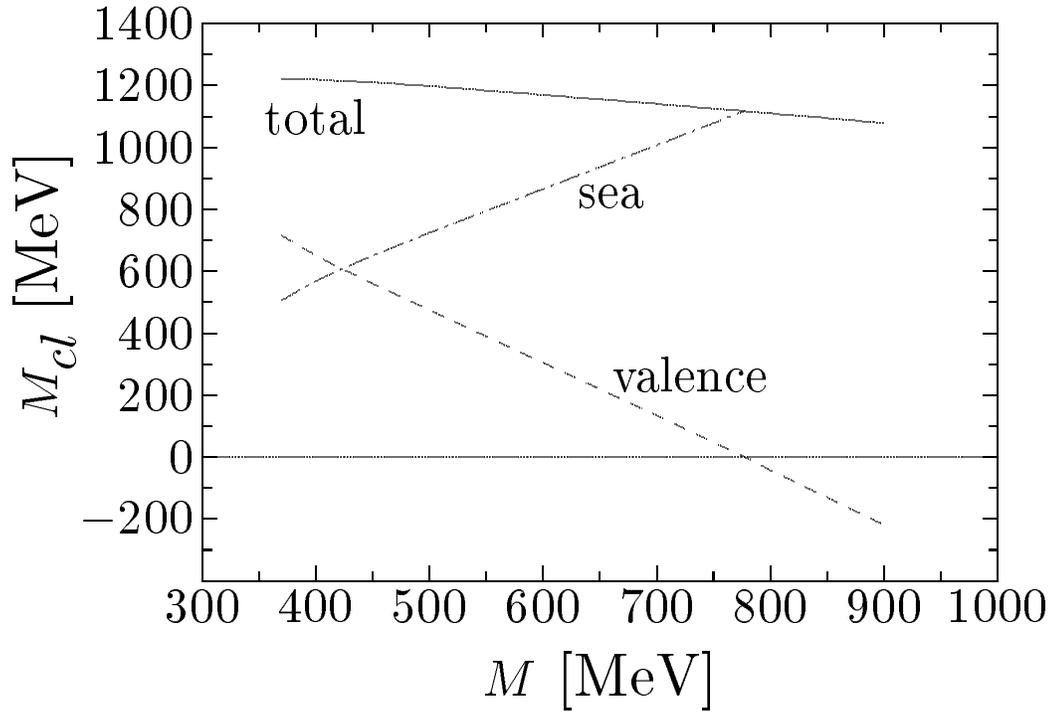}}%\vskip4pt
\caption{Classical energy $M_{cl}$ of the selfconsistent soliton
split in sea and vacuum contributions 
as a function of the constituent quark mass $M$.}
\label{f33}
\end{figure}

Typical results for the energy $M_{cl}$ of the selfconsistent $B=1$
soliton and the separated valence and sea contributions
are presented in fig.\ref{f33} as a function
of the constituent quark mass $M$. The solitonic solutions exist if the $M$ 
exceeds a critical value
$M>M_{cr}\approx 350$ MeV. This cusp behavior is typical for localized
(soliton) solutions of a system of coupled non-linear equations (see e.g.
Lee, 1981 and refs. therein) and has been observed also in other
chiral quark meson models (Birse, 1990 and refs. therein). The soliton
energy shows a rather weak dependence on $M$ as it slightly decreases with
$M$ increasing. However, both the valence and the sea parts depend strongly on
the mass $M$: with increasing $M$ the valence energy decreases and
correspondingly the sea one increases. It is a behavior, similar to the one
obtained with the fixed profile and variation of $R$, and apparently, this is
an indication for a change in the soliton structure.
Below $M_{cr}$  there is no soliton solution and the $N_c$ free quarks
is the most preferable configuration. At $M>M_{cr}$ there is a soliton
which is characterized by a bound valence orbit and polarized Dirac sea.
With increasing $M$ the valence level goes down and at very large values 
$M\sim 1$ GeV approaches the negative
continuum. In the latter case, since the mass $M$ is large enough
($MR\gg 1$) and all negative-energy states are occupied, the baryon
number can be identified via gradient expansion with the winding number
of the chiral field as already discussed in the context of the fixed
profile. Obviously, $\tilde B(U)$ has a non-zero value and it is equal
to the baryon number $B=1$. Thus, the model allows for two different
and to some extent opposite nucleon scenarios. At $M< 600$ MeV the
soliton consists of a well pronounced bound valence level and a polarized
Dirac sea which corresponds to the widely accepted phenomenological valence
picture of the nucleon as valence quarks surrounded by the pion cloud. At much
larger values $M > 1$ GeV the valence level can be considered as a part of
the Dirac sea and we are close to the Skyrme picture of the nucleon as a
topological soliton. In order to make a definite decision which picture is
favored one has to calculate the baryonic observables using $M$ as a 
free model
parameter to reproduce as many as possible experiment data. As
we will see later, fixing the parameters of the model in the vacuum 
to reproduce
the physical values of the pion decay constant and the pion mass, the model is
able to provide an overall good description of the baryon properties for
$M$ between 400 and 450 MeV. Higher $M$ values are excluded. 
It means that in the case of phenomenologically acceptable
values of the parameters of the NJL lagrangian
the saddle point field $U_c$  does not fulfill the condition $MR\gg 1$.
Hence, the gradient expansion cannot be used, and the approximation of the 
non-local NJL effective action by a local one similar to those of the
topological Skyrme-like models is not justified.

\begin{figure}
\centerline{\epsfysize=2.9in\epsffile{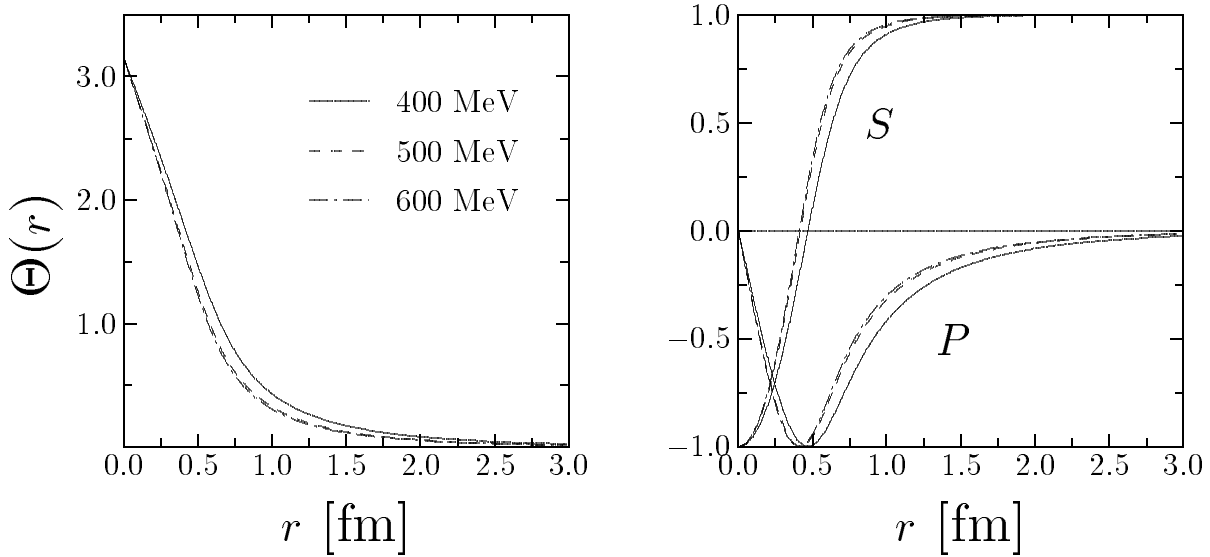}}%\vskip4pt
\caption{Selfconsistent profile function $\Theta$ (left) and the 
selfconsistent scalar $S$ and pseudoscalar $P$
mean fields (right) for three different values of the constituent 
quark mass $M$ and $m_\pi=140$ MeV.}
\label{f34}
\end{figure}

On fig.\ref{f34} one can see the profile function of the
selfconsistent $B=1$  soliton solutions for three values of $M$ as well as
the corresponding mean-field densities (saddle-point meson fields)
$S(r)$ and $P(r)$. As one can see the
actual form of the profile function and the meson fields as function of
the distance $r$ are rather independent of $M$. The typical size of the
soliton is about $0.6$~fm which means that $RM\approx 1$.
The profile function fulfills the boundary conditions (\ref{hgboundcond}) and
(\ref{hgboundcond0}). There is, however,{\em a priori} no reason in this model
to fix $Q_T$ to one because the baryon number $B=1$ is carried by
the valence quarks. This is different from the topological soliton models
(e.g. the Skyrme model), where the baryon number is identified with the winding
number from the very beginning.
There is even no a priori reason why $Q_T$ should be an integer
number, if one leaves mathematical arguments like continuity of the
$\piv$ field in the origin. In contrast to the Skyrme model, where an
integer $Q_T$ is required in order to obtain a finite energy, the total
regularized mass $M_{cl}$ is finite for any value of $Q_T$ in the
present approach. Apparently, we cannot expect that the saddle-point
field $U_c$ describing the baryon with the baryon charge $B=1$
should have the same topological charge $Q_T(U_c)=B$.
Looking for the solution of the stationary equation
(\ref{statcond}) we have to minimize the functional
(\ref{solenergy}) over all fields $U$ which could have in principle
different topological charges. However, the explicit calculations show
that the saddle-point field for the
$B=1$ soliton indeed has the winding number $Q_T=1$.
Diakonov {\it et al.} (1988,1989) and Berg {\it et al.} (1992) have 
investigated the
behavior of the energy $M_{cl}(Q_T)$ of the $B=1$ soliton as a function of
$Q_T$ for a fixed profile form. It turned out, that the minimum of the
soliton energy $M_{cl}$
definitely lies at the point $Q_T=1$, which means  that this feature at least
for the case $B=1$ seems to be a dynamical consequence of the equations of
motion. Unfortunately  the whole analysis is based purely on numerical
arguments and up to now there is no general proof of this fact.

In addition Berg {\it et al.} (1992) also studied selfconsistent solutions
with higher baryon numbers $B>1$.
They found that a local minimum which corresponds to a solitonic solution only
exists if $Q_T \le B$ and in any case these local minima appear for integer
winding number $Q_T$. For example, for the $B=2$ system they found a soliton in
the $Q_T=1$ sector with 2 valence orbitals ($0^+$,$0^-$) occupied if
$M > 370$ MeV. In the $Q_T=2$ sector, solitonic solutions exist if $M> 560$
MeV. The main difference between the two cases is that for
$Q_T=2$ and large mass $M$ both the $0^+$ and the $0^-$ orbital cross the 
zero line and approach
the Dirac sea, whereas in the case of $Q_T=1$
only the $0^+$ orbital comes down and the $0^-$ orbital moves back to the
positive continuum.

Sieber {\it et al.} (1992) performed a detailed analysis of the
NJL model without the chiral circle condition (\ref{chcircle}) and showed
that in contrast
to claims of Meissner Th. {\it et al.} (1989) and Reinhardt and W\"unsch, 
(1989),
at least in case of the proper time regularization scheme, no localized
solitonic solution with a finite energy exists. Using a modified version
of the NJL model, which in addition to the spontaneously broken chiral
symmetry also simulates the {\it anomalous breaking of scale invariance} in
QCD, Meissner Th. {\it et al.} (1993b) and Weiss {\it et al.} (1993a) have 
shown that in this
case stable solitonic solutions exist without the non-linear constraint which
could be considered also as a physical motivation for the chiral circle 
condition (\ref{chcircle}) in the present model.

\vskip1cm
3.6 \underline{Semiclassical quantization of the soliton}
\vskip4mm

In the previous section we have shown that
in the leading order of the large $N_c$ approximation
the calculation of the nucleon mass reduces to solving the saddle 
point equation (\ref{solenergy}). Apparently, if we want to compute the 
$1/N_c$ corrections to it we have to take
into account the quantum fluctuations of the chiral field
$U(x)$ around this classical saddle point solution $U_c(x)$.
Among all possible fluctuations a special role is played by those
which do not change the classical soliton energy. Since
the hedgehog soliton is not invariant under translations, space
rotations and isotopic transformations (although it is invariant
under the simultaneous space and isotopic rotations),
applying these symmetry transformations to the saddle point solution
we can construct a whole family of new solutions which correspond to one and 
the same classical energy $M_{cl}$. We remind that
we had an analogous situation in the vacuum sector where in the chiral
limit there also was a family of vacuum solutions (\ref{u}) with a
spontaneous breakdown of the chiral symmetry. However, since all vacuum
solutions reflect the fact that the chiral
symmetry is spontaneously broken, in the vacuum sector we can choose
one of the vacuum solutions and forget about the others. In the
soliton sector the classical hedgehog solution cannot be
associated directly with the nucleon physical state. Indeed, the physical
nucleon state should have proper quantum numbers like spin, isospin
momentum.  Such quantum states can be obtained as quantum superposition of
classical states with different space and isotopic orientations and with
different space locations. The corresponding techniques are
well known in the non-relativistic many-body physics (see Ring and Shuck, 1980)
and they are also used in the field theory. In particular,  the cranking
(semiclassical quantization) scheme is successfully applied for the
quantization of the Skyrmion soliton (Adkins {\it et al.}, 1983) as well as 
in the chiral sigma
model with valence quarks (Cohen and Broniowski 1986). However, in the present
model we have a more complicated case because of both the non-locality of the
effective action (fermion determinant) and the ultraviolet divergences which
require regularization.

The symmetry of the hedgehog solution leads to the equivalence
of the space-rotational and of the isotopic zero modes,
so that we are left with three
rotational and three translational zero modes.
In order to restore the rotational, isotopic and translational symmetries,
we have to treat the degrees of freedom related to these
symmetries on the quantum level. Even if we treat the meson fields in the
leading order of the $1/N_c$ approximation (no meson loops),
in order to obtain states with good quantum numbers, we have to include
the zero modes in our considerations. In
contrast to other normal modes the zero-mode fluctuations are not small. Since 
they cannot be treated in a perturbative way, one should integrate over
these fluctuations exactly in terms of path integrals. In fact, the
semiclassical quantization scheme for the soliton of the NJL model, including
polarized Dirac sea, is elaborated in (Diakonov {\it et al.}, 1988). The 
problems arising in this scheme due to the regularization are considered at 
first in (Reinhardt, 1989).
\vskip1cm

\underline{Rotational zero modes and the quantum rotational corrections to the
soliton energy.}
In this part we restrict ourselves to consider only the rotational zero
modes needed to construct the nucleon states
with certain spin and isospin eigenvalues.
To this end we restrict the path integral
in (\ref{PiNPI}) to the meson fields of the form:
\begin{equation}
U(\tau,\vec x)= R(\tau) U_c({\vec x}) R^\dagger(\tau)\,,
\label{rotating-soliton}
\end{equation}
where $R(\tau)$ is a unitary time-dependent $SU(2)$ orientation matrix of the
soliton and $U_c$ is the stationary meson field configuration. Now we are left
with a path integral over time dependent matrices $R$
\begin{equation}
\Pi_N(T)= \Gamma^{\{f\}}_N\Gamma^{\{g\}\ast}_N
\int\D R\prod_{i=1}^{N_c}
\langle 0,T/2 |
\frac{1}{D(RU_cR^\dagger)} |0,-T/2 \rangle_{f_i g_i}
\e^{-S(RU_cR^\dagger)}\,.
\label{PiNPIR0}
\end{equation}
For ansatz (\ref{rotating-soliton})
the operator $D(U)$ (\ref{D-U-definition})
can be rewritten as
\begin{equation}
D(U) = R\, [D(U_c)+ R^\dagger \dot R]\, R^\dagger
= R\, [D(U_c)+ i\Omega]\, R^\dagger \,,
\label{D-U-rotating}
\end{equation}
where the dot stands for the derivative with respect to the euclidean time 
$\tau$. In (\ref{D-U-rotating}) we introduced a hermitian angular velocity 
matrix
\begin{equation}
\Omega=-iR^\dagger \dot R=-{i\over 2}\Sp(R^\dagger \dot R\tau^a)\tau^a\equiv
{1\over 2}\Omega_a\tau^a\,.
\label{omega}
\end{equation}
With Dirac operator (\ref{D-U-rotating}) the effective action reads (before
regularization)
\begin{equation}
S(RU_cR^\dagger)=-N_c\Tr\log[D(U_c)+i\Omega]\,.
\label{seffomega}
\end{equation}
In fact, the $D(U_c)+i\Omega$ corresponds to the body-fixed
frame of the soliton in which the quark fields are transformed as
\begin{equation}
\Psi\longrightarrow R(t)\Psi\qquad \hbox{and}\qquad
\Psi^\dagger\longrightarrow \Psi^\dagger R^\dagger(t).
\nonumber\end{equation}
Similarly, the quark propagator in the background meson field $U$ can be
presented in the form
\begin{equation}
\bra{x}\,\frac 1{D(U)}\, \ket{x^\prime}_{fg}\,=\,\,\bra{x}R(x_4)
\frac 1{D(U_c)+i\Omega}\,R^\dagger(x^\prime_4) \ket{x^\prime}_{fg}\,\,.
\label{Eq10b}\end{equation}

In order to perform the integration over $R$ in (\ref{PiNPIR0})
we treat the angular
velocity $R^\dagger \dot R$ of the soliton as a small quantity and neglect
its derivatives. In fact, as we will see later this is justified by
$1/N_c$ arguments. Therefore, in the local expansion in
powers of the angular velocity and its derivatives
we can restrict to the terms
\begin{equation}
N_c \Tr \log [D(U_c)+i\Omega]
= N_c \Tr \log [D(U_c)]
- {I_{sea}\over 2} \int d\tau \Omega_a^2 + \ldots\,.
\label{moment-of-inertia-definition}
\end{equation}
The coefficient
$I_{sea}$ is the Dirac sea contribution to the  moment of inertia of
the soliton. If we forget about the regularization, after some straightforward
calculations we get
\begin{eqnarray}
I_{sea}\delta_{ab}&=&{1\over 4T}N_c
\Tr\Bigl[{1\over D(U_c)}\tau^a
{1\over D(U_c)}\tau^b\Bigr]\nonumber\\
&=& {1\over 4}N_c
\int_{-\infty}^{+\infty}{\d\omega\over 2\pi} \Tr\left[
{1\over i\omega+h(U_c)}\tau^a{1\over i\omega+h(U_c)}\tau^b\right]
\nonumber\\
&=&{1\over 2}N_c\sum_{\epsilon_m<0\atop \epsilon_n>0}{\bra n \tau^a\ket m
\bra m\tau^b\ket n\over  \epsilon_n-\epsilon_m}\,,
\label{seamomentinertianonreg}
\end{eqnarray}
Here, similarly to the baryon number the trace is calculated explicitly in
terms of quark matrix elements. Since the sea contribution to the moment of
inertia originates from the effective action, it is divergent and should be
regularized. It is done (Reinhardt, 1989) using the proper-time regularized
expression for the real part (\ref{e229}) and the expansion
\begin{eqnarray}
\e^{{\hat A}+{\hat B}} =& e^{\hat A}&\, + \,
\int\limits^1_0 \d\alpha\,\e^{\alpha {\hat A}}\ {\hat B}\
\e^{(1-\alpha) {\hat A}}\nonumber\\
&\quad +& \, \int\limits^1_0 \d\beta\,\int\limits^{1-\beta}_0 \d\alpha\,
\e^{\alpha {\hat A}}\ {\hat B}\
\e^{\beta {\hat A}}\ {\hat B}\
\e^{(1-\alpha-\beta) {\hat A}} + ...
\label{FSD}
\end{eqnarray}
The final regularized expression for the soliton moment of
inertia reads
\begin{equation}
I_{sea}\delta_{ab}={1\over 2}N_c\sum_{n\neq m}
{\cal R}_I^\Lambda(\epsilon_m, \epsilon_n)
\bra n \tau^a\ket m \bra m\tau^b\ket n \,,
\label{momentinertiareg}
\end{equation}
where the proper-time regulator is given by (Reinhardt, 1989)
\begin{equation}
{\cal R}_I^\Lambda(\epsilon_m, \epsilon_n) =\frac{1}{4\sqrt{\pi}}
\int\limits_{1/\Lambda^2}^{\infty}\frac{du}{\sqrt{u}}\left(\frac{1}{u}
\frac{ \e^{-u\epsilon_n^2} - \e^{-u\epsilon_m^2} }
{\epsilon_m^2 - \epsilon_n^2 }
- \frac{\epsilon_n \e^{-u\epsilon_n^2}
+\epsilon_m \e^{-u\epsilon_m^2} }
{\epsilon_m + \epsilon_n }
\right) \,.
\label{regmomentinertia}
\end{equation}
For the quark propagator using its spectral representation and treating
$\Omega$ as a perturbation we obtain
\begin{equation}
\bra {0,T/2}\frac 1{D(U_c)+i\Omega}\ket {0,-T/2}^{N_c}
\mathop{\sim}_{T\to \infty}\exp
\left(-N_c\epsilon_{val}T-{I_{val}\over 2}
\int d\tau \Omega_a^2\right)\,.
\label{1/Dexpansion}
\end{equation}
Analogously, $I_{val}$ is the valence quark contribution to the moment of
inertia
\begin{equation}
I_{val}\delta_{ab}={1\over 2}N_c\sum_{n\neq val}
{\bra {val}\tau^a\ket n\bra n\tau^b\ket {val}\over\epsilon_n-\epsilon_{val}}\,.
\label{valmomentinertia}
\end{equation}
It originates from the quark propagators and is not
affected by the regularization. The total moment of inertia of the soliton is
now
\begin{equation}
I=I_{val}+I_{sea}\sim N_c\,.
\label{solmomentinertia}
\end{equation}
Combining (\ref{valmomentinertia}) with the non-regularized $I_{sea}$
(\ref{seamomentinertianonreg})
\begin{equation}
I\delta_{ab}={1\over 2}N_c\sum_{\epsilon_m \le val\atop \epsilon_n> val}
{\bra n \tau^a\ket m \bra m\tau^b\ket n\over  \epsilon_n-\epsilon_m}\,,
\label{solmomentinertianonreg}
\end{equation}
we recover the Inglis formula for the moment of inertia in the
cranking approximation known from the many-body theory (Ring and Schuck, 1980).

Now, using expansions (\ref{moment-of-inertia-definition}) and
(\ref{1/Dexpansion}) we arrive
at a functional integral over time-dependent orientation matrices $R(\tau)$
with an action quadratic in $\Omega$. The quadratic action is
in fact the action of a symmetric rotator and hence, the path integral reduces
to the euclidean evolution matrix element for the corresponding quantum
rotator:
\begin{equation}
\int\limits_{R(-T/2)=R_1}^{R(T/2)=R_2}
{\cal D}R \exp \left(  - \frac{I}{2}\int\limits_{-T/2}^{T/2} d\tau \Omega_a^2
\right)
= \bra{R_2} e^{-H_{rot}T} \ket{R_1}\,.
\label{path-integral-result}
\end{equation}
Here we fixed the boundary conditions in the path integral at
$\tau=\pm T/2$ postponing the integral over $R(\pm T/2)$
for later. The quantum hamiltonian $H_{rot}$
acts in the Hilbert space of functions $\psi(R)$
where $R$ is the $SU(2)$ orientation matrix of the soliton
and in the rhs of (\ref{path-integral-result})
we have the evolution matrix element in this $R$ representation.
From quantum mechanics we know that the hamiltonian of the
symmetric rotator can be expressed through the squared
angular momentum:
\begin{equation}
H_{rot} = {J_a^2\over 2I}={T_a^2\over 2I}\,,
\label{h-rot}
\end{equation}
where the action of generators $J_a$ and $T_a$
on the wave function $\psi(R)$ is as follows
\begin{equation}
\exp(i\omega_a T_a) \psi (R)
= \psi \left( \exp(-i\omega_a \tau_a/2 ) R \right)\,,
\label{SU2-isospin}
\end{equation}
\begin{equation}
\exp(i\omega_a J_a) \psi (R)
= \psi \left(R \exp(i\omega_a \tau_a/2 )  \right)\,.
\label{SU2-spin}
\end{equation}
Comparing these formulas with the form of the rotating soliton
$RU_cR^\dagger$ we see that the left generator $T_a$
is nothing else but the isospin operator, whereas
$J_a$ is the spin one. We stress that the identity (\ref{path-integral-result})
corresponds to the canonical quantization rule (in the euclidean time)
\begin{equation}
\Omega_a \to-i{J_a\over I}
\label{quantization-rule}
\end{equation}

The eigenfunctions of the operators $J^2=T^2, J_3, T_3$ can be expressed
through the Wigner $D$-functions as follows
\begin{equation}
\psi_{JJ_3TT_3}(R) = (-1)^{T+T_3}\sqrt{2T+1}\,D^{T=J}_{-T_3,J_3}(R)\,.
\label{D-functions}
\end{equation}
Obviously these functions are also eigenstates of the hamiltonian
$H_{rot}$ (\ref{h-rot})
\begin{equation}
H_{rot} \psi_{JJ_3TT_3}(R)
={J(J+1)\over 2I} \psi_{JJ_3TT_3}(R)\,.
\label{hrotfunction}
\end{equation}

The path integral over the orientation
matrices $R$ with the local action quadratic
in $\Omega$ is computed exactly in (\ref{path-integral-result})
although this local quadratic form of the action is an approximation
to the non-local rotational action (\ref{moment-of-inertia-definition})
justified in the large $N_c$ limit. Indeed, we are interested
in nucleon rotational excitations with $J=T$ of order of $O(N_c^0)$. For such 
excitations the angular velocity of the semiclassical trajectories, which
saturate our path integral in the large $N_c$ limit, is
$\Omega_a\sim 1/I\sim 1/N_c$

The above argumentation that the left and right generators
$T_a$ and $J_a$ can be interpreted as spin and isospin operators
is based rather on physical intuition than
on a consistent derivation. In fact,
the spin and isospin of the nucleon state created by current
$J_N^\dagger$ are determined by the tensor $\Gamma^{\{g\}*}_N$
on the rhs of (\ref{jn}). Let us show how the quantum
numbers $T^2=J^2,T_3, J_3$ of the current $J_N^\dagger$
penetrate into the rotational wave functions of the nucleon.
We can rewrite the quark propagator in the background field of the
rotating soliton $RU_cR^\dagger$ in the form
\begin{equation}
\langle 0,T/2 |
\frac{1}{D(RU_cR^\dagger)} |0,-T/2 \rangle
=
R(T/2)\langle 0,T/2 |
\frac{1}{D(U_c)+i\Omega} |0,-T/2 \rangle R^\dagger(-T/2)\,.
\label{propagator-R}
\end{equation}
Inserting this into (\ref{PiNPIR0}) we arrive at the following product
\begin{equation}
\sum\limits_{t_1^\prime\ldots t_{N_c}^\prime}
\Gamma^{(t_1^\prime,j_1),\dots (t_{N_c}^\prime,j_{N_c})}_{JJ_3TT_3}
\prod_{i=1}^{N_c}R_{t_1^\prime,t_1}(T/2)
\dots R_{t_{N_c}^\prime,t_{N_c}}(T/2)
=
\sum\limits_{T_3^\prime}
\Gamma^{(t_1,j_1),\dots (t_{N_c},j_{N_c})}_{JJ_3TT_3'}
D^T_{T_3T_3'}[R(T/2)]\,.
\label{solrotfunction}
\end{equation}
Here we have taken into account that
tensor $\Gamma^{\{f\}}_N$ contains Clebsh-Gordan coefficients
which relate the isospins of separate quarks $t_i$ to the isospin $T$
of the nucleon, so that the isospin rotation of $N_c$
quarks is equivalent to the corresponding isospin rotation of
the nucleon isospin index $T_3$. Next, tensor
$\Gamma^{\{f\}}_N$ also contains spin indices of separate quarks that
are contracted with spin indices of the quark propagators
(\ref{propagator-R}).
At large euclidean time separation $T$, the quark propagator is saturated
by the valence  quark contribution (\ref{eval}),
so that effectively the spin and isospin indices
of $\Gamma^{\{f\}}_N$ become contracted with the indices of the
$N_c$ valence wave functions. Using the fact that the valence level has the
grand spin $G=0$ one can see that after some manipulations the valence wave
functions contracted with
the function $D^T_{T_3T_3'}(R)$ in (\ref{solrotfunction})
result in
\begin{equation}
(-1)^{J+J_3} D^T_{T_3,-J_3}[R(T/2)]
= (-1)^{T+T_3} D^{T\ast}_{-T_3,J_3}[R(T/2)]
\end{equation}
which up to a normalization factor is identical with the rotational wave
function of the final state $\psi_{JJ_3TT_3}^\ast[R(T/2)]$ (\ref{D-functions}).
In a similar way, one obtains the wave function
$\psi_{JJ_3TT_3}[R(T/2)]$ of the initial nucleon. Note that
in eq. (\ref{path-integral-result}) we fixed the boundary conditions
for $R(-T/2)=R_1$ and $R(T/2)=R_2$. Now we can turn back to
(\ref{PiNPIR0}) and perform the integral over $R_1$, $R_2$.
We get a rather simple result
\begin{eqnarray}
&&\int\D R \psi_{JJ_3TT_3}^*[R(T/2)]\, \psi_{JJ_3TT_3}^{}[R(-T/2)]
\exp\left\{-{I\over 2} \int d\tau \Omega^2
\right\}\nonumber\\
&=&\int\d R\, \psi_{JJ_3TT_3}^*(R)
\,\psi_{JJ_3TT_3}^{}(R)\,\e^{-{J(J+1)\over 2I}T}
\equiv\bra{JJ_3TT_3}\e^{-H_{rot}T}\ket{JJ_3TT_3}=
\e^{-{J(J+1)\over 2I}T}\,.
\label{Rint}
\end{eqnarray}
From the asymptotics of the nucleon correlation function
\begin{equation}
\Pi_N(T)\mathop{\sim}_{T\to \infty}
\exp\Biggl[-\Bigl(M_{cl}+{J(J+1)\over 2I}\Bigr)T\Biggr]\,.
\label{PiNPIR}
\end{equation}
we obtain the quantized soliton energy
\begin{equation}
E_J=M_{cl}+{J(J+1)\over 2I}\,,
\label{EJ}
\end{equation}
which contains the familiar rotational corrections.
In this formula the classical contribution is $M_{cl}=O(N_c)$
and the rotational correction is of order $1/I=O(N_c^{-1})$.
On the other hand, there exist $O(N_c^{0})$
corrections coming from the non-zero modes which were
neglected in our derivation. These non-zero mode
corrections are parametrically
larger than the $O(N_c^{-1})$ rotational corrections.
However the non-zero mode corrections do not depend on the
spin and isospin of the nucleon state so that the non-zero mode corrections
cancel in mass splittings. In particular, in the
semiclassical quantization scheme the nucleon-delta mass difference is simply
given by the rotational corrections
\begin{equation}
E_{N\Delta}=E_{3/2}-E_{1/2}={3\over 2I}\sim {1\over N_c}\,.
\label{EDN}
\end{equation}
We get the expected results that in large $N_c$ limit the nucleon and Delta are
degenerate in mass.

\begin{figure}
\centerline{\epsfysize=2.8in\epsffile{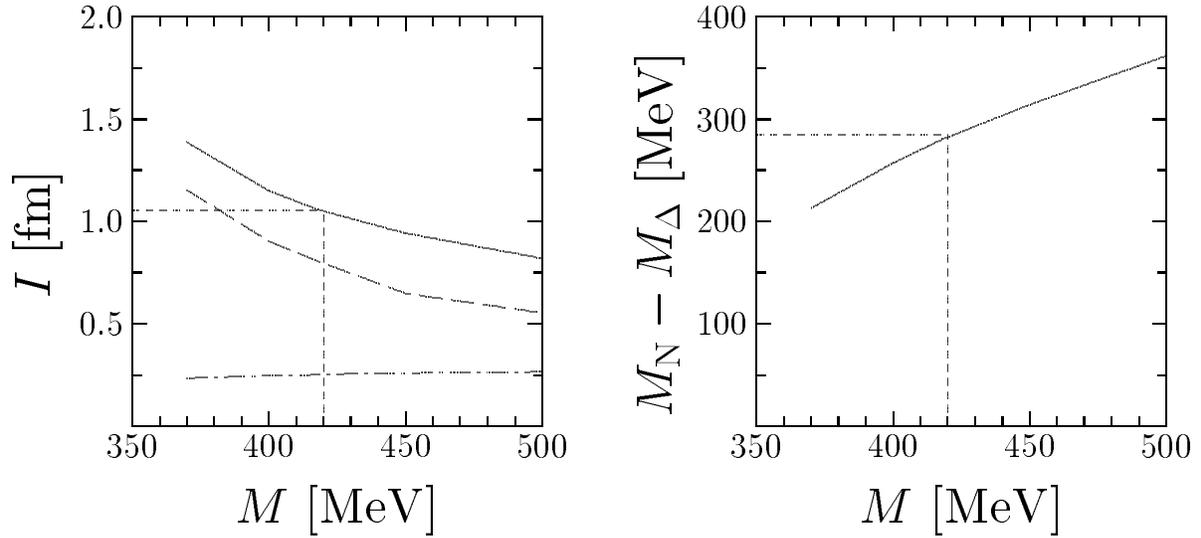}}%\vskip4pt
\caption{Moment of inertia and its valence (dashed) and sea (dash-dotted) parts
(left), and the nucleon-Delta mass splitting (right) for the selfconsistent 
solution and $m_\pi =140$ MeV in dependence of the constituent quark mass $M$. 
The experimental nucleon-Delta mass difference is also shown.}
\label{f41}
\end{figure}

The moment of inertia has been calculated numerically by Wakamatsu and Yoshiki,
(1991) and Goeke {\it et al.} (1991). The results are shown in fig.~\ref{f41} 
as a
function of the constituent quark mass $M$. One notices a clear decrease of the
moment of inertia with $M$ increasing. At $M\approx 420$ MeV the experimental
nucleon-Delta mass difference of 294 MeV can be reproduced
(see eq.~(\ref{EDN}). In this region, the valence quarks dominate and the sea 
contribution to $I$ is about 20 \%. 
The $N-\Delta$ mass difference increase with $M$ and as can be seen the  
values for $M\approx 420$ MeV are favored.

\vskip1cm
\underline{Translational zero modes.}
So far we have considered only the rotational zero modes.
However, the hedgehog
soliton $U_c$ breaks not only the rotational and but also the
translational symmetries. In order to take
into account the translational zero modes, we consider the ``moving soliton''
extending the path integral over all time-dependent meson fields of the form:
\begin{equation}
U(x)= U_c({\vec x-\vec X(\tau) })=T_{X(\tau)}
U_c(\vec x )T^\dagger_{X(\tau)}\,.
\label{moving-soliton}
\end{equation}
Here $T_X$ is a unitary translation operator which corresponds to a 
translation $X$.
Analogously to (\ref{D-U-rotating}) it enters the Dirac operator
\begin{equation}
D(U) = T_X\, [D(U_c)+ T^\dagger_X \dot T_X]\, T^\dagger_X
= T_X\, [D(U_c) - \dot {\vec X}\cdot\vec\nabla]\, T^\dagger_X \,,
\label{D-T-translation}
\end{equation}
where dot stands for the derivative with respect to the euclidean time
$\tau$. Further, the effective action can be transformed to
\begin{equation}
S(T_XU_cT^\dagger_X)=-\Tr\,\log\,[D(U_c)-\dot X_k\partial_k]\,.
\label{seffomegaX}
\end{equation}
and expanded in $\dot X_k$
\begin{equation}
N_c\Tr\,\log\,[D(U_c)-\dot X_k\partial_k]
= N_c \Tr \,\log \,[D(U_c)]
-{M_{sea}\over 2}\int d\tau {\dot X}^2 +\ldots\,,
\label{mass-of-inertia-definition}
\end{equation}
For the quark propagator the expansion leads to
\begin{equation}
\bra {0,T/2}T_X\frac 1{D(U_c)-\dot X_k\partial_k}T^\dagger_X\ket {0,-T/2}^{N_c}
\mathop{\sim}_{T\to \infty}\e^{-N_c\epsilon_{val}T-{M_{val}\over 2}
\int d\tau {\dot X}^2}\,.
\label{val-1/Dexpansion}
\end{equation}
It can be shown (Pobylitsa {\it et al.}, 1992) that
\begin{equation}
M_{sol}=M_{val}+M_{sea}\sim N_c\,.
\label{inertial-mass}
\end{equation}
coincides with the soliton mass $M_{cl}$. The integral over $X$ with an
action quadratic $\dot X$ can be easily computed
and now we arrive at the following expression for the nucleon mass
\begin{equation}
E_J=M_{cl}+{J(J+1)\over 2I}+{P^2\over 2M_{sol}}\,.
\label{EJP}
\end{equation}
As it is expected the large $N_c$ limit leads to a non-relativistic
approximation.

One realizes that although the model is able
to reproduce the $N-\Delta$ mass difference, the masses itself are by 
25\% too large. The
reason is that the hedgehog soliton as a mean-field solution breaks the
translational and rotational symmetry and because of that the classical mass
$M_{cl}$
contains spurious contributions of the center-of-mass motion which should be
subtracted. The problem is well-known in the many-body theory
(Ring and Schuck, 1980) and
analogously to so called RPA scheme one expects two main subtraction terms, 
both of order of $O(N_c^0)$:
\begin{equation}
\Delta E=-{<p^2>\over 2 M_{sol}}-{<j^2>\over 2 I}\,,
\label{RPAcorr}
\end{equation}
where $p$ and $j$ are the linear and angular momentum operators, and $M_{sol}$
and  $I$ are the mass and the moment of inertia of the soliton . A rather
crude estimate of the first term can be done in an simple oscillator potential
model. 
An estimate of the spurious
contributions using directly the formula {(\ref{RPAcorr})} and calculating 
$<p^2>$ and $<j^2>$ for the soliton reduces the nucleon mass by more than 30~\%
(Pobylitsa {\it et al.}, 1992). In the Skyrme model this question has been 
investigated quite in detail by including pion loop effects (Moussallam, 1993, 
Holzwarth, 1994, Holzwarth and Walliser, 1995). The final value for the nucleon
 mass comes out close to experiment. 

Despite the problem of the spurious contributions to the
soliton energy and the related overestimation of the masses has attracted a
lot of attention (Pobylitsa {\it et al.}, 1992, Weigel {\it et al.}, 1995b,
Broniowski {\it et al.}, 1995) a strict derivation and a reliable quantitative
estimate of those contributions in the present model is still missing.
The reason is that in order to solve this problem rigorously  one has to
include the meson fluctuations.
In the path integral formalism it means that we have to go beyond the present
large $N_c$ saddle-point treatment of the path integral (classical meson
fields). Note that although using the
semiclassical quantization scheme we are able to assign proper spin and
isospin number to the soliton, now the ansatz (\ref{rotating-soliton}) is not
enough. The quantum meson contributions to the effective action, which we
neglect as parametrically small (of order $O(N_c)$ and hence suppressed by 
$1/N_c$) in the present
considerations, have to be taken into account. They are given by
the third term in (\ref{lnZ}) ${1\over 2}\Tr\log\Bigl[{\delta^2 S\over\delta
\phi\delta \phi}\Bigr]$. However, including this term one encounters the
following problems:
\begin{itemize}
\item The functional trace over the meson fields includes new divergences and
one needs a new cutoff in order to make the theory finite. It should be done in
a consistent way for both the vacuum and the soliton.
\item The meson-loop term should be evaluated directly in euclidean space-time,
since the used proper-time regularization makes the analytical continuation to
Minkowski space-time ill defined (Broniowski {\it et al.}, 1995). The usual
dispersion relations for the Green's function are not valid and hence, one
cannot use the convenient particle-hole description as it is nevertheless 
done in (Weigel {\it et al.}, 1995b).
\end{itemize}

%\newpage
%\vskip1cm
3.7 \underline{Nucleon matrix elements of quark currents}
\vskip4mm

Most of the nucleon observables are related to the evaluation of the
nucleon matrix element ${\bra N(p^\prime)}j\ket{N(p)}$ of various quark 
currents $j$. In this subsection, we will demonstrate how this
matrix element can be written in terms of path integrals and can be evaluated
by using the $1/N_c$ expansion in the semiclassical quantization scheme.

We start with a matrix element of an arbitrary quark current
$\Psi^\dagger\hat O \Psi$, where $\hat O$ is some matrix with spin and
isospin indices. Analogously to the nucleon correlation function
(\ref{PiNPI}), the
nucleon matrix element can be expressed as an euclidean functional integral:
\begin{eqnarray}
&&\langle N^\prime({\vec p}^\prime) | \Psi^\dagger(0) \hat O \Psi(0) |
N({\vec p}) \rangle
\mathop{=}_{T \to+\infty } \frac 1{Z^\prime} \int \d^3 x\, \d^3x^\prime\,
\e^{- i{\vec p}^\prime {\vec x}^\prime +
i{\vec p} {\vec x}}\nonumber \\
&&\int{\cal D}U \int {\cal D}\Psi \int {\cal D}\Psi^\dagger
J_{N^\prime}(\vec x^\prime,T/2) \Psi^\dagger(0)\hat O \Psi(0) 
J_N^\dagger(\vec x, -T/2) \e^{- \int d^4 z  \Psi^\dagger D(U) \Psi }\,.
\label{form-factor-integral}
\end{eqnarray}
Here the nucleon state $|N\rangle$ is created by the nucleon
current $J_N^\dagger$ (\ref{jn}). The constant $Z^\prime$ in
eq.~(\ref{form-factor-integral}) is chosen, so that the nucleon states obey
the non-relativistic normalization condition:
\begin{equation}
\langle N ({\vec p}^\prime) |  N ({\vec p}) \rangle
= (2\pi)^3 \delta^{(3)}({\vec p}^\prime -{\vec p}) \, .
\end{equation}

\begin{figure}
%\centerline{\epsfig{file=rev_fig8.ps,bbllx=80pt,bblly=480pt,bburx=480pt,bbury=760pt,rheight=4cm,rwidth=8cm,clip=}}
%\centerline{\epsfysize=6in\epsffile{rev_fig8.ps}}%\vskip4pt
\xslide{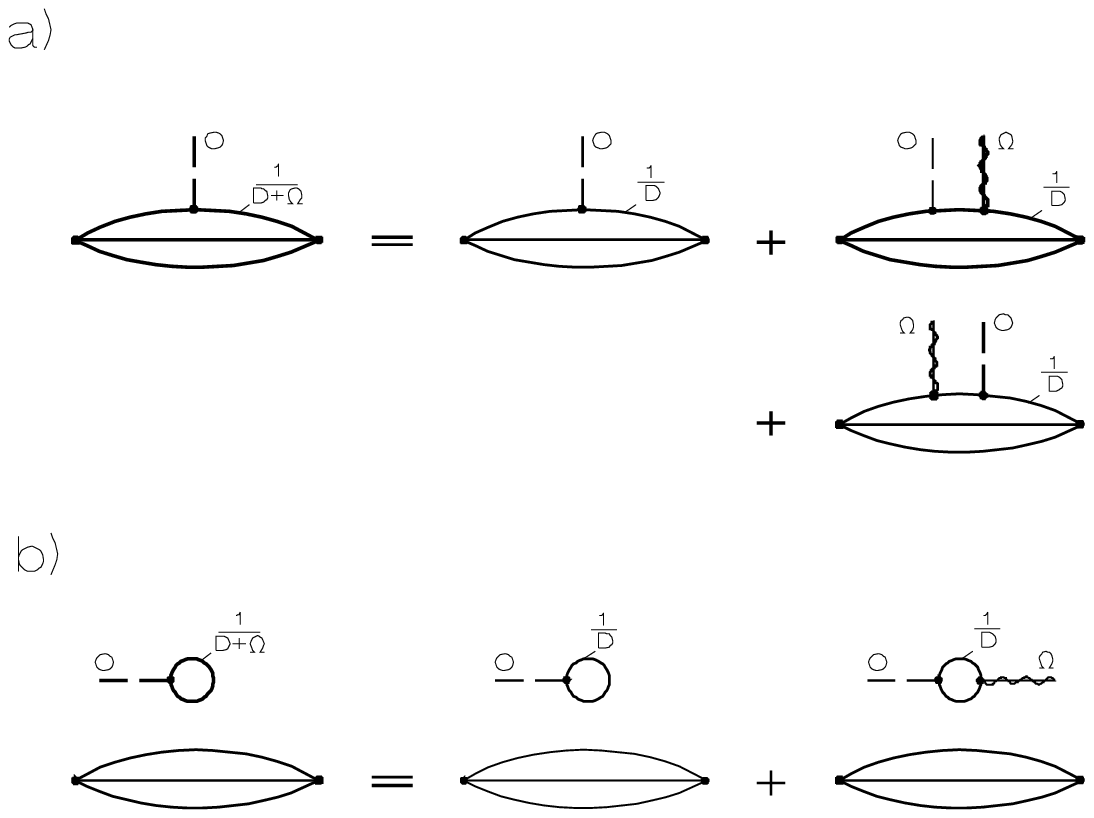}{10cm}{80}{480}{480}{760}
\caption{Diagrams corresponding to the expansion in $\Omega$ of the
current matrix element:
a) the valence contribution and b) the Dirac sea contribution.}
\label{Figr1}
\end{figure}

Integrating out the quarks in (\ref{form-factor-integral})
the result is naturally split in valence and Dirac sea parts (see
Fig.\ref{Figr1}):
\begin{equation}
\langle N^\prime ( {\vec p}^\prime) | \Psi^\dagger \hat O \Psi | N({\vec p}) \rangle
= \langle N^\prime ({\vec p}^\prime) | \Psi^\dagger \hat O \Psi | N({\vec p})
\rangle^{sea}
+ \langle N^\prime ({\vec p}^\prime) | \Psi^\dagger  \hat O \Psi | N ({\vec p})
\rangle^{val}\,,
\end{equation}
where
\begin{eqnarray}
&&  \langle N^\prime ({\vec p}^\prime) | \Psi^\dagger\hat O \Psi | N ({\vec p})
\rangle^{sea}
\mathop{=}_{T\to+\infty}N_c  \frac 1{Z^\prime} \int \d^3 x\, \d^3x^\prime\,
\e^{- i{\vec p}^\prime {\vec x}^\prime + i{\vec p}
{\vec x}} \nonumber \\
&&
\times \int{\cal D}U\,
\Gamma_{N^\prime}^{\{f\}}\Gamma_N^{\{g\}\ast}
\prod\limits_{k=1}^{N_c}
\langle x^\prime,T/2| \frac{1}{D(U)}
|x,-T/2\rangle_{f_kg_k}
\Sp\Bigl\{\hat O_{dd^\prime} \langle 0,0 |
\frac{-1}{D(U)} | 0,0\rangle_{d^\prime d}\Bigr\}
%\nonumber\\
%&&\times
\e^{N_c {\rm Tr}\log [D(U)]} \,
\label{sea-contribution}
\end{eqnarray}
and
\begin{eqnarray}
&&\langle N^\prime ({\vec p}^\prime) | \Psi^\dagger \hat O \Psi(0) |
N ({\vec p})\rangle^{val} \mathop{=}_{T\to+\infty}N_c
\frac 1{Z^\prime} \int \d^3 x \d^3x^\prime \e^{- i{\vec p}^\prime {\vec x}^\prime +
i{\vec p} {\vec x}} \nonumber \\
&&\times
 \int {\cal D}U \,
\Gamma_{N^\prime}^{\{f\}}\Gamma_N^{\{g\}\ast}
\prod\limits_{k=2}^{N_c}
\langle \vec x^\prime,T/2 | \frac{1}{D(U)} |\vec x,-T/2
\rangle_{f_k g_k} \nonumber \\
&& \times
\langle x^\prime,T/2 | \frac{1}{D(U)} |  0,0 \rangle_{f_1d}
 \hat O_{dd^\prime}\langle  0,0|  \frac{1}{D(U)} | \vec x, -T/2
\rangle_{d^\prime g_1} \e^{N_c{\rm Tr}\log [D(U)]} \,.
\label{valence-contribution}
\end{eqnarray}

As in the case of the nucleon correlation function, the integral over the meson
fields can be done in the saddle-point approximation. Since the hedgehog
stationary solution $U_c$ breaks the rotational and translational symmetries
and we need nucleon states of both good spin and isospin numbers and a well
defined linear momentum, we have to consider
both the rotational and the translational zero modes. To this end we extend
the path integral over all time-dependent meson fields of the form:
\begin{equation}
U(x)= R(\tau) U_c({\vec x-\vec X(\tau) }) R^\dagger(\tau)\,,
\label{rotating-moving-soliton}
\end{equation}
and follow the same steps as in the case of the nucleon correlation function.
In  the leading order in $1/N_c$, we obtain (neglecting regularization)
for the sum of the level and sea contributions to the matrix element
\begin{eqnarray}
&& \bra{N'(\vec p')} \Psi^\dagger \hat O \Psi(0) \ket{N(\vec p)} = N_c
\int d^3x e^{i({\vec p}' - {\vec p}){\vec x}}
\nonumber\\
&&\times\int dR \,
\psi_{J'J_3'T'T_3'}^\ast(R) \sum_{\epsilon_n \leq \epsilon_{val}}
\Phi_n^\dagger({\vec x}) R^\dagger\hat O R \Phi_n({\vec x}) 
\psi_{JJ_3TT_3}(R)\,
\label{form-factor-leading-result}
\end{eqnarray}
a simple expression with a clear physical meaning: in the mean-field
approach the nucleon consists of quarks occupying all the levels
with the energy $\epsilon_n \leq \epsilon_{val}$.
In deriving (\ref{form-factor-leading-result}) we took into account the zero
modes of the soliton. The rotational motion of the soliton
results in the rotational wave functions of the soliton
$\psi_{JJ_3TT_3}(R)$ and the isospin rotation of the matrix
$\hat O\to R^\dagger\hat O R$. Similarly, the integral over $x$ and
$\exp(i(\vec p^\prime-\vec p) \vec x)$ appear due to the translational zero 
modes.
In the large $N_c$ approach the mass of the nucleon is $O(N_c)$,
whereas the momentum transfer in the nucleon matrix element
is $\vec q=\vec p^\prime-\vec p=O(N_c^0)$, so that
we work in the non-relativistic limit $|\vec q|\ll M_{cl}$.

Expression (\ref{form-factor-leading-result}) corresponds
to the leading order of $1/N_c$ expansion. In the subleading order
in $1/N_c$, there appear various corrections to this leading-order result.
Below we concentrate on the $1/N_c$ corrections connected to the
rotational motion of the soliton. One of the reasons
why we do it is that for some quantities the leading result
vanishes identically and the non-vanishing contributions
come from the rotational corrections.

Note that the rotational matrix element in (\ref{form-factor-leading-result})
is trivial if the quark current $\Psi^\dagger \hat O \Psi$ is isosinglet,
since in this case $R^\dagger \hat O R=\hat O$ and the integral over $R$
in (\ref{form-factor-leading-result}) reduces to the normalization matrix
element. In the case of the isovector current, we replace in the
above formulas:
\begin{equation}
\hat O \to O^a= \tilde O \tau^a\,,
\end{equation}
where we separate explicitly the isospin part of $O^b$.
Hence, we have
\begin{equation}
R^\dagger\, O^a\, R = D_{ab}(R)\,\tilde O \tau^b\,,
\label{Dab}
\end{equation}
where function $D_{ab}(R)$ is defined by
\begin{equation}
D_{ab}(R) = \frac 12\,\Sp(R^\dagger \tau^a R \tau^b)\,.
\label{Dab-definition}
\end{equation}
We see that the rotational part of the matrix element
(\ref{form-factor-leading-result}) reduces to the calculation
of the integral
\begin{equation}
\int{\cal D}R\,\psi_{N^\prime}^*[R(T/2)]\,
D_{ab}[R(0)]\,\psi_{N}[R(-T/2)]\, \e^{-{I\over 2}\int d\tau
\Omega^2}=\int\d R\,\psi_{N^\prime}^*(R)\,
D_{ab}(R)\,\psi_{N}(R) \e^{ - \frac{J(J+1)}{2I}T}
\,.
\label{DDabD0}
\end{equation}
In the left hand side, we have written the
corresponding path integral
over the time dependent orientation matrices $R(\tau)$
which appears in the systematic derivation of eq.
(\ref{form-factor-leading-result}) from the path integral formulas
(\ref{sea-contribution}), (\ref{valence-contribution}).

Now we turn to the rotational corrections to formula
(\ref{form-factor-leading-result}).
As it was explained earlier the angular velocity of the soliton
is parametrically small ($\Omega=O(1/N_c)$) and we can expand in it
\begin{eqnarray}
&&<\vec x,\tau|\frac 1{D(U_c)+i\Omega}|\vec x,\tau>=
<\vec x,\tau|\frac 1{D(U_c)}|\vec x,\tau>
\nonumber\\
&&
-\int\d^3x^\prime\d\tau^\prime<\vec x,\tau|\frac 1{D(U_c)}|\vec x^\prime,\tau^\prime> i\Omega(\tau^\prime)
<\vec x^\prime,\tau^\prime|\frac1{D(U_c)}|\vec x,\tau>
+\ldots\,.
\label{Dexpansion}
\end{eqnarray}
Inserting this expansion into the general formulas
(\ref{sea-contribution}), (\ref{valence-contribution}),
we arrive at the path integral containing both angular velocity
$\Omega[R(\tau)]$ and the D-function $D_{ab}[R(0)]$.
As long as we are working with the path integral approach
these two quantities are $c$-numbers. However, if we compute
this path integral and express it in terms of the operator approach,
we arrive at non-commuting collective operators
and the result generally depends on the ordering of these operators.
This point has provoked a lot of discussion
(Schechter and Weigel, 1995a,b), (Christov {\it et al.}, 1995a), 
(Wakamatsu, 1995) and it is often misunderstood. The reason is that the 
canonical quantization rule is considered independently of the path integral. 
Indeed, using some
``{\em naive}'' ordering of the collective variables $\Omega$ and
$D_{ab}$ (e.g. Wakamatsu and Watabe (1993) kept the order in which
these variables appear after the expansion (\ref{Dexpansion})) one faces the
problem of violation of the Pauli principle and of the $G$-parity symmetry
(Schechter and Weigel, 1995a,b). Such a problem does not exist using the 
$1/N_c$ expansion in the present semiclassical quantization scheme 
(Christov {\it et al.}, 1994, 1995a). In this scheme, as it was
already shown, the canonical quantization rule is a consequence of the exactly
calculated path integral over $R$ with the action quadratic in $\Omega$ 
and hence, it must be considered in the context of this path integral.
It is well known that the general formula connecting the path integral
and the operator approaches leads to time-ordering
of the operators which correspond to the time-dependent
$c$-number quantities of the path integral.
Therefore, using the quantization rule (\ref{quantization-rule})
we arrive at the following result for the rotational path integral
appearing in the formula for the rotational corrections
to the nucleon matrix element of a quark current:
\begin{eqnarray}
&&\int{\cal D}R\,\psi_{N}^*[R(T/2)]\,
D_{ab}[R(0)]\,\Omega^c[R(\tau)]\psi_{N}[R(-T/2)]\,
\e^{-{I\over 2}\int d\tau \Omega^2}\nonumber\\
&&
=-{i\over I}\int\d R\,\psi_{N}^*(R)\,
[\vartheta(-\tau)\,D_{ab}(R)\,J_c+\vartheta(\tau)\,J_c\,D_{ab}(R)]
\,\psi_{N}(R)
\e^{-{J(J+1)\over 2I}T}\,,
\label{DDabD1} \end{eqnarray}
sandwiched between the nucleon rotational wave functions. Using the spectral
representation of the quark propagator (\ref{spectral-representation}) we can
integrate over $\tau$ and finally get for the quark valence contribution
\begin{eqnarray}
&&  \langle N^\prime({\vec p}^\prime) |
\Psi^\dagger O^a \Psi | N({\vec
p}) \rangle^{val}
= N_c\int \d^3 x\e^{i({\vec p}^\prime - {\vec p}) {\vec x}}
\int\d R \, \psi_{N^\prime}^\ast(R)\Biggl\{
\Bigl[\Phi_{val}^\dagger({\vec x}) \, O^b \,\Phi_{val}({\vec
x})\Bigr]\,D_{ab}(R)
\nonumber\\
&&
+\frac{1}{2I} \sum\limits_{\epsilon_n\neq \epsilon_{val}}
 \frac{1}{\epsilon_{val} - \epsilon_n}
\Bigl\{ [\Phi_n^\dagger({\vec x})\, O^b\,\Phi_{val}({\vec x})]
\langle val |\tau^c| n \rangle[\theta(\epsilon_n) \, J_c \,D_{ab}(R)+
\theta(-\epsilon_n)\,D_{ab}(R) J_c ]
\nonumber \\
&&
+ [\Phi_{val}^\dagger({\vec x}) \,O^b \,\Phi_n({\vec x})] \,
\langle n |\tau^c| val \rangle [\theta(\epsilon_n) D_{ab} J_c+
\theta(-\epsilon_n)J_c \,D_{ab}(R)] \Bigr\} \Biggr\}
\psi_N(R)
\label{valence-contribution-result}
\end{eqnarray}

In the case of the sea part of the matrix element
(\ref{sea-contribution}) we have an additional problem with the divergences
of the model. The integrand in the r.h.s.
of eq.~(\ref{sea-contribution}) contains the fermionic determinant
Det$[\,D(U)]$ and the related generally divergent
matrix element
\begin{equation}
\Sp \Bigl[\hat O \langle 0 |  \frac{-1}{D(U)} | 0 \rangle\Bigr]
=\frac{\delta}{\delta\xi(0)}\Tr\log D_\xi
\Bigr|_{\xi=0}\,.
\label{xi-source}
\end{equation}
where
\begin{equation}
D_\xi = D(U_c) + i\Omega - \xi R^\dagger \hat O R \,.
\end{equation}
As we already stated only the real part of the fermion determinant is
divergent and needs to be regularized whereas the imaginary part is finite and
it is not regularized.

Let us first consider the sea part of the matrix element of a particular
current which contributes only to the imaginary part of $\Tr\log D_\xi(U)$. In
this case the matrix element given by (\ref{xi-source}) is finite and does
not contain any regulator. Following the same steps as for the valence
contribution, namely the expansion
(\ref{Dexpansion}) keeping the terms up to linear order in $\Omega$ and making
use of the identities (\ref{omega}) and  (\ref{Dab}), we get for the
sea part
\begin{eqnarray}
&&\langle N^\prime({\vec p}^\prime) | \Psi^\dagger O^a \Psi | N({\vec
p}) \rangle^{sea}
= N_c
\int \d^3 x \e^{i({\vec p}^\prime - {\vec p}) {\vec x}} \int \d R \,
\psi_{N^\prime}^\ast(R)
\Biggl\{ \sum\limits_{\epsilon_n<0}
\Phi_n^\dagger({\vec x}) \, O^b \,\Phi_n({\vec x})D_{ab}(R)
\nonumber\\
&&
+\frac{1}{2I}\sum\limits_{\epsilon_n>0\atop \epsilon_m<0}\frac{1}
{\epsilon_n-\epsilon_m}
%\nonumber\\
%&&\times
\Bigl\{ J_c D_{ab}(R)\left[\Phi_n^\dagger({\vec x}) \, O^b \,
\Phi_m({\vec x})\right]\langle m |\tau^c| n \rangle  +
\left[\Phi_m^\dagger({\vec x}) \,O^b  \,\Phi_n({\vec x})\right] \,
 J_c D_{ab}(R)\Bigr\}\langle n |\tau^c| m \rangle  \Biggr\}
\nonumber\\
&&
\times\psi_N(R)\,.
\label{sea-contribution-result}
\end{eqnarray}
Adding the valence contribution (\ref{valence-contribution-result})
to the sea part (\ref{sea-contribution-result}),
we obtain the final results for the matrix element of a given current whose
sea part originates from the imaginary part of the effective action:
\begin{eqnarray}
&&\langle N^\prime({\vec p}^\prime) | \Psi^\dagger O^a \Psi | N({\vec
p}) \rangle = N_c \int \d^3 x
\e^{i({\vec p}^\prime - {\vec p}) {\vec x}}
 \int \d R \, \psi_{N^\prime}^\ast(R)
\Biggl\{ \sum\limits_{\epsilon_n \leq \epsilon_{val}}
\Phi_n^\dagger({\vec x}) \,O^b \,\Phi_n({\vec x})\,D_{ab}(R)
\nonumber\\
&&
- \frac{1}{2I} \sum\limits_{\epsilon_n>\epsilon_{val} \atop
\epsilon_m\leq \epsilon_{val}}
 \frac{1}{\epsilon_n - \epsilon_m}
\Bigl\{[\Phi_n^\dagger({\vec x}) \, O^b\,\Phi_m({\vec
x})] \, \langle m |\tau^c| n \rangle J_c D_{ab}(R)
 + [\Phi_m^\dagger({\vec x}) \,O^b \,\Phi_n({\vec x})] \,
\langle n |\tau^c| m \rangle \,D_{ab}(R) J_c \Bigr\} \Biggr\}
\nonumber\\
&&
\times\psi_N(R)\,.
\label{form-factor-result}
\end{eqnarray}
It is easy to see that expression (\ref{form-factor-result}) has the same
structure as the sea contribution but with the valence quark included
into the occupied states.

In the case that the matrix element (\ref{xi-source}) comes from the
divergent real part of $\Tr\log D_\xi(U)$
it must be regularized. Actually, in order to make it
finite it is sufficient to start with the proper-time regularized
real part
\begin{eqnarray}
&&\mbox{Re}[\Tr \log D_\xi(U)]_{reg}\equiv
\frac 12 \Tr\log[D_\xi^\dagger(U) D_\xi(U)]_{reg}
\nonumber\\
&&=-\frac 12 \int\limits_{1/\Lambda^2}^{\infty}\frac{du}{u} \, \Tr
\left[ \e^{-uD_\xi^\dagger D_\xi}
-\e^{-uD_0^\dagger D_0}\right]\,.
\label{proper-time-regularization}
\end{eqnarray}
in the evaluation of the matrix element
\begin{eqnarray}
&&\Sp \Bigl[\hat O \langle x |  \frac{-1}{D(U)} | x \rangle\Bigr]
=\frac{\delta}{\delta\xi(x)} \mbox{Re} \,
(\Tr\log D_\xi)_{reg}
\nonumber\\
&&= -\frac 12\, \Sp \, R^\dagger(x_4)\hat O R(x_4)
\langle x | \e^{-D^\dagger D/\Lambda^2} D^{-1} | x \rangle
 - \frac 12 \, \Sp \, R^\dagger(x_4)\hat O^\dagger R(x_4)
\langle x |(D^\dagger)^{-1} \e^{-D^\dagger D/\Lambda^2} | x \rangle \,.
\label{variational-derivative-regularized}
\end{eqnarray}
Here $D$ stands for
\begin{equation}
D=D_{\xi=0} = \partial_\tau + h(U_c) + i\Omega \,
\end{equation}
Note that we assume $\xi(x)$ to be real but
the matrix $\hat O$ is not necessarily hermitian. After quantization
(\ref{quantization-rule}) of the time-ordered collective
operators and using the spectral representation for the
quark propagator (\ref{spectral-representation}) we get
the regularized sea contribution in the form
\begin{eqnarray}
&&  \langle N^\prime({\vec p}^\prime) | \Psi^\dagger O^a \Psi | N({\vec p})
\rangle_{reg}^{sea}
= N_c
\int \d^3 x
\e^{i({\vec p}^\prime - {\vec p}) {\vec x}}
 \int dR \, \psi_{N^\prime}^\ast(R)
%\nonumber\\
%&&
\biggl\{ \sum\limits_n \frac {1+\eta}2{\cal R}_1(\epsilon_n) \,
\Phi_n^\dagger({\vec x}) \, O^b \,\Phi_n({\vec x})\,D_{ab}(R)
\nonumber\\
&&
+ \frac{1}{4I} \sum\limits_{m,n}
\Bigl[ \,
[\Phi_m^\dagger({\vec x}) \,O^b \,\Phi_n({\vec x})]
\langle n |\tau^a| m \rangle \,-\eta
[\Phi_n^\dagger({\vec x}) \,O^b \,\Phi_m({\vec x})] \,
\langle m|\tau^c| n \rangle \Bigr] \,
\nonumber\\
&&
{}_\times\Bigl[{\cal R}_2^{(+)}(\epsilon_m,\epsilon_n)\,  J_c D_{ab}(R)
+ {\cal R}_2^{(-)}(\epsilon_m,\epsilon_n)\, D_{ab}(R)J_c\Bigr] \biggr\}
\psi_N(R) \, .
\label{sea-regularized}
\end{eqnarray}
The coefficient $\eta=1$ for hermitian matrix $\hat O$ and $\eta=-1$ for
anti-hermitian one:
\begin{equation}
\hat O^\dagger = \eta \hat O \,.
\label{eta-definition}
\end{equation}
The regularization functions are given by
\begin{equation}
{\cal R}_1(\epsilon_n)=-\epsilon_n
\int\limits_{-\infty}^{\infty}\frac{\d\omega}{2\pi}\, 
\frac{\e^{[-(\omega^2+\epsilon_n^2)/\Lambda^2]}}{\omega^2+\epsilon_n^2}
\,
\end{equation}
and
\begin{eqnarray}
&&{\cal R}_2^{(\pm)}(\epsilon_m,\epsilon_n)=
\int\limits_{-\infty}^{\infty}\frac{\d\omega}{2\pi}
\int\limits_{-\infty}^{\infty}\frac{\d\omega^\prime}{2\pi} \,
 \frac{1}{\pm i(\omega - \omega^\prime) +0}
\nonumber\\
&&\times
\biggl\{ \frac{\e^{-[\omega^2+\epsilon_n^2]/\Lambda^2}}
{(i\omega + \epsilon_n)(i\omega^\prime + \epsilon_m)}
%\nonumber\\
%&&
-\frac{1}{\Lambda^2}
\left[1+\frac{i\omega - \epsilon_n}{i\omega^\prime + \epsilon_m}  \right]
\int_0^1\d\alpha
\e^{-[\alpha(\omega^2+\epsilon_n^2)
+(1-\alpha)({\omega^\prime}^2+\epsilon_m^2)]/\Lambda^2}
\biggr\}.
\end{eqnarray}

As can be seen in the present scheme, the matrix element of a given current is
factorized in a quark and collective part. The latter determines the
spin-flavor structure. The particular form of the matrix element
depends on the hermitian properties of the current $\hat O$ and of the
symmetry of the quark matrix element under $m \leftrightarrow n$.

\newpage
\section{Nucleon properties in the SU(2) NJL model}

Using the formalism developed in the previous section, we are in a
position to calculate the nucleon observables. We start with the
electromagnetic form factors of the nucleon
and the related static properties like radii and magnetic moments.
The $E2/M1$ ratio for the $\gamma N \rightarrow \Delta$ transition is also
considered. Further, the axial properties of the nucleon, namely the weak axial
and pseudoscalar form factors and the axial and induced pseudoscalar
coupling constants,  are examined. The results for the pion nucleon form
factor and the related strong pion nucleon coupling constant are also
presented. We discuss the PCAC and the Goldberger-Treiman
relation and consider the nucleon electric polarizability. 
We make some prediction for the tensor charges which is
related to the quark transverse structure function.

%\newpage
%\vskip1cm
4.1 \underline{Electromagnetic Properties}
\vskip4mm

The nucleon electromagnetic Sachs form factors are related to the matrix
element of the electromagnetic current
\begin{equation}
j^\mu = \Psi^\dagger\gamma^0\gamma^\mu \hat Q\Psi
\label{em-current}
\end{equation}
in the standard way:
\begin{equation}
\langle N^\prime (p^\prime) | j^0(0) | N(p) \rangle = G_E(q^2)
\delta_{J_3^\prime J_3} \,, 
\label{Sachs-electric-form-factor}
\end{equation}
\begin{equation}
\langle N^\prime (p^\prime) | j^k(0) | N(p) \rangle = \frac{i}{2
M_N}
\varepsilon^{klm} (\tau^l)_{J_3^\prime J_3} q^m
 G_M(q^2)\,,
\label{Sachs-magnetic-form-factor}
\end{equation}
Here $M_N=938$ MeV is the physical nucleon mass, $Q$ is the quark
charge matrix
\begin{equation}
\hat Q=\left( \begin{array}{cc}
         2/3  & 0 \\
         0  & -1/3     \end{array}\right) = \frac{1}{6} + \frac{\tau^3}{2}
\, .
\label{quark-charge-matrix}
\end{equation}
and
\begin{equation}
q=p^\prime-p \,
\end{equation}
is the momentum transfer. For convenience, we also will use the notation 
$Q^2=-q^2$. Accordingly, the isoscalar and isovector parts of the form factors
are defined by
\begin{equation}
G_{E(M)} = \frac12 G_{E(M)}^{T=0} + T_3 G_{E(M)}^{T=1} \,.
\label{isoscalar-isovector}
\end{equation}

In order to examine the divergent structure of the nucleon matrix element of
the current $j^\mu$ it is enough to consider the regularized
determinant with a small external source $\xi$
(\ref{xi-source}). In the case of the electromagnetic form factors
this external source is an electromagnetic field $A_\mu$ coupled to
the current
$j^\mu$:
\begin{equation}
D(U) - \xi \hat O \quad \to \quad
D(U, \hat QA_\mu) = \gamma_0[\gamma_\mu(\partial_\mu - i \hat Q A_\mu)
+ MU^{\gamma_5}]\, .
\end{equation}
The proper-time regularization in the Euclidean space
(\ref{proper-time-regularization}) contains operator $D^\dagger D$, and
in order to preserve the gauge invariance
the Euclidean electromagnetic field $A_\mu$ must be consider real.
Since the complex conjugate euclidean
Dirac matrices $\gamma_\mu^\ast$ are  connected
to $\gamma_\mu$ by some unitary transformation $V$ (\ref{V}),
it is easy to show that
\begin{equation}
\Tr\log D(U, \hat QA_\mu)=\Tr\log [(V\tau^2) D(U,\hat QA_\mu) (V\tau^2)^{-1}] =
\Tr\log [D(U,\hat Q^\prime A_\mu)]^\ast\,.
\label{D-D-ast}
\end{equation}
Here
\begin{equation}
\hat Q^\prime =  - \frac{1}{6} + \frac{\tau^3}{2}\,,
\label{Q-prime}
\end{equation}
and the asterisk stands for the complex (not hermitian!) conjugation.
Equality (\ref{D-D-ast}) shows that the
isoscalar electromagnetic form factors originate from the
imaginary part of $\Tr\log D(U, \hat QA_\mu)$, whereas the isovector form
factors are generated by the real part of $\Tr\log D(U, \hat QA_\mu)$.
Since the imaginary part of $\Tr\log D(U,\hat QA_\mu)$ is ultraviolet finite,
for the isoscalar form factors we use directly
the non-regularized expression (\ref{form-factor-result}), whereas the sea
contribution to the isovector form factors must be regularized
(\ref{sea-regularized}).

Under the transformation $V\tau^2$ (\ref{D-D-ast}) we also have
\begin{equation}
(V\tau^2) \gamma_\mu (V\tau^2)^{-1}  = \gamma_\mu^T,
\quad
(V\tau^2) \tau^a (V\tau^2)^{-1} = - (\tau^a)^\ast = -(\tau^a)^T,
\end{equation}
\begin{equation}
(V\tau^2) h (V\tau^2)^{-1} = h^\ast = h^T\,,
\end{equation}
where the superscript $T$ stands for the transposition in both the
matrix indices and the coordinate space. The last equation shows that
the eigenfunctions of $h$ can be chosen in such a way that
\begin{equation}
(V\tau^2) \Phi_m({\vec x}) = \Phi_m^\ast({\vec x})\,.
\end{equation}
Hence, the following relations hold:
\begin{equation}
[\Phi_n^\dagger({\vec x}) \gamma_0\gamma_\mu \Phi_m({\vec x})]
\langle m |\tau^a| n \rangle
= \eta\, [\Phi_m^\dagger({\vec x})  \gamma_0\gamma_\mu \Phi_n({\vec x})]
\langle n | \tau^a| m \rangle  \,,
\label{gamma-transposition}
\end{equation}
and
\begin{equation}
[\Phi_n^\dagger({\vec x}) \gamma_0\gamma_\mu\tau^a \Phi_m({\vec x})]
\langle m | \tau^a| n \rangle
 = -\eta\,[\Phi_m^\dagger({\vec x})\gamma_0\gamma_\mu\tau^a
\Phi_n({\vec x})] \langle n | \tau^a| m \rangle
\label{gamma-tau-transposition}
\end{equation}
where the values for $\eta$ are determined by the hermitian properties of
$i\gamma_0\gamma_\mu\tau^a$:
\begin{equation}
\eta
=\left\{ \begin{array}{ll}
         -1  & \mbox{if }   \mu=0  \\
          1  &\mbox{if }   \mu=1,2,3  .   \end{array} \right.
\label{eta-definition-2}
\end{equation}

The rotational matrix elements, which we need,
can be easily computed with the relations
\begin{equation}
\langle T_3^\prime J_3^\prime  |J_{a}| T_3 J_3 \rangle = \frac 12
\,\delta_{T_3^\prime T_3}\,(\tau^a)_{J_3^\prime J_3} \,,
\end{equation}
\begin{equation}
\langle T_3^\prime J_3^\prime  |D_{ab}(R)| T_3 J_3 \rangle = - \frac 13 \,
(\tau^a)_{T_3^\prime T_3}\, (\tau^b)_{J_3^\prime J_3} \,.
\end{equation}

Now using the symmetry properties of the quark matrix element we can
apply the general formulas for the nucleon matrix
elements derived in the previous section to the case of the
electromagnetic current (\ref{em-current}) for evaluation of the nucleon
electromagnetic form factors (\ref{Sachs-electric-form-factor}).

For the isoscalar electric form factor we obtain a simple result:
\begin{eqnarray}
  G_E^{T=0}(q^2)= \frac{N_c}{3}\int \d^3 x
\e^{i{\vec q} {\vec x}} \Bigl\{\sum\limits_{\epsilon_n \leq
\varepsilon_{val}} \Phi_n^\dagger({\vec x}) \Phi_n({\vec x})
- \sum\limits_{\epsilon_n^{(0)} <0}
\Phi_n^{(0)\dagger}({\vec x}) \Phi_n^{(0)}({\vec x}) \Bigr\}\,,
\label{form-factor-electric-isosinglet}
\end{eqnarray}
where the vacuum contribution is subtracted. Since the rotational corrections
exactly vanish this form factor contains only
leading order terms . The baryon number
is given by the form factor at $q^2=0$:
\begin{eqnarray}
B=G_E^{T=0}(0)=\int \d^3 x \Bigl\{\sum\limits_{\epsilon_n \leq
\epsilon_{val}} \Phi_n^\dagger({\vec x}) \Phi_n({\vec x})
- \sum\limits_{\epsilon_n^{(0)} <0}
\Phi_n^{(0)\dagger}({\vec x}) \Phi_n^{(0)}({\vec x}) \Bigr\}=\int \d^3 x
\Phi_{val}^\dagger({\vec x}) \Phi_{val}({\vec x})=1\,.
\label{baryon-number-electric-isosinglet}
\end{eqnarray}
As it was mentioned in the previous chapter it originates from the
imaginary part of the effective action and is determined by the valence
quarks.

For the electric isovector form factor we obtain:
\begin{eqnarray}
&&
  G_E^{T=1}(q^2)
=  \frac{N_c}{6I} \,
\int \d^3 x
\e^{i \vec q\vec x}
 \biggl\{
   \sum\limits_{m,n}
 {\cal R}_I^\Lambda(\epsilon_m,\epsilon_n)
[\Phi_m^\dagger({\vec x}) \,   \tau^a  \, \Phi_n({\vec x})] \,
\langle n |\tau^a| m \rangle
\nonumber\\
&&
-  \sum\limits_{\epsilon_n\neq \epsilon_{val}}
\frac{1}{\epsilon_{val} - \epsilon_n}
[\Phi_n^\dagger({\vec x}) \,   \tau^a  \,
\Phi_{val}({\vec x})] \,
\langle val |\tau^a| n \rangle
\biggl\}\,,
\label{electric-isovector-form-factor}
\end{eqnarray}
where the proper-time regulator ${\cal R}_I^\Lambda$ is the same
as in the case of the moment of inertia (\ref{regmomentinertia}).
Hence, the isovector electric form
factor has a proper normalization (electric charge):
\begin{equation}
G_E^{T=1}(0) = 1\,.
\end{equation}
It should be mentioned that in contrast to the isoscalar electric form factor
the isovector one vanishes in leading order ($\Omega^0\sim N_c^0$) and
the only
non-zero contribution comes from the rotational corrections (next to
leading order terms in $\Omega$).

As in the case of the isovector electric form factor, the isoscalar
magnetic one also vanishes in the leading order in $\Omega$, and we find
non-vanishing contributions only from the $1/N_c$ rotational
corrections:
\begin{equation}
G_M^{T=0}(q^2) = \frac{N_c M_N}{6I} \, \varepsilon^{kaj} \,
\frac{iq^j}{|q^2|}
\int \d^3x \e^{i{\vec q}{\vec x}}
\sum\limits_{\scriptstyle \epsilon_n > \epsilon_{val}
\atop \epsilon_m \leq \epsilon_{val}   }
\frac{[\Phi_m^\dagger({\vec x})\gamma^0\gamma^k \Phi_n({\vec x})]
\, \langle n |\tau^a | m \rangle}
{\epsilon_m - \epsilon_n}   \,.
\label{magnetic-isoscalar-form-factor}
\end{equation}

The calculations of the magnetic isovector form factor are more
involved and the final expression
\begin{eqnarray}
G_M^{T=1}(q^2) &=& \frac{N_c{\cal M}_N}{3} \, \varepsilon^{kbj} \,
\frac{iq^j}{|q^2|}\int \d^3x \e^{i{\vec q}{\vec x}}
%\nonumber \\
%&&
\Biggl\{
[\Phi_{val}^\dagger({\vec x})\gamma^0\gamma^k\tau^b \Phi_{val}({\vec x})]
 - \sum\limits_n {\cal R}_{M1}^\Lambda(\epsilon_n) \,
[\Phi_n^\dagger({\vec x})\gamma^0\gamma^k\tau^b  \Phi_n({\vec x})]
\nonumber \\
&&-\frac{i}{2I} \varepsilon_{abc}\Biggl[ \sum\limits_{n,m}
{\cal R}_{M2}^\Lambda(\epsilon_m,\epsilon_n) \,
[\Phi_m^\dagger({\vec x})\gamma^0\gamma^k \tau^a \Phi_n({\vec x})]
\,\langle n |\tau^c |m\rangle
\nonumber \\
&&-\sum\limits_{\epsilon_n \neq \epsilon_{val}}
\frac{\mbox{sign}(\epsilon_n) }
{\varepsilon_{val} - \epsilon_n} \,
[\Phi_n^\dagger({\vec x})\gamma^0\gamma^k \tau^a \Phi_{val}({\vec x})]
\,\langle val |\tau^c | n \rangle
\Biggr]\Biggr\} \,
\label{isovector-magnetic-form-factor}
\end{eqnarray}
includes leading order terms (the first two terms in the rhs)  as well
as rotational corrections. Accordingly, two different regularization 
functions appear:
\begin{equation}
{\cal R}_{M1}^\Lambda(\epsilon_n) = - {\cal R}_1(\epsilon_n)
=\frac{\epsilon_n}{2\sqrt{\pi}}
\int\limits_{1/\Lambda^2}^{\infty}\frac{\d u}{\sqrt{u}}
\e^{-u\epsilon_n^2}
\,,
\end{equation}
\begin{eqnarray}
{\cal R}_{M2}^\Lambda(\epsilon_m,\epsilon_n) =
 \frac{1}{4\pi} \int_0^1 \frac{\d\beta}{\sqrt{\beta(1-\beta)}} \,
\frac{(1-\beta)\epsilon_m - \beta\epsilon_n}
{(1-\beta)\epsilon_m^2 + \beta\epsilon_n^2} \,
\e^{-[(1-\beta)\epsilon_m^2 + \beta\epsilon_n^2]/\Lambda^2} \,.
\end{eqnarray}

Eqs.
(\ref{form-factor-electric-isosinglet}),
(\ref{electric-isovector-form-factor}),
(\ref{magnetic-isoscalar-form-factor}) and
(\ref{isovector-magnetic-form-factor})
are our final expressions for the electromagnetic
isoscalar and isovector form factors in the NJL model.
According to (\ref{isoscalar-isovector})
the proton and neutron form factors are expressed in terms of the isoscalar
and isovector form factors as follows
\begin{equation}
G_{E(M)}^p = \frac12 [G_{E(M)}^{T=0} +  G_{E(M)}^{T=1}] \,,
\end{equation}
\begin{equation}
G_{E(M)}^n = \frac12 [G_{E(M)}^{T=0} -  G_{E(M)}^{T=1}]\,.
\label{FFPNb}
\end{equation}

\begin{figure}
\centerline{\epsfysize=5.5in\epsffile{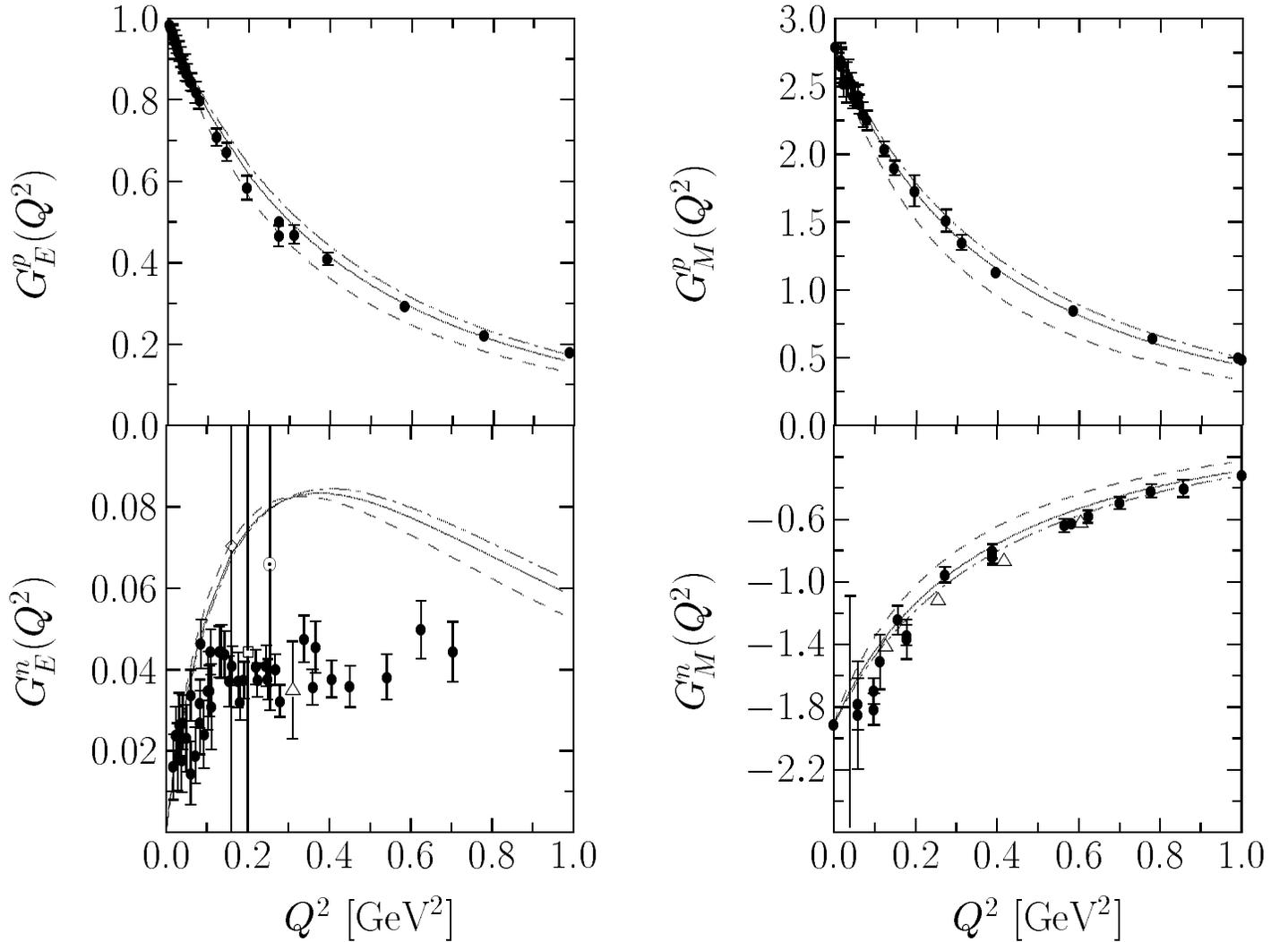}}%\vskip4pt
\caption{Electric (left) and magnetic (right) form
factors of the nucleon as functions of $Q^2$ for three different values of 
$M$: 370 (dashed line), 
420 (solid line) and 
450 (dash-dotted line) MeV in the SU(2)-NJL model 
(Christov {\it et al.}, 1995). The experimental data in the case of the 
electric form factor are from (H\"ohler {\it et al.}, 1976) for the proton 
and from (Platchkov {\it et al.}, 1990, Eden {\it et al.}, 1994, Meyerhoff 
{\em et al.}, 1994) 
for the neutron. For the magnetic form factors they are from (H\"ohler 
{\em et al.}, 1976, Bruins {\em et al.}, 1995). 
The theoretical magnetic form factors are normalized to the experimental 
values of the magnetic moments.}
\label{Figr2}
\end{figure}

The nucleon electric and magnetic form factors
are displayed in figs.\ref{Figr2} a) and b). The theoretical curves
(Christov {\it et al.}, 1995b) are given for three different values of the
constituent quark mass, $370, 420$ and $450$ MeV.
The magnetic form factors are normalized to the experimental
values of the corresponding magnetic moments at $Q^2=0$.

With one exception, the neutron electric form factor, 
all other form factors agree with the experimental data quite well
for the constituent quark mass around $420$ MeV. 
It should be noted, however, that the neutron electric form factor,
which deviates from the experimental data, is most sensitive
to the used approximations and numerical errors.  According to the
formula (\ref{FFPNb}) the form factor has been calculated as a
difference of the electric isoscalar and electric isovector form
factors. Both form factors are of order of one, whereas
the resulting neutron form factor has experimental
values of order $0.04$, {\it i.e.} about $4\%$ of the value of its
components. It means that the numerical accuracy together with the used
large $1/N_c$ approximation behind the model
are strongly magnified and could in principle be an explanation for the
deviation from the experimental data. It should be also stressed that the
experimental data (see Platchkov {\it et al.}, 1990), available for
the neutron electric form factor, are strongly model
dependent and a different N - N potential used
in the analysis of the data can lead to an enhancement of the
experimental numbers by more than 50 \%. 

The mean squared radii and the magnetic moments are obtained from the 
form factors:
\begin{equation}
\langle r^2 \rangle^{T=0,1}_{E(M)} = -{6\over G_{E(M)}^{T=0,1}}\,
{\d G_{E(M)}^{T=0,1}\over \d q^2}\, \Bigg\vert_{q^2=0} \ ,
\label{ELRAD} \end{equation}
\begin{equation}
\mu^{T=0,1} = G_M^{T=0,1} (q^2) \Big\vert_{q^2=0} \ .
\label{MAGMOM} \end{equation}
and the results (Christov {\it et al.}, 1995b) for three values of the  quark
constituent mass,
370, 420 and 450 MeV, and pion mass $m_{\pi}=140$ MeV are
presented in table~\ref{Tabl1}. The value $\sim 420$ MeV is favored
which agrees with the conclusion drawn from the form factors.
With the exception of the neutron electric squared radii, to
which remarks similar to the case of the neutron electric form factor
are valid, the contribution of the valence quarks is dominant. However,
the contribution of the Dirac sea is non-negligible and it varies within
the range 15 -- 40\%. As can be seen, the numerical results for
the nucleon $N$--$\Delta$ mass splitting,
the mean squared proton, isovector electric radii as well as the
$q$-dependence of the proton electric and magnetic and the neutron
magnetic form factor differ from the experimental data by no more than
about $\pm 5\%$. Finally, for the magnetic moments the results are by
25--30\% below their experimental values. However, the experimental ratio
$\mu_p/\mu_n$ is almost exactly reproduced which is not the case of other
chiral models of nucleon.

\begin{table}
\caption{ Nucleon observables calculated in the SU(2) NJL model 
with the physical pion mass.}
\vbox{\offinterlineskip
\hrule height1pt
\halign{&\vrule width1pt#&
 \strut\quad\hfill#\hfill\quad
 &\vrule#&\quad\hfill#\hfil\quad
 &\vrule#&\quad\hfill#\hfil\quad
 &\vrule#&\quad\hfill#\hfil\quad
 &\vrule#&\quad\hfill#\hfil\quad
 &\vrule#&\quad\hfill#\hfil\quad
 &\vrule#&\quad\hfill#\hfil\quad
 &\vrule width1pt#&\quad\hfil#\hfil\quad\cr
height6pt&\omit&&\multispan{11}&&\omit&\cr
& &&\multispan{11} \hfill {\bf Constituent\ Quark\ Mass} \hfill &&
&\cr
height4pt&\omit&&\multispan{11}&&\omit&\cr
&\hfill {\bf Quantity}\hfill &&\multispan3 \hfill 370\ MeV\hfill
&&\multispan3
\hfill 420   MeV\hfill
&&\multispan3
\hfill 450  MeV\hfill
&&\hfill {\bf Exper.}\hfill &\cr
height3pt&\omit&&\multispan3 && \multispan3 &&
\multispan3 &&\omit&\cr
&\omit&&\multispan{11}\hrulefill&&\omit&\cr
height3pt&\omit&&\omit&&\omit&&\omit&&\omit&&\omit&&\omit&&\omit&\cr
&\hfill    &&\hfill total \hfill &&\hfill sea \hfill &&\hfill total
\hfill &&\hfill sea \hfill && \hfill total \hfill  &&\hfill sea \hfill  &&
&\cr 
height3pt&\omit&&\omit&&\omit&&\omit&&\omit&&\omit&&\omit&&\omit&\cr
\noalign{\hrule height1pt}
height3pt&\omit&&\omit&&\omit&&\omit&&\omit&&\omit&&\omit&&\omit&\cr

& \hfill $ <r^2>_{T=0}$ [fm$^2$] \hfill &&
0.63  &&
0.05  &&
0.52  &&
0.07  &&
0.48  &&
0.09  &&
0.62  &\cr
height3pt&\omit&&\omit&&\omit&&\omit&&\omit&&\omit&&\omit&&\omit&\cr
\noalign{\hrule}
height3pt&\omit&&\omit&&\omit&&\omit&&\omit&&\omit&&\omit&&\omit&\cr

& \hfill $ <r^2>_{T=1}$ [fm$^2$] \hfill &&
1.07  &&
0.33  &&
0.89  &&
0.41  &&
0.84  &&
0.45  &&
0.86 &\cr
height3pt&\omit&&\omit&&\omit&&\omit&&\omit&&\omit&&\omit&&\omit&\cr
\noalign{\hrule}
height3pt&\omit&&\omit&&\omit&&\omit&&\omit&&\omit&&\omit&&\omit&\cr

& \hfill $ <r^2>_p$ [fm$^2$] \hfill &&
 0.85  &&
 0.19  &&
 0.70  &&
 0.24  &&
 0.66  &&
 0.27  &&
 0.74  &\cr
height3pt&\omit&&\omit&&\omit&&\omit&&\omit&&\omit&&\omit&&\omit&\cr
\noalign{\hrule}
height3pt&\omit&&\omit&&\omit&&\omit&&\omit&&\omit&&\omit&&\omit&\cr

& \hfill $ <r^2>_n$ [fm$^2$]\hfill  &&
--0.22   &&
--0.14   &&
--0.18   &&
--0.17   &&
--0.18   &&
--0.18   &&
--0.11   &\cr
height3pt&\omit&&\omit&&\omit&&\omit&&\omit&&\omit&&\omit&&\omit&\cr
\noalign{\hrule}
height3pt&\omit&&\omit&&\omit&&\omit&&\omit&&\omit&&\omit&&\omit&\cr

& \hfill $\mu_{T=0}\hfill$ [n.m.]  &&
0.68  &&
0.09  &&
0.62  &&
0.03  &&
0.59  &&
0.05  &&
0.88  &\cr
height3pt&\omit&&\omit&&\omit&&\omit&&\omit&&\omit&&\omit&&\omit&\cr
\noalign{\hrule}
height3pt&\omit&&\omit&&\omit&&\omit&&\omit&&\omit&&\omit&&\omit&\cr

& \hfill $ \mu_{T=1}$ [n.m.] \hfill &&
 3.56 &&
 0.77 &&
 3.44 &&
 0.97 &&
 3.16 &&
 0.80 &&
 4.71        &\cr
height3pt&\omit&&\omit&&\omit&&\omit&&\omit&&\omit&&\omit&&\omit&\cr
\noalign{\hrule}
height3pt&\omit&&\omit&&\omit&&\omit&&\omit&&\omit&&\omit&&\omit&\cr

& \hfill $ \mu_p $ [n.m.] \hfill &&
  2.12 &&
  0.43 &&
  2.03 &&
  0.50 &&
  1.86 &&
  0.43 &&
  2.79        &\cr
height3pt&\omit&&\omit&&\omit&&\omit&&\omit&&\omit&&\omit&&\omit&\cr
\noalign{\hrule}
height3pt&\omit&&\omit&&\omit&&\omit&&\omit&&\omit&&\omit&&\omit&\cr
& \hfill $ \mu_n $ [n.m.] \hfill &&
 --1.44 &&
 --0.34 &&
 --1.41 &&
 --0.47 &&
 --1.29 &&
 --0.38 &&
 --1.91    &\cr
height3pt&\omit&&\omit&&\omit&&\omit&&\omit&&\omit&&\omit&&\omit&\cr
\noalign{\hrule}
height3pt&\omit&&\omit&&\omit&&\omit&&\omit&&\omit&&\omit&&\omit&\cr

& \hfill $ |\mu_p/\mu_n|$ \hfill  &&
  1.47  &&
  ---   &&
  1.44  &&
  ---   &&
  1.44  &&
  ---   &&
  1.46           &\cr
height3pt&\omit&&\omit&&\omit&&\omit&&\omit&&\omit&&\omit&&\omit&\cr
\noalign{\hrule}
height3pt&\omit&&\omit&&\omit&&\omit&&\omit&&\omit&&\omit&&\omit&\cr
& \hfill $ <r^2>^\mu_p$ [fm$^2$]  &&
 1.08  &&
 0.32  &&
 0.66  &&
 0.28  &&
 0.56  &&
 0.25  &&
 0.74  &\cr
height3pt&\omit&&\omit&&\omit&&\omit&&\omit&&\omit&&\omit&&\omit&\cr
\noalign{\hrule}
height3pt&\omit&&\omit&&\omit&&\omit&&\omit&&\omit&&\omit&&\omit&\cr

& \hfill $ <r^2>^\mu_n$ [fm$^2$]  &&
--1.17   &&
--0.51   &&
--0.65   &&
--0.31   &&
--0.52   &&
--0.24   &&
--0.77   &\cr
height3pt&\omit&&\omit&&\omit&&\omit&&\omit&&\omit&&\omit&&\omit&\cr
\noalign{\hrule}
height3pt&\omit&&\omit&&\omit&&\omit&&\omit&&\omit&&\omit&&\omit&\cr

& \hfill $ M_{\Delta}-M_N$ [MeV]\hfill &&
 213  &&
 ---  &&
 280  &&
 ---  &&
 314  &&
 ---  &&
 294  &\cr
height3pt&\omit&&\omit&&\omit&&\omit&&\omit&&\omit&&\omit&&\omit&\cr
\noalign{\hrule }
height3pt&\omit&&\omit&&\omit&&\omit&&\omit&&\omit&&\omit&&\omit&\cr

& \hfill $ g_A  $\hfill &&
 1.26  &&
 0.08  &&
 1.21  &&
 0.11  &&
 1.13  &&
 0.06  &&
 1.26   &\cr
height3pt&\omit&&\omit&&\omit&&\omit&&\omit&&\omit&&\omit&&\omit&\cr
\noalign{\hrule height1pt} 
}}
\label{Tabl1}
\end{table}

The magnetic isovector form factor is the
only one which includes non-zero contributions in both leading ($N_c^0$) and
next to leading order ($1/N_c$) rotational corrections. However, the
numerical calculations show that the ($1/N_c$) rotational corrections do not
affect the $q^2$-dependence (slope) of the form factor but rather the
value at the origin, $G_M^{T=1}(0)\,=\,\mu^{T=1}$ which is the isovector
magnetic moment. It is not surprising, since (as can be seen from
eq.(\ref{isovector-magnetic-form-factor})) in both the leading and next
to leading order terms the shape of the wave
functions $\Phi_n$ determines the $q^2$-dependence of the form factor,
whereas the value at $q^2=0$ depends on the particular matrix elements
included. In leading order the isovector magnetic moment is
strongly underestimated (Wakamatsu and Yoshiki, 1991). The enhancement for
this quantity due to the $1/N_c$ corrections is of order $(N_c+2)/N_c$
(Christov {\it et al.}, 1994) and it improves considerably the agreement with
experiment. It is also important to note that since the $1/N_c$ rotational
corrections have the same spin-flavor structure like the leading term,
they do not violate the consistency condition of Dashen and
Manohar (1993) derived in the large-$N_c$ of QCD. The
latter means that other $1/N_c$ corrections (e.g. meson loops) should
contribute to both the leading and next to leading orders.

\begin{figure}
\centerline{\epsfysize=3.in\epsffile{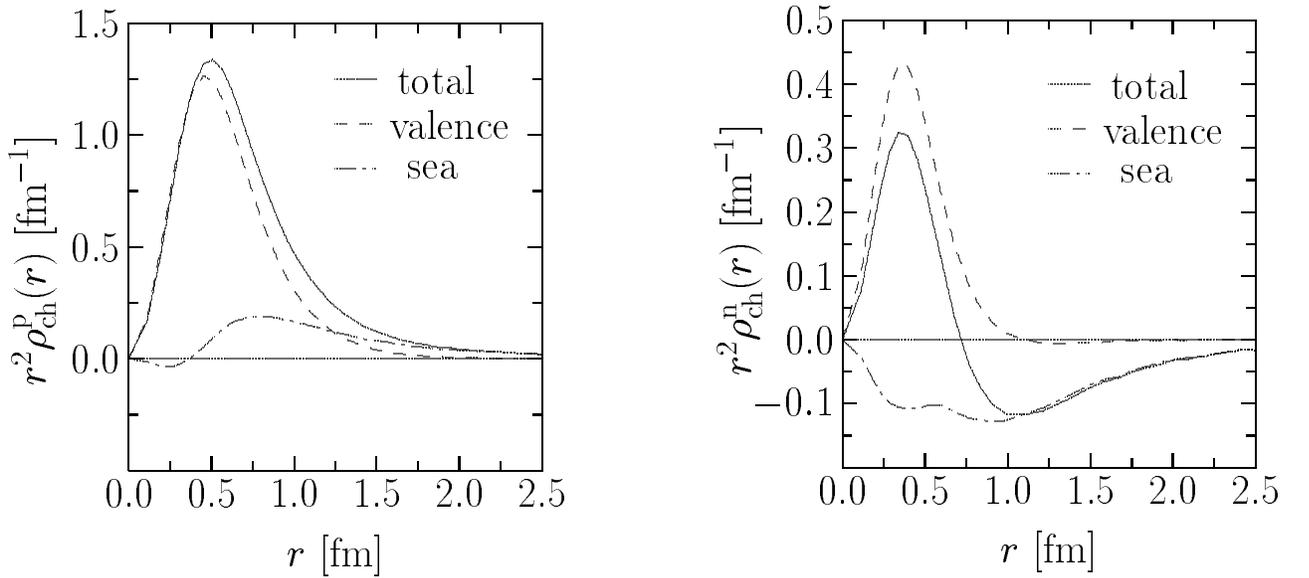}}%\vskip4pt
\caption{ Charge density distribution of the proton (lower) and
neutron (upper) for the constituent quark mass $M = 420$ MeV taken from 
(Christov  {\it et al.}, 1995b).}
\label{Figr10}
\end{figure}

\begin{figure}
\centerline{\epsfysize=4.0in\epsffile{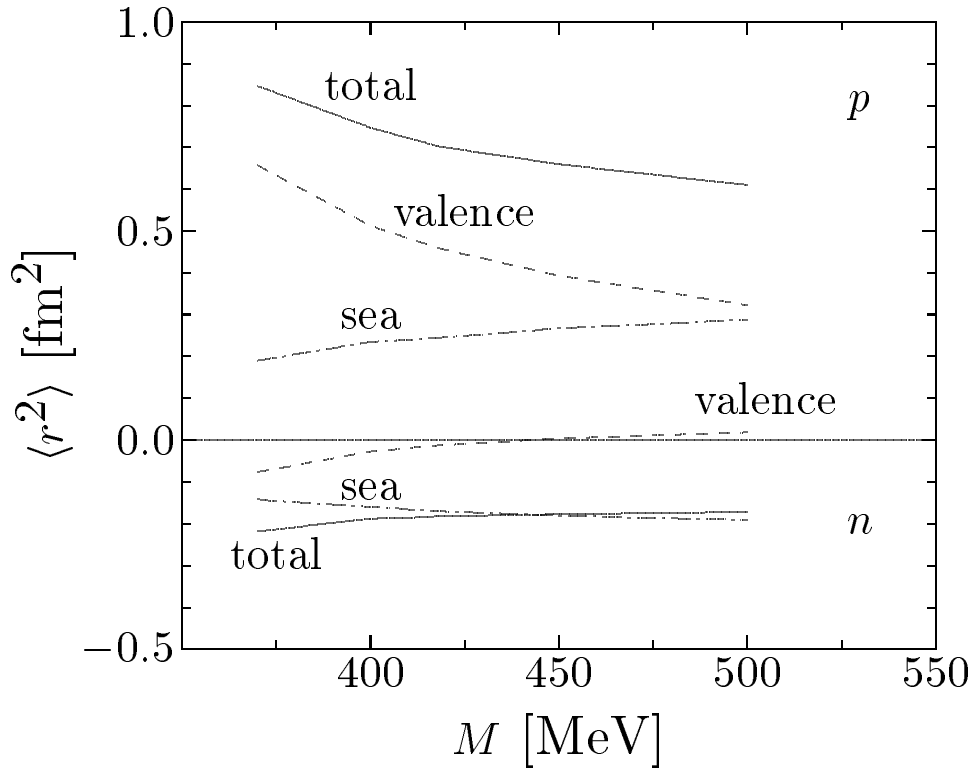}}%\vskip4pt
\caption{Electric charge radii of proton and neutron
as functions of the constituent quark mass $M$ in the SU(2) NJL model. 
The valence and sea parts are marked by the dashed and dash-dotted lines, 
respectively (Christov {\it et al.}, 1995b).}
\label{Figr9}
\end{figure}

In fig.\ref{Figr10} we plot the typical proton and neutron charge
distributions from (Christov {\it et al.}, 1995b) which is similar to those 
of Wakamatsu (1992) and Gorski {\it et al.} (1992) for the 
constituent quark mass $M\,=\,420\,$ MeV. For the
proton we have a positive definite charge distribution completely dominated
by the valence contribution whereas in the case of the neutron the Dirac sea
is dominant. In accordance with the well accepted phenomenological
picture, one realizes a positive core coming
from the valence quarks and a long negative tail due to the polarization
of the Dirac sea. Using the gradient expansion the latter can be expressed
in terms of the dynamical pion field -- pion cloud. It is reflected on
the proton and neutron charge radii shown on fig.\ref{Figr9}.
In the case of proton the sea contribution is about 30\% whereas for the
neutron charge radius the negative sea part is dominating and the valence
contribution is negligible.

The theoretical values of the static nucleon properties but with the physical
pion mass set to zero can be found in (Christov {\it et al.}, 1995b) and the 
corresponding electric form factors in (Gorski {\it et al.}, 1992). The
chiral limit ($m_{\pi}\to 0$)
mostly influences the isovector charge radius. In fact, as it is
expected (B\'eg and Zapeda 1972, Adkins {\it et al.}, 1983), the isovector
charge radius diverges in chiral limit and the calculations with zero
pion mass confirms it. Another quantity strongly influenced by the
chiral limit is the neutron electric form factor.
For the $m_{\pi}\to 0$ the discrepancy from the experiment is
by almost a factor two larger than in the case $m_{\pi}\neq 0$.
The other observables differ in the chiral limit by about 30\%.
The reason is that in chiral limit one observes much larger
contributions from the sea effects, up to 50\% of the total value.
One concludes that using the physical pion mass one is able to get
a much better agreement with the experimental data.

\vskip1cm
4.2 \underline{$E2/M1$ ratio for the $\gamma N \rightarrow \Delta$
transition}
\vskip4mm

Here, we consider the ratio of electric quadrupole to
magnetic dipole amplitude $E2/M1$ ratio for the process
$\gamma + N \rightarrow \Delta$. It is a quantity of a
current interest, since it is sensitive to a presence of charge
deformations in the baryon structure and can be measured experimentally.
In a very recent $\pi^{0(+)}$-photoproduction experiment (Beck {\it et al.},
1995) performed at MAMI, Mainz the direct model-independent estimate of
the ratio $E2/M1 = (-2.4 \pm 0.2)\%$ indicates a presence of an oblate
charge deformation in the nucleon or/and delta, and as such it imposes a
strong constraint for the effective models of baryon structure. Explicitly 
one has
\begin{equation}
\frac{E2}{M1} \equiv \frac{1}{3} \
\frac{M^{E2}(\vec{k}, \lambda = +1 \ ; \
p(J_3 = -\frac{1}{2}) \rightarrow \Delta^+(J_3 = +\frac{1}{2}))}
{M^{M1}(\vec{k}, \lambda = +1 \ ; \
p(J_3 = -\frac{1}{2}) \rightarrow \Delta^+(J_3 = +\frac{1}{2}))}
\label{deftr}
\end{equation}
of the electric quadrupole amplitude $M^{E2}$ to the magnetic dipole one
$M^{M1}$. Here $k$ is the momentum  of a photon of helicity $\lambda$
in the $\Delta$ rest frame:
\begin{equation}
k = \frac{M_\Delta^2 - M_N^2}{2 M_\Delta}\,.
\label{restk}
\end{equation}
Both amplitudes are related to the corresponding matrix element of the
isovector current:
\begin{equation}
j^\mu_3(z) = \bar{\Psi}(z) \gamma^\mu \frac{\tau_3}{2} \Psi(z)\,.
\end{equation}
For the electric quadrupole amplitude $M^{E2}$, according to the
Siegert's theorem~(see for instance (Eisenberg and Greiner, 1970)), one can
use the zero
component $j^0_{T=1}$ as well the space component $\nabla\cdot\vec j_{T=1}$.
Similarly to (Wirzba and Weise, 1987), we decide to express the amplitude 
$M^{E2}$
via the $N-\Delta$ transition matrix element of $j^0_{T=1}$ (charge
density) of the current $j^\mu_{T=1}$:
\begin{equation}
M^{E2}(\vec{k}, \lambda ; N \rightarrow \Delta) =
\sqrt{15 \pi} \int d^3x \ \langle\Delta \vert j^0_3(\vec{x}) |
N\rangle Y_{2 \lambda} (\hat{x}) j_2(kx) ,
\label{mel}
\end{equation}
The reason is that in the large $N_c$ treatment this quantity can be
calculated (Christov {\it et al.}, 1995b) directly in terms of quark
matrix elements, whereas the matrix element of $\nabla\cdot\vec j_{T=1}$ is
suppressed additionally by $1/N_c$ and in order to evaluate it one should use
supplementary the saddle-point equation (\ref{eqsmotion}) for which
the ansatz (\ref{rotating-soliton}) is apparently not a solution.

The amplitude $M^{M1}$ is directly related to the $N-\Delta$ transition
matrix element of the space components $j^k_3$:
\begin{equation}
M^{M1}(\vec{k}, \lambda ; N \rightarrow \Delta) =
- \lambda {3\over 2} \int d^3x \langle\Delta | (\hat x \times
\vec j_3)_\lambda|N \rangle_{N\Delta} j_1(kx) .
\label{mml}
\end{equation}
The matrix elements the current $j^\mu$ are calculated in a way similar
to those of the nucleon electromagnetic form factors.

In the particular case of $\lambda = +1$,  $J_3^p = -\frac{1}{2}$
$J_3^\Delta = +\frac{1}{2}$, we have
\begin{equation}
M^{E2} = \frac{15 \sqrt{3}}{4} \int dr r^2 j_2(kr)\rho^{E2}_{N
\Delta}(r)\,,
\label{me2}
\end{equation}
and
\begin{equation}
M^{M1}= - 3 \int dr r^2 j_1(kr) \rho^{M1}_{ N \Delta}(r)\,,
\label{mm1}
\end{equation}
respectively.
The corresponding transition  charge density $\rho^{E2}_{N \Delta}(x)$
has a more complicated structure (Watabe {\it et al.}, 1995a) including a
spherical harmonics tensor $Y_{2\mu}$ which acts on the quark wave
function as a projector for the charge deformation.

In the $k \, r \ll 1$
approximation, one can relate the ratio $E2/M1$ to the electric
quadrupole $N\Delta$ transition moment:
\begin{equation}
\frac{E2}{M1} = \frac{1}{2} M_N\, k\,
\frac{< Q_{zz} >_{N\Delta}}{\mu_{N\Delta}}\,,
\label{rtlw}
\end{equation}
where $ Q_{zz} >_{N\Delta}$ is the electric quadrupole
transition moment and $\mu_{N\Delta}$ is the transition magnetic moment.

In table~\ref{Tabl2}, the results (Watabe {\it et al.}, 1995a)
for the ratio $E2/M1$ as well as for some related observables, namely
the isovector charge m.s.radius, the $N-\Delta$ transition magnetic moment
$\mu_{N\Delta}$, the $N-\Delta$ mass difference, and the quadrupole
electric transition moment $\langle Q_{zz}\rangle_{N\Delta}$, are given 
for three different
values of the constituent quark mass $M$ in comparison with the experiment. 
For the constituent quark mass $M$ around 420 MeV
the $E2/M1$ ratio is between $-2.5\%$ and $-2.3\%$  quite in
agreement with the experiment estimate (Beck {\it et al.}, 1995).
From the relation (\ref{rtlw}) we
also present an estimate for the electric quadrupole transition moment
$< Q_{zz} >_{N\Delta}=-0.026$ using the experimental values for
$E2/M1=-2.4\pm 0.2$ and $\mu_{N\Delta}=3.3$ which
is not far from the model prediction $-0.02$. This negative non-zero
value indicates a charge deformations of oblate type in the nucleon
or/and  delta structure. As can be seen  from table~\ref{Tabl2}
the dominant contribution to the $< Q_{zz} >_{N\Delta}$ in the NJL
model comes from the Dirac sea. It means that the main charge deformation
is due to the polarized Dirac sea, whereas the
valence quarks are almost spherically distributed. Since using the
gradient expansion, the polarization of the Dirac sea can be expressed
in terms of the dynamical pion field -- pion cloud, one can think of
the nucleon or/and of the delta as consisting of an almost spherical
valence quark core surrounded by a deformed pion cloud. One should note, 
however, that in the present approach the Delta is a stable particle and has 
no width which is in fact a result of the large $N_c$ treatment. The 
implications of this approximation are discussed by Cohen (1995).

We remind that for the same values of the constituent quark
mass $M\approx 420$  MeV the nucleon properties (including also the
nucleon form factors) are reproduced fairly well. The only
exception is the
$N-\Delta$ transition magnetic moment which similar to the isovector
magnetic moment is underestimated by 25\%.
The results for other observables in table~\ref{Tabl2} show an overall
good agreement with the experiment.
%which, however, is not the case of
%the Skyrme model calculations (Wirzba and Weise, 1987)

\begin{table}
\caption{Ratio $E2/M1$ and some related observables, calculated in
the SU(2) NJL model (Watabe {\it et al.}, 1995a)  for three different values 
of the 
constituent mass $M=400, 420$ and $450$ MeV, compared with experimental values.
%The Skyrme model results (Wirzba and Weise, 1987) are also presented.
}
\label{Tabl2}
\begin{tabular}{|c|c|c|c|c|c|c|c|}\hline
 & \multicolumn{6}{|c|}{{\bf Constituent quark mass $M$}} & \\ 
{\bf Quantity} &
\multicolumn{2}{|c|}{400 MeV} &
\multicolumn{2}{|c|}{420 MeV} &
\multicolumn{2}{|c|}{450 MeV} &
{\bf Exp} \\ 
\cline{2-7}
 & total & sea & total & sea & total & sea &  \\ \hline
$< r^2 >_{I=1}  \ [\fm^2]$ &
0.88 & 0.35 & 0.84 & 0.37 & 0.79 & 0.41 & 0.86 \\   \hline
$\mu_{\Delta N}  \ [n.m.]$ &
2.34 & 0.57 & 2.28 & 0.58 & 2.20 & 0.60 & 3.33 \\   \hline
$M_\Delta - M_N  \ [\MeV]$ &
255  &      & 278  &       & 311  &      & 294  \\  \hline
$< Q_{zz} >_{\Delta N}$ \ [fm$^2$] &
$-$0.020 & $-$0.014 & $-$0.020 & $-$0.015 & $-$0.021 & $-$0.016 &
$-0.026$\\  \hline
$E2/M1$\ [\%]& $-$2.19 & & $-$2.28& &$-$2.42& &$-2.4\pm$0.2\\ \hline
\end{tabular}
\end{table}

\vskip1cm
4.3 \underline{Axial properties}
\vskip4mm

The axial and pseudoscalar form factors reflect the nucleon structure
seen by the weak probes, and as such provide an important test for any
effective model of the nucleon. Most of the theoretical efforts are
concentrated on the axial whereas the pseudoscalar one is believed to be
generally understood in the context
of the chiral symmetry.

The general decomposition of the nucleon matrix element of the axial
current
\begin{equation}
A^a_\mu(x)=\Psi^\dagger(x)\gamma_0\gamma_\mu
\gamma_5\frac{\tau_a}2\Psi(x)
\label{axial-current}
\end{equation}
in terms of the axial $G_A$ and
pseudoscalar form factor $G_p$ reads
\begin{equation}
\langle  N(p^\prime,\xi^\prime)|
A^a_\mu(0)
|N(p,\xi)\rangle
= {\bar u}(p^\prime,\xi^\prime)\Bigl[ G_A(q^2)\gamma_\mu + {G_p(q^2)\over
2M_N}q_\mu \Bigr]\gamma_5 \frac{\tau_a}2 u(p,\xi)\,.
\label{Eq:axial}
\end{equation}
where $\xi$ stands for both spin and isospin.

In the large $N_c$ limit the time component of the
axial current is suppressed by $1/N_c$ compared to the space one. It
means that in the present large $N_c$ treatment the matrix element of
the space component of the axial current, which we use for the
evaluation of the weak form factors, is ``enhanced''.

In the calculations, one makes use of
the general expressions (\ref{valence-contribution-result}) and
(\ref{sea-regularized}) for the nucleon matrix element derived in
the previous section. As for the case of the isovector magnetic current
the Dirac sea contribution comes from the real part of the effective
action and needs regularization. Also $i\gamma_0\gamma_\mu\tau^a$ is
hermitian ($\eta=1$) and the quark matrix elements have the same
symmetry properties under $n \leftrightarrow m$ as those in the
isovector magnetic case. Thus, the matrix element of the space
components of the axial current $A^k_3$ includes leading order
terms~$\sim \Omega^0$ as well as next to leading order ones
$\sim\Omega$ ($1/N_c$)
\begin{eqnarray}
&&\bra{N(p^\prime)} A^k_3(\vec x)\ket{N(p)}=-N_c\int\d^3x\,\e^{iqx}
\Biggl\{\Bigl(\Phi^\dagger_{val}({\vec x})
\,\gamma^0\gamma^k\gamma_5\tau_3\,\Phi_{val}({\vec
x})\Bigr)
\nonumber\\
&&
-\sum\limits_n {\cal R}^\Lambda_{M1}(\epsilon_n)
\Bigl(\Phi^\dagger_n({\vec x}) \gamma^0 \gamma^k\gamma_5 \tau_b
\Phi_n({\vec x})\Bigr)
+\frac i{2\Theta}\varepsilon^{cb3} \Biggl[\sum\limits_{n\neq val}
\sign({\epsilon_n})\frac {\Bigl(\Phi^\dagger_{val}({\vec
x})\gamma^0\gamma^k\gamma_5\tau_b\Phi_n({\vec
x})\Bigr)\bra{n}\tau_c\ket{val}}{\epsilon_n -
\epsilon_{val}}\nonumber\\
&&-\sum\limits_{n,m}{\cal
R}^\Lambda_{M2}(\epsilon_n,\epsilon_m)\Bigl(\Phi_m^{\dagger}({\vec x})
\gamma^0 \gamma^k \gamma_5\tau_b\Phi_n({\vec x})\Bigr)
\bra{n}\tau_c\ket{m}\Biggl]\Biggr\}\,,
\label{MEAC}\end{eqnarray}
where the first and
third terms are valence quark contributions in leading and next to
leading order in angular velocity, respectively. The other
terms represent the divergent Dirac sea part.

Using result (\ref{MEAC}) after some
straightforward calculations one can express the axial and
pseudoscalar form factors in terms of two axial densities $A_0(r)$ and
$A_2(r)$ (Watabe {\it et al.}, 1995b)
\begin{equation}
G_A(q^2) = \frac{M_N}{E} \ \int r^2 dr \,
[j_0(qr) A_0(r) -j_2(qr) A_2(r)] \,.
\label{GAff}
\end{equation}
and
\begin{equation}
G_P(q^2) = -{M_N^2\over E(E+M_N)}\,\int r^2 dr \,
j_0(qr) A_0(r) -24 {M_N^2\over q^2}(1+{M_N\over 2E})\,
\int r^2 \mbox{d}r j_2(qr) A_2(r) \,,
\label{GPff}
\end{equation}
where
\begin{equation}
A_0(r)=\langle  N|A^3_3|N\rangle \,,
\label{A0}
\end{equation}
\begin{equation}
A_2(r)=\langle  N|{1\over 3}A^3_3-A_3^ix^ix^3|N\rangle\,.
\label{A2}
\end{equation}
In the above expressions $M_N$ and $E$ are the nucleon mass and
energy $E=\sqrt{M_N^2+\vec{q}^2/4}$ in the Breit frame, respectively.

\begin{figure}
\centerline{\epsfysize=4.0in\epsffile{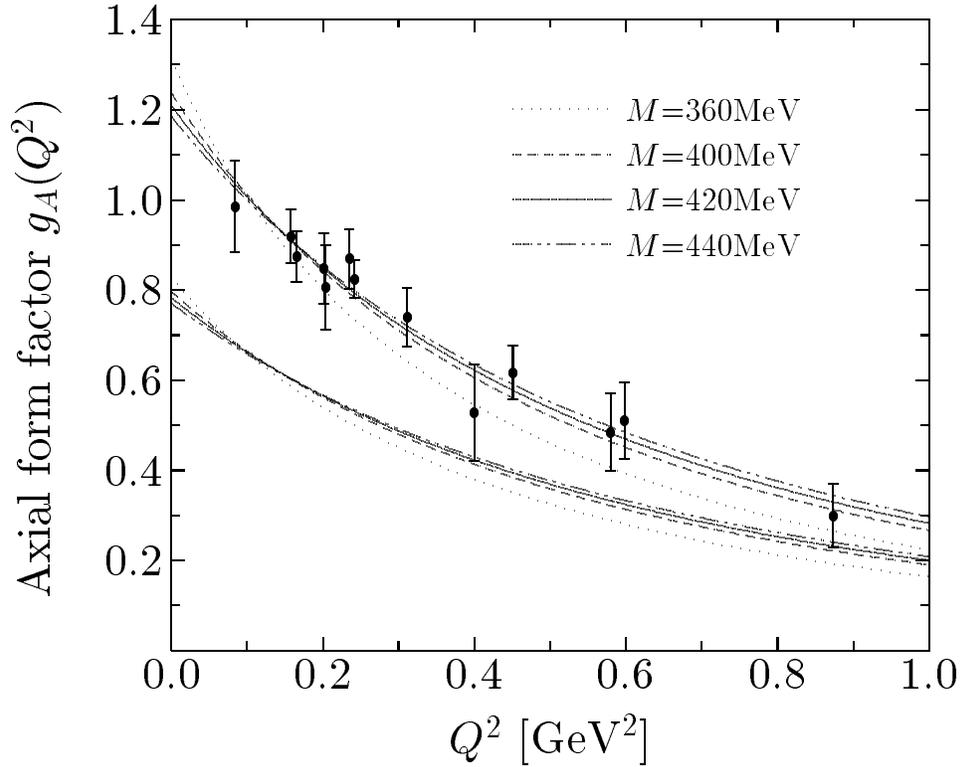}}%\vskip4pt
\caption{Axial form factor in the SU(2) NJL model 
(Watabe {\it et al.}, 1995b) in leading 
(lower curves) and next to leading order (upper curves) in
angular velocity in comparison with the experimental
data (Baker {\it et al.}, 1981, Kitagaki {\it et al.}, 1983).}
\label{Figr53}
\end{figure}

\vskip1cm
\underline{Axial form factor and the axial-vector coupling constant.}
The results for the axial form factor
(Watabe {\it et al.}, 1995b) are shown in fig.\ref{Figr53} for four
different values of the constituent mass $M$, 360, 400, 420 and
440 MeV. The main contribution to the axial form factor in the leading
as well as in the next to leading order in angular velocity comes from
the valence quarks and the Dirac sea contribution is almost negligible.
As for the other nucleon properties a value for constituent mass
around 420 MeV is preferred. For this mass value the theoretical curve
agrees well with the experimental data without any scaling.
On fig.\ref{Figr53} we also show the leading order results
(Meissner Th. and Goeke, 1992).
As can be seen the two groups of curves differ in both  magnitude and
slope. The dipole masses (see table~\ref{Tabl53}) for the axial form
factors, calculated in leading order, are slightly larger. 

The axial charge radius calculated as
\begin{equation}
\langle r^2 \rangle_A =
- \frac{6}{G_A(0)} \ \frac{\mbox{d} G_A(Q^2)}{\mbox{d} Q^2}
\Bigg\vert_{Q^2=0}
= \frac{1}{G_A(0)} \int r^4 \mbox{d}r \ [A_0(r)+\frac{2}{5}A_2(r)]
+\frac{3}{4M_N^2} \,.
\label{sqradi}
\end{equation}
is presented in Table~\ref{Tabl53} together with the values
from the dipole fit (in
brackets) in comparison with the estimate from the experimental dipole
fit. As can be seen the numbers directly calculated from the
corresponding
densities $A_0$ and $A_2$ are very close to those of the dipole fit
which is an indication that even at small $q^2$ the dipole fit is a good
approximation to the theoretical curves.

The axial-vector coupling constant is given by
\begin{equation}
g_A=G_A(0)=2\int\d^3x<N|A^3_3(x)|N>\,.
\label{ga}
\end{equation}
which means that
only the density $A_0$ contributes (\ref{GAff}). The strong
underestimation of $g_A$ in most of the chiral models of nucleon is a
problem which have attracted a lot of attention and theoretical efforts
in both the Skyrme-like models (see Holzwarth, 1994 and references
therein) and the NJL model (Wakamatsu and Watabe, 1993,
Blotz {\it et al.}, 1993c, Christov {\it et al.}, 1994, 1995a,b, 
Wakamatsu 1995, Schechter and Weigel, 1995). Wakamatsu and Watabe (1993) made 
an important
observation that since after the canonical quantization, the collective
operators do not commute, it may lead to non-zero rotational corrections
to $g_A$ which could resolve the problem of the underestimation of $g_A$
in the leading order in $\Omega$ (Wakamatsu and Yoshiki, 1991, Meissner Th. 
and Goeke, 1991). However, in their calculations, the non-zero result is
due to a chosen order
of the collective operators being not justified by path integrals or
many-body techniques. As a consequence their expression violate the
$G$-parity symmetry (Schechter and Weigel, 1995) and the Pauli
principle (even though, numerically, this Pauli
violating contribution turns out to be only a tiny fraction of the valence
contribution), and the rotational ($1/N_c$) corrections from the Dirac
sea are missing. As it was already mentioned in chapter 3 such a problem
does not exist in the present semiclassical quantization scheme
(Christov {\it et al.}, 1994) and it can be easily checked directly from
(\ref{MEAC}) (Christov {\it et al.}, 1995a).

\begin{figure}
%\centerline{\psfig{file=ganjl.ps,height=6cm,bbllx=80,bblly=480,bburx=480,bbury%=760,clip=,angle=90}}
\centerline{\epsfysize=4.in\epsffile{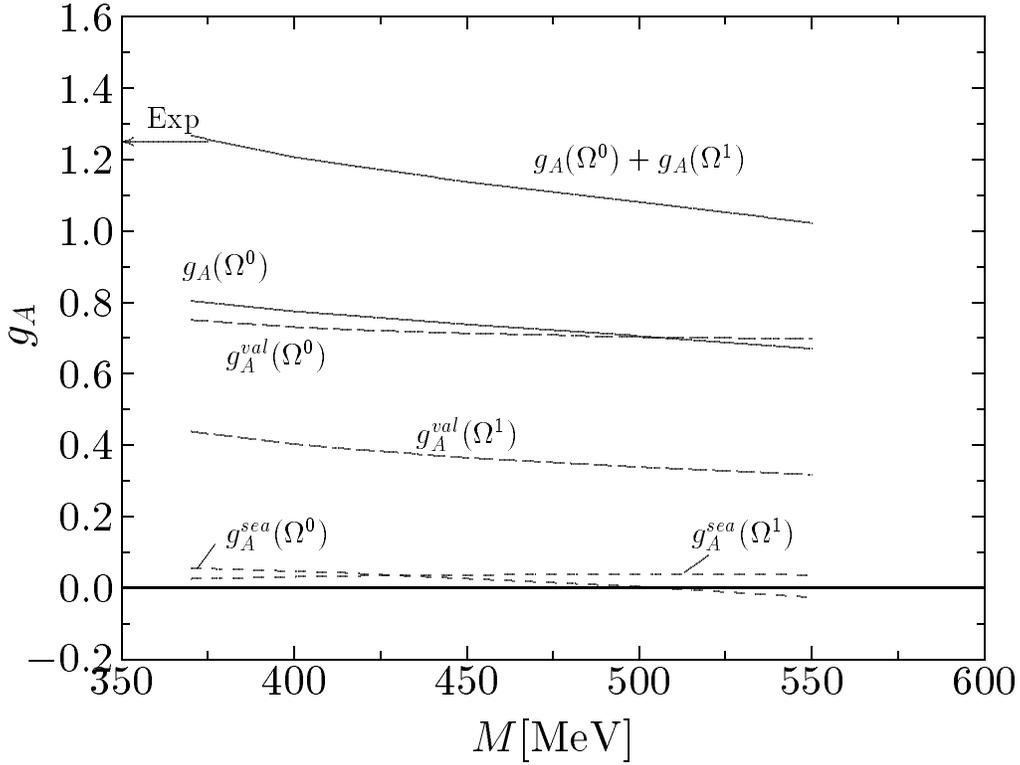}}%\vskip4pt
%\xslide{ganjl.ps}{6cm}{80}{480}{480}{760}
\caption{Axial coupling constant $g_A$ evaluated in the leading
($\Omega^0$) and next to leading order $\Omega^0+\Omega^1$ in the SU(2) NJL 
model as a function of the constituent quark mass $M$ 
(Christov {\it et al.}, 1994).}
\label{Figr56}
\end{figure}

On fig.\ref{Figr56} we present the results for the axial-vector coupling
constant in leading and next to leading order in angular velocity $\Omega$
as a function of the constituent mass $M$ (Christov {\it et al.}, 1994). As
can be seen, in both the leading and next to leading order the valence
part dominates and the Dirac sea contribution is almost negligible.
The enhancement due to the $1/N_c$ rotational corrections improves
considerably the agreement with experiment and for the
constituent quark mass of about 420 MeV the experimental value of
$g_A$ is almost exactly reproduced. It is interesting to notice that this
enhancement numerically is very close to $\frac {N_c+2}{N_c}$.

\begin{table}
\caption{Axial properties of the nucleon, calculated in
the NJL model (Watabe {\it et al.}, 1995b) for four different values of the 
constituent mass $M=360, 400, 420$ and $440$ MeV, compared with experimental
values. The obtained dipole mass for the axial form factor and the
axial m.s.radius are given in both leading and next to leading order in angular
velocity. The radius is also calculated from the
dipole fit (in brackets). The induced pseudoscalar coupling constant
$g_P$ is also presented.}
\label{Tabl53}
\begin{center}
\begin{tabular}{|c|c|c|c|c|c|} \hline
 &\multicolumn{4}{|c|}{{\bf Constituent quark mass $M$ [MeV]}} & \\
{\bf Quantity} & 360 & 400 & 420 & 440 &{\bf Exp} \\ \hline
 $M_A^{\Omega^0}$ [GeV] & $0.91 $ & $1.01 $ & $1.04 $ & $1.072 $ & 
$1.05^{+0.12}_{-0.16}$ \\ \hline
 $M_A^{\Omega^0+\Omega^1}$ [GeV] & $0.85 $ & $0.96 $ & $0.99 $ &
 $1.03 $ & $1.05^{+0.12}_{-0.16}$ \\ \hline
 $< r^2 >_A$ [${\rm fm}^2$] &
 0.70 (0.65) & 0.51 (0.51) & 0.46 (0.47) & 0.43 (0.44) &
 $0.42^{+0.18}_{-0.08}$ \\ \hline 
 $g_A^{\Omega^0}$ &
 0.83 & 0.80 & 0.78 & 0.77 & 1.26     \\ \hline
 $g_A^{\Omega^0+\Omega^1}$ & 1.31 & 1.24 & 1.21 & 1.18 & 1.26\\ \hline
 $g_P^{\Omega^0+\Omega^1}$  & 6.09 & 6.05 & 6.01 &6.00 & $8.6\pm 1.9$\\ \hline
\end{tabular}
\end{center}
\end{table}

In the present large $N_c$ treatment, only the rotational $1/N_c$ corrections 
coming from the rotational zero modes are taken into account .
Apparently, they are only a part of the existing $1/N_c$ corrections and
the inclusion of all $1/N_c$ corrections could in principle changes
the obtained results. In the present approach the only source of
additional $1/N_c$
corrections is the term ${1\over 2}\Tr\log\Bigl[{\delta^2 S\over\delta
\phi\delta \phi}\Bigr]$ in (\ref{lnZ}) which represents the meson quantum
(loop) effects. One generally expects that the meson loops would
affect mostly the contribution of the polarized Dirac sea (represents
the meson clouds) and not much that of the valence quarks. It means
that for the physical
quantity which are dominated by the valence quarks like the axial vector
coupling constant one should not expect large changes due to the $1/N_c$
corrections from the meson loops. Additionally, the rotational $1/N_c$
corrections obey $G$-parity symmetry and fulfill the consistency
condition of Dashen and Manohar  (1993) derived in the large-$N_c$ of QCD.
The latter means that the other $1/N_c$ contributions
should contribute to both the leading and next to leading order terms,
and the enhancement due to the rotational corrections will
survive in a strict next to leading order in $1/N_c$.

\begin{figure}
\centerline{\epsfysize=4.0in\epsffile{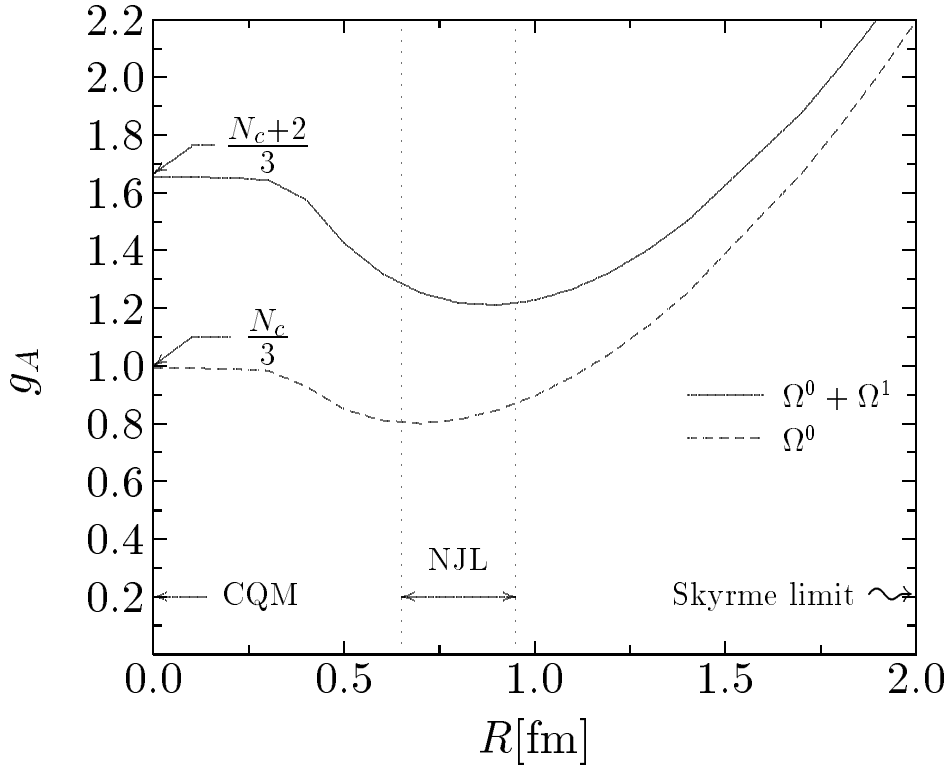}}%\vskip4pt
\caption{Axial vector coupling constant $g_A$ as a function of the
soliton size $R$ in the leading ($\Omega^0$) and next to leading
order $\Omega^0+\Omega^1$ for a fixed profile function in the SU(2) NJL model
(Praszalowicz {\it et al.}, 1995).}
\label{Figr57}
\end{figure}

Praszalowicz {\it et al.}, (1995) studied the behavior of $g_A$ in the limit 
of both
small and large soliton size $R$. We remind that at $MR\gg1$ the NJL soliton
shows a structure similar to the topological soliton of the Skyrme model. At
small $R$ limit the contribution of the Dirac sea vanishes and we have
a physical picture of three quarks of constituent mass $M$ similar to
the one of the constituent quark model (CQM).
The results of (Praszalowicz {\it et al.}, 1995) are shown in 
fig.\ref{Figr57}. One
immediately sees that only in the case with the rotational $1/N_c$
corrections included, the results ${N_c+2\over 3}$ for $g_A$ of CQM can
be reproduced. It should be
stressed that since in the NJL model the nucleon appears as a non-trivial bound
state of valence quarks and polarized Dirac sea, and the CQM is based on pure
symmetry assumptions concerning the wave functions, this
agreement is by no means obvious. Apparently, the semiclassical quantization 
with  expansion up to first order in the rotational velocity for the 
observables and up to the second order for the energy provides a consistent 
and powerful method to describe the rotational degrees of freedom.
In the large $R$ limit, the rotational $1/N_c$
corrections exactly vanish and we recover the Skyrme case of no rotational
$1/N_c$ corrections.

\begin{figure}
\centerline{\epsfysize=4.0in\epsffile{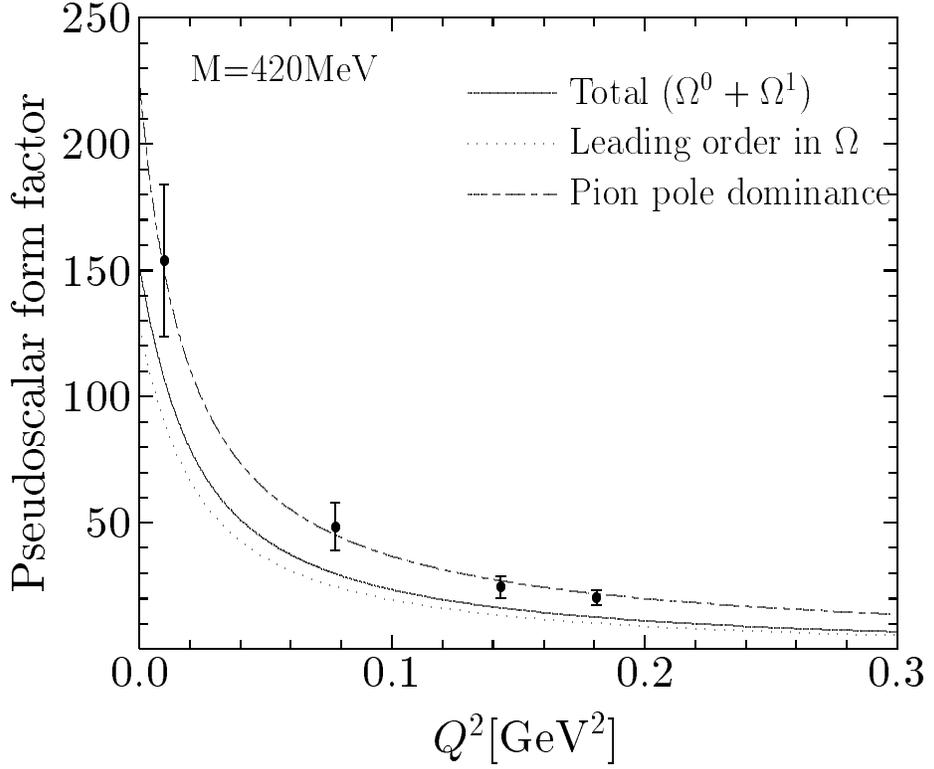}}%\vskip4pt
\caption{Pseudoscalar form factor of the SU(2) NJL model in leading and 
next to leading order
in angular velocity in comparison with the experimental
data (Choi {\it et al.,} 1993, Esaulov {\it et al.,} 1978). 
The pion pole dominance fit (dash-dotted line) is also presented.}
\label{Figr54}
\end{figure}

\vskip1cm
\underline{Pseudoscalar form factor.}
For the pseudoscalar form factor the chiral symmetry
predicts (Amaldi {\it et al.}, 1979) to be entirely dominated by the pion pole
and as such to be very sensitive to the pion content of the
nucleon. The model result for $M=420$ MeV is shown in
fig.\ref{Figr54}. The experimental points are from the first exclusive
experiment (Choi {\it et al.}, 1993) determining independently both
weak form factors. The muon capture point (Esaulov {\it et al.}, 1978) is 
included as well. We also present the pion
pole dominance prediction (dashed line). As can be seen, despite the fact
that the magnitude of $G_P$ is underestimated, its $q^2$-behavior is in a
good  agreement with both the experimental points and the pion pole
dominance. As it is expected, the pseudoscalar form factor is entirely
dominated by the polarized Dirac sea (the main contribution comes from $A_2$)
and the valence contribution is almost negligible. We remind that using the
gradient expansion the polarization of the Dirac sea can
be expressed in terms of the dynamical pion field -- the pion cloud. It means
that even in the approximation of no meson loops the model is able to account
for an essential part of the pion dynamics relevant for the nucleon structure.
It should be also noted that the pseudoscalar form factor is very
insensitive to the particular value of $M$ and that the enhancement due
to the rotational corrections in the case of $G_P$  is much smaller
than for the axial one. The latter feature is related to the fact that the  
$G_P$ is dominated by the Dirac sea.
Since the rotational corrections are $\sim 1/I$ (see eq.(\ref{MEAC})) and
the moment of inertia $I$ is dominated by the valence quarks, one expects
that these corrections in the case of Dirac sea should be much smaller than in
the case of the valence contribution.

The calculated values of the induced pseudoscalar
coupling constant defined at the muon capture point
\begin{equation}
g_P={M_\mu\over 2M_N}G_P(q^2=-0.88 M_\mu^2) 
\label{gP}
\end{equation}
are presented in table~\ref{Tabl53}.
As experimental value we take the value at the muon capture
point directly from the experiment (Esaulov {\it et al.}, 1978).
As can be seen, the model calculations
yield numbers for $g_P$ which are slightly outside the experimental error bars.
If we use the estimate of Bernabeu $g_P=8.2\pm 2.1$ (Bernabeu, 1978),
extracted from the same experimental data (Esaulov {\it et al.}, 1978), the
theoretical predictions are even better. Apparently, despite the
agreement in the $q$-dependence on fig.\ref{Figr54} this underestimation 
shows that some
physics is still missing in the present model picture. Indeed,
in the approximation of large $N_c$, used in the calculations, the meson
quantum (loop) effects are not taken into account. Since $g_P$ is
dominated by the Dirac sea, this could be the reason (Bernard {\it et al.},
1994) for the underestimation of $g_P$.

\vskip1cm
4.4 \underline{Pion nucleon form factor}
\vskip4mm

We parametrize the pion nucleon vertex in terms of the pion nucleon
form factor $G_{\pi NN}$ and relate it to the matrix element of the
pseudoscalar density:
\begin{equation}
\langle
N(p^\prime,\xi^\prime)|\bar\Psi i\gamma_5\tau^a\Psi|N(p,\xi)\rangle
= {m_\pi^2 f_\pi\over m_0}{\bar u}(p^\prime,\xi^\prime)i\gamma_5
\frac{\tau_a}2 u(p,\xi) {G_{\pi NN}(q^2)\over m_\pi^2 - q^2} \,,
\label{Gpnndef}
\end{equation}
where $q^2={p^\prime}^2-p^2$ and the denominator $m_\pi^2 -q^2$
represents the pion pole. The factor ${m_\pi^2 f_\pi\over m_0}$ ensures
that the definition of $G_{\pi NN}$ is for a physical pion field:
\begin{equation}
\pi^a= {m_0\over m_\pi^2 f_\pi} \bar \Psi i \gamma_5
\tau^a  \Psi\,,
\label{piPCAC}
\end{equation}
which is consistent with (\ref{e29c}).

Using the same techniques as before, we evaluate the pion nucleon form
factor $G_{\pi NN}$ and the final results reads
\begin{equation}
G_{\pi NN}(q^2)={2M_N\over q}{m_\pi^2 f_\pi\over m_0}{N_c\over 3}
(m_\pi^2 - q^2)\, \int \d^3xj_1(qr)<N|\gamma^0\gamma_5x^3\tau^3|N>\,.
\label{pNN-form-factor}
\end{equation}
The pseudoscalar density $<N|\Psi\gamma^0\gamma_5x^3\tau^3 \Psi |N>$ has the 
following structure:
\begin{eqnarray}
&&<N|\Psi\gamma^0\gamma_5x^3\tau^3 \Psi|N>={1\over 3}[\Phi_{val}^\dagger({\vec
x})\gamma^0\gamma_5\tau^b {x^b\over r}
\Phi_{val}({\vec x})] - \sum\limits_n {\cal R}_{M1}^\Lambda(\epsilon_n) \,
[\Phi_n^\dagger({\vec x})\gamma^0\gamma_5\tau^b {x^b\over r}  \Phi_n({\vec x})]
\nonumber \\
&&
-\frac{i}{6I} \varepsilon_{abc}\Biggl[ \sum\limits_{n,m}
{\cal R}_{M2}^\Lambda(\epsilon_m,\epsilon_n) \,
[\Phi_m^\dagger({\vec x})\gamma^0\gamma_5\tau^a {x^b\over
r}\Phi_n({\vec x})]\,\langle n |\tau^c |m\rangle
\nonumber \\
&&
-\sum\limits_{\epsilon_n \neq \epsilon_{val}}
\frac{\mbox{sign}(\epsilon_n) }
{\varepsilon_{val} - \epsilon_n} \,
[\Phi_n^\dagger({\vec x})\gamma^0\gamma_5 \tau^a {x^b\over r}
\Phi_{val}({\vec x})]\,\langle val |\tau^c | n \rangle
\Biggr] \,,
\label{pNN-density}
\end{eqnarray}
including both leading and next to leading terms.

The pion nucleon coupling constant $g_{\pi NN}(0)$ at $q^2=0$ is given by
\begin{equation}
g_{\pi NN}(0)=G_{\pi NN}(0)=2M_N {m_\pi^2 f_\pi\over m_0}
\int\d^3x<N|\bar\Psi\gamma_5x^3\tau^3\Psi |N>\,,
\label{gpnn}
\end{equation}
and it differs slightly (at the level of about 5\%) from the physical pion 
nucleon coupling constant $g_{\pi NN}(m_\pi^2)$ defined via the residue at the 
pole in (\ref{Gpnndef}).
 
\begin{table}
\caption{Monopole cutoff mass and $g_\pi N N$ coupling constant in the SU(2) 
NJL model}
\label{Tabl54}
\begin{center}
\begin{tabular}{|c|c|c|c|c|} \hline
 &\multicolumn{3}{|c|}{\bf $M$ [MeV]}&  \\
{\bf Quantity} & 370 & 420 & 450& {\bf Phen} \\ \hline
$\Lambda_\pi$ [MeV] & 835 & 1086 & 1242 & $\approx 0.9$ GeV\\ \hline
$g_{\pi N N}^{(\Omega^0)}(0)$ & 8.1 & 7.8 & 7.5 & 13.1\\ \hline
$g_{\pi N N}^{(\Omega^0+\Omega^1)}(0)$ & 10.2 & 10.3 & 10.3& 13.1 \\ \hline
\end{tabular}
\end{center}
\end{table}

\begin{figure}
\centerline{\epsfysize=4.0in\epsffile{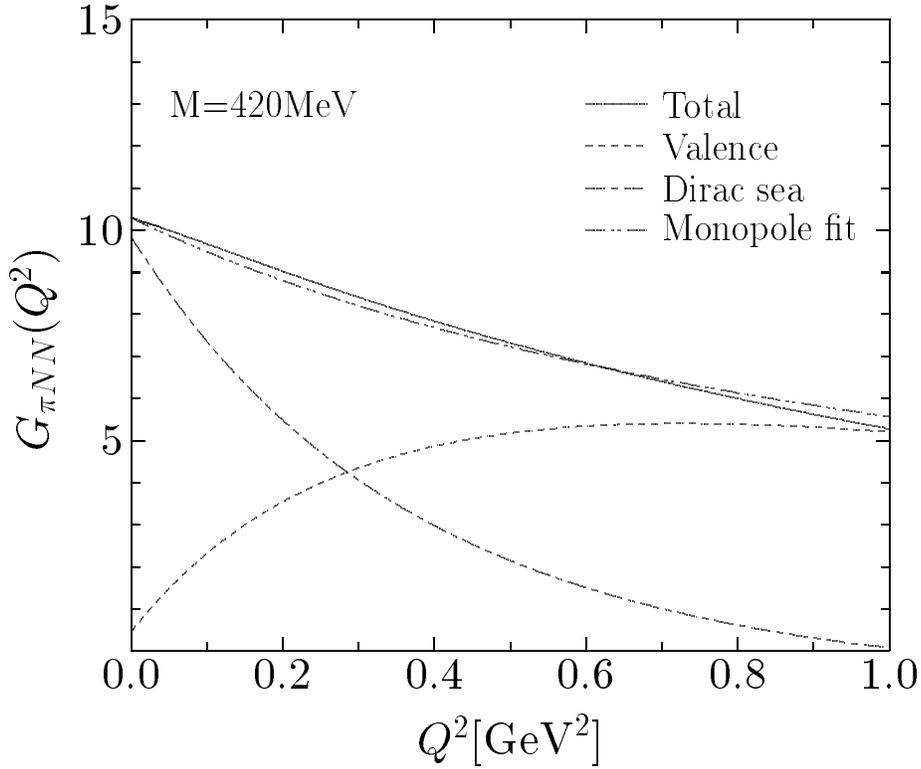}}%\vskip4pt
\caption{Pion nucleon form factor calculated in the next to leading
order in angular velocity for $M=420$ MeV in the SU(2) NJL model. The 
valence (dashed line) and Dirac
sea (dash-dotted line) contributions as well as the monopole fit 
are presented.}
\label{Figr58}
\end{figure}

The calculated pion form factor is presented on fig.\ref{Figr58}
for $M=420$ MeV together with separated valence and Dirac sea contributions.
We plot also the corresponding monopole fit:
\begin{equation}
G_{\pi NN}(q^2)=g_{\pi NN}
{\Lambda_{mon}^2-m_\pi^2\over\Lambda_{mon}^2-q^2}\,,
\label{monople-fit}
\end{equation}
with $g_{\pi NN}\equiv g_{\pi NN}(m_\pi^2)$ being the on-shell pion coupling 
constant. 
The monopole fit with a monopole cutoff mass of order of 1 GeV is commonly 
used for phenomenological parametrizations of the pion nucleon vertex. The 
extracted values for the monopole cutoff mass are given in table~\ref{Tabl54}. 
As can be seen, the calculations
deviate from the monopole fit. Although the form factor is a
smooth monotonically decreasing function of the momentum transfer $q^2$, the
valence and the sea contributions have rather different and to some extent
opposite behavior. The sea contribution increases rapidly with $q^2\to 0$ and
hence, entirely dominates at small $q^2$. Keeping in mind that in the model the
polarized Dirac sea plays the role of the pion cloud, it can be easily
understood as a dominance of the pion-pion resonance interaction
at small $q^2$. On the other hand, at larger $q^2$ this interaction is
suppressed and the valence part contribution, almost negligible at
small $q^2$, increases with $q^2$ and becomes dominant.

The obtained values of $g_{\pi NN}(0)$ evaluated in leading as well as in the 
next to leading order in $\Omega$ are presented in table \ref{Tabl54}. 
As in the case of the axial coupling constant $g_A$, the strong pion nucleon 
one $g_{\pi NN}$ is also 
strongly underestimated in leading order. However, despite the enhancement due 
to the rotational $1/N_c$ corrections, in the next to leading order it is 
still underestimated by 20 \%. Similarly to $g_P$ the strong constant 
$g_{\pi NN}$  is dominated by the Dirac sea,
and the meson loop effects ($1/N_c$ corrections) could be of importance.

\vskip1cm
4.5 \underline{PCAC and the Goldberger-Treiman relation}
\vskip4mm

Because of the non-zero current mass $m_0$ the axial current is only
partially conserved (\ref{e29c}) (PCAC). According to the PCAC hypothesis its
divergence is related to the pion field
\begin{equation}
m_\pi^2 f_\pi\pi^a(x)=\partial^\mu  A_\mu^a
= m_0 \bar \Psi i\gamma_5\tau^a \Psi\,,
\label{piPCAC1}
\end{equation}
which we have already used in the case of the pion nucleon vertex
(\ref{Gpnndef}). However, as we discuss below, this identity is not necessary 
valid in the case of nucleon matrix element
\begin{equation}
q^\mu <N(p^\prime)| A_\mu^a|N(p)>
= m_0<N(p^\prime)| \bar \Psi i\gamma_5\tau^a \Psi|N(p)> =
m_\pi^2 f_\pi<N(p^\prime)|\pi^a|N(p)>\,,
\label{NNpiPCAC}
\end{equation}
evaluated in a particular model.

In the present model, the nucleon appears as a large $N_c$ soliton
(localized mean-field solution) and the zero modes are taken into account in 
order to assign proper
quantum numbers to the soliton solution. In the many-body theory, it is
known as {\it projection after variation} (Ring and Schuck, 1980). In 
evaluating (\ref{NNpiPCAC}) two problems arises: First, the term 
$q^0\langle A_0\rangle$ is by two orders in $1/N_c$ suppressed compared to 
$q^i\langle A_i\rangle$. Hence, one cannot expect the NJL model to describe 
$q^0\langle A_0\rangle$ properly. Second, one also cannot evaluate 
$q^i\langle A_i\rangle$ systematically. This is due to the fact that one needs 
the Dirac equation 
(\ref{Diracequation}) in order to replace $i\gamma_i\partial_i\Phi_n$ by 
$\epsilon_n-\gamma_0 MU_c^{\gamma_5}$ where $U_c$ is the selfconsistent 
solution of large 
$N_c$ saddle-point equation (\ref{eqsmotion}). This, however, is inconsistent, 
since 
$\epsilon_\alpha-MU_c$ is evaluated without rotational $1/N_c$ corrections, 
whereas the rhs of (\ref{NNpiPCAC}) are evaluated including those 
corrections. Actually the mismatch
between the lhs and rhs of (\ref{NNpiPCAC}) tells something about the 
theoretical errors we do by taking from the $1/N_c$ corrections only those 
which correspond to the rotational zero modes. This mismatch may also be 
seen in connection with a virial theorem used in the chiral
sigma model by Birse and Banerjee (1985), Birse (1986, 1990) and 
Fiolhais {\it et al.} (1988).

Let us estimate to what extend the violation of the equality (\ref{NNpiPCAC})
will affect the axial coupling constant $g_A$. In the case of a finite pion 
mass, $g_A$ (\ref{ga}) can be rewritten as
\begin{equation}
g_A=2\int\d^3x<N|A^3_3(\vec x)|N>=-2\int\d^3x x^3\partial^\mu 
<N|A^3_\mu(\vec x)|N>\,.
\label{ga1}
\end{equation}
Using the PCAC (\ref{piPCAC}) we can obtain another estimate $g_A^\pi$
from the pion field (or from the pseudoscalar density):
\begin{equation}
g_A^\pi=m_\pi^2 f_\pi\int\d^3x x^3<N|\pi^3(\vec x)|N>= m_0\int\d^3x
<N|\bar\Psi i x^3\gamma_5\tau^3  (\vec x)\Psi|N>
\label{ga-pi}
\end{equation}
In the case that the equality (\ref{NNpiPCAC}) is valid, we would have
$g_A^\pi=g_A$.  The numerical calculations, however, show that
$g_A^\pi$ is by 15 \% lower than $g_A$ calculated directly
from the matrix element of the axial current. The reason for this
discrepancy can be easily seen if one calculates explicitly
$<N|\partial_\mu A^3_\mu(\vec x)|N>$ making use of (\ref{Diracequation}) with
(\ref{SPTc}):
\begin{equation}
g_A-g_A^\pi=N_cM\int\d^3x \bigl[ <N|\bar\Psi i x^3 \gamma_5
\tau^3\Psi |N> S_c(r)-<N|\bar\Psi\Psi |N> P_c(r)\bigl]\,.
\label{ga-gapi}
\end{equation}
where the scalar $S_c(r)$ and the pseudoscalar $P_c(r)$ mean-field densities 
are selfconsistent solutions of large $N_c$ 
saddle-point equation (\ref{eqsmotion}):
\begin{equation}
S_c(r)=M\cos\Theta_c(r)\qquad \mbox{and}\qquad
P_c(r)=M\sin\Theta_c(r)\,.
\label{SPTc}
\end{equation}

Using that the pseudoscalar density (\ref{pNN-density}) contains
non-zero both leading and next to leading terms
\begin{equation}
<N|\bar\Psi i\gamma_5x^3\tau^3 \Psi |N>=P_c(r)+
<N|\bar\Psi i\gamma_5x^3\tau^3\Psi  |N >^{(\Omega^1)}\,,
\label{pdens}
\end{equation}
whereas the scalar density is given only by the leading
order term $S_c(r)$, and also that both $S_c(r)$ and $P_c(r)$ are solutions of 
(\ref{eqsmotion}), we get
\begin{equation}
g_A-g_A^\pi=N_cM\int\d^3x
<N|\bar\Psi i\gamma_5x^3\tau^3\Psi|N>^{(\Omega^1)}S_c(r)> 0\,.
\label{ga-gapi1}
\end{equation}
Obviously, in the leading order ($\Omega^0$) the two estimates, $g_A$ and
$g _A^\pi$, coincide and the PCAC equality (\ref{NNpiPCAC}) is valid.
However, in the next to leading order they differ and it is probably
due to the fact that no rotational $1/N_c$ corrections are present in the
saddle-point equations. In the present scheme the PCAC relation 
(\ref{NNpiPCAC}) is fulfilled within 15~\% ``measured'' in terms of 
the axial coupling constant. 

Alkofer and Weigel (1993) proposed as a solution to this problem to
use the rhs of (\ref{ga-gapi}) as a new ``equation of motion'' which
respects the PCAC. Making use of the fact that the moment of inertia
$I(\Theta)$ is a functional of the profile function $\Theta$ they
solved this new equation together with (\ref{Diracequation}) iteratively
and found a new selfconsistent solution which satisfies PCAC. In fact,
in their calculations following the incomplete treatment of Wakamatsu and 
Watabe (1993), they
disregarded the Dirac sea contribution. The full calculations, however,
including the Dirac sea as well, show no solution, which is not
surprising, since this new ``equation of motion'' does not correspond to
any stationary point of the effective action.

Apparently, as argued above the present scheme based essentially on 
the large $N_c$ approximation is not very appropriate for evaluation of
quantities like the divergence of the axial current which requires the use 
of the equations of motion. Hence, on theoretical grounds, the estimate 
$g_A$ calculated directly from the matrix element of the space component of 
the axial current is preferred to $g_A^\pi$. Accordingly, 
the numbers of fig.\ref{Figr53} correspond to $g_A$ and not to $g_A^\pi$.   

On the other hand, it is not an easy task to generalize the
present large $N_c$ treatment in order to include the zero-mode
rotational corrections in the saddle-point equation. It would correspond
to {\it projection before variation} and a deviation from the hedgehog
structure already for the saddle-point solution would be inevitable. It
would also lead to conceptual problems related to the consistency of
$1/N_c$ expansion which we have used so far. Therefore, the most
general solution would be to go consistently beyond the large $N_c$
approximation including the quantum meson (loop) effects from the
very beginning.

Closely related to the PCAC is the Goldberger-Treiman (GT) relation. It makes 
an important link between the strong $g_{\pi NN}$ and the axial vector $g_A$ 
coupling constants and the pion decay constant $f_\pi$. In fact, this 
relation is valid in nature within some percents. Let us
assume that the PCAC equality (\ref{NNpiPCAC}) is valid and use
(\ref{ga-pi}) to evaluate $g_A$. Inserting it
in (\ref{gpnn}) one gets
\begin{equation}
g_A={g_{\pi NN}(0)f_\pi\over M_N}\,.
\label{GTNR}
\end{equation}
which differs from the GT relation in $g_{\pi NN}(0)$
taken at $q^2=0$ and not at the pole $q^2=m_\pi^2$. In fact, in the model 
$g_A$ differs from $g _A^\pi$ by 15 \% which means 
that with the estimate $g_A$ the relation (\ref{GTNR}) is fulfilled at the 
15\% level and apparently, due to the approximations used, the present model 
scheme cannot describe the Goldberger-Treiman discrepancy observed in 
nature.

\newpage
\vskip0.5cm
4.6 \underline{Tensor charges}
\vskip4mm

The nucleon tensor charges $g_T^{(a)}$ ($a=0,3$ corresponds to the
singlet and isovector charges, respectively) are defined as the nucleon
forward (zero momentum transfer) matrix element of the tensor current:
\begin{equation}
\langle N(p)| \bar\Psi  \sigma_{\mu\nu} {\tau^a\over 2}\Psi |N(p) \rangle =
g_T^{(a)} \bar u(p)\sigma_{\mu\nu}{\tau^a\over 2} u(p)\,,
\label{tensor_charge_definition}
\end{equation}
where $\sigma_{\mu\nu}=\frac i2[\gamma_\mu, \gamma_\nu]$ and $\tau(0)\equiv
{\bf 1}$. Although the tensor charge like other nucleon charges
is a fundamental quantity, which characterizes the nucleon,
too little is known about its values and its implication to the nucleon
structure. The reason is that the tensor charge is difficult to access
experimentally, since there is no experimental probes coupled directly
to the tensor current. However, as it is shown by Jaffe and Ji, (1991)
the nucleon tensor charge is related to the first moment of the
transversity quark distribution $h_1(x)$:
\begin{equation}
\int\limits_0^1 \d x (h_1^{(a)}(x)-\bar{h}_1^{(a)}(x))=g_T^{(a)}\,,
\label{first_moment}
\end{equation}
where $a=0,3$ stands for the singlet and isovector transversity quark
distributions, respectively. This sum rule gives a hope to gain some
experimental information about the tensor charge.
It should also be noted that the
tensor charge depends on the renormalization-scale dependent. However,
the dependence on the
normalization point is rather weak
\begin{equation}
g_T^{(f)}(\mu^2)=
\Biggl(\frac{\alpha_s(\mu^2)}{\alpha_s(\mu_0^2)}\Biggr)^{\frac{4}{29}}
g_T^{(f)}(\mu_0^2),
 \end{equation}
which means that the initial value of the normalization point is not
of big importance.

In the non-relativistic limit using the identity
$\sigma_{ik}=\varepsilon_{ikj}\gamma_0\gamma_j\gamma_5$ one can easily
see that the nucleon matrix element of the tensor current has a
structure similar to those of the axial current:
\begin{equation}
g_T^{(0)}= {N_c\over 6 I}\sum_{nm}R_T^\Lambda(\epsilon_n,\epsilon_m)
<n|\tau^a|m><m|\gamma_0\sigma_a|n>=\alpha N_c^0\,,
\end{equation}
\begin{equation}
g_T^{(3)}= N_c \sum_{\epsilon_n\leq val}<n|\gamma_0\vec
\sigma\cdot\vec\tau|n>+
{N_c\over 9}{1\over 2I} \sum_{\epsilon_n> val\atop \epsilon_m\leq
val}{1\over\epsilon_n-\epsilon_m}
<m|\gamma_0[\vec \sigma\times\vec\tau]_a|n><n|\tau^a|m>
= \beta N_c+ \delta N_c^0
\end{equation}
In contrast to the case of the axial current, here the singlet tensor
charge diverges and needs regularization:
\begin{equation}
{\cal R}_T^\Lambda(\epsilon_n,\epsilon_m)=-{1\over 2\sqrt{\pi}}\int
{\d u\over \sqrt{u}}{\epsilon_n\e^{-u\epsilon_n^2}
+\epsilon_m\e^{-u\epsilon_m^2}\over \epsilon_n+\epsilon_m}\,,
\end{equation}
whereas the isovector tensor charge is finite and is not regularized.
In large $N_c$ limit $g_T^{(0)} \sim N_c^0$ and
$g_T^{(3)} \sim N_c$, which is
the same as in case of the non-relativistic quark model. In fact, the
isovector charge, similar to the axial vector coupling constant,
contains non-zero contributions from both the leading ($N_c$) and next
to leading terms ($N_c^0$) and the consistency condition of Dashen
and Manohar (1993) is also fulfilled.

The model prediction for the nucleon tensor charges for $M=420$~MeV
(Kim {\it et al.}, 1995d):
\begin{equation}
g_T^{(3)}\approx 1.45, \qquad  g_T^{(0)}\approx 0.69\,,
\label{ten_num}
\end{equation}
or
\begin{equation}
g_T^{(u)}\approx 1.07, \qquad  g_T^{(d)}\approx -0.38
\end{equation}
are close to those in bag model
(Jaffe and Ji, 1991) and not so far (within the error bars)
from the QCD sum rule estimates of (He and Ji, 1995).

\vskip1cm
4.7 \underline{Electric polarizability of the nucleon}
\vskip5mm

So far we have considered nucleon observables which are related to matrix 
element of a single quark current. In this subsection, we discuss a more 
complicated case of an physical quantity which is related to a matrix element 
of two quark currents, namely the nucleon electric polarizability.

Recent measurements  of the electric, $\alpha$, and magnetic, $\beta$, 
polarizabilities of the nucleon (Federspiel {\em et al.}, 1991, Zieger 
{\em et al.}, 1992, Schmiedmayer {\em et al.}, 1991) 
narrowed considerably the experimental uncertainties in
these observables, and were accompanied by a number of theoretical studies.

The electric polarizability is given by the change of the nucleon
energy in an external electric field and it can be expressed in terms of the 
correlator of two nucleon currents~(\ref{PiNPI}). Apart from the regularization
the explicit calculations give for the leading-$N_c$ contribution to $\alpha$ 
\begin{eqnarray} 
\label{eq:pol}
\alpha^{(0)} 
&=& \frac 12  {N_c}\;e^2 \langle N \mid D^{(1)}_{a 3} D^{(1)}_{b 3} 
\mid N\rangle
\sum_{\epsilon_n >\epsilon_{val} \atop \epsilon_m \leq\epsilon_{val} }
{\langle m |\tau^a z | n \rangle\langle n|\tau^b z| m \rangle
\over \epsilon_n - \epsilon_m} ,
\end{eqnarray}
where $|N\rangle$ is the collective nucleon state, $D^{(1)}_{a 3}$ 
is an element of a Wigner $D$-matrix, 
and $| n \rangle$ and $\epsilon_n$ denote the eigenstates and eigenvalues
of the Dirac hamiltonian.
This formula can be split into the valence and sea parts.
The sea contribution is divergent and includes a cut-off function. The 
vacuum contribution has to be subtracted as well.
Finally, evaluating the collective matrix element one obtains the
leading-$N_c$ valence and sea
contributions to the electric polarizability of the nucleon in the form
\begin{equation} 
\label{eq:polval}
\alpha^{(0)}_{\rm val} ={N_c\over 6}\;e^2 \sum_{n \ne {\rm val}}
{\langle {\rm val} |\tau^a z | n \rangle\langle n|\tau^a z| {\rm val} \rangle
\over \epsilon_n - \epsilon_{\rm val}}\,,
\end{equation}
\begin{equation} 
\label{eq:polsea}
\left.
\alpha^{(0)}_{\rm sea}=
{N_c\over 6}\;e^2 \sum_{n \ne m} {\cal R}^\Lambda_I(\epsilon_m,\epsilon_n)
{\langle m |\tau^a z | n \rangle\langle n|\tau^a z| m \rangle}
- \right. {\rm vac},
\end{equation}
where ${\cal R}^\Lambda_I(\epsilon_m,\epsilon_n)$ is the
proper-time regularization function of
the moment of inertia $I$~(\ref{regmomentinertia}). 
The subtracted vacuum contribution in (\ref{eq:polsea}) has the same
structure as the first term, but the free quark states and
eigenvalues are used. 

The rotational $1/N_c$ correction $\alpha^{(1)}$ is given by a 
triple sum over quark states (Nikolov {\em et al.}, 1994). The valence 
part $\alpha^{(1)}_{\rm val}$ has been calculated exactly, whereas the 
sea part $\alpha^{(1)}_{\rm sea}$ has been approximated by the leading term 
in the gradient expansion $\alpha^{(1)}_{\rm sea,lowest}$.  

\begin{figure}
\xslide{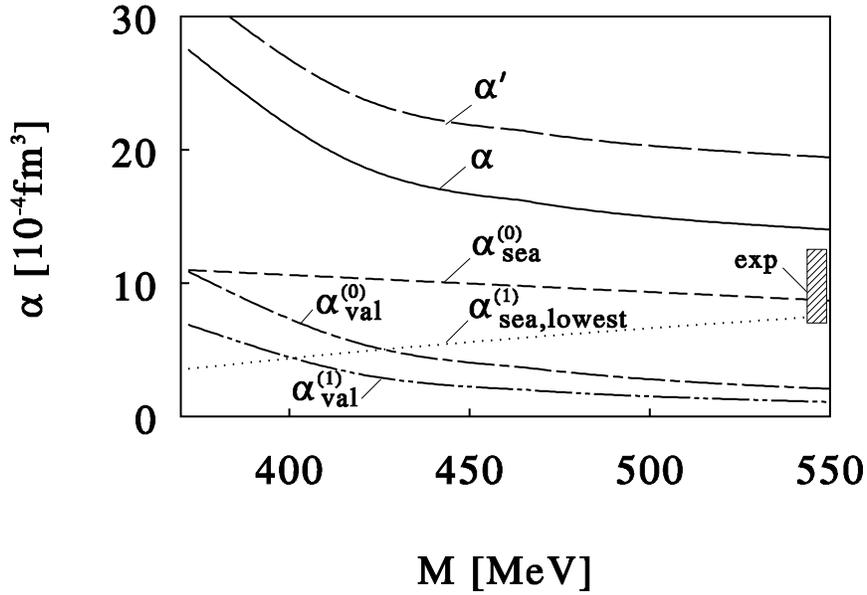}{8cm}{110}{305}{530}{595}
\caption{Electric polarizability of the nucleon of the SU(2) NJL model 
(Nikolov {\em et al.}, 1994)
as a function of the constituent quark mass $M$. The $\alpha^\prime$ indicates 
the direct result of the NJL model, the $\alpha$ incorporates corrections due 
to $N-\Delta$ mass splitting. The polarizability of the nucleon is defined
as $\alpha=(\alpha_p+\alpha_n)/2$.}
\label{fig:pol}
\end{figure}
In fig.~\ref{fig:pol} we present the results (Nikolov {\em et al.}, 1994) for 
the average electric polarizability of the proton and neutron, 
\mbox{$\alpha \equiv \frac{1}{2} (\alpha_p + \alpha_n)$} obtained
in the NJL model as
a function of the constituent quark mass $M$.
The different contributions as well as the total nucleon electric
polarizability are presented. The long-dashed line corresponds
to $\alpha^\prime =  \alpha^{(0)}_{\rm sea}+\alpha^{(0)}_{\rm val}
+ \alpha^{(1)}_{\rm sea,lowest}+\alpha^{(1)}_{\rm val}$ and this is the direct
result of the NJL model. The solid
line shows the total $\alpha$, roughly corrected for the effects of the
$N$-$\Delta$ mass splitting.
For values of $M$ in the physically relevant range 400--450~MeV the
changes of $\alpha$ with  $M$
are small.  The present results are in qualitative agreement with the results 
of other models (Weiner and Weise, 1985, Broniowski and Cohen, 1993) with 
explicit pion degrees of freedom. 

The sea contributions clearly dominate over the valence ones both on leading 
order in $N_c$ and for the rotational $1/N_c$ corrections.   
These contributions are very well reproduced
by the leading two terms in the gradient expansion and the first
term is the so called ``seagull'' contribution to $\alpha$,
discussed in many previous papers. 

The inclusion of rotational corrections has important phenomenological
consequences. The dominant contribution to the electric polarizability is
obtained from pion tail effects, and is proportional
to the axial constant $g_A^2$. Since the rotational corrections have been 
shown (Christov {\em et al.,} 1994) to be crucial for reproducing the 
experimental value of   
$g_A$ in the model, it is important to take into account these
rotational corrections for the electric polarizability. The calculation 
shows that indeed the rotational corrections give a sizeable contribution
of about 50--60\% of the leading-$N_c$ result.
After including approximately
the corrections due to the $N$-$\Delta$ mass splitting, 
the theoretical value is closer to experiment 
than in other studies in soliton models. For the typical choice of the 
constituent quark mass \mbox{$M = 420\ \rm{MeV} $} we obtain
\mbox{$\alpha^\prime \simeq 25 \times 10^{-4}\ {\rm fm}^3$}
and \mbox{$\alpha \simeq 19 \times 10^{-4}\ {\rm fm}^3$},
compared to the experimental value $\alpha_{\rm exp}=9.6 \pm 1.8 \pm 2.2
\times 10^{-4}\ \rm fm^3$ (Federspiel {\em et al.}, 1991,
Zieger {\em et al.}, 1992, Schmiedmayer {\em et al.}, 1991). 
The model prediction is, however, still too large indicating
that the inclusion of other $1/N_c$ effects (such as e.g. meson loops)
is necessary. In other approaches like e.g. the chiral perturbation
theory with nucleons (for review see (Bernard {\it et al.}, 1995)) the  
pion loops provide the main effect. In our model, parts of those meson effects
are taken into account already in the leading $1/N_c$ order due to the presence
of the mean-field. The effects of meson loops in the next to leading order 
in $1/N_c$ remains to be investigated. In particular,
one hopes that the meson-loop $1/N_c$ effects related 
to the zero modes would lead to a reduction in the case of the electric 
polarizability.

\newpage

\section{SU(3)-NJL model}

In the previous chapters, we have reviewed the Nambu--Jona-Lasinio
model with scalar isoscalar and pseudoscalar isovector
couplings. The extension of this model to the larger symmetry of SU(3)
introduces not only the kaon and eta mesons to the model but
offers in the baryon sector, which is the main concern of the present
article, the possibility to calculate the properties of strange baryons as well
as the contributions of strange quarks to the properties of the nucleon and
Delta. In fact, at present these effects are in the case of the spin structure
functions or the strange vector form factors under
vivid discussion.

We start with a very short review of the mesonic sector
which is already
extensively discussed in the reviews of Klevansky (1992),
Vogl and Weise (1991) and Hatsuda and Kunihiro (1994). After performing the 
bosonization and
saddle-point approximation we fix the parameters of the
model by reproducing pion and kaon properties and have
again the constituent quark mass of the non-strange quarks
as the only free parameter.
Actually, the extension from SU(2) to SU(3) is formally not trivial,
since the current mass of the strange quark, $m_s$, is noticeably
larger than that of the up and down quarks, $m_u$ and $m_d$.
This corresponds to a non-negligible explicit breaking
of the SU(3)-symmetry which plays an important role in
the baryon sector.
We remind that the semiclassical treatment of the NJL model
is justified in the large $N_c$ limit. In the SU(3) case,
apart from the $1/N_c$ expansion one has to deal also with the expansion in
powers of $m_s$. These expansions in two small parameters $1/N_c$
and $m_s$ are very sensitive to which limit is being taken
first:  $N_c\to \infty$ or $m_s\to 0$. This leads to certain
difficulties in the collective quantization of the soliton rotation.
We discuss the perturbative approaches (Weigel {\it et al.,} 1992, Blotz 
{\it et al.,} 1992, 1993a), which in view of explicit 
calculations of baryonic observables are most far developed at the present 
time. Special emphasis then is put on the possible representations of the
collective hamiltonian.
After a discussion of the mass spectra of the hyperon multiplets in
SU(3) we review baryonic observables like the electromagnetic
form factors, sigma terms, axial vector
couplings, spin properties of the proton and the Gottfried sum rule.
Predictions will be done for the nucleon  strange form factors.

Besides the perturbative approach one can also consider the hyperons as 
bound states of kaon mesons in the background field of the SU(2) soliton. This 
approach has been invented by the Callan and Klebanov (1985) and has been 
applied in the chiral sigma model by McGovern and Birse (1990) and 
in the case of the NJL model by Weigel {\it et al.} (1994).  

\vspace{1cm}
5.1 \underline{Extension of the NJL model to SU(3) flavors}
\vspace{4mm}

An extension of the SU(2) flavor NJL lagrangian (1)
to SU(3) can be done by a direct generalization ($\tau^a\to\lambda^a$)
of the SU(2) case with scalar and pseudoscalar couplings.
Then the simplest SU(3) form of a four-fermion interaction with scalar and 
pseudoscalar couplings reads
\beq  {\cal L}
       = {\bar \Psi}(x) \left[
      i \dsl - {\hat m} \right] \Psi(x)
       + {G\over 2 }       \sum_{a=0}^8
       \left[
         \left(
        {\bar \Psi}(x)  \lambda^a \Psi (x) \right)^2
              +
         \left( {\bar \Psi}(x)
         \lambda^a i \gamma_5  \Psi (x) \right)^2
                    \right]
           \label{gl1}
\eeq
where the $\lambda_a$ are the Gell-Mann matrices with
$\Sp\lambda^a \lambda^b=2\delta_{ab}$
and the sum over the $\lambda^a$ matrices includes
$\lambda^0=\sqrt{2\over 3} {\bf 1}_3$ as well.
This form of the couplings is demanded by requiring
a chiral $SU(3)_R\otimes~SU(3)_L$ symmetry.
The mass matrix in \qeq{gl1}
is now enlarged by the strange quark mass
\beq    {\hat m} =
         \left( \ba{ccc}  m_u & 0 & 0     \\
                    0 & m_d & 0     \\
                    0 & 0   & m_s   \ea
               \right)
            =  {\bar m}_0 {\bf 1} +
               \lambda^3 m_3 +
               \lambda^8 m_8
               \label{gl1a}
\eeq
where ${\bar m}_0=(m_u+m_d+m_s)/3$,
$m_3=(m_u-m_d)/2$ and $m_8=(m_u+m_d-2m_s)/(2\sqrt{3})$.
Note that in contrast to SU(2), where one could write
a minimal lagrangian with the singlet scalar field only and a
triplet of pion
fields, the  rank 2 of the SU(3) group compared with rank 1 of SU(2)
together with chiral symmetry
requires the presence of the whole nonet of scalar and pseudoscalar
couplings.
Using integral identities similar to eq. (15) the bosonized form
of (\ref{gl1}) reads
\beq
         {\cal L}' = {\bar \Psi}(x) \left(
         i \dsl - {\hat m}
        - \sum_{a=0}^8 \left(  \sigma_a \lambda^a
        + i\gamma_5  \pi_a \lambda^a \right)
        \right) \Psi(x)
        - { 1 \over 2G}
     \sum_{a=0}^8   \left[  \sigma_a^2 +\pi_a^2 \right]
       -       {\cal L}^\prime_A [\sigma_a,\pi_a]\,.
\eeq
Here we have added "by hand" an extra term
${\cal L}^\prime_A [\sigma_a,\pi_a]$. The necessity
of this additional term is related to the U$_A(1)$ problem.
Besides $SU(3)_R\otimes~SU(3)_L\otimes~U_V(1)$ symmetry, the
lagrangian ${\cal L}$ (\ref{gl1}) has in addition the
$U_A(1)$ symmetry. On the other hand, it is well know that in QCD
the $U_A(1)$ symmetry is broken by the anomaly. As a consequences
the $\eta'$ meson remains massive even in the chiral limit.
This means the effective low energy theory should contain some term
${\cal L}^\prime_A [\sigma_a,\pi_a]$
explicitly breaking the $U_A(1)$ symmetry. In fact, we use the presence of 
${\cal L}^\prime_A$ as an argument to consider $\eta^\prime$  massive and 
not as a Goldstone boson. For the fixing of the parameters of the NJL model 
only Goldstone bosons are needed. In the baryonic sector, we will use the 
chiral circle (\ref{gl18a}) and hence the detailed structure of 
${\cal L}^\prime_A$ is irrelevant for our considerations.

After integrating over the quark fields
the partition function in Euclidean space is given by
\beq   Z = \int {\cal D} \sigma_a {\cal D} \pi_a
e^{-S (\sigma,\pi) }
       \label{gl20}
\eeq
where
\beq      S (\sigma,\pi)=-N_c {\rm Tr}\  \log D(\sigma,\pi)
           +  {1\over 2G} \int d^4x
            \sum_{a=0}^8  \left( \sigma_a^2 + \pi_a^2 \right)
             +  \int d^4x
         {\cal L}^\prime_A [\sigma_a,\pi_a]
            \label{gl21}
\eeq
with
$D(\sigma,\pi)=\partial_\tau + h$ and
\beq  h = -i\gamma_0\gamma^k\partial_k +
         \gamma_0 {\hat m}
            + \gamma_0
       \sum_{a=0}^8 \left(
       \sigma_a \lambda^a  +
      i \gamma_5  \pi_a \lambda^a \right) \,.
\eeq

\vspace{10mm}
5.2 \underline{Vacuum solutions and fixing of parameters}
\vspace{4mm}

In the vacuum we make again a saddle point approximation
similar to eq. (28)
\beq  { { \delta   S(\sigma,\pi)    \over \delta \sigma_a }
      = 0 \linie}_{   \sigma_a=\sigma_{ac}\atop \pi_a=\pi_{ac}
            } \ \ \
      { { \delta    S(\sigma,\pi)    \over \delta  \pi_a }
       = 0
          \linie}_{   \sigma_a=\sigma_{ac}\atop \pi_a=\pi_{ac}
               } \,.
\eeq
For three flavors there are the $\sigma_0$,
$\sigma_3$ and $\sigma_8$ fields which can have non-vanishing
vacuum expectation values. Therefore, instead of
eq.~(\ref{gap}) one has
in general one gap equation for each flavor degree of freedom.
Neglecting isospin breaking (${\mbar} =m_u=m_d$) in the following the
non-trivial gap equations read
\bea     {1 \over G}   \sigma_1 - 8 N_c  \sigma_1
            I_1(\sigma_1)   &=&  {  {\mbar} \over G}
\nn
         {1 \over G}   \sigma_2 - 8 N_c  \sigma_2
            I_1(\sigma_2)   &=&  { m_s \over G}
        \label{gl13}
\eea
where we set  $\sigma_1=\sqrt{2\over 3}\sigma_{0c}+{1\over\sqrt{3}}
\sigma_{8c}+{\mbar} $
and  $\sigma_2=\sqrt{2\over 3}\sigma_{0c}-{2\over\sqrt{3}}
\sigma_{8c}+m_s$.
Deriving these equations we neglected the contribution of
the ${\cal L}^\prime_A [\sigma_a,\pi_a]$.

The spontaneous breaking of chiral symmetry leads now to the appearance
of constituent quark masses for the non-strange and strange
sector according to
\beq   M_u = M_d = \sigma_1,\ \ \ \ \  M_s=\sigma_2 \,.  \eeq
This resembles two independent Dirac equations with mass
gaps $M_u$ and $M_s$.
In addition, there are two quark condensates
\bea     <{\bar u}u>\> = \> <{\bar d}d> &=&
         - {1\over V_4}\, \frac12 { \partial \over \partial  {\mbar} }
                   Z
                  \nn
         < {\bar s}s > &=&
         - {1\over V_4} {  \partial \over \partial  m_s   }
                  Z
           \label{gl15}
\eea
which become equal in the chiral limit and actually are
of the order of $-(250\,\MeV)^3$.
Using the gap equations  eqs. (\ref{gl13}) the eqs. (\ref{gl15})
can be cast into the form
\bea    <{\bar u}u+{\bar d}d> &=&  -{1\over G} \left( M_u-{\mbar}
         \right)
     \nn
     < {\bar s}s > &=&  -{1\over 2G} \left( M_s - m_s  \right) \,.
        \label{gl16}
\eea
Similarly to the definition of the pion decay constant
in eq.~(\ref{fpi-final}),
we can define the corresponding vacuum matrix element
of the axial vector current $A_\mu^a,\>a=4,5,6,7$ with
the physical kaon fields $\pi_a,\>a=4,5,6,7$. As a result one
obtains (Schneider {\it et al.}, 1995)
\bea   f_\pi   &=&  M_u  g_\pi  Z_p(q^2=-m_\pi^2)  \nn
       f_K     &=&  {M_s+M_u\over 2}  g_K    Z_p(q^2=-m_K^2)
         \label{gl17}
\eea
where  $g_K$ can be determined from the corresponding
eq.~(\ref{couplconst}).
In the chiral limit, from the \qeq{gl16} and
the gap equation \queq{gl13} it follows that
\beq       f_\pi = f_K  =  M_u  g_\pi  Z_p(q^2=0)
\eeq
and
\beq   {m_K^2 \over m_\pi^2} =
        { m_s + m_0  \over  2 {\mbar}  } \,.
 \eeq
For finite quark current masses the kaon decay constant deviates from the pion
decay constant but however it is still underestimated by 10 \%.

The masses of the pion and kaon are determined from
the two-point meson functions (50) 
\bea     m_\pi^2 &=&  {  {\mbar}  \over
                 G M_u Z_p(q^2=-m_\pi^2)       }
                \nn
         m_K^2   &=&  \left(  {  m_s      \over  M_s}
                   + {  {\mbar}   \over  M_u}
                    \right)
              {1\over  2  G  Z_p(q^2=-m_K^2)       }
                + \Delta^2_{M_s}
%                + ( M_s - M_u )^2
           \label{gl18}
\eea
with $\Delta_{M_s} = M_s - M_u$.

\vspace{10mm}
\underline{Fixing of the parameters.}
The SU(3) NJL model in the present form contains four unknown parameters:
the coupling constant $G$, the current quark masses $m_0$ and $m_s$, and the
regularization cut-off $\Lambda$.
First, instead of $G$ the constituent quark mass $M_u$ is used as parameter
which is related to $G$ by the first gap equation (\ref{gl13}).
Second, for a given $M_u$ the $m_0$, $m_s$ and $\Lambda$ are fixed to
reproduce $f_\pi = 93$ MeV, $m_\pi = 139$ MeV and the kaon mass
$m_K = 496$ MeV.
The $M_u$ is not fixed in the meson sector but in the baryon sector, in order
to reproduce best the experimental data of the nucleon.
This results, as in SU(2), in preferred values of $M_u$ about 420 MeV.
The strange constituent mass $M_s$ is not an independent quantity but  
determined by $M_u$ according to the second gap equation (\ref{gl13}).

In contrast to the SU(2) case, where the average current quark 
mass $m_0$ is rather small and the results do not depend on its particular 
value fixed by the physical pion mass, 
in the SU(3) case, the explicit chiral symmetry breaking due to the strange 
current mass $m_s\approx 175$ MeV (Bijnens {\it et al.}, 1995)\footnote{We 
note that for the strange current mass we use the value at the same 
$\overline{\mbox{MS}}$ scale $\mu=1$ GeV (of order of the model cutoff) 
as in the case of the quark condensate.} will be treated 
perturbatively up to the second order corrections in $m_s$. Apparently 
we need to reproduce the value of the strange quark current mass properly.  
Actually, if one follows the above scheme and uses the proper time 
regularization (\ref{e229}) with 
$\phi (u)= \theta ( {1\over \Lambda^2} - u )$, the 
absolute values of $m_u$, $m_d$ and $m_s$ come out about two times larger 
than expected.  To that end we make use 
essentially of the freedom to choose the particular form of the function 
$\phi (u)$ in the proper-time
regulator (\ref{e229}) in order to adjust the quark condensate and the current
quark masses to their phenomenologically accepted values.
In our practical calculations we use 
\begin{eqnarray}
\phi(u) = c    \theta ( u - \Lambda_1^{-2} ) +
         (1-c) \theta ( u - \Lambda_2^{-2} )
\end{eqnarray}
in which two additional parameters $c$ and $\Lambda_2$ appear.
The $c$ is used to obtain a value for $m_0$ of 6.1 MeV.
The $\Lambda_2$ is chosen as small as possible in order to facilitate the
numerical effort in the calculation of the soliton.

\vspace{8mm}
5.3 \underline{Restriction to Goldstone modes}
\vskip4mm

Up to now we have considered the meson sector of SU(3) NJL in terms of
independent fields $\sigma^a$ and $\pi^a$ with $a=0,1,\dots,8$.
The parameters of the model are fixed by pion and kaon properties.
Like in the SU(2) case the model as such does not support solitonic solutions 
and, again, one has to subject the field to the condition of the chiral circle.
In analogy to (\ref{u}) this is defined as
\beq       \sigma_a(x) \lambda^a
              + i \pi_a(x) \lambda^a
           =
           M  U^{(3)}_c (\vx)
           =  M e^{ i \pi_a' \lambda^a }
            \label{gl18a}
\eeq
where $U_c^{(3)^\dagger} (\vx)U^{(3)}_c(\vx)=1$ similar to
 the case of SU(2) chiral fields.
The sum goes over $a=0,1,\dots,8$ and $M$ is given by $M=M_u$ even in the case
of $M_s \neq M_u$.
This is consistent with the so called trivial embedding of SU(2) into SU(3)
discussed in the next subsection.

The condition of the chiral circle corresponds to a clear separation between
the Goldstone degrees of freedom,
i.e. pions, kaons and eta, and non-Goldstone ones like sigmas.
Since we are interested in describing the ground state properties of the
nucleon and the hyperons, the degrees of freedom corresponding to heavy mesons
are of less importance and hence, will be ignored.
With the $U_c^{(3)}(x)$ of eq.(\ref{gl18a}) in the solitonic sector, the 
present model is identical to the SU(3) version of the Chiral quark soliton
model of Diakonov {\em et al.} (1988). This model is based on the instanton 
model of the QCD vacuum (Diakonov and Petrov, 1986). We remind that
the effective quark lagrangian derived by Diakonov and Petrov (1986) from
the instanton model of the QCD vacuum has a form
of a `t Hooft-like interaction which explicitly breaks the U$_A(1)$
symmetry. Its bosonization leads to meson fields of the form (\ref{gl18a}).

\vspace{10mm}
5.4 \underline{Trivial embedding of SU(2) hedgehog soliton into SU(3)}
\vspace{5mm}

By analogy with the SU(2) case we expect that the low-lying
baryons can be described as rotational excitations of some
classical soliton solution. The spectrum of these
baryons in the real world imposes certain restrictions on the symmetry
of the SU(3) soliton. It was suggested by Witten (1983)
that one should consider the trivial embedding of the
SU(2) hedgehog soliton into SU(3)
\beq   U^{(3)}_c (\vx)
          =\left( \ba{ccc}  U^{(2)}_c(\vx)  & \linie &  0 \\
            \hline
           0 & \linie &  1   \ea
\right)
\label{c51}  \eeq
where
$U^{(2)}_c(\vx)=\cos\Theta(r)+ i{\vec\tau}{\hat
x}\sin\Theta(r)$
coincides with the SU(2) stationary meson field configuration
$U_c(\vx)$ of eq.~(\ref{Uhedgehog}).
The advantage of this ansatz for the saddle point solution
is that its rotational excitations naturally describe the particles of
the SU(3) octet and decuplet.

If one neglects the quark masses then there is a continuous set
of saddle point solutions $R^\dagger U^{(3)}_c(\vx)R$
where $R$ is an arbitrary SU(3) matrix. Similarly to the SU(2)
case, in the path integral scheme we have to take into account
the path integral over time dependent fields of the form
\beq
       U^{(3)}(\tau,\vx)=  R(\tau)  U^{(3)}_c(\vx)
       R^\dagger(\tau) \,.
\label{SU-3-rotation}
\eeq
After this path integral is computed one arrives at the baryon rotational
wave functions $\psi(R)$ depending on the SU(3) orientation matrix $R$.
As in the SU(2) case
the SU(3) flavor generators and the spin operator
acting in the space of this rotational wave functions are nothing else but
the generators of the left and right rotations of the matrix $R$.
By analogy with (\ref{SU2-isospin}), (\ref{SU2-spin}) we obtain
\begin{equation}
\exp(i\omega_a T_a) \psi (R)
= \psi \left( \exp(-i\omega_a \lambda_a/2 ) R \right)\,,
\label{SU3-isospin}
\end{equation}
\begin{equation}
\exp(i\omega_a J_a) \psi (R)
= \psi \left(R \exp(i\omega_a \lambda_a/2 )  \right)\,.
\label{SU3-spin}
\end{equation}
Here $T_a$ are eight SU(3) flavor generators. As for the right
generators $J_a$, the components with $a=1,2,3$ have the meaning of the
spin operator. The other components of $J_a$ have no direct physical
meaning. Note that under the rotation associated with $J_8$
the field $U^{(3)}(\tau,\vx)$ does not change. This means that the
parametrization of the rotating soliton by the SU(3)
matrix $R$ contains "non-physical" degrees of freedom.
As it will be seen later this leads to a constraint on the allowed
eigenvalues of $J_8$.

\vspace{8mm}
5.5 \underline{Expansion in angular velocity}
\vskip4mm

Restricting the path integral over $U(x)$ in ((\ref{PiNPI}) to meson 
configurations of the form $U^{(3)}(\tau,\vx)$
(\ref{SU-3-rotation}) we obtain by analogy with (\ref{PiNPIR0})
\beq    \Pi_B(T) =
        \Gamma_B^{\{f\}}  \Gamma_B^{\{g\}*}
        \int {\cal D} R \prod_{i=1}^{N_c}
       \langle0,T/2\mid
       {1 \over D(RU^{(3)}_cR^\dagger) }
       \mid0,-T/2\rangle_{f_ig_j}  e^{-S(RU^{(3)}_cR^\dagger)}
          \label{gl29}
\eeq
where the $\Gamma_B^{\{f\}}$ carry the quantum numbers of the hyperons
and
the Dirac operator in the body-fixed frame of the soliton is now
given by:
\beq     D(U^{(3)}_c(\tau,\vx))
        =  R \left[ \partial_\tau +  h_0
           +      h_1 + i \Omega
           \right] R^\dagger \,.
\label{gl30}
\eeq
Here $h_0$ is the single particle hamiltonian
\bea        h_0     &=&   -i\gamma_0 \gamma^k \partial_k
             + M_u \gamma_0  U^{(3)\gamma_5}_c
                +  \gamma_0  m_0   {\bf 1}
           \label{gl32}
\eea
which includes the two flavor hamiltonian  eq.~(\ref{hU})
embedded in  SU(3). In contrast to SU(2) case in \qeq{gl30} we have 
additionally the term
\beq  h_1 =  \gamma^0
          \left( {\Delta_{M_s}\over 3}  {\bf 1}     -
           {\Delta_{M_s}\over \sqrt{3}   }    D_{8a}^{(8)}(R(\tau))\lambda_a
          \right)
       \label{gl33}
\eeq
which appears due to the fact,
that the mass matrix ${\hat m}$
does not commute
with the rotation matrix $R(\tau)$.
We also have defined
\beq
\Delta_{M_s} = M_s - M_u\,.
\eeq
Below we consider $h_1$ as a perturbation. Note that there is a freedom
of the redistribution of the singlet part of the strange quark mass
between $h_0$ and $h_1$. Different
saddle points corresponding to this freedom have been studied by
Schneider {\it et al.} (1995).
In the following we will discuss
the case where the saddle point is determined by using
hamiltonian \qeq{gl32} and $h_1$ is treated as a perturbation.

Expanding the correlation function \qeq{gl29} in powers of $\Omega$
and $\Delta_{M_s}(m_s)$ up to the
second order in the rotational velocity we get
for the propagators
\bea  & &    \prod_{i=1}^{N_c}
       \langle0,T/2\mid
       {1 \over D(RU^{(3)}_cR^\dagger) }
       \mid0,-T/2\rangle_{f_ig_i}
       {}_{\zeichen }
       \nn  & &
       \exp{\left( - N_c \epsilon_{val}(U) T
       - \half I_{ab}^{val} \int d\tau  \Omega_{ }^a \Omega_{ }^b
       - { N_c \over 2 \sqrt{3} }
                                  i \int d\tau \Omega_{ }^8
             - \int d\tau  L_{val}[\Delta_{M_s}]
            \right)}
           \label{gl35}
\eea
as well as from the effective action
\beq
    \exp( -S(R U^{(3)}_c R^\dagger)  )
     = \exp{\left( -  T E_{sea}(U)
     - \half I_{ab}^{sea} \int d\tau  \Omega_{ }^a \Omega_{ }^b
         - \int d\tau  L_{sea}[\Delta_{M_s}]
          \right)}
           \label{gl36}
\eeq
where $L[\Delta_{M_s}]=L_{sea}[\Delta_{M_s}]+L_{val}[\Delta_{M_s}]
      = O(\Delta_{M_s})$ and will be discussed below.
The tensor of the moments of inertia $I_{ab}$ is diagonal but
in contrast to the SU(2) case its components are different
\beq       I_{ab} =  \left\{   \ba{l}
              I_1,\> a=b=1,2,3    \\
              I_2, \> a=b=4,5,6,7  \\
             0, \>  {\rm otherwise.}    \ea  \right.
         \label{gl37}
\eeq
The second important difference from the SU(2) case
is the linear term in $\Omega_{ }^8$ in (\ref{gl35}).
In Skyrme type models, this expression is proportional to the topological
winding number of the chiral field which in these models coincides
with the baryon number. In the present non-topological chiral
model the term with $\Omega_{ }^8$ is due to the discrete valence level
in the Dirac spectrum.
The presence of the $\Omega^8$ term is a reflection of the fact that right
rotations associated with the generator $J_8$
have no effect on the chiral field.
Therefore, the generator $J_8$ corresponding the
rotational velocity $\Omega^8$ is constrained.
From the term linear in $\Omega^8$ in eq.~(\ref{gl35}) we see that
\beq
J_8 = - \frac{N_c}{2\sqrt 3} \,.
\label{J-8-constraint}
\eeq
By analogy with the hypercharge
\beq
Y=(2/\sqrt{3})T_8\,,
\eeq
one can introduce  the "right" hypercharge
\beq
Y_R=(2/\sqrt{3})J_8\,,
\eeq
in terms of which the constraint (\ref{J-8-constraint}) can be rewritten
in the form
\beq
Y_R = - N_c/3 \,.
\eeq
This in turn transforms into a constraint for the possible
representations of SU(3). Actually, one can show that
only representations with zero triality survive. These are
the octet and decuplet representations with spin $1/2$ and $3/2$
respectively.

By analogy with the case of the SU(2) case the effective
rotational lagrangian leads to the following quantization rules
\beq        J_a         =  \left\{
              \begin{array}{ll}
             i { \Omega_{ }^a  I_1},& a=1,2,3     \\
             i { \Omega_{ }^a  I_2},& a=4,5,6,7   \\
           {\sqrt{3}\over 2}Y_R=-{N_c \over 2\sqrt{3}}   ,& a=8 \,.
              \end{array}
                 \right.
         \label{gl37a}
\eeq
Here, the components $J_a$ with $a=1,2,3$
correspond to the spin operator.
Using these quantization rules we obtain
a collective hamiltonian from \qeqs{gl35}{gl36} in the form
\beq   H_{coll}^{(0)}  =
             \half \left( { 1 \over I_1}
          -   { 1 \over I_2}  \right)
          {\cal C}_2(SU(2))
              + { 1 \over 2 I_2}
           {\cal C}_2(SU(3))
            - {3\over 8 I_2} \,.
             \label{gl38}
\eeq
Here
\begin{equation}
{\cal C}_2(SU(3)) = \sum\limits_{a=1}^8 J_a^2 = \sum\limits_{a=1}^8 T_a^2
\end{equation}
is the quadratic Casimir operator of the SU(3) group and
\begin{equation}
{\cal C}_2(SU(2)) = \sum\limits_{a=1}^3 J_a^2
\end{equation}
is the Casimir operator of the SU(2) subgroup corresponding to spin
generators $J_a$ $(a=1,2,3)$.

The upper index $(0)$ in $H_{coll}^{(0)}$ indicates that we are working in the
chiral limit without any mass corrections.
Those will be incorporated later and will yield additional terms
$H_{coll}^{(1)}$ and $H_{coll}^{(2)}$.

The Hamiltonian $H_{coll}^{(0)}$ commutes with all flavor SU(3) generators
$T_a$. Therefore the eigenstates of
$H_{coll}^{(0)}$ can be classified
according to the irreducible representations of SU(3).
Below we use the well known $(p,q)$ parametrization the SU(3)
representations in which the octet is $(1,1)$ and the decuplet is
$(3,0)$.

Since in our model the spin $J_a$ and flavor $T_a$ operators
are realized as the generators of right and left rotations
of the same orientation matrix $R$, the representation $(p,q)$ should be the
same for $T_a$ and for $J_a$. Apart from the representation $(p,q)$
the eigenstates of $H_{coll}^{(0)}$ are also characterized by the quantum 
numbers corresponding to the following set of commuting operators:
\begin{eqnarray}
&&
Y = \frac{2}{\sqrt{3}} T_8,\qquad
\sum\limits_{a=1}^3 T_a^2, \qquad T_3 \,,
\nonumber
\\
&&
Y_R = \frac{2}{\sqrt{3}} J_8,\qquad
\sum\limits_{a=1}^3 J_a^2, \qquad J_3  \,.
\end{eqnarray}
The corresponding wave functions are
\begin{equation}
\psi_{(Y,T,T_3)(Y_R,J,J_3)}^{(p,q)}(R) = \sqrt{\mbox{dim}(p,q)}\,
(-1)^{Y_R/2+J_3}D_{(Y,T,T_3)(-Y_R,J,-J_3)}^{(p,q)\ast}(R)
\label{wavefunc}
\end{equation}
where the asterisk stands for the complex conjugation and
$\mbox{dim}(p,q)$ is the dimension of the representation $(p,q)$.

With these wave functions
the Casimir operator has the following spectrum
\begin{equation}
{\cal C}_2(SU(3))  \psi_{Y,T,T_3;Y_R,J,J_3}^{(p,q)}
=\frac13(p^2+q^2+3(p+q)+pq)  \psi_{(Y,T,T_3)(Y_R,J,J_3)}^{(p,q)} \,,
\label{C-SU3-action}
\end{equation}
\begin{equation}
{\cal C}_2(SU(2))  \psi_{(Y,T,T_3)(Y_R,J,J_3)}^{(p,q)}
= J(J+1) \psi_{(Y,T,T_3)(Y_R,J,J_3)}^{(p,q)}    \,.
\label{C-SU2-action}
\end{equation}

In the chiral limit presently considered, the
members of each multiplet are degenerate in energy.
Using  (\ref{C-SU3-action}), (\ref{C-SU2-action})
we find the eigenvalues of the rotational hamiltonian (\ref{gl38})
which leads to the following result
for the energy of the center of the multiplets, defined as the average
mass of all the members,
\beq   E_{p,q,J} =
         M_{cl}
          + \half \left( { 1 \over I_1}
        -  { 1 \over I_2}  \right)
               J(J+1)
            + { 1 \over 6 I_2}
           \left( p^2 +q^2 +3(p+q)+pq \right)
          - {3\over 8 I_2}\,.
           \label{gl39}
\eeq
Note that constraint (\ref{J-8-constraint})
fixes the spin of the particles of the $(p,q)$ multiplet.
The analog of the nucleon delta splitting (\ref{EDN})
has to be interpreted in SU(3) as
octet and decuplet-splitting, i.e.
the energy difference between the centers of the
octet and decuplet,
and follows as
\beq  E_{8-10} = E_{3,0,3/2} -   E_{1,1,1/2}  =
                 { 3 \over 2 I_1 } \,.
\eeq
Formally this coincides with $E_{N\Delta}$ in SU(2), however, the
phenomenological value for the difference of the average octet and decuplet
masses is now $E_{8-10}^{ph}=230\MeV$.

\vspace{8mm}
5.6 \underline{Strange mass terms of the collective lagrangian}
\vspace{5mm}

In order to calculate the splitting {\it within} the multiplets we have
to include the SU(3) symmetry breaking terms due to $h_1$ in  
(\ref{gl33}). Note that we deal with two quantities
which may be considered as small quantities. The first one is the rotational
velocity $\Omega$, which is of the order of $1/N_c$.
The second one is the mass difference $\Delta_{M_s}=M_s-M_u$.
Assuming both corrections to be roughly of the same order we use 
systematic expansions up to the second order.

Expanding the baryon correlator \qeq{gl29} and considering
sea and valence contributions together, one obtains
for $L[\Delta_{M_s}]$ in
\qeq{gl35} and \qeq{gl36}
\bea       L[\Delta_{M_s}]   &=&
           L_1 [\Delta_{M_s}]
       +   L_2 [\Delta_{M_s},\Omega ]  \nn & &
       +   L_3 [\Delta_{M_s}^2]
             + O \left( \Delta_{M_s}^m \Omega^n
                  \right)\ \ \ m+n\ge 3 \,.
\eea
The first term in this expansion
\beq      L_1 [\Delta_{M_s}]
         =
           { \sigma \over {\mbar} }
            { \Delta_{M_s} \over  3  }
         \left( 1 - D_{88}^{(8)}(R(\tau)) \right)
\eeq
contains the SU(2) sigma
term $\sigma$, which in the present
model is given by
\beq  \sigma = {\mbar}
    { {  \partial M_{cl}[{\mbar}  ] \over
           \partial {\mbar} }  \linie}_{{\mbar} =0} \,.
\eeq
Then we have in the next order  $O\left(
\Omega\Delta_{M_s}\right)$
\bea        L_2[ \Delta_{M_s},\Omega]
                    &=&
            { \Delta_{M_s} \over \sqrt{3} }
                2  K_1 \sum_{i=1}^3
         D_{8i}^{(8)}(R(\tau))
                   \Omega_i(\tau)    \nn & &
                      +    { \Delta_{M_s} \over \sqrt{3} }
          2 K_2  \sum_{a=4}^7   D_{8a}^{(8)}(R(\tau))
                   \Omega_a(\tau) \,,
\eea
where $K_1$ and $K_2$ similar to $I_1,I_2$ are related to tensor
$K_{ab}$ 
\beq       K_{ab} =  \left\{
             \begin{array}{ll}
             K_1, & a=b=1,2,3    \\
             K_2, &a=b=4,5,6,7  \\
             0,\  & {\rm otherwise.}
             \end{array}
\right.
              \label{gl42}
\eeq
The latter can be split in valence
\beq     K_{ab}^{val}  =
        { N_c \over 2}
       \sum_{n\neq m} {
       \langle val \mid \lambda^a \mid n\rangle
      \langle n   \mid \lambda^b \gamma_0 \mid val \rangle
           \over \epsilon_n - \epsilon_{val} }\,.
            \label{gl43}
\eeq
and sea part
\beq     K_{ab}^{sea}  =
        { N_c \over 4}
       \sum_{\epsilon_n>0\atop \epsilon_m<0} 
       {\langle m   \mid \lambda^a \mid n\rangle
      \langle n   \mid \lambda^b \gamma_0 \mid m \rangle
           \over \epsilon_n - \epsilon_m  }
              \label{gl44}
\eeq
In contrast to the sea contributions of $I_1$ and $I_2$ that of $K_{ab}$ 
originates from the imaginary part of the effective euclidean action. 
Hence, it 
is ultraviolet finite and needs no regularization. One should note that in 
Skyrme approaches with vector mesons such contributions from the effective 
action are obtained only by adding the gauged Wess-Zumino term 
(Park and Weigel, 1991).

From the quadratic corrections in the symmetry breaking we obtain
\bea       L_3 [\Delta_{M_s}^2]
             &=&
            \left( { \Delta_{M_s} \over \sqrt{3} }
                           \right)^2
                2 N_1 \sum_{i=1}^3       D_{8i}^{(8)}(R(\tau))
                     D_{8i}^{(8)}(R(\tau))    \nn & &
                 +       \left( { \Delta_{M_s} \over \sqrt{3} }
                           \right)^2
                2 N_2 \sum_{a=4}^7   D_{8a}^{(8)}(R(\tau))
                     D_{8a}^{(8)}(R(\tau))     \nn & &
      +     \left(   { \Delta_{M_s} \over \sqrt{3}  }  \right)^2
                 {2\over 3}
              N_0
                 \left( 1 - D_{88}^{(8)}(R(\tau)) \right)
                 \left( 1 - D_{88}^{(8)}(R(\tau)) \right)\,,
                  \label{gl41}
\eea
where $N_1$ and $N_2$ are diagonal elements of
tensor $N_{ab}$
\beq
  N_{ab} =  \left\{   \ba{l}
     N_1,\qquad a=b=1,2,3    \\
     N_2,\qquad a=b=4,5,6,7  \\
     N_0,\qquad a=b=8   \,.       \ea  \right.
 \label{gl46}
\eeq
It consists of a valence part
\beq     N_{ab}^{val}  =
        { N_c \over 2}
       \sum_{n\ne val} {
       \langle val \mid \lambda^a\gamma_0 \mid n\rangle
      \langle n   \mid \lambda^b \gamma_0 \mid val \rangle
           \over E_n - E_{val} }
            \label{gl48}
\eeq
and a regularized sea contribution
\beq     N_{ab}^{sea}  =
        { N_c \over 4}
       \sum_{n,m}
       \langle m   \mid \lambda^a \gamma_0 \mid n\rangle
      \langle n   \mid \lambda^b \gamma_0 \mid m \rangle
           {\cal R}_N^\Lambda(\epsilon_n,\epsilon_m) \,.
              \label{gl49}
\eeq
The regularization function ${\cal R}_N^\Lambda$
is given by
\beq    {\cal R}_N^\Lambda (\epsilon_n,\epsilon_m)
        = { 1 \over 2 \sqrt{\pi} }
        \int_0^\infty
       { d u \over \sqrt{u} } \phi(u)
      \left[ { \epsilon_n e^{-u \epsilon_n^2} -
               \epsilon_m e^{-u \epsilon_m^2}  \over
               \epsilon_n - \epsilon_m  }    \right] \,.
\eeq
Before we can study the mass splittings from the collective
hamiltonian resulting from \qeq{gl41}, one has to remember that
in the presence of the linear terms in the rotational
velocity $\Omega$, the generators of the group are modified
according to
\beq        J_a         =  \left\{
                   \ba{ll}
             i { \Omega_{ }^a  I_1}
              -2   { \Delta_{M_s}\over \sqrt{3} }
                        K_1     D_{8a}^{(8)}(R(\tau))
         ,& a=1,2,3     \\
             i { \Omega_{ }^a  I_2}
              -2   { \Delta_{M_s}\over \sqrt{3} }
                        K_2     D_{8a}^{(8)}(R(\tau))
              ,& a=4,5,6,7   \\
           -{N_c \over 2\sqrt{3}} = {\sqrt{3}\over 2} Y_R,
                  & a=8 \,.
                     \ea         \right.
         \label{gl50}
\eeq

\vspace{8mm}
5.7 \underline{Hyperon  splittings in linear order - Sum rules}
\vspace{4mm}

Here, we consider the hyperon splitting of the octet and
decuplet in linear order in $\Delta_{M_s}$ and
neglect the isospin breaking, i.e.
set $m_u=m_d=m_0$. As shown in the meson sector in linear order of
$\Delta_{M_s}$ the difference $M_s-{\barM}$ equals $m_s$. Now the collective 
Hamiltonian
$$H_{coll} = H_{coll}^{(0)} +  H_{coll}^{(1)}$$
includes a new term
\bea    H_{coll}^{(1)}   &=&
% ------------m^1
                 { \sigma   \over {\mbar} } {m_s\over 3}
\left(
              1  -   D_{88}^{(8)}(R)    \right)
% ----------------m R ---
        +    {2m_s \over \sqrt{3} }
            \left(    {K_1 \over I_1}  -{K_2\over I_2} \right)
           \sumi     D_{8i}^{(8)}(R)    J_i
          \nn & &      - {m_s} {K_2\over I_2}  Y
       +
             {N_c m_s \over 3}  {K_2\over I_2}   D_{88}^{(8)}(R) \,.
\label{hamsymm2}   \eea
The moments of inertia
and the matrix elements of the Wigner functions can be found in
(Blotz {\it et al.}, 1993a).

Although $H_{coll}^{(1)}$ contains three different operators
$Y,D_{88}^{(8)}(R)$ and $\sumi D_{8i}^{(8)}(R)J_i$,
the mass splitting can be parametrized in terms the two quantities
\beq   \Delta^{hyp} = {1\over 3} {m_s\over {\mbar} } \sigma
                 +m_s \left( 2{K_2\over I_2} - 3{K_1\over I_1}
                 \right)
\eeq
and
\beq    \delta^{hyp} = m_s {K_1\over I_1} \,.  \eeq
For the masses of the octet and decuplet
baryons, calculated relative to the mass of the  $\Sigma^*$, one gets 
(Blotz {\it et al.}, 1993a)
\newpage
\beq
\begin{array}{lllrlrll}
&\Delta m_N &=  &-\frac{3}{10}&\Delta^{hyp} &-&\delta^{hyp} &-
E_{8-10}
                                                            \\
&\Delta m_\Lambda &=  &-\frac{1}{10}&\Delta^{hyp} & &   &-
E_{8-10}
                                                            \\
&\Delta m_\Sigma &=  &\frac{1}{10}&\Delta^{hyp} & &    &- E_{8-10}
                                                            \\
&\Delta m_\Xi &=  &\frac{1}{5}&\Delta^{hyp} &+&\delta^{hyp}   &-
E_{8-10}
                                                            \\
&\Delta m _\Delta &=  &-\frac{1}{8}&\Delta^{hyp} &-&\delta^{hyp}  &
                                                            \\
&\Delta m_{\Sigma^*}&= &0           &  &       &
                                                            \\
&\Delta m_{\Xi^*}&=   &\frac{1}{8}&\Delta^{hyp}    &+&\delta^{hyp}   &
                                                            \\
&\Delta m_\Omega &=   &\frac{1}{4}&\Delta^{hyp}    &+&2\delta^{hyp} &
                \label{c6pl}                                \\
\end{array}
\eeq
The reason that we consider the mass spectrum relative to the
$\Sigma^*$ is that the mass of the $\Sigma^*$ almost
corresponds to the center of the mass of the decuplet.
From the formulas \queq{c6pl}      above, it follows  the mass
relation of Guadagnini (1984)
\beq
     m_{\Xi^*}-  m_{\Sigma^*} +   m_N ={1\over 8} (11m_\Lambda - 3
      m_\Sigma)  \,.
\label{c635}     \eeq
Here
it is interesting to note that this sum rule can be obtained in
the pseudoscalar Skyrme model only by introducing the   hypercharge
operator by hand with some unknown coefficients. In the present approach, 
this term arises naturally \queq{hamsymm2}.

Furthermore, one obtains also the Gell-Mann -- Okubo  relations
\beq
   {3\over 4}m_\Lambda+{1\over 4}m_\Sigma={1\over 2}(m_N+m_\Xi)
\label{c636a}
\eeq
and
\beq
   m_\Omega-m_{\Xi^*}=m_{\Xi^*}-m_{\Sigma^*}=m_{\Sigma^*}-m_{\Delta}\,.
\label{c636b}
\eeq
It should be noted that these relations are mass sum rules which rely
basically on the fact, that the SU(3) flavor symmetry breaking part of
the strong interaction can be treated in first order perturbation
theory. Their validity shows in some way the consistency of
the quantization procedure. On the other hand the values of the
hyperon masses themselves depend on the dynamics of the present model
and this goes beyond the perturbative sum rule approach.

\vspace{8mm}
5.8 \underline{Hyperon splitting in second order - Wave functions}
\vspace{4mm}

In the present scheme, the corrections in second order of the strange 
quark mass
have three different origins. The first one results from the expansion of the
baryon correlator in $\Delta_{M_s} = M_s-M_u$.
The second one is due to the quantization formulas
\qeq{gl50}, which replace the angular velocities in
\qeq{gl41} by the corresponding generators.
Accordingly, the collective hamiltonian contains not only 
\bea    H_{coll}^{(1)}   &=&
% ------------m^1
       { \sigma   \over {\mbar}  } { \Delta_{M_s} \over 3}
\left(
        1    -      D_{88}^{(8)}(R)      \right)    \nn & &
% ----------------m R ---
    +    {2 \Delta_{M_s} \over \sqrt{3} }  {K_1 \over I_1}  \sumi
              D_{8i}^{(8)}(R)    J_i
      +  {2 \Delta_{M_s} \over \sqrt{3}} {K_2\over I_2} \suma
                  D_{8a}^{(8)}(R)   J_a
 \label{gl80}
\eea
%--------quantization
but also second order terms in the $\Delta_{M_s}(m_s)$
\bea
      H_{coll}^{(2)}    &=&
              2 {\Delta_{M_s}^2\over 3}   {K_1^2 \over I_1}  \sumi
    D_{8i}^{(8)}(R)      D_{8i}^{(8)}(R)
             +2 {\Delta_{M_s}^2\over 3}  {K_2^2\over I_2} \suma
          D_{8a}^{(8)}(R)     D_{8a}^{(8)}(R)
             \nn
% -----m^2 -------
   & &           - 2    N_1  {\Delta_{M_s}^2\over 3}    \sumi
                               D_{8i}^{(8)}(R)
                               D_{8i}^{(8)}(R)
                           - 2  N_2    {\Delta_{M_s}^2\over 3}  \suma
               D_{8a}^{(8)}(R)   D_{8a}^{(8)}(R)
                   \nn & &
          -  2    N_0    {\Delta_{M_s}^2 \over 3}
         \left(  1 -  D_{88}^{(8)}(R)  \right)^2  \,.
         \label{gl81}
\eea
To study the third contribution in second order
of the symmetry breaking we recall the perturbation theory for
our system of $H_{coll}^{(0)}+H_{coll}^{(1)}+H_{coll}^{(2)}$.
For the energy of a baryon $B$ in the representation $[n]$, which has to
be the lowest dimensional one where the quantum numbers of the baryon
appears, we can write in increasing order of the perturbation:
\bea  E_{B,n}^{(0)}&=& \langle B,n \mid  H_{coll}^{(0)} \mid B,n \rangle
       \label{gl100}  \\
      E_{B,n}^{(1)}&=&  \langle B,n \mid H_{coll}^{(1)} \mid B,n
                        \rangle
            \label{gl101}  \\
      E_{B,n}^{(2)}&=&
       \langle B,n \mid H_{coll}^{(2)} \mid B,n
       \rangle
            +
     \sum_{n\ne n'}
  { \mid \langle B,n' \mid H_{coll}^{(1)} \mid B,n \rangle \mid^2 \over
         E_{B,n}^{(0)}    -   E_{B,n'}^{(0)} }  \,.
     \label{gl102}
\eea
Here, $E_{B,n}^{(m)}$ is the energy of baryon of quantum
numbers $B$ in the representation $[n]$ with strangeness corrections
of the order $m$ taken into account. In the lowest order the energies 
$E_{B,n}^{(0)}$
\qeq{gl100} are degenerate for each multiplet and correspond to the mass
center of the multiplet. In the next order $E_{B,n}^{(1)}$
include linear strange mass contributions from the
matrix elements of $H_{coll}^{(1)}$ between the unperturbed
wave functions $\mid B,n \rangle$.
This removes the degeneracy for the zero-order results $E_{B,n}^{(0)}$.
The second-order contributions  $E_{B,n}^{(2)}$ \qeq{gl102} 
are from $H_{coll}^{(2)}$ between the wave functions
of lowest order as well as from the last term in \qeq{gl102}.
There the summation is over all representations $[n']\ne[n]$.
In fact, this term appears since the baryon wave-functions
of the collective hamiltonian $H_{coll}^{(0)}+H_{coll}^{(1)}$
after diagonalization in the space of all $\mid B,n \rangle$
are no longer pure octet or decuplet states.
Because the operator  $H_{coll}^{(1)}$ commutes
with  $Y,T,T_3$ and $J$ operators,
there are in addition only higher representations like $[{\bar 10}]$
of $[27]$ in the case of octet baryons  and
$[27]$ and $[35]$ in the case of decuplet ones.
These higher
representations enter the baryon wave functions
implicit
in \qeq{gl102}
perturbatively
in first order of the symmetry breaking and therefore
the total contribution to the energy in \qeq{gl102}
is quadratic in the symmetry breaking.

Actually, if the baryon is a member of the
octet, the only non-vanishing matrix elements  can be parametrized by
the quantum numbers of the baryons according to
\beq   \mid
        \langle B,{\bar{10}}\mid H_{coll}^{(1)} \mid  B,8\rangle
           \mid^2
        = {1\over 40 } \left(
          \alpha +\gamma\half\right)^2   \left(
             Y+{1\over 4} Y^2 + T(T+1)   \right)
\label{pert2}\eeq
and
\beq  \mid
     \langle B,{{27}}\mid H_{coll}^{(1)}  \mid B,8\rangle
              \mid^2
        = {1\over 100 } \left(
          \alpha -\gamma{1\over 6} \right)^2  \left(
     9 + {5\over 2} \left( T(T+1)-{1\over 4}Y^2 \right) - {7\over 4}Y^2
          \right)
\label{pert3}  \eeq
with the abbreviations
\bea
\alpha &=& -{1\over 3} {\Delta_{M_s} \over {\mbar}  } \sigma
           + {K_2\over I_2} \Delta_{M_s}   \nn
\beta &=& - \Delta_{M_s} {K_2\over I_2}    \nn
\gamma &=& 2 \Delta_{M_s}  \left( {K_1\over I_1} - {K_2\over I_2} \right)
\,.
\label{pert4}
\eea
Similarly, if the baryon is in the decuplet, the only non-vanishing
matrix elements are given by
\beq    \langle B,27 \mid H_{coll}^{(1)} \mid  B,10\rangle
        =    {1\over 16}   \left(
          \alpha +\gamma{5\over 6} \right)^2
          \left( 1 + {3\over 4} Y + {1\over 8} Y^2  \right)
\label{pert5}\eeq
and
\beq    \langle B,35 \mid H_{coll}^{(1)} \mid  B,10\rangle
        =       {5\over 112}        \left(
          \alpha -\half \gamma     \right)^2
        \left(  1- {1\over 4}  Y  - {1\over 8} Y^2  \right) \,.
\label{pert6}\eeq

We will consider expectation values
of external current operators, for which we need also the wave function
$\mid B\rangle$
itself
in the next to leading order of the symmetry breaking.
Then
the wave functions can be written in first order perturbation theory as
\beq   \mid B\rangle =     \mid n,B\rangle   +
       \sum_{n\ne n'}      \mid n',B\rangle       {
        \langle B,n'\mid H_{coll}^{(1)} \mid  B,n\rangle \over
         E_{B,n}^{(0)}   -   E_{B,n'}^{(0)}    } \,.
\eeq
Furthermore, the wave functions for the octet baryons,
which will be the case of special interest in the next
sections, in linear approximation in $\Delta_{M_s}$ read
\bea   \mid     B\rangle  &=&   \mid  B,8\rangle  -
        {I_2\over 6\sqrt{5}} \left( \alpha+ \half\gamma \right)
       \left( Y+{1\over 4}Y^2 + T(T+1) \right)
        \mid B,{\bar{10}} \rangle   \nn
        & & - {2 I_2 \over 50}
        \left( \alpha -{1\over 6}\gamma \right)
       \left(
       \sqrt{6}  Y^2 + 3(1-{7\over 8}Y^2)-\half T(T+1)  \right)
      \mid B,{{27}}    \rangle
\label{wavefun} \eea
The expectation values of these current operators, corresponding to some
observable quantity, can be obtained consistently by sandwiching
the operators, calculated up to the linear order in the symmetry
breaking, between the unperturbed wave functions (\qeq{wavefunc})
and then adding the matrix element of the operators, calculated now in
zeroth order of the symmetry breaking, between the
perturbed wave function (\queq{wavefun}).
In this scheme, one can calculate consistently the same type
of corrections from the symmetry breaking to the
expectation values of these current operators
as it was previously in the case of the energy. However, because of technical
reasons, current expectation values are always calculated with
linear rotational and linear strangeness corrections.

\vspace{8mm}
5.9 \underline{Yabu-Ando diagonalization method }
\vspace{5mm}

In contrast to the former perturbative methods of evaluating the
collective hamiltonian either between the unperturbed
eigenfunctions (\ref{wavefunc}) or the first order perturbated ones 
(\ref{wavefun}), in the
case of the wave-function corrections, Yabu and Ando
(1988) developed a method
for treating the baryonic wave functions to all orders in  $m_s$
or in our scheme in $\Delta_{M_s} = M_s-M_u$.
Without reviewing technical details, which can be found
in the original work (Yabu and Ando, 1988),
the method is based on diagonalizing the collective hamiltonian
$H_{coll}^{(0)} + H_{coll}^{(1)} + H_{coll}^{(2)}$ in the
basis, spanned by the eigenfunctions of $H_{coll}^{(0)}$.
Therefore, the wave functions for the octet or decuplet baryons
after the diagonalization do not only have an admixture
of the representations mentioned above, but contain representations
of arbitrary high dimensions. Of course, if the symmetry breaking
can be considered as small, these higher representations are
also suppressed in the wave functions.
Nevertheless, this method is able to calculate the strange
mass corrections from the wave function to all orders.
On the other hand, as we have seen above, there are
contributions of the same order from expanding the
baryon correlator itself. Therefore it seems to be more
consistent to calculate all kinds of strange mass corrections
up to the second order instead of calculating only one
of these corrections up to all orders and the others
up to the second. The fact that the latter ones are calculated only
up to the second order is a technical problem to which
no solution is seen.

However, apart from these theoretical considerations
a numerical comparison between the Yabu-Ando treatment
and the perturbative treatment of the wave function correction
up to the second order has been performed recently
(Blotz {\it et al.}, 1995a). As can be seen clearly in this
investigation both methods lead to almost identical
results for the mass splittings. This means that
higher order corrections, at least for the wave function,
$O\left(m_s)^3\right)$ do not play a
significant role.

\vskip1cm
5.10 \underline{Results: Mass splittings in the SU(3) NJL model}
\vspace{4mm}

Several authors have performed numerical calculations for the hyperon mass 
splittings. First order perturbative results can be found in (Weigel 
{\it et al.,} 1992, Blotz {\it et al.,} 1992, 1993a). These papers present 
also the numbers obtained in the Yabu-Ando approach which has been compared to
the the second order results by (Blotz {\it et al.,} 1995a). The inclusion of 
the second order corrections in the strange mass improves the agreement with 
the experiment and the corresponding second order results from (Schneider 
{\it et al.,} 1995) are presented in the Table~\ref{bms}. 

It is also interesting to see to what extent the numbers of table \ref{bms} 
fulfill the the Gell-Mann -- Okubo mass relations. If the nucleon mass in the 
NJL model is put to 938 meV one obtains for ${3\over 4}m_\Lambda+{1\over 4}
m_\Sigma-{1\over 2}(m_N+m_\Xi)$ a value of $-9.2$ MeV (from table~\ref{bms}) 
and $+6.7$ MeV for the experimental masses. Both values of these violations 
are on the scale of 1 \% in the baryon masses, nevertheless one should note 
the different sign. For the baryon 
decuplet (\ref{c636b}) 
the three terms in (\ref{c636b}) are $138, 149\mbox{ and }150$ MeV from table 
\ref{bms} compared to $139, 149\mbox{ and }152$ 
MeV for the experimental masses. Apparently, both the experiment and the 
SU(3) NJL model calculations fulfill the ll-Mann -- Okubo mass relations 
equally well.

\begin{table}
\caption{Hyperon mass splittings in second order perturbative calculations in 
the SU(3) NJL model (Schneider {\it et al.,} 1995).} 
\label{bms}
\begin{center}
\begin{tabular}{|c|c|c|}  \hline
$\mbox{Baryon}$ & $M_B-M_N$ [MeV] 
& $M_B^{exp}-M_N^{exp}$ [MeV] \\
\hline 
$N$           &   0 &   0 \\\hline
$\Sigma$      & 224 & 254 \\\hline
$\Xi$         & 354 & 379 \\\hline
$\Lambda$     & 149 & 177 \\\hline
$\Delta$      & 288 & 293 \\\hline
$\Sigma^\ast$ & 438 & 445 \\\hline
$\Xi^\ast$    & 587 & 594 \\\hline
$\Omega$      & 725 & 733 \\\hline
\end{tabular}
\end{center} 
\end{table}

\begin{figure}
\centerline{\epsfysize=3.5in\epsffile{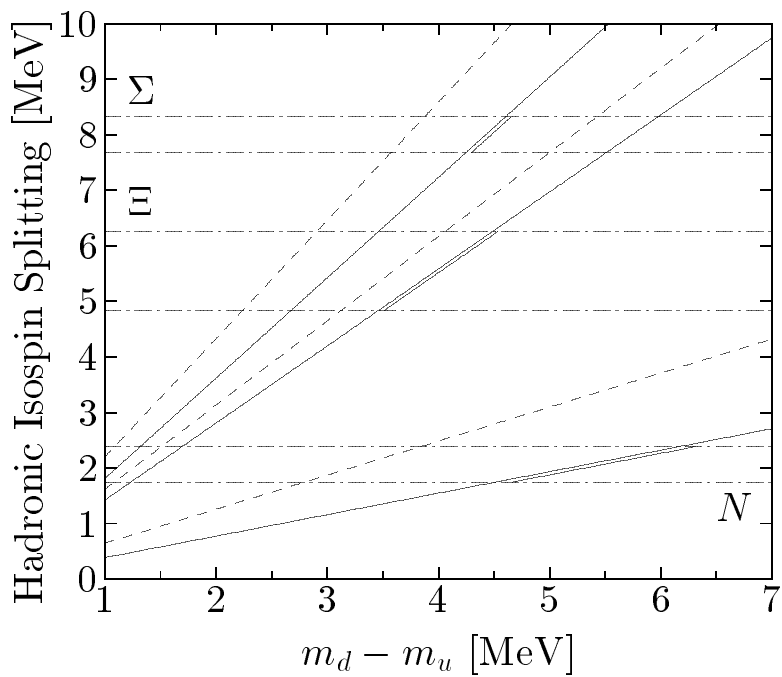}}%\vskip4pt
\caption{Hadronic part of the octet isospin splittings 
$N(n-p)$, $\Xi(\Xi^--\Xi^0)$, $\Sigma(\Sigma^--\Sigma^+)$.
The SU(3)-NJL calculations are presented by solid lines (with $m_s$-
corrections) and dashed lines (without $m_s$-corrections).
The experimental values including error are indicated by dashed-dotted 
lines. The double lines indicate the regions where the SU(3)-NJL calculations 
with $m_s$-corrections overlap with the experimental numbers.}
\label{isospl}
\end{figure}

The above mass splittings of the baryons belonging to the octet and decuplet 
are due to the finite value of $m_s$. It is also interesting to calculate the 
mass splittings within the isospin multiplets. It consists of the hadronic 
and the electromagnetic part
\begin{eqnarray}
\Delta M_B = \left( \Delta M_B \right)_h + \left( \Delta M_B \right)_e\,.
\end{eqnarray}
The origin of this is theoretically clear, since the electromagnetic part is
due to the baryon electromagnetic self-energy and the hadronic part is due to
the current mass difference $\delta m _{ud}=m_d-m_u$.
Gasser and Leutwyler (1982) quote the following values for the experimental
estimates of the electromagnetic parts of the octet baryons (in MeV):
\begin{eqnarray}
\left( n - p \right)_e = -0.76 \pm 0.3\,, \ \ \
\left( \Sigma^- - \Sigma^+ \right)_e = 0.17 \pm 0.3 \,, \ \ \
\left( \Xi^- - \Xi^0 \right)_e = 0.86 \pm 0.3\,.
\end{eqnarray}
These values have to be subtracted from the measured mass differences
\begin{eqnarray}
\left( n - p \right)_{exp} = 1.29 \,,\ \ \
\left( \Sigma^- - \Sigma^+ \right)_{exp} = 8.07 \pm 0.09 \,,\ \ \
\left( \Xi^- - \Xi^0 \right)_{exp} = 6.4 \pm 0.6\,,
\end{eqnarray}
and one obtains for the hadronic parts
\begin{eqnarray}
\left( n - p \right)_h = 2.05 \pm 0.3 \,,\ \ \
\left( \Sigma^- - \Sigma^+ \right)_h = 7.89 \pm 0.3 \,,\ \ \
\left( \Xi^- - \Xi^0 \right)_h = 5.5 \pm 0.7\,.
\end{eqnarray}
Unfortunately, these simple estimates do not exist for the baryon decuplets.
In chiral quark models, the hadronic part of the mass splitting is related to
the current quark mass difference
$\Delta m_{ud} = m_d - m_u$
and in Skyrme type models to the corresponding boson fields.
Actually, in the SU(2) NJL the hadronic
part of the neutron-proton mass difference vanishes identically in the leading order in $N_c$.
One way to cure this disease is to enlarge the symmetry group.
Jain {\it et al.} (1989) used a U(2) $\otimes$ U(2) extension of the Skyrme
model with pseudoscalar and vector fields.
They predicted, however, a n-p hadronic mass difference which was 35$\%$ too
small. Some increase was obtained by the same authors in the case of U(3) 
$\otimes$ U(3).
Similar results were obtained in the chiral bag model (Park and Rho, 1989),
where quark degrees of freedom are explicitly taken into account.
The symmetry breaking in the Skyrme model has recently successfully been 
reinvestigated by Walliser (1993).

In contrast to some of these estimates, the SU(3) NJL calculations of 
Praszalowicz
{\it et al.} (1993) and Blotz {\it et al.} (1994) show good agreement with the
experimental data not only for the n-p system but also for
$\Sigma^- - \Sigma^+$ and $\Xi^- - \Xi^0$.
In these calculations, the hadronic part of the isospin splittings is evaluated
perturbatively to orders $O(m_s^2)$ and $O(m_s \Delta m_{ud})$.
The results for a constituent mass of $M=420$MeV they obtained are
shown in fig.~\ref{isospl}.
As can be seen, for a common value of $\Delta m_{ud}$ of
4.4MeV the experimental data are reproduced within the error bars.
If one ignores strange mass corrections, the difference $\Delta m_{ud}$ 
reduces to about
3.5MeV. Obviously the SU(3) NJL offers a satisfactory description for the 
isospin mass differences in the octet.
Predictions for the decuplet can be found in (Praszalowicz {\it et al.,} 1993).

\newpage
\section{Baryon properties in the SU(3) NJL model}

In the previous chapter, we have presented the collective
quantization in SU(3) case and showed that the SU(3) model provides a
successful description of the mass splittings of the SU(3) baryon
octet and decuplet. From the mass splittings for the
constituent quark mass $M=420\;\mbox{MeV}$ has been chosen and
this value will be used throughout all calculations, which we will present in 
this subsection, so that there will be no more free adjustable parameter.  
Since the SU(3)-formalism is a generalization of the SU(2)
NJL model, though being quite involved and complicated,
we will not repeat the details of the calculations but concentrate on
the results for the static properties as well as for various form factors of
the SU(3) octet baryons.  We will consider the contributions up to the
first order in both $\Omega$ and $\Delta_{M_s}(m_s)$. 

The investigations in SU(3) case has two significant meanings:
First, stimulated by recent experiments, it is now an important issue of hadron
physics to understand the role of the  strangeness in the nucleon structure.
Second, to study the properties of the other members
of the SU(3) multiplets.

We will start with studying the
electromagnetic properties of the SU(3) octet baryons.

\vspace{1cm}
6.1 \underline{Electromagnetic properties}
\vspace{4mm}

The electromagnetic Sachs form factors $G_{E} (q^2)$ and
$G_M (q^2)$
\begin{equation}
\langle B^\prime (p^\prime) | j^0(0) | B(p) \rangle = G_E(q^2)
\delta_{J_3^\prime J_3} \,, \label{Sachs-electric-form-factor-B}
\end{equation}
\begin{equation}
\langle B^\prime (p^\prime) | j^k(0) | B(p) \rangle = \frac{i}{2
M_N}
\varepsilon^{klm} (\tau^l)_{J_3^\prime J_3} q^m
 G_M(q^2)\,,
\label{Sachs-magnetic-form-factor-B}
\end{equation}
are related to the matrix element of the time and space components of the
electromagnetic current
\begin{equation}
j_\mu (x) \;=\; \bar{\Psi}(x) \gamma_\mu \hat{Q}\Psi(x)\,.
\label{elmag-current-su3}
\end{equation}
Here $ M_N$ denotes the experimental nucleon mass and $\hat{Q}$ is the quark
charge matrix
\begin{equation}
\hat{Q}\;=\;\left( \begin{array}{ccc} \frac{2}{3} & 0 & 0 \\
0 & -\frac13 & 0 \\     0 & 0 & -\frac13 \end{array} \right)
\end{equation}
The $q$ is the four momentum transfer and we also use $Q^2 = -q^2$.
Hence, the electromagnetic current $j_\mu$ can be decomposed
in third and eighth  SU(3) octet currents
\begin{equation}
j_\mu\;=\; j^{(3)}_{\mu} \;+\; \frac{1}{\sqrt{3}} j^{(8)}_{\mu}
\end{equation}
with
\begin{equation}
j^{(3)}_{\mu} = \frac12 \bar{\Psi}\gamma_\mu \lambda^3 \Psi \qquad \mbox{and}
\qquad j^{(8)}_{\mu} = \frac12 \bar{\Psi}\gamma_\mu \lambda^8 \Psi .
\label{Eq:emcurrent2}
\end{equation}

The matrix element of the electromagnetic current are evaluated according
to the technique developed in chapter 3 (for details, see 
(Kim {\em et al.}, 1995a)).

Fig.~\ref{Figr18} shows the electromagnetic form factor of the nucleon for 
three different values of the constituent quark mass $M$ 
(Kim {\em et al}. 1995a). As can be seen, the model results
agree very well with the presented experimental data. In particular, the
$q$-dependence of the form factors are reproduced rather well. The deviations
from the experimental data, seen in the case of the magnetic form factors, is
due to a slight underestimation of both proton and neutron magnetic moments.
It should be also mentioned that the electromagnetic form factors depend weakly
on the particular values of the constituent quark mass if this is chosen  
between 400 and 450 MeV. As in the case of SU(2) version, values of the 
constituent mass 
around $M=420$ MeV seem to be preferable. In the case of the magnetic form 
factors on fig.~\ref{Figr18}, one should note that the results are presented 
without any additional normalization to the experimental values of the 
magnetic moments. 

\begin{figure}
\centerline{\epsfysize=5.5in\epsffile{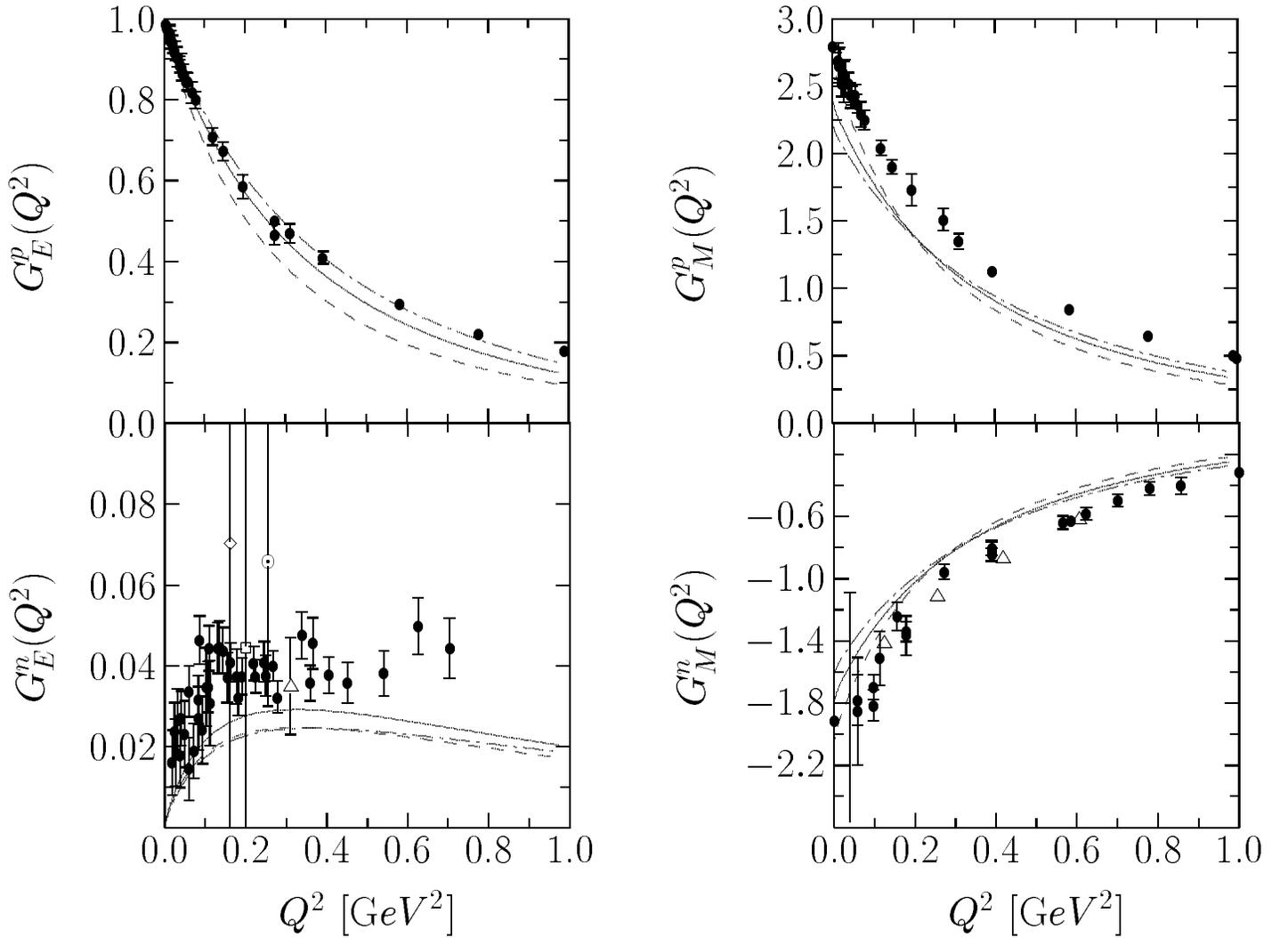}}%\vskip4pt
\caption{Electric (left) and magnetic (right) form
factors of the nucleon as functions of $Q^2$ evaluated in the SU(3) NJL model 
for three different values of $M$: $M=370 \mbox{MeV}$ (dashed line), 
$M=420\mbox{MeV}$ (solid line) and 
$M=450\mbox{MeV}$ (dash-dotted line) (Kim et al., 1995a). 
For the strange mass $m_s=180 \;\mbox{MeV}$ is used.
The experimental data are from
(H\"ohler {\em et al.}, 1976, Platchkov {\em et al.}, 1990, Eden  {\em et al.},
 1994, Meyerhoff {\em et al.}, 1994, Bruins {\em et al.}, 1995). 
The results for the magnetic form factors are presented without any scaling.}
\label{Figr17}
\end{figure}
\begin{figure}
\centerline{\epsfysize=3.8in\epsffile{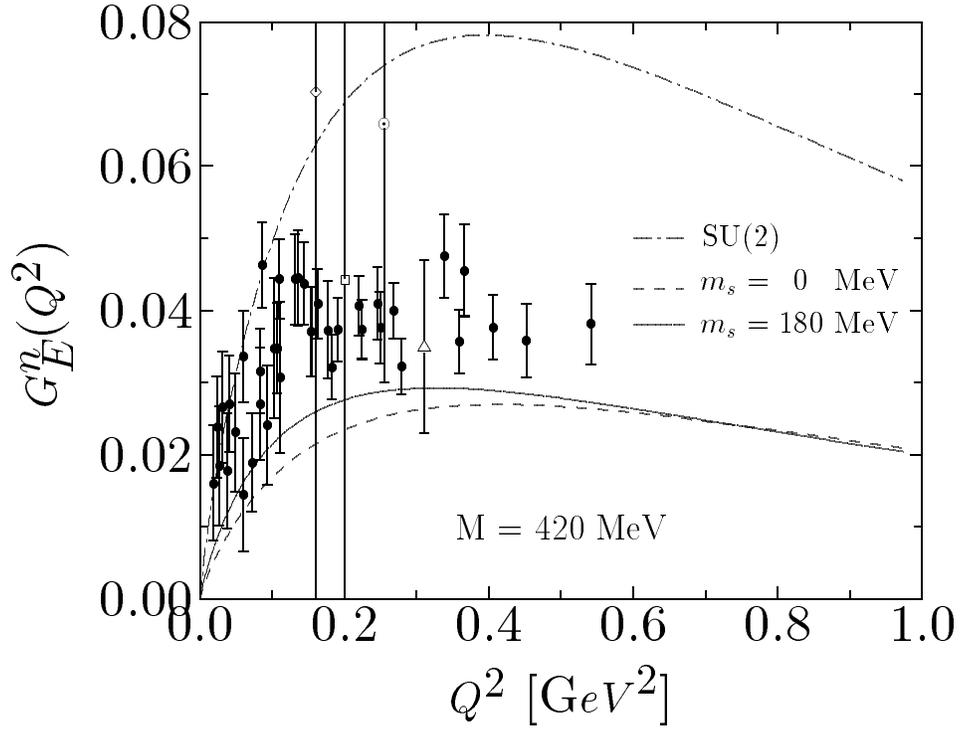}}%\vskip4pt
\caption{Neutron electric form factor as a function of $Q^2$  
evaluated in the SU(3) NJL model (Kim et al., 1995a) 
with $m_s=180 \;\mbox{MeV}$ (solid line)
and without $m_s$ corrections (dashed line) in comparison with the SU(2)
result (dash-dotted line). The experimental data are the same as in 
fig.~\protect{\ref{Figr17}}.}
\label{Figr18}
\end{figure}

\begin{figure}
\centerline{\epsfysize=4.in\epsffile{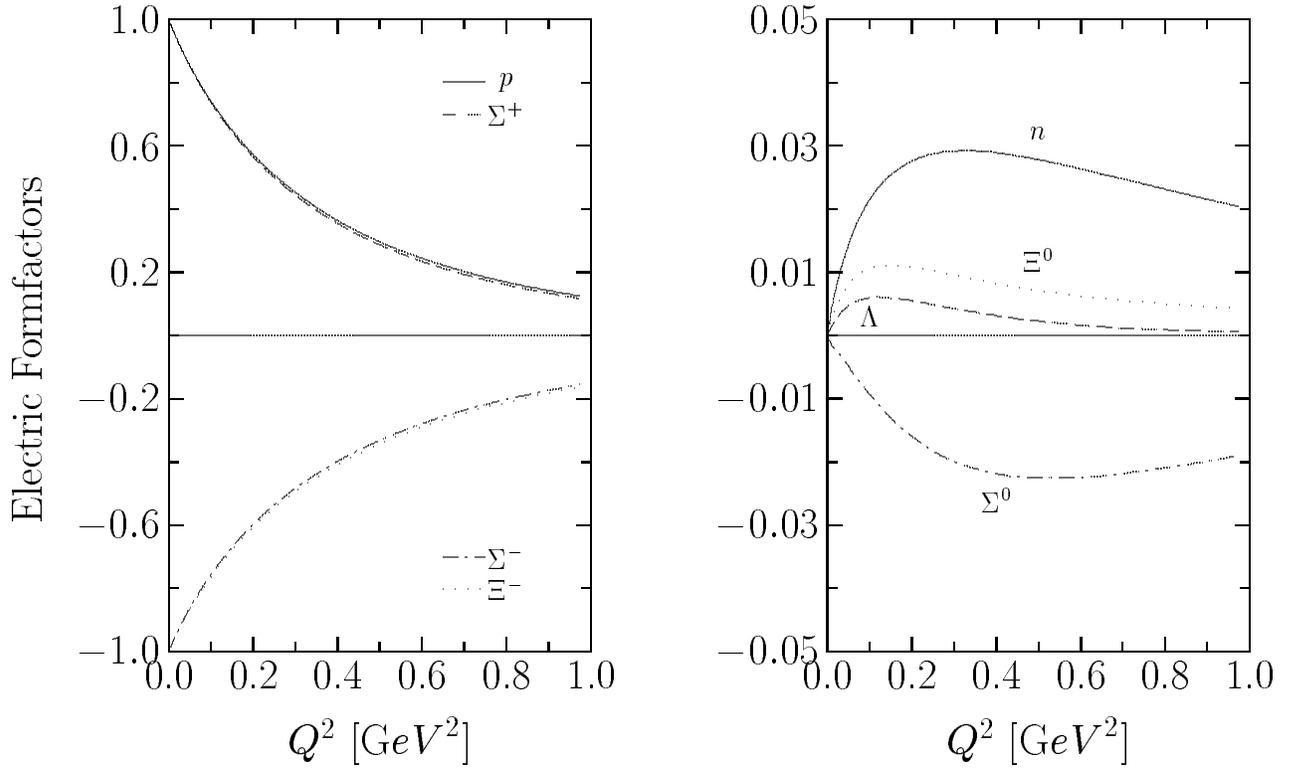}}%\vskip4pt
\caption{Electric form factors of charged (left) and neutral 
(right) octet baryons as a function of $Q^2$  in the SU(3) NJL model 
for $M=420$ MeVwith $m_s$ corrections taken into account 
(Kim {\em et al}., 1995a).}
\label{Figr19}
\end{figure}

On fig.~\ref{Figr19} we plot again the neutron electric form factor with and 
without the strange mass corrections compared to the experimental data as 
well as to the
results of SU(2) case. As can be seen, the $m_s$ corrections influence the
results at low momentum transfer and in particular, the neutron charge radius.
It should be also noted that in SU(3) case the results
for the neutron electric
form factors are quite different from those of the SU(2) version and are in a 
better agreement with experiment. In fact,
this difference can be understood looking at the structure of the quark charge
matrix. In contrast to SU(2) case where  $\hat{Q}= \frac16 (T=0)
+ \frac{\tau_3}{2} (T=1)$, i.e. the isoscalar part is a constant, in SU(3) it
includes $\lambda_8$, $\hat{Q}=\frac{\lambda_8}{2\sqrt{3}} (T=0) +
\frac{\lambda_3}{2} (T=1)$. Thus the isoscalar part of electromagnetic current
(\ref{elmag-current-su3}) is not invariant under rotation $R$.
This leads to a reduction of the electric isoscalar form factor in SU(3)
and results in considerably smaller numbers compared to those of the SU(2)
model.

\begin{figure}
\centerline{\epsfysize=4.in\epsffile{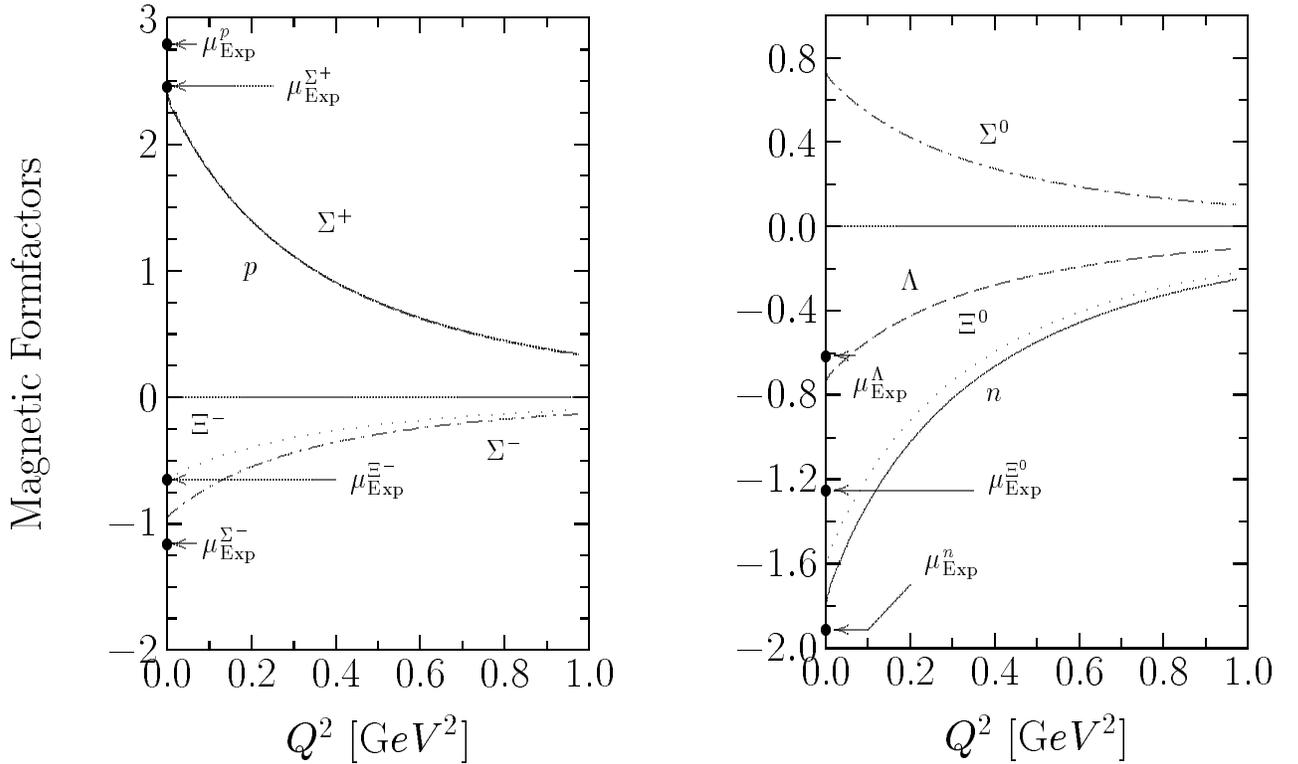}}
\caption{The same as in fig.~\protect{\ref{Figr19}} but for magnetic form 
factors (Kim {\em et al}., 1995a). The experimental
values for the magnetic moments are also shown (full circles).}
\label{Figr20}
\end{figure}

On fig.~\ref{Figr19} and fig.~\ref{Figr20} we present the electromagnetic form factors for the SU(3) octet hyperons from (Kim {\em et al}., 1995a).


Without $m_s$ corrections, for all momentum transfers the form factors fulfill 
the relations
\begin{eqnarray}
G^{p}_{E,M} & = & G^{\Sigma^{+}}_{E,M},\;\;
G^{\Sigma^{-}}_{E,M}=G^{\Xi{-}}_{E,M}, \nonumber \\
G^{n}_{E,M} & = & G^{\Xi^{0}}_{E,M},\;\;
G^{\Lambda}_{E,M} \;=\; -G^{\Sigma^{0}}_{E,M}\,,
\label{Eq:uspin}
\end{eqnarray}
which follow from the $U$-spin symmetry.

In table~\ref{emff1} we present the magnetic moments and the
charge and magnetic squared radii of the SU(3) octet baryons. The results
agree with the experimental data within $20\%$.  It should be mentioned that
Bae and McGovern (1996) made a $\chi^2$ analysis of the
hyperon magnetic moments for effective theories and found that
the present NJL model provides the best description among the hedgehog 
approaches to these quantities. It should be mentioned that the SU(3) version 
provides a better description of the nucleon magnetic moments compared to the 
SU(2) one.

\begin{table}
\caption{Electromagnetic properties of the octet baryons in the SU(3) NJL 
model (Kim {\em et al}. 1995e).}
\label{emff1}
\begin{center}
\begin{tabular}{|c|c|c|c|c|c|c|}    \hline
$\mbox{Baryons}$ &
$<r^2>_e[\mbox{fm}^2]$ & exp. &
$\mu_B[n.m.]$ & exp. &
$<r^2>_m[\mbox{fm}^2]$ & exp. \\
\hline
$p$          & $~0.78$ & $~0.74$ & $~2.39$ & $~2.79$ & $0.70$ & $0.74$ \\
$n$          & $-0.09$ & $-0.11$ & $-1.76$ & $-1.91$ & $0.78$ & $0.77$ \\
$\Lambda$    & $-0.04$ & --      & $-0.77$ & $-0.61$ & $0.70$ & --     \\
$\Sigma^{+}$ & $~0.79$ & --      & $~2.42$ & $~2.46$ & $0.71$ & --     \\
$\Sigma^{0}$ & $~0.02$ & --      & $~0.75$ & --      & $0.70$ & --     \\
$\Sigma^{-}$ & $-0.75$ & --      & $-0.92$ & $-1.16$ & $0.74$ & --     \\
$\Xi^{0}$    & $-0.06$ & --      & $-1.64$ & $-1.25$ & $0.75$ & --     \\
$\Xi^{-}$    & $-0.72$ & --      & $-0.68$ & $-0.65$ & $0.51$ & --     \\
\hline
\end{tabular}
\end{center}
\end{table}
\newpage
\vspace{1cm}
6.2 \underline{Strange vector form factors of the nucleon and related
observables}
\vspace{4mm}

We now proceed to discuss the strange vector form factors which
recently have
attracted much attention.  The success of the model in describing the
electromagnetic properties of the nucleon encourages one to apply
it for studying also the strange vector form factors. Experimentally,
although several experimental proposals have been suggested and a
series of
experiments is in progress (Musolf {\em et al.}, 1994), there is still
no experimental result for the strange vector form factors available.
From the side of the theory the strange form factors of the nucleon
have been
studied within different model schemes. Using pole fit analyses based on
dispersion relations, Jaffe (1989) first calculated the strange form
factors and gave estimates for the squared strange radius and magnetic
moment of the nucleon.
These quantities have been also studied in the SU(3) Skyrme model
without (Park {\em et al.}, 1991) and with vector mesons (Park and
Weigel, 1992), in a kaon-cloud model including kaon-loops
(Musolf and Burkardt, 1994) and also based on the vector dominance
(Cohen {\it et al.}, 1993). Forkel {\em et al.}, (1994) have
evaluated the strange vector form factors using different models, in
particular, a vector dominance model with $\omega-\phi$-mixing and a
kaon cloud. Hammer {\em et al.},
(1996) updated the Jaffe's results using a new dispersion theoretical
analysis of the nucleon electromagnetic form factor (Mergell {\em et
al.,} 1996). In a very recent work, using some QCD equalities for the
baryon current elements Leinweber (1995) estimated the strange
magnetic moment of the nucleon. In the NJL model, including both the $m_s$ 
and the rotational $1/N_c$ corrections the strange vector
form factors have been studied recently by Kim {\em et al.}, (1995c). 

The information of the strange vector form factors in the nucleon is
contained in the quark matrix elements:
\begin{equation}
\langle N(p^\prime)| J^{s}_{\mu} | N(p) \rangle =
\langle N(p^\prime)| \bar{s} \gamma_\mu s | N(p) \rangle.
\label{Eq:mat}
\end{equation}
The strange quark current $j^{s}_{\mu}$ can be expressed in terms
of the baryon current $j^{B}_{\mu}$ and the hypercharge current
$j^{Y}_{\mu}$:
\begin{equation}
j^{s}_{\mu}\;=\; \bar{s} \gamma_{\mu} s
=\frac{1}{N_c} \bar{q} \gamma_\mu q - \frac{1}{\sqrt{3}}
\bar{q} \gamma_\mu \lambda_8 q\equiv j^{B}_{\mu}-j^{Y}_{\mu}
\end{equation}
Here,  the sign convention of Jaffe (1989) is used.

As in case of the electromagnetic form factors, Sachs isoscalar and
isovector form factors $G^{s}_{E}(q^2)$ and~$G^{s}_{M} (q^2)$
\begin{eqnarray}
\langle N^\prime(p^\prime) |J^{s}_{0}(0) | N(p) \rangle & = &
G^{s}_{E} (q^2) \nonumber \\
\langle N^\prime(p^\prime) | J^{s}_{i}(0) |  N(p) \rangle & = &
\frac{1}{2M_N} G^{s}_{M} (q^2) i
\epsilon_{ijk} q^j \langle s^\prime| \sigma_k |s \rangle\,
\label{Eq:gm}
\end{eqnarray}
can be defined related to the matrix element of the time and space
components of the strange current, respectively. However, in contrast
to the isoscalar electric
form factor, which at $q^2=0$ determines the baryon number $B=1$,
the strangeness of the nucleon is zero, $G^{s}_{E}(0)=0$.

\begin{figure}
\centerline{\epsfysize=3.5in\epsffile{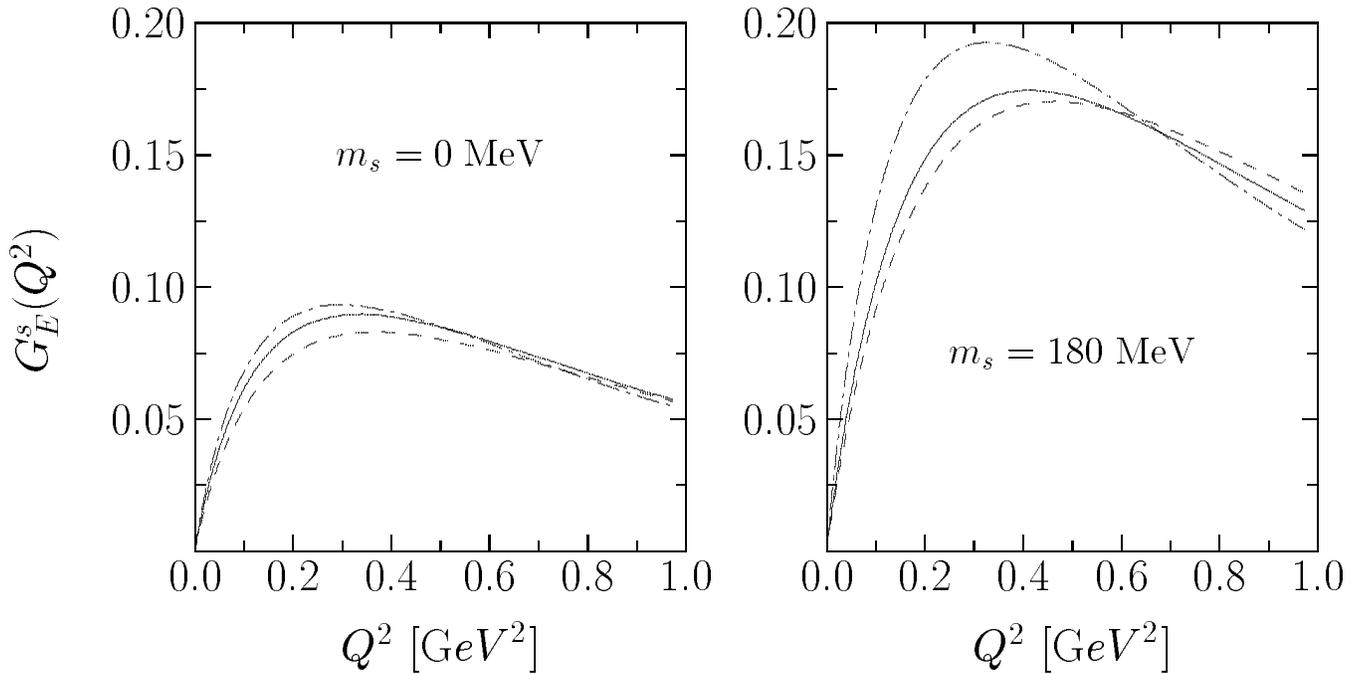}}
\caption{Strange electric form factor $G^{s}_{E}$ evaluated in the SU(3) 
NJL model (Kim {\em et al.,} 1995c)
as functions of $Q^2$ without $m_s$ (left) and with $m_s$
corrections (right) for different values of the constituent quark mass $M$:
$370$ MeV (dash-dotted line), 420 MeV (solid line) and 450 MeV (dashed
line).}
\label{Figr21}
\end{figure}

\begin{figure}
\centerline{\epsfysize=4.in\epsffile{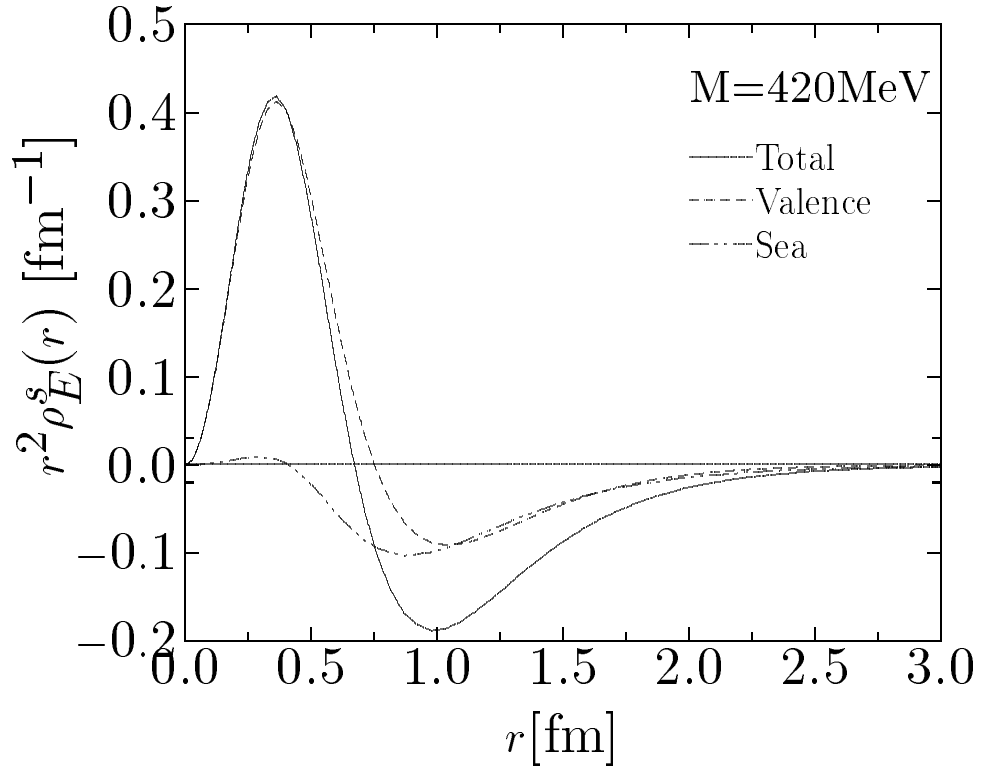}}%\vskip4pt
\caption{Strange electric densities split in valence and Dirac sea
contributions in the SU(3) NJL model (Kim {\em et al.}, 1995c). 
The sign convention of Jaffe (1989) is used.}
\label{Figr22}
\end{figure}

Fig.~\ref{Figr22} shows the strange electric form factor $G^{s}_{E}$ with
the constituent quark mass $M$ between 370 and 450 MeV calculated in the 
NJL model (Kim {\em et al.}, 1995c). As can be seen,
the strange electric form factor $G^{s}_{E}$ depends weakly on $M$ but
it is rather sensitive to the $m_s$ corrections. The inclusion of these 
strange mass corrections leads to a strong enhancement of more
than 70 \% of the form factor at finite momentum transfers.
Similarly to Forkel {\em et al.}, (1994), the strange electric form factor
$G^{s}_{E}$ in the NJL model  (Kim {\em et al.}, 1995c), has a sign opposite 
to the one of the pole-fit analyses of Jaffe, (1989) and Hammer 
{\em et al.}, (1996).

The strange electric squared radius is given by
\begin{equation}
\langle r^2\rangle_s \;=\; -6\frac{d G^{s}_E (Q^2)}{dQ^2}
\left.\right|_{Q^2=0}  .
\end{equation}
and the results from different model calculations are summarized in
table~\ref{sff2}. 

\begin{table}
\caption{Comparison of the strange magnetic moments
and strange electric squared radii calculated in different models.
In the SU(3) NJL calculations a constituent mass $M=420$ MeV is used 
(Kim {\em et al.}, 1995c).}
\label{sff2}
\begin{center}
\begin{tabular}{|c|c|c|c|}   \hline
models&$\mu_s[\mu_N]$
&$\langle r^2\rangle^{Sachs}_{s}[\mbox{fm}^2]$&references \\
\hline
pole-fit & $-0.31 \pm 0.09$ & $0.14\pm 0.07$ & Jaffe 89 \\
pole-fit & $-0.24 \pm 0.03$ & $0.19\pm 0.03$ & Hammer 95 \\
kaon cloud & $-(0.31\to 0.40)$
&$-(0.027\to 0.032)$& Musolf \& Burkardt 94 \\
VMD  &$-(0.24\to 0.32)$&$-0.04$ & Forkel 94\\
Skyrme&$-0.13$&$-0.11$& Park  91 \\
Skyrme with VM &$-0.13$&0.05& Park $\&$ Weigel 92 \\
QCD equalities &$-0.75\pm 0.3$&& Leinweber 95 \\
NJL ($\Omega^0+\Omega^1$) &$-0.45$&$-0.35$  & Kim  95 \\
\hline
\end{tabular}
\end{center}
\end{table}

In the lack of experimental data, it is useful to compare the
results of different model calculations presented in table~\ref{sff2}.
In fact, only two approaches, pole-fit analysis (Jaffe, 1989, Hammer
{\em et al.}, 1996) and NJL with $m_s$ and rotational corrections (Kim
{\em et al.}, 1995c), predict clear non-zero values for the strange
radius but of opposite signs which simply reflect the opposite signs of
the corresponding strange electric form factors. It should be stressed
that both approaches incorporate a proper description of the nucleon
electric form factors and the corresponding charge radii. 

In fig.~\ref{Figr22} we show the strange electric density 
from (Kim {\em et al.}, 1995c).
As in the case of the neutron charge densities (see fig.~\ref{Figr10}), this 
density has a negative tail which is due to the
contribution of the polarized Dirac sea.  In fact, the shape of this
tail is determined by the behavior of quark wave function at large
distances. Because of the embedding ansatz (\ref{c51}) it is
simply given by the pion mass, $\exp(-m_\pi r)$. On the other hand, it
is generally expected that the tail of the strange electric density
should be related to the kaon cloud with asymptotics 
$\exp(-m_K r)$. Hence, one can expect some overestimation
for the quantities related to the strangeness in the NJL approach
for which the contribution of the Dirac sea is dominant. In
particular, this {\it caveat} applies to the strange radius. 
As one can see in 
fig.~\ref{Figr22} the present model yields a valence contribution to the 
strange charge distribution localized in the interior of nucleon. In terms of 
a baryon-meson picture this corresponds to a $\Lambda$ or $\Sigma$ core 
coupled to the corresponding kaon cloud.  

\begin{figure}
\centerline{\epsfysize=3.5in\epsffile{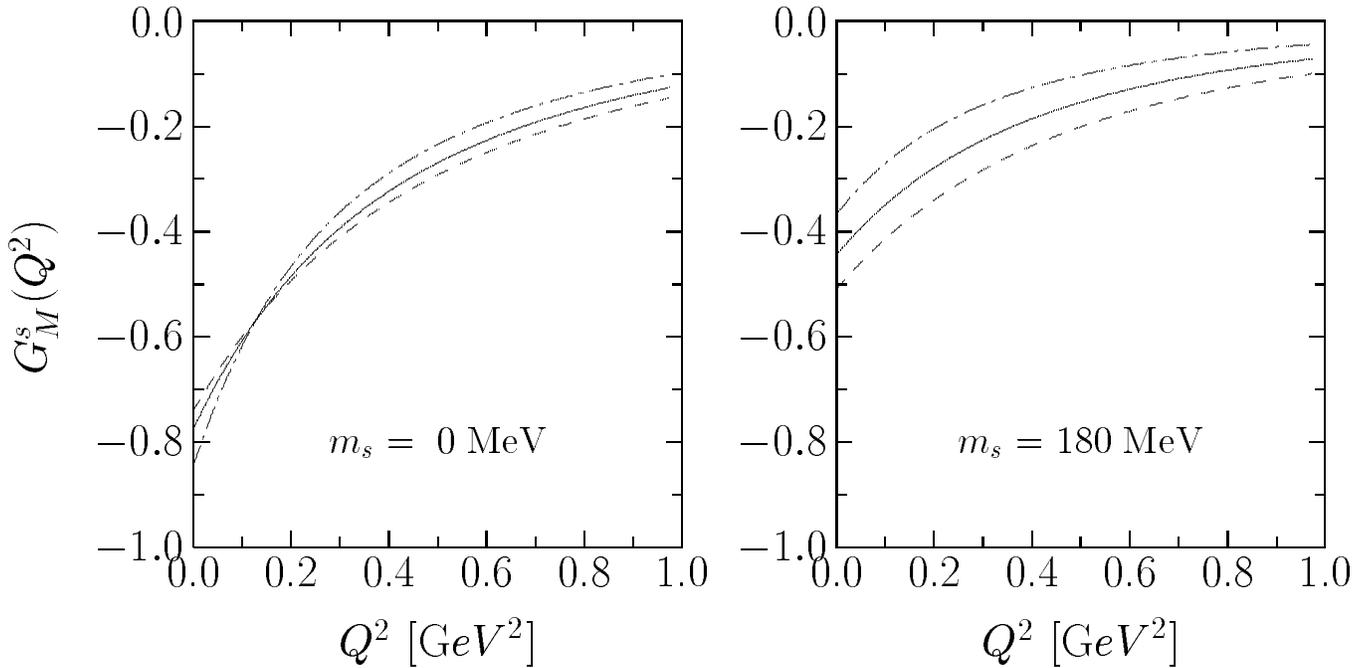}}%\vskip4pt
\caption{Strange magnetic form factor $G^{s}_{M}$ 
(Kim {\em et al.,} 1995c)
as functions of $Q^2$ evaluated in the SU(3) NJL model without 
$m_s$ (left) as well as with $m_s$
corrections (right) for different values of the constituent quark mass:
$M=370$ MeV (dash-dotted line), 420 MeV (solid line) and 450 MeV (dashed
line).}
\label{Figr23}
\end{figure}

\begin{figure}
\centerline{\epsfysize=4.in\epsffile{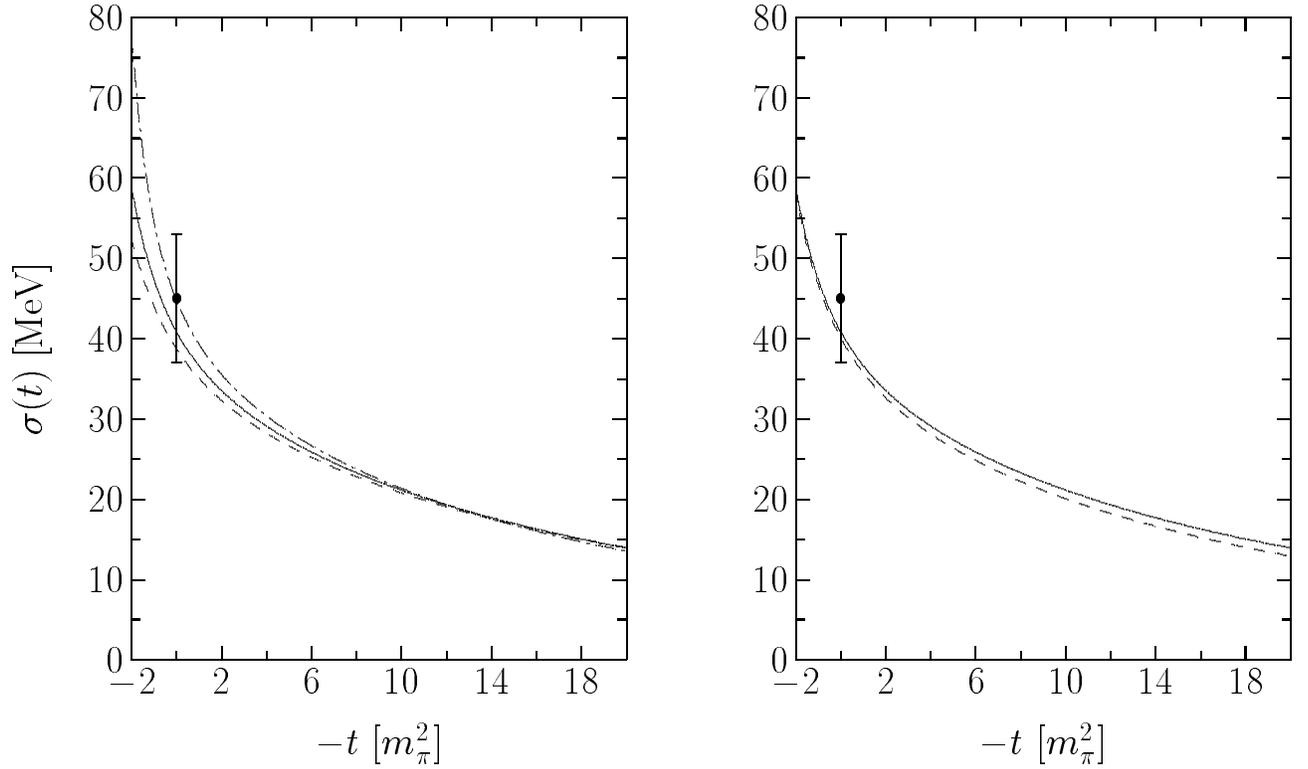}}%\vskip4pt
\caption{Scalar form factor as a function of $Q^2$ evaluated
in the SU(3) NJL model (Kim {\it et al.}, 1996): On the left for different 
values of the constituent up and down quark mass 370 MeV (dotted line),
420 MeV (solid line) and 450 MeV (dashed line) and with $m_s$ corrections, 
and on the right, for $M=420$ MeV and with (solid line) 
and without (dashed line) $m_s$ corrections. The empirical value for 
$\sigma(0)=45\pm 8$ MeV of (Gasser {\em et al.}, 1991) is also presented.}
\label{Figr24}
\end{figure}

The strange magnetic form factor (Kim {\em et al.}, 1995c) is plotted on
fig.~\ref{Figr24}. As in the case in the case of the strange electric form
factor the
magnetic one shows a weak dependence on the constituent mass whereas the
inclusion of the $m_s$ corrections leads to a common enhancement of more
than 50~\%. The results are similar to those of Jaffe (1989), Hammer
{\em et al.} (1996) and Forkel {\em et al.} (1994).

The strange magnetic moments for different models are given in
table~\ref{sff2}. All presented model calculations
predict a negative sign and similar values for $\mu^s$. We remind the
reader that both the pole-fit analyses (Jaffe, 1989, Hammer 
{\em et al.}, 1996) and NJL model (Kim {\em et al.}, 1995c) provide a proper
description of both proton and neutron magnetic form factors and the
corresponding magnetic moments. 

There are also some results of strange form factors using the NJL model 
performed by Weigel {\em et al.}, (1995a). The authors have used 
non-selfconsistent profiles and have ignored the rotational $1/N_c$ 
corrections. In addition, there seem to be also some purely numerical 
shortcomings of the calculations of Weigel {\em et al.}, (1995a).

\vspace{1cm}
6.3 \underline{Scalar Form Factor and $\pi N$ sigma term}
\vspace{4mm}

The scalar form factor $\sigma (t)$ is defined as
\begin{equation}
\sigma (t) \;=\; m_0
\langle N(p') |\bar{u}u + \bar{d}{d} | N(p) \rangle
\label{Eq:sigma1}
\end{equation}
with $m_0=(m_u+m_d)/2\simeq 6\;\mbox{MeV}$. The mass $m_0$ is fixed using the 
additional free parameter in the proper-time regularization integral 
(\ref{e229g}). Here we use the usual notation $t$ for the square of the 
momentum transfer.
 
Because of a factor-of-two discrepancy between empirical value for the
$\pi N$ sigma term ($\Sigma=64\pm 8$ MeV) (Koch, 1982) from the
analyses of $\pi N$ data
and naive estimates of the $\sigma$-term ($\sigma\simeq 25$ MeV) from the
baryon mass spectrum (Cheng, 1976), which contradicts to the prediction of the
low-energy theorem of current algebra, there have been a lot of discussion in
the literature (see (Gasser {\it et al.}, 1991, Gasser and Leutwyler, 1991) 
and references therein) devoted to this problem. 
Donoghue and Nappi (1986) suggested that the discrepancy is due to the 
presence of strange quarks in the nucleon, i.e.
$\langle N| \bar{s}s |N\rangle\neq 0$. However, using both the Skyrme
model and bag model they got unlikely large $\langle N| \bar{s}s
|N\rangle$ contribution (almost $30\%$) to the quark condensate in the
nucleon. Gasser {\it et al.} (1991) reanalysed the $\sigma$ term
carefully taking into account all currently available $\pi N$
scattering data. They concluded a stronger $t$-dependence of the scalar
form factor $\sigma(t)$ resulting in
$$\Delta_\sigma=\sigma(2m^{2}_{\pi})-\sigma (0)\simeq 15 \;\mbox{MeV}\,,$$
and a larger value for sigma term $\sigma(0) = 45 \pm 8 \mbox{MeV}$
which means that the value at the Cheng-Dashen point
$\sigma(2m^{2}_{\pi})\equiv\Sigma \simeq 60
\mbox{MeV}$. This value differs from the estimates from the baryon spectrum and
it is due to both non-zero  strange current mass and the strange content 
$\langle N|\bar ss|N\rangle$ in the quark condensate. For 
$y=2\langle N| \bar{s}s |N\rangle/\langle N| \bar{u}u + \bar{d}d |N\rangle$, 
the strange contribution to the $\sigma$ term, Gasser {\it et al.} (1991) 
suggested a value of about 0.2, which means that the strange contribution to 
$\langle N|\bar uu+\bar dd + \bar ss|N\rangle$ should be about 10 \% and 
hence much smaller than suggested by Donoghue and Nappi (1986).

Fig.~\ref{Figr24} shows the scalar form factor as a function of
the constituent quark mass $M$ (Kim {\it et al.}, 1996).
As can be seen, the scalar form factor is very sensitive to the  
the particular value of $M$ only in the time-like region close to the 
Cheng-Dashen point. For $M=420\;\mbox{MeV}$ both the empirical estimate 
$\sigma(0)=45\pm 8$ MeV and the deduced $t$ 
dependence $\Delta_\sigma= 15.2\pm 0.4$ MeV as well as the slope at zero 
momentum transfer (strange scalar ms radius  
$\langle r^2\rangle^{S}_{N} \simeq 1.6 \mbox{fm}^2$)
(Gasser {\it et al.}, 1991) are reproduced. The 
actual values of (Kim {\it et al.}, 1996) are $\sigma= 41$ MeV, 
$\Delta_\sigma=18.2 MeV$ and $\langle r^2\rangle^{S}_{N}=1.5~\mbox{fm}^2$. 
It should be noted that the $m_s$ corrections do not influence $\sigma (t)$ 
much. However, the $m_s$ corrections play an important role of reducing 
strongly the strangeness contribution $\langle N | \bar{s} s | N \rangle$.  
With the $m_s$ corrections taken into account,
Kim {\it et al.} (1996) obtain $y=0.27$ in case of  $M=420\mbox{MeV}$. 
This qualitatively agrees with the value $y\simeq 0.2$ of Gasser {\em et al.} 
(1991), whereas without the $m_s$ corrections $y$ it is strongly 
overestimated $y=0.48$. 

\vspace{1cm}
6.5 \underline{Axial charges of the nucleon}
\vspace{4mm}

The axial charges of the nucleon are defined as
forward matrix elements of the axial current
\begin{eqnarray}
\langle N(p)| \bar{\psi} \gamma_\mu \gamma_5 \lambda^a \Psi | N(p) \rangle = 
g_A^{(a)} \bar{u}(p) \gamma_\mu \gamma_5 u(p) ,
\end{eqnarray}
where $a=0,3,8$ and $\lambda^3$, $\lambda^8$ are Gell-Mann
matrices, $\lambda^0$ is in this context 
$3 \times 3 $ unit matrix and $u(p)$ is
a standard Dirac spinor.

The flavor non-singlet axial charges
$g^{(3)}_A$ and $g^{(8)}_A$ can be measured in the experiments on
hyperon semileptonic decays. In particular, the $g^{(3)}_A$
is known with good accuracy from the neutron $\beta$-decay data
(Particle Data Group 1994):
\begin{eqnarray}
g_A^{(3)}\equiv g_A = 1.257 \pm 0.0028 . 
\label{g3_num}
\end{eqnarray}
In order to extract $g^{(8)}_A$ from experiment an additional assumption of 
flavor $SU(3)$ symmetry is usually used for analysing the data on hyperon 
$\beta$-decays. In particular, Hsueh {\it et al.} (1988) give the following 
estimate
\begin{eqnarray}
g_A^{(8)} = 0.34 \pm 0.02 . 
\label{g8_num}
\end{eqnarray}
and for a recent discussion of the applicability of flavor $SU(3)$
symmetry see (Lichtenstadt and Lipkin, 1995).
The flavor singlet axial charge $g^{(0)}_A$ cannot be obtained directly from 
hyperon weak decay experiments, but it can be
related to the first moment of the polarized quark distribution
in the proton. The latter is defined as
\begin{eqnarray}
g^p_1 (x) &=& \frac{1}{2} \sum_{q=u,d,s} e_q^2 \left[
q_\uparrow (x) - q_\downarrow (x) 
+ \bar{q}_\uparrow (x) - \bar{q}_\downarrow (x) 
\right] \nonumber \\
&=& \frac{1}{2} \sum_{q=u,d,s} e_q^2 \Delta q (x).
\label{g1}
\end{eqnarray}

In the parton model, the moment $\Delta q = \int_0^1 \d x \Delta q(x)$ is 
related  to the light quark (of flavor $q$) contribution to the proton spin.
The distribution (\ref{g1}) has been measured in deep
inelastic scattering experiments. Using (\ref{g1}) and the 
relations
\begin{eqnarray}
g_A^{(3)} &=&\Delta u - \Delta d \nonumber \\
g_A^{(8)} &=& \frac{1}{\sqrt{3}} 
\left( \Delta u + \Delta d - 2 \Delta s \right) \nonumber \\
g_A^{(0)} &=&\Delta u + \Delta d + \Delta s = \Delta \Sigma.
\end{eqnarray}
the moments of separate polarized quark
distributions $\Delta u, \Delta d$ and $\Delta s$ as well as the singlet 
axial charge $g_A^{(0)}$
can be obtained from the experimentally measured moment $\int_0^1 \d x g_1^p$
and axial charges $g_A^{(3)}$ and $g_A^{(8)}$.
The 1988 EMC result (Ashman {\it et al.}1988) suggested that
the flavor singlet axial charge $g^{(0)}_A$ is close to zero
which indicates that none of the proton spin is carried by its constituent
quarks. In addition the EMC data disagreed with the assumption
$\Delta s=0$.
This has sometimes been called the ``proton spin crisis". This EMC
result was rather surprising from the point of view of the constituent
quark model (CQM), which would suggest $g^{(0)}_A=1$ and $\Delta s=0$.
Soon after the first data became available it was demonstrated
by Brodsky {\it et al.} (1988) that in the Skyrme model the
singlet axial charge is suppressed by $1/N_c$ in comparison with
CQM and this was suggested as a solution of the `` spin crisis". 

\begin{figure}
\centerline{\epsfysize=4.in\epsffile{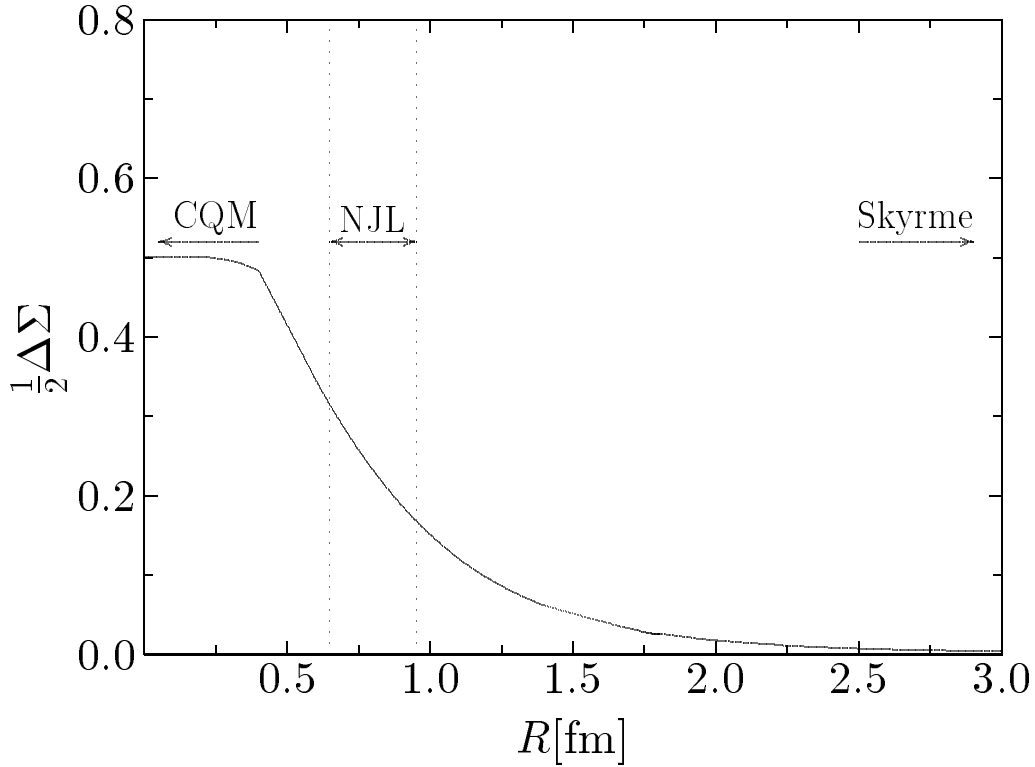}}%\vskip4pt
\caption{Proton spin fraction carried by light quarks as a function of the
soliton size $R$ for a fixed profile function of the SU(2) NJL model.}
\label{g0}
\end{figure}

In the NJL model, the $g^{(0)}_A$ has a contribution which originates from the 
imaginary part of the effective action and hence it has no counterpart in the 
Skyrme model (Blotz {\it et al.} 1993b). An analogous contribution is also 
pertinent to the Skyrme model with explicit vector mesons 
(Park and Weigel, 1991), though its origin and particular
value differ from that of NJL model. It should be noted that in the large 
$N_c$ limit this NJL contribution is non-vanishing. However, it is 
suppressed in  large soliton size $R$ (or large constituent quark mass)
limit  $g^{(0)}_A\sim 1/(MR)^4$. In fig.~\ref{g0}
the proton spin fraction carried by light quarks
 $\frac12 \Delta \Sigma = \frac12 g_A^{(0)} $ in dependence of  
$MR$ is plotted. At small $MR$, corresponding to CQM, it is close to one 
half whereas it is
rapidly vanishing with increasing $MR$, so that in the limit of 
large $MR$ the Skyrme model result $g^{(0)}_A=0$ is reproduced.
This figure illustrates how the NJL model in some sense interpolates
between two models of baryons, namely the constituent quark model and 
the Skyrme model. Indeed, if $MR$ is small, the valence level is situated 
just below the threshold $M$, and the picture resembles 
the non-relativistic constituent quark model. In the opposite case of
$MR \gg 1$ the NJL picture of nucleon appears to be ideologically close
to that of the Skyrme--Witten topological soliton. Actually, the truth 
seems to be just between these two cases, because in the NJL model for the 
selfconsistent solution, which provides an overall good description of the 
baryon properties, $M R\approx 1$ .
 
\begin{table}
\caption{Axial constants and the light quark contributions to the proton spin 
calculated in the SU(3) NJL model (Blotz {\it et al.,} 1993b, 1996) compared 
with the experimental values taken from the analysis of Ellis and 
Karliner (1995).} 
\label{axial-charges} 
\begin{center}
\begin{tabular}{|c|c|c|c|c|c|c|}\hline
 & $g_A^{(3)}$ & $g_A^{(8)}$ & $g_A^{(0)}$ &
 $\Delta u$ & $\Delta d$ & $\Delta s$ \\
\hline 
NJL SU(3) & $1.38$ & $0.30$ & $0.37$ & 
$0.90$ & $-0.48$ & $-0.05$ \\\hline
exp.      & $1.26$ & $0.34 \pm 0.02$ & $0.31 \pm 0.07$ & 
$0.83 \pm 0.03$ & $-0.43 \pm 0.03$ & $-0.10 \pm 0.03$ \\
\hline  
\end{tabular}
\end{center}
\end{table}
 
The results of the selfconsistent calculations of the axial charges 
$g_A^{(3)}$,
$g_A^{(8)}$ and $g_A^{(0)}$ in the three flavor NJL model are summarized
in table~\ref{axial-charges} (Blotz {\it et al.} 1993b, 1996a).
These results are obtained in the $SU(3)$ NJL model with both linear in 
$m_s$ and rotational $1/N_c$ corrections taken into account with a constituent 
quark mass $M=420$~MeV.
 
As can be seen from the table, the values of axial charges calculated
in the three flavor NJL model are in a good agreement with experimental 
values from
hyperon $\beta$ decays ( Hsueh {\it et al.} 1988) and recent
analysis of experimental data on polarized deep inelastic scattering 
(Ellis and Karliner, 1995).

In QCD, the value of $g_A^{(0)}$ is related to the nucleon matrix
element of the topological charge, 
$\langle N | F\widetilde F | N \rangle$, by the $U(1)_A$ anomaly. 
Recently, using the instanton model of QCD vacuum, a method has been 
developed to represent 
systematically QCD gluon operators as constituent quark 
operators in the NJL model derived from the instanton model.
(Diakonov {\em et al.}, 1995b). It was shown that this method fully 
preserves the $U(1)_A$ anomaly. Thus, in the context of the instanton
vacuum, the result $g_A^{(0)} = 0.37$ of (Blotz {\em et al.}, 1993b)
can be regarded as a consistent estimate of the nucleon matrix 
element of $F\widetilde F$.

\vspace{1cm}
6.6 \underline{Gottfried sum rule}
\vspace{4mm}

The Gottfried sum rule, which is the unpolarized analogue of the
Bjorken sum rule for the polarized structure functions,
was reexamined in the recent NMC experiment
(Arneondo {\it et al.}, 1994) and the result of the measurement is again in
contradiction to the naive quark model expectation.
The Gottfried sum is defined as
\beq  S_G = \int_0^1 \d x {1\over x} \left( F_2^p (x) - F_2^n (x) \right)
      \label{gott1}
\eeq
where the $F_2^{(p/n)} (x)$ are   the unpolarized structure
functions.

\begin{table}[b]
\label{tabgott3}
\caption{NJL estimates of (Wakamatsu 1992) and of (Blotz {\it et al.}, 1995b),
respectively,
for the Gottfried sum $S_G$  for SU(2) and SU(3). The values are given for 
$M=420$ MeV in the chiral limit as well as with $m_s$ corrections.}  

\begin{center}
\begin{tabular}{|c|c|c|c|c|}\hline
$M$ [MeV]   & $S_G^{SU(2)}$ & $S_G^{SU(3)}$ no $m_s$ & $S_G^{SU(3)}$ with $m_s$
&Exp\\ \hline
%  372  &              0.2228 &   0.2448 &  0.2469  \\
 420  &              0.2091 & 0.2336  &  0.2341 & $ 0.221\pm~0.008$ \\\hline
\end{tabular}
\end{center}
\end{table}

In the language of the parton model and in terms of quark
distribution functions:
\bea  S_G &=& \int_0^1 \d x \sum_{i=u,d}   e_i^2  \left(
               q_i^p + {\bar q}_i^p - q_i^n - {\bar q}_i^n \right)
           \label{gott1a} \\
       &=&
            \int_0^1 \d x {1\over 3}\left( u + {\bar u} - d - {\bar d}
\right)                \label{gott2}
\eea
where in the last line the SU(2) flavor symmetry is used.
The $u,{\bar u},d$ and ${\bar d}$ are the quark distribution functions 
of  proton. Using standard quark sum rules for the valence part
of the proton
\beq  \int \d x (u - {\bar u})=2,\ \ \int \d x (d - {\bar d})=1 , \eeq
the Gottfried sum  \queq{gott2} becomes
\beq  S_G = {1\over 3} +{2\over 3} \int \d x \left(
           {\bar u}^p - {\bar d}^p \right)
      \label{go7}
\eeq
where the second part in rhs of (\ref{go7}) is related to the flavor asymmetry 
of the sea. If it is flavor symmetric one obtains the Gottfried sum rule
$$S_G = {1\over 3}.$$
However, the recent NMC data (Arneondo {\it et al.}, 1994) leads to a 
different value
\beq  S_G = 0.221 \pm 0.008  \label{go8}   \eeq
which suggests that the Dirac sea carries some flavor asymmetry.

Assuming a particular ansatz (Wakamatsu, 1992) for the integrated 
distributions (second term in (\ref{go7})) Wakamatsu (1992) in SU(2) and Blotz 
{\em et al.} (1996b) in SU(3) version analyzed the Gottfried sum rule in terms
of the flavor asymmetry of the Dirac sea in the NJL model.
The obtained estimations are given in table~\ref{tabgott3}. In the SU(3) case 
the $m_s$ corrections are also included (Blotz {\it et al.}, 1995b). 
It should be noted that in the NJL model the deviation from of $S_G=1/3$ 
occurs even in the chiral limit.

\newpage
\section{Summary}

The present article reviews the theory and applications of the 
Nambu--Jona-Lasinio model with two and three flavors for the description
of low-mass baryons. The model corresponds to an effective
chiral theory where quarks interact with selfconsistently generated
bosonic mean fields. After concentrating on Goldstone fields (pion,
kaon and eta) and after semiclassical quantization of the collective
rotational degrees of freedom the model coincides with the Chiral
quark soliton model of Diakonov, Petrov and Pobylitsa obtained from QCD
assuming an instanton liquid for the QCD vacuum.  After adjusting
the parameters of the model in the meson sector to the pion
decay constant and the pion and kaon masses, the constituent
quark mass is left as the only free parameter. This parameter is tuned
in the baryonic sector so that the model using a 
hedgehog form for the meson fields reproduces best the
wealth of experimental data with a value of the constituent quark mass 
$M$ around 420 MeV. With this uniquely defined parameter set 
the calculations are
performed for various radii, magnetic moments, mass splittings
of the baryon octet and decuplet, and axial and spin properties
of the nucleon. Particular emphasis is put on electromagnetic and
axial form factors. Predictions are made for strange form factors
and tensor charges. Without any readjustment of the parameters, a very good
agreement between the theory and nearly all low-energy experimental
data is obtained. Therefore, one can consider the present
model as a quite useful tool to describe the ground-state properties of the
octet and decuplet baryons.
 
For the parameter set associated with a constituent quark mass of 420 MeV,
the practical outcome of the calculations can be summarized as
follows. SU(2) case: Within 15~\% we reproduce the proton charge
radius, the axial vector coupling constant, the pion-nucleon sigma term,
the delta-nucleon splitting, the E2/M1-ratio, the electric proton form factor 
and the momentum dependence of the the magnetic form factors of proton
and neutron. The nucleon magnetic moments deviate by 25~\% from
the experiment. The electric neutron form factor and the corresponding
charge radius appear to be overestimated by 50~\%. Predictions
are given for tensor charges. SU(3) case: Within 15~\% we reproduce
the proton charge radius, the axial vector coupling constant, the
pion nucleon sigma term, the delta-nucleon splitting, the mass
splitting within the octet and decuplet, the isospin splittings
within the octet. The electric proton form factor and the magnetic
form factors of proton and neutron and the corresponding magnetic
moments deviate also less than 15~\%
from experiment. The electric
neutron form factor and the neutron charge radius are by 20~\%
too small compared to the experimental data. The flavor singlet
and octet axial constants come out within the experimental errors.
These quantities are connected with the "spin crisis" of the proton.
Predictions are given for strange electric and magnetic form factors
of the nucleon and for electromagnetic form factors of octet and
decuplet baryons. The corresponding magnetic moments agree well with
experiment.
 
The present model reproduces most of the relevant
low-energy observables of the nucleon and of various hyperons.
It is important to note that this agreement is rather good
and hence the model might be used to obtain some insight into
the microscopic structure of the baryons. The structure of the baryons emerges 
to be based on chiral symmetry breaking as a dominating mechanism
and the important resulting forces are those between quarks
and Goldstone bosons. Confining forces, being absent in the
model, seem not to play a relevant role for the baryonic ground
states. Also vector mesons and scalar mesons are apparently not needed.
This simple picture of the interactions is associated with a simple
picture of the microscopic structure of the baryons. They appear
as composed of three valence quarks in a bound state together with a 
continuum of the polarized Dirac sea. In fact, the contribution of the
polarized Dirac sea to the baryon observables is at about the 30~\% level. 
It should be stressed that the polarized Dirac sea contribution is absolutely 
necessary since it causes the valence quarks and hence the whole baryon to be 
bound. The baryon number of this system is given by
the occupation of the valence level and not by the
topological properties of the meson fields, as it is in 
the Skyrme model. For the parameter values used, an approximation of the 
present non-local NJL effective action by a local Skyrme-type action is not 
justified. Thus, the present model picture of the 
baryon with clearly distinguished valence quarks differs from the topological 
soliton and hence, the present approach does not support topological baryon 
models like the Skyrme one.  
 
The above considerations are all made within the non-linear version of
the Nambu-Jona-Lasinio model (on the chiral circle) being in this form 
equivalent to the Chiral quark soliton model. There are
several problems still within this model, which need further
consideration. One consists in the undesired asymptotic form 
$\exp(-m_\pi r)$ of the meson profile in the strange direction (kaon field)
as it turns out in the present treatment of the strange
mass corrections. This affects the applicabilty of
the formalism to those ``strangeness'' quantities which strongly depend on 
the tail of the polarized Dirac sea, as e.g. strange radii. Another problem 
is related to the present large $N_c$ treatment in which only the
$1/N_c$ corrections originating from the zero modes are taken into account.  
This question is closely connected with the fact that the absolute 
masses of the of the baryons come out noticeably too high and the 
meson-loop corrections, being presently very much discussed in other chiral 
effective theories, are needed to provide a remedy for this. In the present 
model it seems to be quite a formidable task, both formally and numerically.
 
There is also an interesting
field of applications of the model that is related to
gluonic degrees of freedom in baryons. For example, the evaluation of matrix 
elements of
operators related to higher twist contributions to moments of nucleon
structure functions has recently attracted much attention. The
Nambu--Jona-Lasinio model as such, based on a four quark
interaction, cannot be used here since it does not contain
gluon degrees of freedom. However, the non-linear version
of NJL model, having been bosonized and restricted to 
Goldstone modes by invoking the chiral circle, corresponds
directly to the effective quark meson action obtained in the 
instanton liquid model of QCD. One therefore
has a link between the description of baryons, as it is
presented in this review, and the instanton configuration of
the gluon field. Making use of that, the present approach has the potentiality
systematically to represent gluon operators as effective quark
operators, so that baryon matrix elements of gluon operators
can be evaluated using the techniques presented in this review 
(Diakonov {\it et al.}, 1995). In this way useful information for the 
understanding of nucleon structure functions may be obtained.

{\large \bf Acknowledgements}:
The authors are indebted to W.Broniowski, D.Diakonov, V.Petrov, 
M.Pra\-szalowicz, G.Ripka and H.Weigel for useful discussions. 
The suggestions and criticism of M.Polyakov and a contribution of 
E.Nikolov, while preparing the article, were most
welcome. A.B. and P.P acknowledge grants from the Alexander von
Humboldt foundation. The work has been supported partially by the
Deutsche Forschungsgemeinschaft, the Bundesministerium f\"ur Bildung
und Wissenschaft, the COSY-Project J\"ulich, the Volkswagen-Foundation,
the INTAS-93-0283 Contract (P.P.) and the Bulgarian Science Foundation 
under contract $\Phi$--32 (C.V.C.).

\newpage
\vskip2cm
\parindent20pt
\parskip0pt
\hskip3cm{\bf REFERENCES}
\bigskip

%----- A -----
\noindent Adkins, G. S., C. R. Nappi and E. Witten (1983). 
{\it Nucl.Phys.}, {\bf B228}, 552. \\
Ahrens, L. A., {\it et al.} (1987).
{\it Phys.Rev.}, {\bf D35}, 785. \\
Aitchison, I. J. R. and C. M. Fraser (1984). 
{\it Phys.Lett.}, {\bf B146}, 63. \\
Aitchison, I. J. R. and C. M. Fraser (1985a). 
{\it Phys.Rev.}, {\bf D31}, 2605. \\
Aitchison, I. J. R. and C. M. Fraser (1985b). 
{\it Phys.Rev.}, {\bf D32}, 2190. \\
Alkofer, R. and H. Reinhardt (1990).
{\it Phys.Lett.}, {\bf B244}, 461. \\
Alkofer, R., H. Reinhardt and H. Weigel (1994). {\it Phys.Rep.}, {\bf 265} 139\\
Alkofer, R., H. Reinhardt, H. Weigel and U. Z\"{u}ckert (1992).
{\it Phys.Lett.}, {\bf B298}, 132. \\
Alkofer, R. and H. Weigel (1993).
{\it Phys.Lett.}, {\bf B319}, 1. \\
Alkofer, R. and I. Zahed (1990). 
{\it Phys.Lett.}, {\bf B238}, 149. \\
Amaldi, E., S. Fubini and G. Furlan (1979). In: Springer tracts in modern 
physics, 

{\it Pion-electroproduction}, Vol.83. Springer-Voerlag. \\
Arneodo M. {\it et al.} (1994). {\it Phys.Rev.}{\bf D50}, 1\\
%----- B -----
Bae, M.-S. and J. A. McGovern (1996).{\it J.Phys.} {\bf G22}, 199. \\
Baker, N. J., {\it et al.} (1981).
{\it Phys.Rev.}, {\bf D23}, 2499. \\
Balachandran, A. P., F. Lizzi, V. G. J. Rodgers and A. Stern (1985). 
{\it Nucl.Phys.}, {\bf B256}, 525. \\
Ball, R. (1989). 
{\it Phys.Rep.}, {\bf 182}, 1. \\
Ball, R. D., S. Forte and J. Tigg (1994).
{\it Nucl.Phys.}, {\bf B428}, 485. \\
Bass, S. D. (1994a).
{\it Phys.Lett.}, {\bf B329}, 358. \\
Bass, S. D. (1994b).
Cavendish Preprint HEP 94/11., nucl-th/9408023. \\
Beck, R. (1995).
private communication. \\
B\'eg, M. A. B., and A. Zepeda (1972).
{\it Phys.Rev.}, {\bf D6}, 2912. \\
Beise, E. J. and R.D. McKeown (1991).
{\it Comm.Nucl.Part.Phys.}, {\bf 20}, 105. \\
Berg, D., F. Gruemmer, K. Goeke and E. Ruiz Arriola,  (1992). 
{\it J.Phys.}, {\bf G18}, 35. \\
Bernabeu, J. (1978).
{\it Nucl.Phys.}, {\bf A374}, 593c. \\
Bernard, V. (1986). 
{\it Phys.Rev.}, {\bf D34}, 1601. \\
Bernard, V., R. Brockmann, M. Schaden, W. Weise and E. Werner (1984). 
{\it Nucl.Phys.}, {\bf A412}, 349. \\
Bernard, V., U.-G. Meissner and I. Zahed (1987). 
{\it Phys.Rev.}, {\bf D36}, 819. \\
Bernard, V., R. L. Jaffe and U.-G. Meissner (1988). 
{\it Nucl.Phys.}, {\bf B308}, 753. \\
Bernard, V. and U.-G. Meissner (1988). 
{\it Nucl.Phys.}, {\bf A489}, 647. \\
Bernard, V. and Ulf-G. Meissner (1991). {\it Phys.Lett.}, {\bf B266}, 403. \\ 
Bernard, V., N. Kaiser and U.-G. Meissner (1994).
{\it Phys.Rev.}, {\bf D50}, 6899. \\
Bernard, V., N. Kaiser and U.-G. Meissner (1995).
{\it Int.J.Mod.Phys.}, {\bf E4}, 193. \\
Bijnens, J., C. Bruno and E. de Rafael (1993). {\it Nucl.Phys.}, {\bf B390}, 
501. \\ 
Bijnens, J., J. Prades and E. de Rafael (1995). {\it Phys.Lett.}, {\bf B348}, 
226. \\
Birse,  M. C. and M. K. Banerjee (1984). {\it Phys.Lett.}, {\bf B136}, 284. \\
Birse,  M. C. and M. K. Banerjee (1985). {\it Phys.Rev.}, {\bf D31}, 118. \\
Birse,  M. C. (1986). {\it Phys.Rev.}, {\bf D33}, 1934. \\ 
Birse, M. C. (1990). 
In: {\it Progress in particle and nuclear physics, Vol.25} 
(A. Faessler, ed.), pp.1. \\
Blin, A. H., B. Hiller and M. Schaden (1988). 
{\it Z.Phys.}, {\bf A331}, 75. \\
Blotz, A., F. D\"{o}ring, Th. Meissner and K. Goeke (1990). 
{\it Phys.Lett.}, {\bf B251}, 235. \\
Blotz, A., K. Goeke, N. W. Park, D. Diakonov, V. Petrov and P. V. Pobylitsa  
(1992)

{\it Phys.Lett.}, {\bf B287}, 29. \\
Blotz, A., D. Diakonov, K. Goeke, N. W. Park, V. Petrov and P. V. Pobylitsa 
(1993a). 

{\it Nucl.Phys.}, {\bf A555}, 765;\\
Blotz, A., M. Polyakov and K. Goeke (1993b). 
{\it Phys.Lett.}, {\bf B302}, 151. \\
Blotz, A., M. Praszalowicz and K. Goeke (1993c). 
{\it Phys.Lett.}, {\bf B317}, 195. \\
Blotz, A., K. Goeke and M. Praszalowicz (1994).
{\it Acta Phys. Polonics}, {\bf B25}, 1443. \\
Blotz, A., M. Praszalowicz and K. Goeke (1995a).
{\it Phys.Rev.}, {\bf D53}, (1996) 484.  \\
Blotz, A., M. Praszalowicz and K. Goeke (1995b).
{\it Phys.Rev.}, {\bf D53}, (1996) 551. \\
Broniowski, W.\ and T.~D. Cohen, (1993). {\it Phys.\ Rev.}\ {\bf D47}, 299.\\
Broniowski, W., G. Ripka, E. Nikolov and K. Goeke (1995).
{\it Z.Phys}, {\bf A}, (in press).\\
Brodsky, S. J., J. Ellis and M. Karliner (1988).
{\it Phys.Lett.}, {\bf B206}, 309. \\
Bruins, E. E. W., {\it et al.} (1995).
{\it Phys.Rev.Lett.}, {\bf 75}, 21. \\
Buck, A., R. Alkofer and H. Reinhardt (1992). {\it Phys. Lett.}, 
{\bf B 286}, 29.\\
%----- C -----
Cahill R. T. and C. D. Roberts, (1985). {\it Phys.Rev.}, {\bf D32}, 2419.\\
Callan, C. G. and I. Klebanov (1985). 
{\it Nucl.Phys.}, {\bf B262}, 365. \\
Chemtob, M. (1985). 
{\it Nucl.Phys.}, {\bf B256}, 600. \\
Cheng, T. P. (1976).
{\it Phys.Rev.}, {\bf D13}, 2161. \\
Choi, S., {\it et al.} (1993).
{\it Phys.Rev.Lett.}, {\bf 71}, 3927. \\
Christov, Chr. V., A. Blotz, K. Goeke, P. Pobylitsa, V. Petrov, 
M. Wakamatsu and T. Watabe 

(1994). {\it Phys.Lett.}, {\bf B325}, 467. \\
Christov, Chr. V., K. Goeke and P. V. Pobylitsa (1995a).
{\it Phys.Rev.}, {\bf C52}, 425. \\
Christov, Chr. V., A. Z. Gorski, K. Goeke and P. V. Pobylitsa (1995b).
{\it Nucl.Phys.}, {\bf A592}, 513. \\
Chu, M. C., J. M. Grandy, S. Huang and J. W. Negele (1994).
{\it Phys.Rev.}, {\bf D49}, 6039. \\
Cohen, T. D. (1995). {\it Phys.Lett.}, {\bf B359}, 23.\\
Cohen, T. D. and W. Bronjowski (1986). 
{\it Phys.Rev.}, {\bf D34}, 3472. \\
Cohen, T. D., H. Forkel and M. Nielsen (1993). 
{\it Phys.Lett.}, {\bf B316}, 1. \\
Coleman, S. (1985). 
In: {\it Aspects of Symmetry}. Cambridge: University Press. \\
%----- D -----
Dashen, R. and A. V. Manohar (1993). 
{\it Phys.Lett.}, {\bf B315}, 425. \\
Dhar, A. and S. R. Wadia (1984). 
{\it Phys.Rev.Lett.}, {\bf 52}, 959. \\
Dhar, A., R. Shankar and S. R. Wadia (1985). 
{\it Phys.Rev.}, {\bf D31}, 3256. \\
Diakonov, D. I. and V. Yu. Petrov (1984).
{\it Nucl.Phys.}, {\bf B245}, 259. \\
Diakonov, D. I. and V. Yu. Petrov (1986). 
{\it Nucl.Phys.}, {\bf B272}, 457. \\
Diakonov, D. I. and V. Yu. Petrov (1992).
In: {\it Quark cluster dynamics, Lecture notes in physics}, 

Vol.417, pp.288., Springer-Verlag. \\
Diakonov, D., V. Petrov and P. Pobylitsa (1988). 
{\it Nucl.Phys.}, {\bf B306}, 809. \\
Diakonov, D., V. Petrov and M. Praszalowicz (1989). 
{\it Nucl.Phys.}, {\bf B323}, 53. \\
Diakonov, D., M. Poliakov and C.Weiss (1996). 
{\it Nucl.Phys.}, {\bf B}, (in press) \\
Donoghue, J. F., and C. R. Nappi (1986). 
{\it Phys.Lett.}, {\bf B168}, 105. \\
D\"{o}ring, F., A. Blotz, C. Sch\"{u}ren, Th. Meissner, E. Ruiz Arriola and 
K. Goeke (1992a).

{\it Nucl.Phys.}, {\bf A536}, 548. \\
D\"{o}ring, F., E. Ruiz Arriola and K. Goeke (1992b). 
{\it Z.Phys.}, {\bf A344}, 159. \\
D\"{o}ring, F., C. Sch\"{u}ren, E. Ruiz Arriola and K. Goeke (1993). 
{\it Phys.Lett.}, {\bf B298}, 11. \\
D\"{o}ring, F., C. Sch\"{u}ren, E. Ruiz Arriola, T. Watabe and K. Goeke (1995).

Granada Preprint UG-DFM-2/95, (submitted to {\it Nucl.Phys.}, {\bf A}). \\
%----- E -----
Eden, T., {\it et al.} (1994). {\it Phys.Rev.}, {\bf C50}, R1749. \\
Ebert, D. and M. K. Volkov, (1983). {\it Z.Phys.}, {\bf C16}, 205. \\
Ebert, D. and H. Reinhardt (1986). 
{\it Nucl.Phys.}, {\bf B271}, 188. \\
Eguchi, T. (1976). 
{\it Phys.Rev.}, {\bf D14}, 2755. \\
Eisenberg, J. M. and W. Greiner (1970).
In: {\it Excitation mechanisms of the nucleus}.
North-Holland. \\
Ellis, J. and M. Karliner (1995). 
Preprint CERN-TH-95-279, hep-ph/9510402. \\
EMC Collaboration, Ashman, J., {\it et al.} (1988). 
{\it Phys.Lett.}, {\bf B206}, 364. \\
EMC Collaboration, Ashman, J., {\it et al.} (1989). 
{\it Nucl.Phys.}, {\bf B328}, 1. \\
Esaulov, A. S., A. M. Pilipenko and Yu. I. Titov (1978).
{\it Nucl.Phys.}, {\bf B136}, 511. \\
%----- F -----
Federspiel, F.\ {\it et~al.} (1991).
{\it Phys.\ Rev.\ Lett.}\ {\bf 67}, 1511. \\
Ferstl, P., M. Schaden and E. Werner (1986).
{\it Nucl.Phys.}, {\bf A452}, 680. \\
Fiolhais, M., K. Goeke, F. Gr\"ummer, J. N. Urbano (1988). 
{\it Nucl.Phys.}, {\bf A481}, 727. \\ 
Forkel, H., M. Nielsen, X. Jin and T. D. Cohen (1994).
{\it Phys.Rev.}, {\bf C50}, 3108. \\
Forte, S. (1993).
{\it Phys.Rev.}, {\bf D47}, 1842. \\
%----- G -----
Garvey, G. T., W. C. Louis and D. H. White (1993).
{\it Phys.Rev.}, {\bf C48}, 761. \\
Gasser, J., H. Leutwyler and M. E. Sainio (1991). 
{\it Phys.Lett.}, {\bf B253}, 252. \\
Gasser, J. and H. Leutwyler (1982).
{\it Phys.Rep.}, {\bf 87}, 77. \\
Gasser, J., H. Leutwyler and M. E. Sainio (1991).
{\it Phys.Lett.}, {\bf B253}, 260. \\
Gell-Mann, M. and M. L\'evy (1960). {\it Nuovo Cim.}, {\bf 16}, 705.\\
Goeke, K., A. Z. Gorski, F. Gruemmer, Th. Meissner, H. Reinhardt and 
R. W\"{u}nsch (1991).
 
{\it Phys.Lett.}, {\bf B256}, 321. \\
Gorski, A. Z., F. Gruemmer and K. Goeke, (1992).
{\it Phys.Lett.}, {\bf B278}, 24. \\
Guadagnini, E. (1984). {\it Nucl.Phys.}, {\bf B236}, 35. \\
%----- H -----
Hammer, H.-W., U.-G. Meissner and D. Drechsel (1996). 
{\it Phys.Lett.}, {\bf B367}, 323. \\
Hansson, T. H., M. Prakash and I. Zahed (1990). {\it Nucl.Phys.}, {\bf B335}, 
67. \\
Hatsuda, T. (1990). 
{\it Phys.Rev.Lett.}, {\bf 65}, 543. \\
Hatsuda, T. and T. Kunihiro (1994). 
{\it Phys.Rep.}, {\bf 247}, 221. \\
He, H. and X. Ji (1995).
{\it Phys.Rev.}, {\bf D52}, 2960. \\
H\"ohler, G., E. Pietarinen and I. Sabba-Stefanescu (1976).
{\it Nucl.Phys.}, {\bf B114}, 505. \\
Holzwarth, G. and B. Schwesinger (1986). 
{\it Rep.Prog.Phys.}, {\bf 49}, 825. \\
Holzwarth, G. (1994). 
{\it Nucl.Phys.}, {\bf A572}, 69. \\
Holzwarth, G. and H. Walliser (1995). 
{\it Nucl.Phys.}, {\bf A587}, 721. \\
Hsueh, S. Y. {\it et al.} (1988). 
{\it Phys.Rev.}, {\bf D38}, 2056. \\
Huang, S.--Zh. and J. Tjon, {\it Phys.Rev.}, {\bf C 49}, 1702.\\
%----- I -----
Ioffe, B. L. (1981).
{\it Nucl.Phys.}, {\bf B188}, 317, ibid. {\bf B191}, 591(E). \\
Ioffe, B. L. and A. Khodzhamirian (1995).
{\it Phys.Rev.}, {\bf D51}, 3373. \\
%----- J -----
Ishii, N., W. Bentz and K. Yazaki (1993). {\it Phys.Lett.}, 
{\bf B 301}, 165, {\em ibid}.\ {\bf B 318}, 26.\\
Jaffe, R. L. and X. Ji (1991).
{\it Phys.Rev.Lett.}, {\bf 67}, 552. \\
Jaffe, R. L. and C. L. Korpa (1987).
{\it Comm.Nucl.Part.Phys.}, {\bf 17}, 163. \\
Jaffe, R. L. (1989). 
{\it Phys.Lett.}, {\bf B229}, 275. \\
Jain, P., R. Johnson, N. W. Park, J. Schechter and H. Weigel (1989). 
{\it Phys.Rev.}, {\bf D40}, 855. \\
Jaminon, M., G. Ripka and P. Stassart (1989). 
{\it Nucl.Phys.}, {\bf A504}, 733. \\
Jaminon, M., R. Mendez-Galain, G. Ripka and P. Stassart (1992). 
{\it Nucl.Phys.}, {\bf A537}, 418. \\
Jones-Woodward, C. E., {\it et al.} (1991).
{\it Phys.Rev.}, {\bf C44}, R571. \\
%----- K -----
Kahana, S. and G. Ripka (1984). 
{\it Nucl.Phys.}, {\bf A429}, 462. \\
Kahana, S., G. Ripka and V. Soni (1984). 
{\it Nucl.Phys.}, {\bf A415}, 351. \\
Kaplan, D. B. and I. Klebanov (1990).
{\it Nucl.Phys.}, {\bf B335}, 45. \\
Kaplan, D. B. and A. Manohar (1988).
{\it Nucl.Phys.}, {\bf B310}, 527. \\
Kikkawa, K. (1976). 
{\it Prog.Theor.Phys.}, {\bf 56}, 947. \\
Kim, H.-C., A. Blotz, M. Polyakov and K. Goeke (1996a).
{\it Phys.Rev.}, {\bf D53} 4013. \\
Kim, H.-C., M. Polyakov, A. Blotz and K. Goeke (1996b).
{\it Nucl.Phys.}, {\it A598} 379. \\
Kim, H.-C., T. Watabe and K. Goeke (1996c).
Bochum Preprint RUB-TPII-11-95, 11pp.,

hep-ph/9506344), {\it Nucl.Phys.}, {\bf A} (in print)\\
Kim, H.-C., M. V. Polyakov and K. Goeke (1996d).
{\it Phys.Rev.}, {\bf D53} 4715R.\\
Kim, H.-C., A. Blotz, C. Schneider and K. Goeke (1996).
{\it Nucl.Phys.}, {\bf A596}, 415. \\
Kirchbach, M. and D. O. Riska (1991).
{\it Nuovo Cim.}, {\bf A104}, 1837. \\
Kitagaki, T., {\it et al.} (1983).
{\it Phys.Rev.}, {\bf D28}, 436. \\
Kleinert, H. (1976). In: {\it Erice Summer Institute, 
Understanding the fundamental constituents 

of matter} (ed. A. Zichichi), pp.289. Plenum Press, NY. \\
Klevansky, S. P. (1992). 
{\it Rev.Mod.Phys.}, {\bf 64}, 649. \\
Klimt, S., M. Lutz, U. Vogl and W. Weise (1990). 
{\it Nucl.Phys.}, {\bf A516}, 429. \\
Koch, R. (1982).
{\it Z.Phys.}, {\bf C15}, 161. \\
%----- L -----
Lee, T. D. (1981). 
In: {\it Particle Physics and Introduction to Field Theory}. Harwood Chur. \\
Leinweber, D. B. (1995). 
Preprint DOE/ER/40427-27-N95, 20pp, hep-ph/9512319. \\
Lichtenstadt, J. and H. J. Lipkin (1995). 
{\it Phys.Lett.}, {\bf B353}, 119. \\
%----- N -----
Nueber, T., Fiolhais, M., K. Goeke, J. N. Urbano (1993). 
{\it Nucl.Phys.}, {\bf A560}, 909. \\ 
%----- M -----
Mazur, P. O., M. A. Nowak and M. Praszalowicz (1984). 
{\it Phys.Lett.}, {\bf B147}, 137. \\
McGovern, J.A. and M. C. Birse (1990). 
{\it Nucl. Phys.}, {\bf A506}, 392.\\ 
McKay D.W. and H. J. Munczek, (1985). {\it Phys.Rev.}, {\bf D32}, 266. \\
McNamee, P., S. J. Chilton and F. Chilton (1964). 
{\it Rev.Mod.Phys.}, {\bf 36}, 1005. \\
Meissner, Th. and K. Goeke (1991). 
{\it Z.Phys.}, {\bf A339}, 513. \\
Meissner, Th., E. Ruiz Arriola, A. Blotz and K. Goeke (1993a).
Bochum Preprint RUB-TPII-42-93,
 
123pp., hep-ph/9401216. \\
Meissner, Th., G. Ripka, R. W\"{u}nsch, P. Sieber, F. Gruemmer and 
K. Goeke (1993b).
 
{\it Phys.Lett.}, {\bf B299}, 183. \\
Meissner, Th., E. Ruiz Arriola and K. Goeke (1990). 
{\it Z.Phys.}, {\bf A336}, 91. \\
Meissner, Th., E. Ruiz Arriola, F. Gruemmer, H. Mavromatis and K. Goeke (1988).

{\it Phys.Lett.}, {\bf B214}, 312. \\
Meissner, Th., F. Gruemmer and K. Goeke (1989). 
{\it Phys.Lett.}, {\bf B227}, 296. \\
Mergell, P., U.-G. Meissner and D. Drechsel (1996). 
{\it Nucl. Phys.}, {\bf A596}, 367. \\
Meyerhoff, M., {\it et al.} (1994).
{\it Phys.Lett.}, {\bf B327}, 201. \\
Moussallam, B., (1993). {\it Ann.Phys.}, {\bf 225}, 284. \\
Musolf, M. J. and M. Burkardt (1994).
{\it Z.Phys.}, {\bf C61}, 433. \\
Musolf, M. J., {\it et al.} (1994).
{\it Phys.Rep.}, {\bf 239}, 1. \\
%----- N -----
Nambu, Y. and G. Jona-Lasinio (1961a). 
{\it Phys.Rev.}, {\bf 122}, 345. \\
Nambu, Y. and G. Jona-Lasinio (1961b). 
{\it Phys.Rev.}, {\bf 124}, 246. \\
Nelson, T. J. (1967). 
{\it J.Math.Phys.}, {\bf 8}, 857. \\
Nepomechie, R. I. (1985).
{\it PhysRev.}, {\bf D31}, 3291. \\
Nikolov, E.~N., W. Broniowski and K. Goeke (1994). {\it Nucl.\ Phys.}\ 
{\bf A579}, 398\\
NMC Collaboration, Arneondo, M., {\it et al.} (1994).
{\it Phys.Rev.}, {\bf D50}, 1. \\
%----- O -----
%----- P -----
Park, B.-Y. and M. Rho (1989). 
{\it Phys.Lett.}, {\bf B220}, 7. \\
Park, N. W., J. Schechter and H. Weigel (1991). 
{\it Phys.Rev.}, {\bf D43}, 869. \\
Park, N. W. and H. Weigel (1991). 
{\it Phys.Lett.}, {\bf B268}, 155. \\
Park, N. W. and H. Weigel (1992). 
{\it Nucl.Phys.}, {\bf A541}, 453. \\
Particle Data Group (1994).
{\it Phys.Rev.}, {\bf D50}, 1218. \\
Pearce, B. C., K. Holinde and J. Speth (1992). 
{\it Nucl.Phys.}, {\bf A541}, 663. \\
Platchkov, S., {\it et al.} (1990).
{\it Nucl.Phys.}, {\bf A510}, 740. \\
Pobylitsa, P., E. Ruiz Arriola, Th. Meissner, F. Gruemmer, K. Goeke 
and W. Broniowski (1992).

{\it J.Phys.}, {\bf G18}, 1455. \\
Praszalowicz, M. (1985). 
{\it Phys.Lett.}, {\bf B158}, 264. \\
Praszalowicz, M., A. Blotz and K. Goeke (1993). 
{\it Phys.Rev.}, {\bf D47}, 1127. \\
Praszalowicz, M., A. Blotz and K. Goeke (1995).
{\it Phys.Lett.}, {\bf B354}, 415. \\
da Providencia, J., M. A. Ruivo and C. A. de Sousa (1987). 
{\it Phys.Rev.}, {\bf D36}, 1882. \\
%----- Q -----
%----- R -----
Reinhardt, H. (1989). 
{\it Nucl.Phys.}, {\bf A503}, 825. \\
Reinhardt, H. and R. Alkofer (1988). 
{\it Phys.Lett.}, {\bf B207}, 482. \\
Reinhardt, H. and R. W\"{u}nsch (1988). 
{\it Phys.Lett.}, {\bf B215}, 577. \\
Reinhardt, H. and R. W\"{u}nsch (1989). 
{\it Phys.Lett.}, {\bf B230}, 93. \\
Ring, P. and P. Schuck (1980). 
In: {\it The Nuclear Many Body Problem}. Springer Verlag. \\
Ruiz Arriola, E., P. Alberto, J. N. Urbano and K. Goeke (1989). 
{\it Z.Phys.}, {\bf A333}, 203. \\
Ruiz Arriola, E., (1991). {\it Phys.Lett.}, {\bf B253}, 430. \\
%----- S -----
Salam, A. and J. Strathdee (1982). 
{\it Ann.Phys.}, {\bf 141}, 316. \\
Schaden, M., H. Reinhardt, P. A. Amundsen and M. J. Lavelle (1990). 
{\it Nucl.Phys.}, {\bf B339}, 595. \\
Schechter, J. and H. Weigel (1995a).
{\it Phys.Rev.}, {\bf D51}, 6296. \\
Schechter, J. and H. Weigel (1995b).
{\it Mod.Phys.Lett.}, {\bf A10}, 885. \\
Schmiedmayer, J., P. Riehs, J. Harvey, and N. Hill (1991). 
{\it Phys.\ Rev.\ Lett.}\ {\bf 66}, 1015.\\
Schneider, C., A. Blotz and K. Goeke (1995).
Bochum Preprint RUB-TPII-23-95. \\
Sch\"{u}ren, C., F. D\"{o}ring, E. Ruiz Arriola and K. Goeke (1993). 
{\it Nucl.Phys.}, {\bf A565}, 687. \\
Sch\"{u}ren, C., E. Ruiz Arriola and K. Goeke (1992). 
{\it Nucl.Phys.}, {\bf A547}, 612. \\
Shuryak, E. V. (1982).{\it Nucl.Phys.}, {\bf B203}, 93, 116, 140. \\
Shuryak, E. V. (1983).
{\it Nucl.Phys.}, {\bf B214}, 237. \\
Sieber, P., Th. Meissner, F. Gruemmer and K. Goeke (1992). 
{\it Nucl.Phys.}, {\bf A547}, 459. \\
Skyrme, T. H. R. (1961). 
{\it Proc.R.Soc.}, {\bf A260}, 127. \\
Skyrme, T. H. R. (1962). 
{\it Nucl.Phys.}, {\bf 31}, 556. \\
SMC Collaboration, Adeva, B., {\it et al.} (1993). 
{\it Phys.Lett.}, {\bf B302}, 533. \\
SMC Collaboration, Adeva, B., {\it et al.} (1994).
{\it Phys.Lett.}, {\bf B320}, 400. \\
Steininger, K. and W. Weise (1994).
{\it Phys.Lett.}, {\bf B329}, 169. \\
de Swart, J. J. (1963). 
{\it Rev.Mod.Phys.}, {\bf 35}, 916. \\
%----- T -----
Takizawa, M., K. Tsushima, Y. Kohyama and K. Kubodera (1990). 
{\it Nucl.Phys.}, {\bf A507}, 611. \\
Theberge, S., A. W. Thomas and G. A. Miller (1980).  {\it Phys.Rev.}, 
{\bf D22}, 2838.\\
Thomas, A. W. (1983). {\it Adv.Nucl.Phys.}, {\bf 13}, 1.\\
Thompson, A. K., {\it et al.} (1992).
{\it Phys.Rev.Lett.}, {\bf 68}, 2901. \\
't Hooft, G. (1974). 
{\it Nucl.Phys.}, {\bf B72}, 461. \\
't Hooft, G. (1975). 
{\it Nucl.Phys.}, {\bf B75}, 461. \\
't Hooft, G. (1976a). 
{\it Phys.Rev.Lett.}, {\bf 37}, 8. \\
't Hooft, G. (1976b). 
{\it Phys.Rev.}, {\bf D14}, 3432. \\
%----- U -----
%----- V -----
Veneziano, G. (1979). 
{\it Nucl.Phys.}, {\bf B159}, 213. \\
Vogl, U. and W. Weise (1991). 
{\it Prog.Part.Nucl.Phys.}, {\bf 27}, 195. \\
Volkov, M. K., (1984), {\it Ann.Phys.}, {\bf 157}, 282.\\
%----- W -----
Wakamatsu, M. (1992).
{\it Phys.Rev.}, {\bf D46}, 3762. \\
Wakamatsu, M. (1995).
{\it Phys.Lett.}, {\bf B349}, 204. \\
Wakamatsu, M. and H. Yoshiki (1991). 
{\it Nucl.Phys.}, {\bf A524}, 561. \\
Wakamatsu, M. and T. Watabe (1993). 
{\it Phys.Lett.}, {\bf B312}, 184. \\
Walliser, H. (1993). in ``Baryons as Skyrme Solitons'', ed.: G. Holzward,
World Scientific Publ.Comp., Singapore 1993, p.247\\ 
Watabe, T., Chr. V. Christov and K. Goeke (1995a).
{\it Phys.Lett.}, {\bf B349}, 197-203. \\
Watabe, T., Chr. V. Christov and K. Goeke (1995b).
Bochum Preprint RUB-TPII-03-95, 9pp.,

hep-ph/9506440. \\
Watabe, T., H.-C. Kim and K. Goeke (1995c).
Bochum Preprint RUB-TPII-17-95, 13pp.,

hep-ph/9507318. \\
Watabe, T. and H. Toki (1992). 
{\it Prog.Theor.Phys.}, {\bf 87}, 651. \\
Weigel, H., A. Abada, R. Alkofer and H. Reinhardt (1995a).
{\it Phys.Lett.}, {\bf B353}, 20. \\
Weigel, H., R. Alkofer and H. Reinhardt (1992). 
{\it Nucl.Phys.}, {\bf B387}, 638; {\it Phys.Lett.}\ {\bf B284}, 296 \\
Weigel, H., R. Alkofer and H. Reinhardt (1994).
{\it Nucl.Phys.}, {\bf A576}, 477. \\
Weigel, H., R. Alkofer and H. Reinhardt (1995b).
{\it Nucl.Phys.}, {\bf A582}, 484. \\
Weiner, R.\ and W. Weise (1985). {\it Phys.\ Lett.}\ {\bf B159}, 85\\
Weiss, C., R. Alkofer and H. Weigel (1993a). 
{\it Mod.Phys.Lett.}, {\bf A8}, 79. \\
Weiss, C., A. Buck, R. Alkofer and H. Reinhardt (1993b).  {\it Phys.Lett.}, 
{\bf B 312}, 6.\\
Wess, J. (1972). 
{\it Acta Phys.Austr.}, {\bf 10}, 494. \\
Wess, J. and B. Zumino (1971). 
{\it Phys.Lett.}, {\bf B37}, 95. \\
Wirzba, A. and W. Weise (1987).
{\it Phys.Lett.}, {\bf B188}, 6. \\
Witten, E. (1979). 
{\it Nucl.Phys.}, {\bf B156}, 269. \\
Witten, E. (1983). 
{\it Nucl.Phys.}, {\bf B223}, 422. \\
%----- X -----
%----- Y -----
Yabu, H. and K. Ando (1988). 
{\it Nucl.Phys.}, {\bf B301}, 601. \\
%----- Z -----
Zahed, I. and G. E. Brown (1986). 
{\it Phys.Rep.}, {\bf 142}, 1. \\
Zieger, A.\ {\it et~al.} (1992). {\it Phys.\ Lett.}\ {\bf B278}, 34.\\
Z\"{u}ckert, U., R. Alkofer, H. Reinhardt and H. Weigel (1994). 
{\it Nucl.Phys.}, {\bf A570}, 445. \\
\noindent
%
%\end{document}

\end{document}